\documentclass[useAMS,usenatbib]{mn2e}
\usepackage{subfigure}
\usepackage{natbib}
\usepackage{lscape}
\usepackage[dvips]{color}
\usepackage{amssymb}
\usepackage{amsmath}
\usepackage{graphicx}
\usepackage{url} 
\usepackage{ulem} 
\usepackage{setspace}
\usepackage{threeparttable}
\usepackage{multirow}
\usepackage{booktabs}

\title[AGB in high-$z$~galaxies. ]{Revisiting the role of the Thermally-Pulsating Asymptotic Giant Branch phase in high-redshift galaxies}
\author[Capozzi D. et al. 2015] {Diego Capozzi$^{1}$\thanks{E-mail: diego.capozzi@port.ac.uk}, Claudia Maraston$^{1}$, Emanuele Daddi$^{2}$, Alvio Renzini$^{3}$, \and 
Veronica Strazzullo$^{2}$, Raphael Gobat$^{4}$ \\ \\ 
  1 - Institute of Cosmology and Gravitation, University of Portsmouth, Dennis Sciama Building, Burnaby Road,\\
  \quad \, Portsmouth, PO1 3FX, UK\\
  2 - Laboratoire AIM, CEA/DSM-CNRS-Universit\`{e} Paris Diderot, Irfu/Service d'Atrophysique, CEA Saclay, \\
  \quad \, Orme des Merisiers, 91191 Gif-sur-Yvette Cedex, France\\
  3 - INAF - Osservatorio Astronomico di Padova, Vicolo dell'Osservatorio 5, I-35122 Padova, Italy\\
  4 - KIAS , 85 Hoegiro, Dongdaemun-gu, Seoul 130-722, Republic of Korea}

\date{Accepted ;
  Received ; in original form }
\pagerange{\pageref{firstpage}--\pageref{lastpage}} \pubyear{2015}
\begin{document}
\maketitle
\label{firstpage}
\begin{abstract}
We study the debated contribution from thermally pulsing asymptotic giant branch (TP-AGB) stars in evolutionary population synthesis models. We investigate the Spectral Energy Distributions (SEDs) of a sample of 51 spectroscopically confirmed, high-z ($1.3<z_{\rm spec}<2.7$), galaxies using three evolutionary population synthesis models with strong, mild and light TP-AGB. Our sample is the largest of spectroscopically confirmed galaxies on which such models are tested so far. Galaxies were selected as passive, but we model them using a variety of star formation histories in order not to be dependent on this pre-selection.
 
We find that the observed SEDs are best fitted with a significant contribution of TP-AGB stars or with substantial dust attenuation. Without including reddening, TP-AGB-strong models perform better and deliver solutions consistent within $1\sigma$ from the best-fit ones in the vast majority of cases.  
Including reddening, all models perform similarly. 
Using independent constraints from observations in the mid- and far-IR, we show that low/negligible dust attenuation, i.e. $E(B-V)\lesssim 0.05$ , should be preferred for the SEDs of passively-selected galaxies. 
Given that TP-AGB-light models give systematically older ages for passive galaxies, we suggest number counts of passive galaxies at higher redshifts as a further test to discriminate among stellar population models.
\end{abstract}
\section{Introduction}
\label{sec:sec1}
Understanding the star formation histories (SFHs) and the stellar-mass assembly  of galaxies at different cosmic epochs is a fundamental step for constraining galaxy formation and evolution and its theoretical models. Galaxy evolution studies rely on evolutionary population synthesis models (e.g,  \citealp{Bruzual-2003, Maraston-2005}, herefater respectively referred to as BC03 and M05), which are the tool to derive galaxy physical properties like age, SFH and stellar mass, and to predict the Spectral Energy Distribution (SED) of model galaxies from semi-analytic models \citep{Baugh-2006, Tonini-2009, Henriques-2011}. Stellar population models are based on stellar evolution theory and on empirical calibrations for those evolutionary phases whose modelling is still challenging. Stars in the advanced post-main-sequence stellar evolutionary phase called Thermally Pulsing Asymptotic Giant Branch (TP-AGB) are especially difficult to model, given the complex interplay of envelope convection, mixing and mass loss. Different stellar population models incorporate  the contribution of TP-AGB stars in different ways, which over the last years  has resulted in a lively debate  (e.g, \citealp{Maraston-1998, Bruzual-2003, Maraston-2005, Maraston-2006, van-der-Wel-2006, Cimatti-2008, Riffel-2008, Conroy-2010, Kriek-2010, Lyubenova-2010, MacArthur-2010, Melnick-2014, Noel-2013}).

High-redshift galaxies, and in particular those among them in nearly passive evolution, are the ideal laboratories where to test the effect of the TP-AGB phase in stellar population models. This is so because  its contribution is at its maximum for stellar populations $\sim 1$ Gyr old in M05 models (see also \citealp{Marigo-2008}), whereas it is much  smaller in BC03 models.

This test was carried out by \citet{Maraston-2006} (hereafter M06) on a sample of seven high-$z$ ($1.39<z_{\rm spec}<2.67$) galaxies in the Hubble Ultra Deep Field (HUDF), selected for being  nearly passively evolving \citep{Daddi-2005} and investigated the effects of different stellar population models - in particular of the different treatment of TP-AGB stars - on the determination of their physical properties, e.g. ages, SFHs and stellar masses. This was done by fitting (with both BC03 and M05 models) high-quality observed SEDs made up of 14 optical and {\it Spitzer} InfraRed Array Camera (IRAC) photometric bands (corresponding to a rest-frame spectral range from UV  to the $K$ band). 
The redshift range explored in M06 was ideal for assessing the importance of TP-AGB stars in the bolometric and near-IR flux budget of galaxies, because at this cosmic epoch they are dominated by stars with ages in the range $0.2\lesssim t\lesssim2\ {\rm Gyr}$, where the  
models are most discrepant (see M05. Crucial to this exercise was to have galaxies with high-quality data and spectroscopic redshifts.

\citet{Maraston-2006} found that the M05 models were typically able to match well the rest-frame optical and near-IR fluxes of those galaxies without having to invoke reddening.
BC03 models gave comparable good fits only when reddening was included in the fit.This happens because in models with a light TP-AGB the rest-frame near-IR is provided by older, RGB stars and the optical by a young component, whose flux may need to be partly suppressed to allow a good fit to both the optical and the near-IR spectrum.
The reasons for the differences between M05 and BC03 models are extensively discussed in these papers. 

The role of the TP-AGB stars in the models and data was since then investigated in several publications (e.g., 
\citealp{Maraston-2006, Conroy-2010, Kriek-2010, Zibetti-2013, Riffel-2015}, see \citealp{Maraston-2013} for a short review), using different methods compared to M06 and sometimes leading to discrepant conclusions.
At low redshift, \citet{Zibetti-2013} analysed near-infrared spectro-photometric data for a sample of 16 post-starburst selected SDSS galaxies aiming at testing the presence of TP-AGB spectral features as predicted by the M05 models. They found that none of the 16 observed galaxies displayed such features and that all of them showed near-infrared fluxes relative to optical consistent with those predicted by TP-AGB light models (BC03). However, \citet{Riffel-2015} reported on the need of strong TP-AGB (e.g.,  M05) to recover the deep near-{\it IR} spectral features of Seyfert galaxies in the local Universe. Riffel et al. also showed that their galaxies have deep absorptions and argued that the observations by Zibetti et al. did not have enough resolution (300 vs. 1200) nor S/N (30 vs 110 in K) to detect the looked-for spectral features.

At intermediate/high $z$, \citet{Kriek-2010} used galaxies selected as post-starbursts  as calibrators for the TP-AGB phase. In particular, by using photometric redshifts and rest-frame synthetic colours, \citet{Kriek-2010} selected 62 post-starburst galaxies over a wide redshift range and constructed a composite SED to which exponentially declining SFH templates were fitted. They found that BC03 models performed better then M05 ones in reproducing both the rest-frame optical and NIR parts of such composite SED, but noted that this difference is reduced when fitting galaxies individually, especially for high-$z$ galaxies. 

In order to shed some light on these discrepancies, in this paper we take a twofold strategy. First, we perform again the analysis carried out in M06 on the same spectroscopic galaxy sample of 7 passively evolving galaxies, but using the latest photometry and more finely sampled SEDs available in the Cosmic Assembly Near-infrared Deep Extragalactic Legacy Survey (CANDELS, \citealp{Grogin-2011, Koekemoer-2011}) Multi-Wavelength Catalogues \citep{Guo-2013}. This helps with assessing the robustness of the results obtained in 2006. 

In addition, in order to increase our sample of spectroscopically confirmed galaxies, we carry out the same analysis on a sample of 44 passive galaxies at $1.3<z_{\rm spec}<2.1$  selected in the Cosmological Evolution Survey  field (COSMOS, \citealp{Scoville-2007}) and for which photometry is available in 26 broad- and medium-band filters from the COSMOS/UltraVISTA (Visible and Infrared Survey Telescope for Astronomy) catalogue by \citet{Muzzin-2013}. This second step helps with increasing the statistics for such a study and with securing ourselves against possible sample selection biases affecting the 2006 sample of galaxies.

In total we have a sample of 51 galaxies spanning the redshift range $1.3<z_{\rm spec}<2.67$ which constitutes  the largest sample  so far of spectroscopically confirmed galaxies on which models are tested.
Note that, despite their observed-frame colours being located in the ``passive''  area of colour-colour diagrams (e.g., the {\it BzK}, \citealp{Daddi-2004}; the rest-frame {\it UVJ}, \citealp{Williams-2009}), when  fitting models we allow the widest freedom in star formation histories, spanning from passive models to vigorous star formation and dust.

In addition to M05 and BC03 models, we use a modification of the M05 models, where the new models are forced to match the colours of Magellanic Clouds star clusters vs cluster age \citep{Noel-2013}, hereafter M13 models. This leads to a shift  of $\sim300\ {\rm Myr}$ on the onset age of 
TP-AGB-dominated stellar population and to a reduced TP-AGB {\it fuel consumption} of this evolutionary phase with respect to M05 models.

The paper is structured as follows.
Section \ref{sec:sec2} is dedicated to the description of the HUDF and COSMOS samples and their photometry. In Section \ref{sec:sec3} we describe our SED fitting procedure and the revised M13  models that we will be testing here. In Sections \ref{sec:sec4} and \ref{sec:sec5} we present results for both the HUDF and COSMOS samples and assess the statistical consistency among SED fitting solutions obtained via different models, the role of dust attenuation and the physical reliability of fitting solutions. Section \ref{sec:sec6} is dedicated to the study of galaxy physical properties, while in Section \ref{sec:sec7} we discuss our results and the implications of our findings in comparison with those available in the literature. Finally, we draw our conclusions in Section \ref{sec:sec8}. 

Throughout this paper we make use of magnitudes in the AB photometric system and assume a standard cosmology with $H_{0}=70\ {\rm km\ s^{-1}\ Mpc^{-1}}$, $\Omega_{\rm m}=0.3$ and $\Omega_{\rm \lambda}=0.7$.
\begin{table*}
%\begin{tiny}
\begin{center}
\caption{Observed photometry for the HUDF sample in AB magnitudes.}
\label{tab:Table1}
\begin{threeparttable}
\begin{tabular}{cccccccccc}
\hline
  \multicolumn{1}{c}{\bf New ID} &
  \multicolumn{1}{c}{$\mathbf{z_{\rm spec}}$} &
  \multicolumn{1}{c}{\bf U\_CTIO} &
  \multicolumn{1}{c}{\bf U\_VIMOS} &
  \multicolumn{1}{c}{\bf F435W} &
  \multicolumn{1}{c}{\bf F606W} &
  \multicolumn{1}{c}{\bf F775W} &
  \multicolumn{1}{c}{\bf F814W} &
  \multicolumn{1}{c}{\bf F850LP} &
  \multicolumn{1}{c}{\bf F105W} \\
\hline
16273 & 1.39 &  $>$28\tnote{*}          & $>$29\tnote{*}          & 27.2$\pm$0.2   & 26.29$\pm$0.05 &  25.14$\pm$0.02   & 24.79$\pm$0.07 & 24.26$\pm$0.02   & 23.454$\pm$0.007 \\
13586 & 1.55 &  25.68$\pm$0.09          & 25.47$\pm$0.04          & 25.18$\pm$0.02 & 24.69$\pm$0.01 &  23.828$\pm$0.005 & 23.66$\pm$0.02 & 23.103$\pm$0.005 & 22.252$\pm$0.007 \\
10767 & 1.73 &  25.6$\pm$0.1            & 25.47$\pm$0.04          & 25.16$\pm$0.02 & 24.90$\pm$0.01 &  24.553$\pm$0.008 & 24.46$\pm$0.04 & 24.06$\pm$0.01   & 23.38$\pm$0.01   \\
12529 & 1.76 &  $>$28\tnote{*}          & $>$30\tnote{*}	  & 27.5$\pm$0.1   & 26.40$\pm$0.03 &  25.48$\pm$0.02   & 25.33$\pm$0.08 & 24.75$\pm$0.01   & 23.78$\pm$0.02   \\
12751 & 1.91 &  $>$28\tnote{*}          & $>$30\tnote{*}	  & 27.4$\pm$0.2   & 26.14$\pm$0.03 &  25.28$\pm$0.02   & 24.88$\pm$0.06 & 24.35$\pm$0.01   & 23.552$\pm$0.007 \\
12567 & 1.98 &  $>$29\tnote{*}          & $>$30\tnote{*}	  & 29.0$\pm$0.6   & 26.90$\pm$0.05 &  26.30$\pm$0.03   & 25.8$\pm$0.1   & 25.38$\pm$0.03   & 24.65$\pm$0.01   \\
11079 & 2.67 &  $>$28\tnote{*}          & 29.1$\pm$0.6	          & 28.3$\pm$0.2   & 26.96$\pm$0.04 &  26.14$\pm$0.02   & 26.1$\pm$0.1   & 25.81$\pm$0.03   & 25.41$\pm$0.05   \\
\hline
\end{tabular}
\begin{tablenotes}\footnotesize 
\item[*] Upper limit at 1$\sigma$ level.
\end{tablenotes}
\end{threeparttable}
\end{center}
% \end{tiny}
\end{table*}

\addtocounter{table}{-1}
\begin{table*}
%\begin{tiny}
\begin{center}
\caption{Continued.}
\begin{tabular}{ccccccccc}
\hline
  \multicolumn{1}{c}{\bf M06 ID} &
  \multicolumn{1}{c}{\bf F125W} &
  \multicolumn{1}{c}{\bf F160W} &
  \multicolumn{1}{c}{\bf Ks\_ISAAC} &
  \multicolumn{1}{c}{\bf Ks\_HAWK-I} &
  \multicolumn{1}{c}{\bf 3.6 ${\rm \mu m}$} &
  \multicolumn{1}{c}{\bf 4.5 ${\rm \mu m}$} &
  \multicolumn{1}{c}{\bf 5.8 ${\rm \mu m}$} &
  \multicolumn{1}{c}{\bf 8.0 ${\rm \mu m}$} \\
\hline
8238    &  22.974$\pm$0.004 & 22.534$\pm$0.003 & 21.86 $\pm$0.01  & 21.942$\pm$0.004 &  21.31 $\pm$0.01  & 21.390$\pm$0.009 & 21.74 $\pm$0.04  & 22.27$\pm$0.08 \\
4950    &  21.670$\pm$0.004 & 21.188$\pm$0.003 & 20.672$\pm$0.005 & 20.641$\pm$0.001 &  19.987$\pm$0.002 & 19.920$\pm$0.002 & 20.178$\pm$0.008 & 20.69$\pm$0.01 \\
1025    &  22.636$\pm$0.007 & 22.294$\pm$0.006 & 21.83 $\pm$0.02  & 21.849$\pm$0.003 &  21.242$\pm$0.009 & 21.128$\pm$0.007 & 21.18 $\pm$0.02  & 21.54$\pm$0.04 \\
3523    &  23.023$\pm$0.009 & 22.642$\pm$0.008 & 22.21 $\pm$0.02  & 22.254$\pm$0.006 &  21.659$\pm$0.008 & 21.59 $\pm$0.01  & 21.77 $\pm$0.04  & 22.14$\pm$0.06 \\
3650    &  22.645$\pm$0.002 & 22.178$\pm$0.002 & 21.63 $\pm$0.01  & 21.717$\pm$0.003 &  21.187$\pm$0.007 & 21.089$\pm$0.006 & 21.13 $\pm$0.02  & 21.73$\pm$0.04 \\
3574    &  23.646$\pm$0.005 & 23.113$\pm$0.003 & 22.55 $\pm$0.02  & 22.648$\pm$0.006 &  22.02 $\pm$0.02  & 21.97 $\pm$0.01  & 21.90 $\pm$0.04  & 22.50$\pm$0.08 \\
1446    &  24.90 $\pm$0.03  & 23.46 $\pm$0.01  & 22.70 $\pm$0.03  & 22.795$\pm$0.007 &  22.16 $\pm$0.02  & 22.06 $\pm$0.01  & 21.93 $\pm$0.04  & 22.05$\pm$0.05 \\
\hline
\end{tabular}
%\label{tab:Table1}							    
\end{center}
% \end{tiny}
\end{table*}

\section{Galaxy data}
\label{sec:sec2}
\subsection{The Hubble Ultra Deep Field sample}
\label{subsec:subsec2.1}
\begin{figure}
\centering
\includegraphics[width=0.4\textwidth]{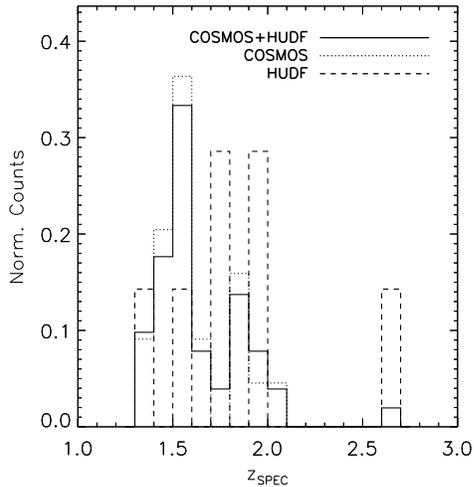}
\caption{Normalised distributions of spectroscopic redshifts for HUDF, COSMOS and combined samples (dashed, dotted and full line, respectively).}
\label{fig:Fig1}
\end{figure}

The seven HUDF galaxies used here are the same as in \citet{Maraston-2006} that were selected as passively evolving according to the $BzK$ criterion of \cite{Daddi-2004}. Spectroscopic redshifts (1.39$z_{\rm spec}<2.67$, see Figure \ref{fig:Fig1}) were obtained by \citet{Daddi-2005} using low-resolution spectra extracted from the {\it Hubble Space Telescope} (HST)  ACS (Advanced Camera for Surveys) grism data taken over the HUDF by the Grism ACS Program for Extragalactic Science (GRAPES) project. Only for galaxy ID=11079, the spectroscopic redshift was measured via the ultra-deep (30 hr integration) VLT (Very Large Telescope)+FORS2 (FOcal Reducer/low dispersion Spectrograph 2) spectroscopy as part of the Galaxy Mass Assembly Ultra-Deep Spectroscopic Survey (GMASS, \citealp{Kurk-2013}).  These galaxies exhibit early-type morphologies on HUDF images and significant Mg$_{\rm UV}$ features \citep{Daddi-2005}, consistent with their UV/optical spectra being dominated by A- or F-type stars.  

In this paper we use the new photometry\footnote{The photometry used in M06 was: i) {\it BViz} ACS; ii) {\it JH} from HST NICMOS (Near Infrared Camera and Multi-Object Spectrometer); iii) {\it JK} from VLT ISAAC (Infrared Spectrometer And Array Camera); iv) {\it VR} from FORS2; v) IRAC channels at 3.5, 4.5, 5.8 and 8.0 $\mu m$.} available in the CANDELS Multi-Wavelength Catalogues  \citep{Grogin-2011, Koekemoer-2011} for  the Great-Observatories-Origins-Deep-Survey (GOODS-SOUTH field \citep{Giavalisco-2004, Guo-2013}, which contains the HUDF. This catalogue is based on CANDELS {\it F160W} detections with  HST's Wide Field Camera 3 (WFC3) and constructed by combining CANDELS data (HST/WFC3 {\it F105W, F125W} \& {\it F160W}) with available public imaging data from UV to mid-IR wavelengths. The CANDELS catalogue consists of images taken in the following filters: i) {\it U} with CTIO Blanco telescope and Visible Multi-Object Spectrograph (VIMOS) on the Very Large Telescope (VLT); ii) {\it F435W, F606W, F775W, F814W} and {\it F850LP} with HST/ACS; iii) {\it F098M, F105W, F125W} and {\it F160W} with HST/WFC3; iv) {\it Ks} with the Infrared Spectrometer and Array Camera (ISAAC) and with the High Acuity Wide field K-band Imager (HAWK-I) on the VLT; v) {\it Spitzer}'s IRAC channels at 3.5, 4.5, 5.8 and 8.0 $\mu m$. The source detection was carried out on the {\it F160W} image by running SExtractor in a two-modes configuration, i.e. hot and cold \citep{Galametz-2013}. This was done in order to make sure to effectively detect  both bright/large (cold mode, i.e. avoiding bright source over-deblending) and  faint/small (hot mode, i.e. pushing SExtractor to detect faint sources close to the image's limiting depth) sources. Photometry in other bands is then measured in dual-image mode. For all {\it HST} bands this was done on Point-Spread-Function (PSF)-matched images, while for other low-resolution bands (whose PSF FWHMs varied by almost a factor of 10), the profile template-fitting package TFIT (developed by the GOODS team) was used to measure the uniform photometry among them. We refer to \citet{Guo-2013} for details on the catalogue, its photometry and on the filters' characteristics (see Table 1 in their article). The fluxes measured (in 16 filters out of 17) for the galaxies studied here are reported in Table \ref{tab:Table1}. 

In addition to having measurements in more filters than in M06 (resulting in a more refined sampling of galaxy SEDs), the uncertainties on the photometry used here are on average smaller than those affecting the data used in M06. This enables us to test whether a more accurate photometry and more finely sampled SEDs deliver significant differences in the fitted galaxy properties compared to the M06 results.

\subsection{The COSMOS sample}
\label{subsec:subsec2.2}
We use a sample of 44 spectroscopically confirmed galaxies with $1.3<z_{\rm spec}<2.1$ (see Figure \ref{fig:Fig1}) in the COSMOS survey. These galaxies were selected as passive via the {\it BzK} criterion and extracted  from the \citet{McCracken-2010} catalogue.  
Galaxy photometry is taken from the ${\it K}_{\rm s}$-selected  COSMOS/UltraVISTA catalogue by \citet{Muzzin-2013}. This catalogue nominally provides PSF-matched aperture photometry in 30 broad- and medium-bands  (see Figure \ref{fig:Fig3}) over most of the COSMOS field, i.e.,  i) GALEX FUV and NUV (see \citealp{Martin-2005}); ii) Subaru/SuprimeCam ${\it g^{\rm +},\ r^{\rm +},\ i^{\rm +},\ z^{\rm +},\ B_{\rm j},\ V_{\rm j}}$ and CFHT/MegaCam ${\it u^{\rm *}}$ \citep{Taniguchi-2007, Capak-2007}; iii) Subaru/SuprimeCam IA427-IA827 optical medium bands \citep{Capak-2007}; iv) UltraVISTA ${\it YJHK_{\rm s}}$ \citep{McCracken-2012}; v) observations at the 3.6, 4.5 5.8, 8.0 and 24 ${\rm \mu m}$ channels from {\it Spitzer}'s IRAC+MIPS cameras (see \citealp{Sanders-2007}). We note that the  {\it i}-band photometry of galaxies $IDs=12\ \&\ 7$ is flagged as ``possibly contaminated'', but their inclusion in the COSMOS sample does not change any of our results. 

Spectroscopic redshifts were measured via the D4000 break and the Mg$_{\rm UV}$ features,  either from spectra taken with the Multi-Object InfraRed Camera and Spectrograph (MOIRCS) at Subaru (17 galaxies from \citealp{Onodera-2012}) or from VLT/VIMOS observations (the remaining 27, ESO programs 086.A-0681 and 088.A-0671, PI Daddi, \citealp{Strazzullo-2015}, Gobat et al., in prep.).  Note that the presence of MgI+II  absorptions at $\lambda\simeq 2800$ \AA\ and/or the D4000 break  further demonstrate  that these galaxies are passive. They are also all undetected in the UV\footnote{For the SED fit we use all available filters, except the 24 ${\rm \mu m}$, whose measurements are available only for 1 galaxy and the theoretical SEDs we use do not model that spectral region.}. The spectroscopic catalogue used here will be presented in a separate paper.

Besides being selected as passively evolving by the $BzK$ criterion, our galaxies also satisfy the rest-frame {\it UVJ} selection criteria \citep{Williams-2009}. As we shall see in Section~ 6, stacked images built on galaxy samples of quiescent galaxies including these same galaxies display hardly any signal in {\it Herschel} maps, thereby supporting our classification as passively evolving galaxies.

We point out  that the optical and NIR photometry in the \citet{Muzzin-2013} catalogue was re-calibrated by applying zero-point offsets in order to improve the quality of photometric redshifts. This re-calibration is somewhat dependent on  the population models  used as templates for measuring photometric redshifts  (in this case BC03 and PEGASE), which could bias the analysis towards  such models. Therefore, unless explicitly stated, our following analysis is carried out on photometry uncorrected for such offsets. This issue will be discussed more thoroughly in Section \ref{subsec:subsec4.4}. 
\section{SED fitting procedure}
\label{sec:sec3}
We carry out SED fitting on the observed SEDs following the same method adopted in M06.
We use a version of the public code HYPERZ \citep{Bolzonella-2000} modified in order to allow to fit SEDs by fixing the galaxy redshift at the spectroscopic value (named HYPERZSPEC). The procedure consists in fitting different model template spectra to the observed data, and evaluating for each of them the reduced  $\chi^{2}$, which is used as a figure of merit to identify the best-fitting model. Once the latter is found, all the characteristics of the stellar populations generating the chosen model (e.g., stellar age, metallicity, SFH), are assigned to the fitted galaxy SED\footnote{It is important to note that HYPERZSPEC does not interpolate on the template grids and that the best-fitting solution should only be regarded as the closest model to the observed data within the given template set. Using a more densely populated template set might result in a different best-fitting solution. Here we use the latest version containing 221 ages per template. \citet{Maraston-2010} and \citet{Pforr-2012} tested whether using 221 model ages instead of the original 51 set in Hyper-Z caused any substantial difference, and the resulting differences were marginal.}. 

In order to be able to make a direct comparison with the results shown in M06, we keep the template setup identical. This is made of 32 sets of theoretical model spectra covering a broad range of SFHs: i) SSP (simple stellar population, i.e. single star burst); ii) exponentially declining SFR ($\tau$-model with $\tau=0.1, 0.3$ and $1\ {\rm Gyr}$); iii) truncated SFR (step-like star formation, i.e. constant for a time interval $\Delta t=0.1, 0.3$ and $1 \ {\rm Gyr}$ since galaxy formation, null afterwards); iv) constant SFR. Each SFH option is calculated for four metallicities (1/5, 1/2, 1 and 2 $Z_{\odot}$). All templates refer to a Salpeter (1955) IMF and each template has an age grid of 221 values within the range of $0.001<t<15\ {\rm Gyr}$.  As in M06, we limit fitted ages to lie within the age of the Universe at the given redshift in a standard cosmology ( in our case $0.001<age<5\ {\rm Gyr}$).
As the minimum magnitude error we conservatively adopt $0.05\ {\rm mag}$ (even when the measurement is nominally more precise).
Finally, only filters with effective wavelengths such that $\lambda_{\rm eff}< \lambda_{\rm max}$, where $\lambda_{\rm max}=25000\ {\rm \AA}$, are used when carrying out the SED fitting\footnote{Filters with $\lambda_{\rm eff}>\lambda_{\rm max}$ are only used when $\lambda_{\rm eff}/(1+z_{\rm spec})<\lambda_{\rm max}$, which happens only for our highest-$z$ galaxy.}. 
We run HYPERZSPEC in two modes: i) assuming no reddening; ii) allowing reddening to vary in a range $0<A_{\rm V}<3$ with a 0.2 step,  for 5 different reddening laws, namely: Milky Way (\citealp{Allen-1976} and \citealp{Seaton-1979}), Large Magellanic Cloud \citep{Fitzpatrick-1986}, Small Magellanic Cloud (SMC, \citealp{Prevot-1984, Bouchet-1985}) and the so-called Calzetti law \citep{Calzetti-2000}. 

With this template setup, we separately carry out the fitting procedure using M05 and BC03-based templates, as in M06, and additionally we experiment with the M13 revised version of M05 models which include a milder contribution from TP-AGB stars (described in the next section). The output stellar population properties are:  age $t$ (i.e., the time since the start of star formation); metallicity $[Z/H]$; SFH; reddening law; reddening $E(B-V)$ and stellar mass $M^{\ast}$.  Stellar mass is calculated by re-normalising the best-fitting model template SED to the observed one and subtracting stellar mass losses (see M98, M05, M06), using a routine from \citet{Daddi-2005}. 

The TP-AGB phase is  controlled by three main physical processes, namely envelope convection that may lead to the so-called hot bottom burning, mixing via the so called third dredge-up and strong mass loss. All such processes are poorly understood in quantitative terms and one has to resort to parametrizations. Yet they all affect the lifetime and luminosity evolution of TP-AGB models, hence the  TP-AGB contribution to the integrated light of stellar populations. 
As a consequence, such  contribution  cannot be calculated from first principles and must be calibrated with suitable data.

In \citet{Maraston-1998} and M05 models, a semi-empirical approach was adopted by using the energetics and optical/near-IR colours of Magellanic Cloud (MC)  star clusters of various ages known to contain TP-AGB stars, thus  fixing the TP-AGB  contribution in stellar population models. Obviously, models constructed in such a way depend on the data used for calibration, the ages assigned to the observed clusters  and their integrated photometry, the latter setting the colours of the integrated models. 

The ages  of the MC star clusters used to measure the fractional contribution of the TP-AGB  to the total bolometric light  were derived from the \citet{Frogel-1990} SWB\footnote{\citet{Searle-1980} provided a classification scheme for rich star clusters into seven types, called SWB types, on the basis of reddening-free parameters derived from integrated {\it uvgr} photometry. \citet{Frogel-1990} surveyed for AGB stars in 39 Magellanic Clouds clusters and associated ages to each SWB type.}-type-age relation, allowing to derive the TP-AGB fuel consumption as a function of cluster age. However, this SWB-type-age relation was calibrated on relatively few cluster ages derived from colour-magnitude diagrams (CMD), some being  now considered underestimated by factors between $\sim40$ and $\sim 80$ per cent, e.g. see \citep{Crowl-2001, Pessev-2008, Milone-2009}, and were based on a Large Magellanic Cloud distance modulus  $\sim0.2$ mag closer than currently measured.
 
\citet{Conroy-2010} performed a similar  calibration, but used newer photometry and average colours and set the clusters on a new age scale resulting in systematically older ages (up to $2\ {\rm Gyr}$ older). As a consequence, in this new framework the M05 models resulted to be off with respect to age (i.e., the TP-AGB contribution was kicking-in at a too young age) and also too red (e.g., in $V-K$) with respect to the average cluster  colours. \cite{Noel-2013}  took the same approach,  performing the age calibration again using optical-to-near-IR photometry, and  assigned ages to 43 MC clusters based on up-to-date CMD measurements from the literature. Then No\"el and collaborators  stacked the clusters  in age bins and determined colours and integrated luminosities. They found that clusters' optical and optical-to-infrared colours become suddenly redder  at an age between $\sim 0.6$ and $\sim 1$ Gyr, compared to younger ones (e.g., $\sim 1$ mag redder in $(V-K)_\circ$, see Figure 3 in \citealp{Noel-2013}), indicative of  a sudden TP-AGB phase transition, as previously argued (\citealp{Maraston-1998}, M05). Clusters were then found slightly bluer at older ages. The shift in the transition age with respect to the M05 calibration was found to be $\sim 300\ {\rm Myr}$. Hence, the conclusions by \citet{Conroy-2010} of a lack of a colour transition and a much later ($\sim 3$~Gyr) development of the TP-AGB were not confirmed. It was instead confirmed that at maximum TP-AGB contribution the M05 models  were $\sim 0.6$ mag too red in  $(V-K)_\circ$, lying  at the reddest edge of the variance with respect to the average cluster colours. 
 
The calibration in \cite{Noel-2013} was used by C. Maraston to re-calibrate the M05 models accordingly. The new ``M13" models have been  calculated by setting to zero the TP-AGB fuel consumption for ages younger than $0.6\ {\rm Gyr}$, and by reducing the TP-AGB fuel consumption by $\sim~50$~per cent at $\gtrsim 0.6\ {\rm Gyr}$,
with respect to the original M05 calibration. As it can be seen in Figure 3 of \citet{Noel-2013}, the M13 models go perfectly through the average data in the $(V-K)_\circ$ vs. age plot. The same models have also been argued to better match a sample of post-starburst galaxies at low redshift ($z\sim~0.5$) by \citet{Melnick-2013,Melnick-2014}.

\begin{figure*}
\centering
\includegraphics[width=0.48\textwidth]{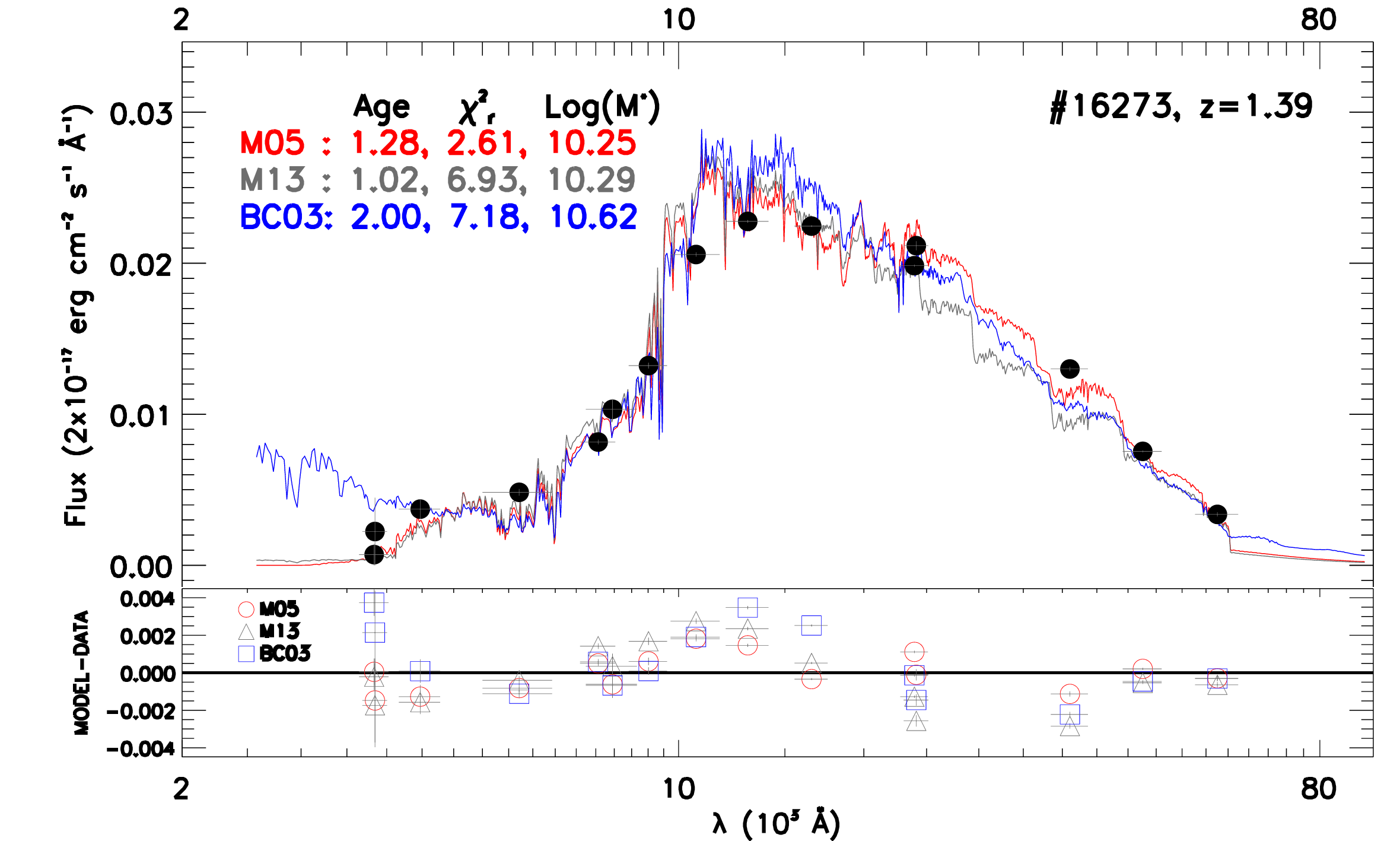}
\includegraphics[width=0.48\textwidth]{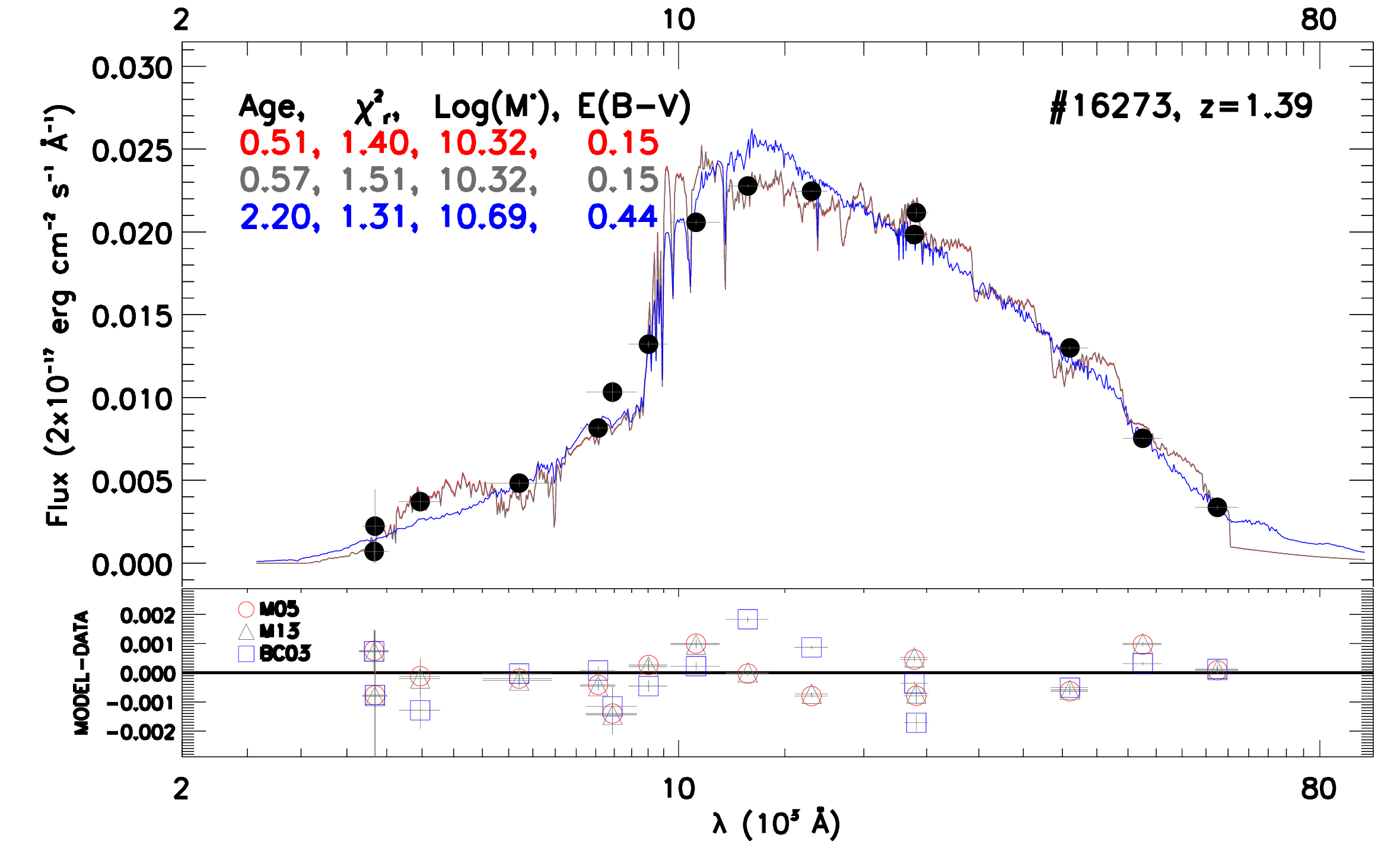}
\includegraphics[width=0.48\textwidth]{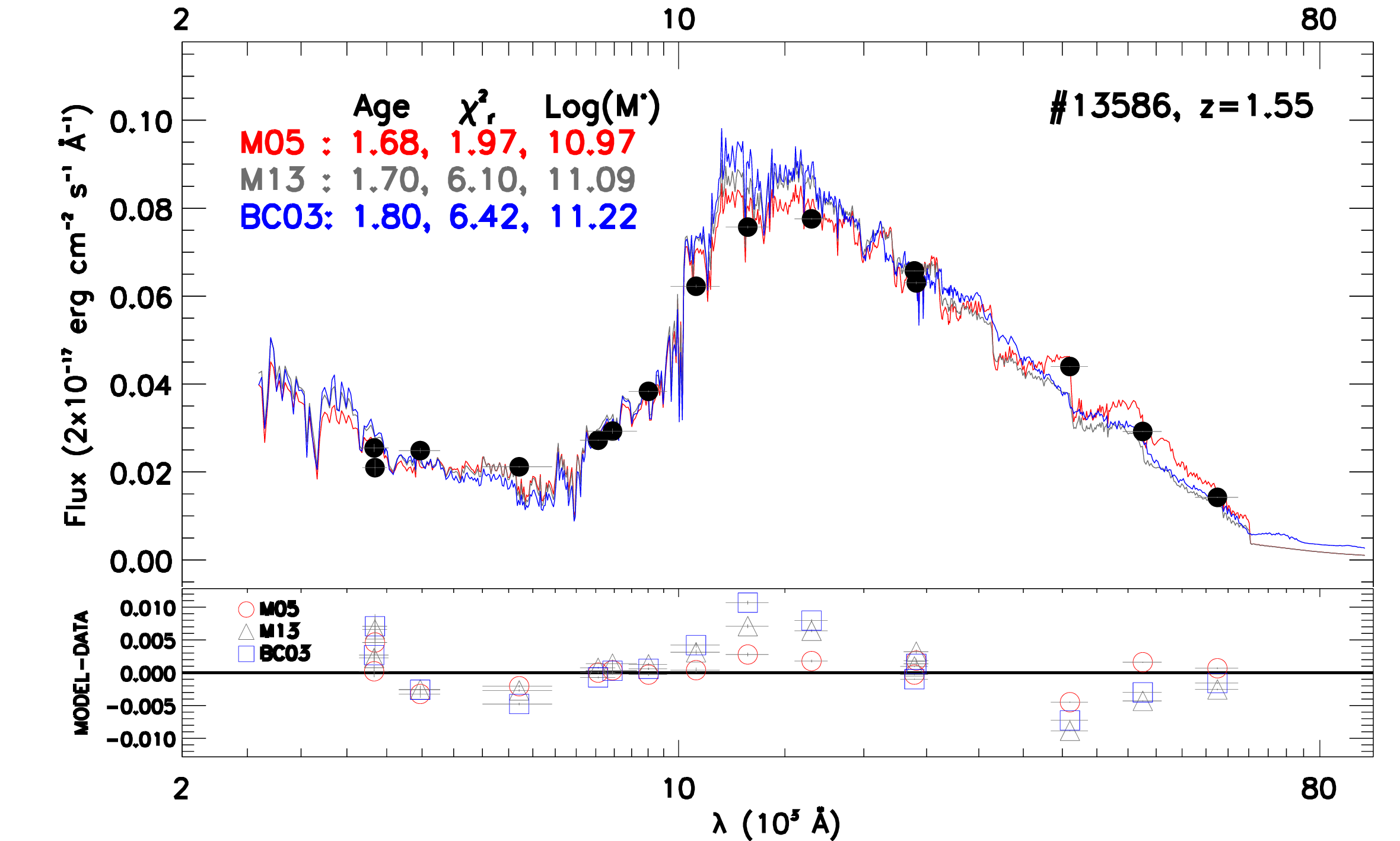}
\includegraphics[width=0.48\textwidth]{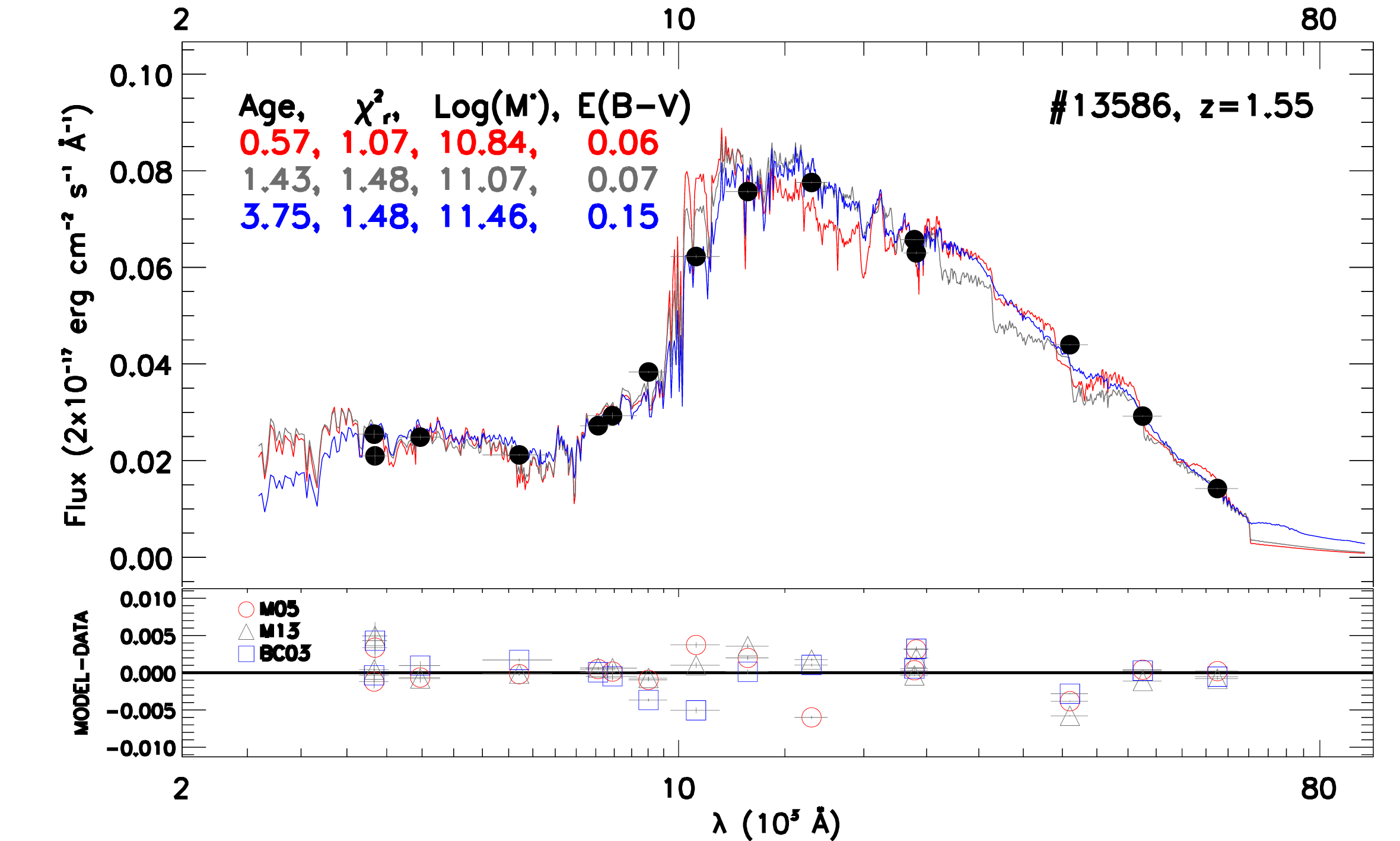}
\includegraphics[width=0.48\textwidth]{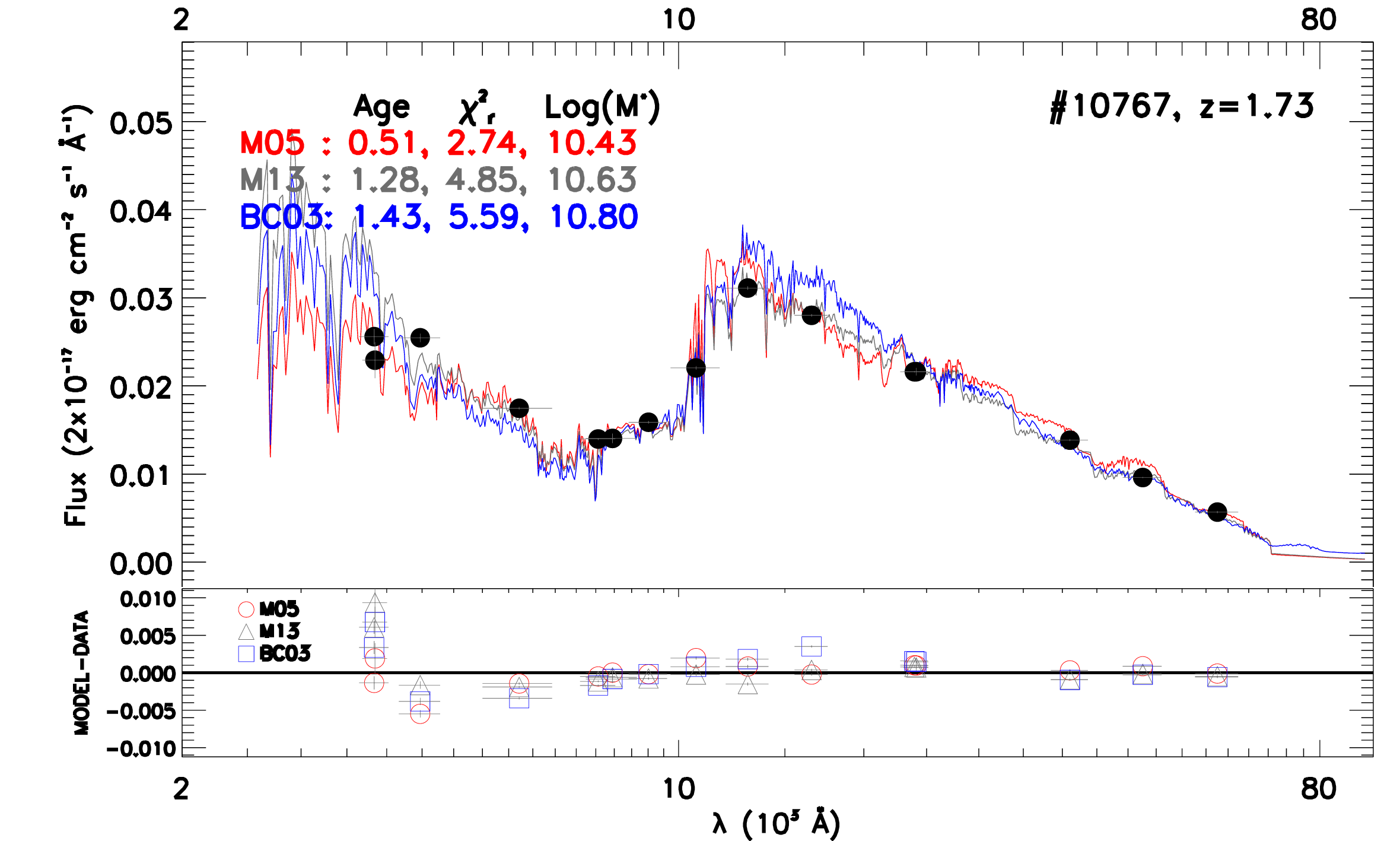}
\includegraphics[width=0.48\textwidth]{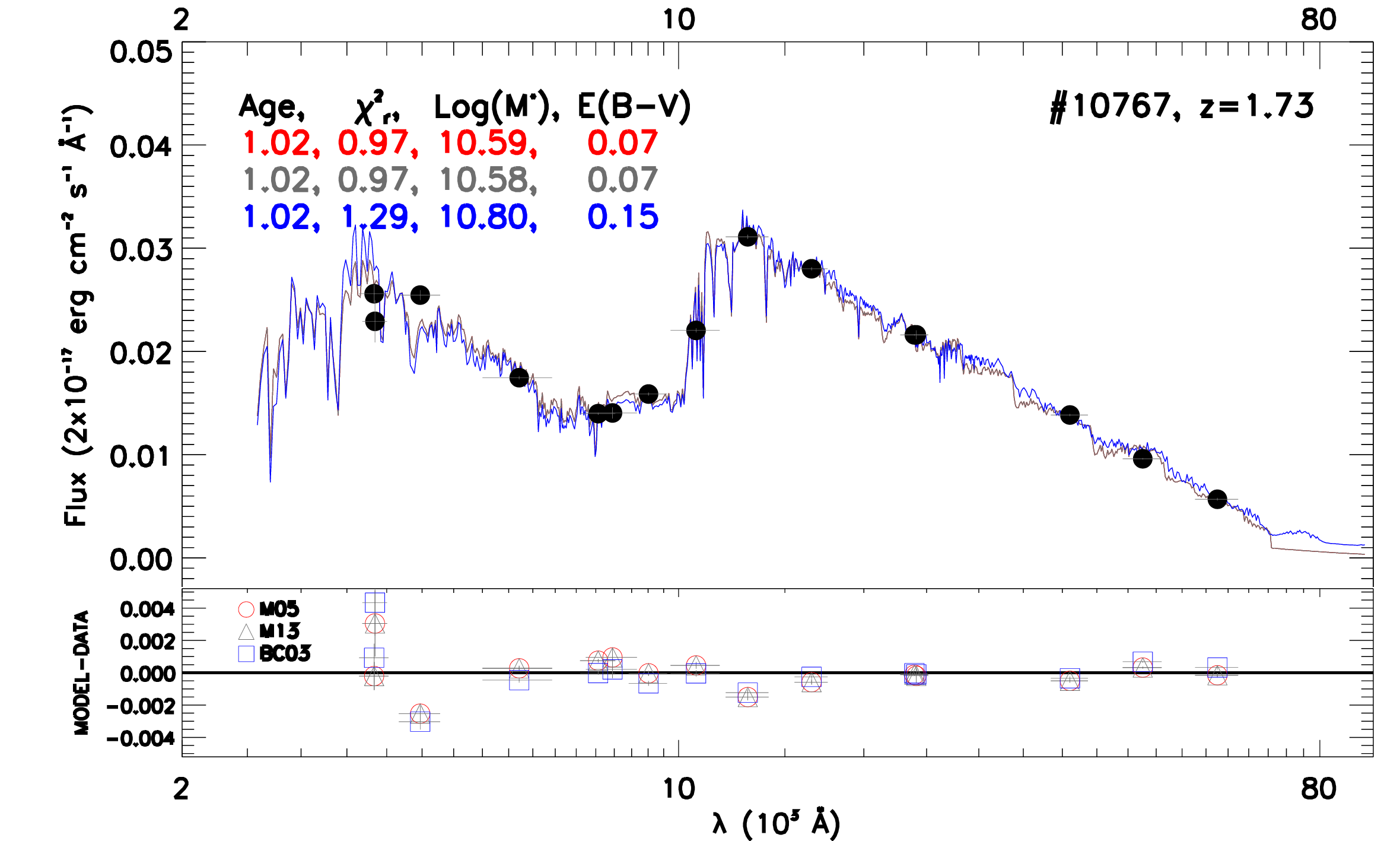}
\includegraphics[width=0.48\textwidth]{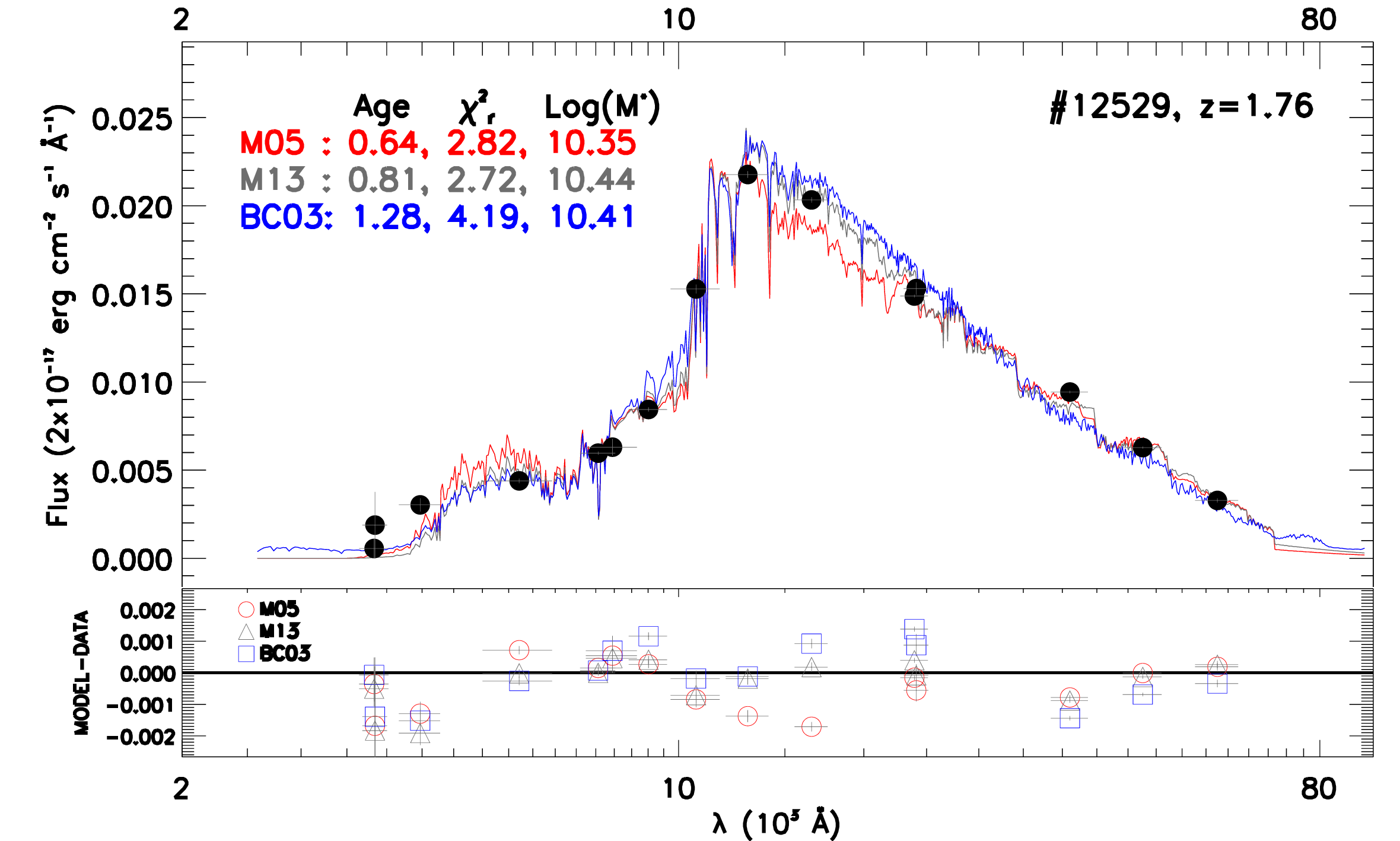}
\includegraphics[width=0.48\textwidth]{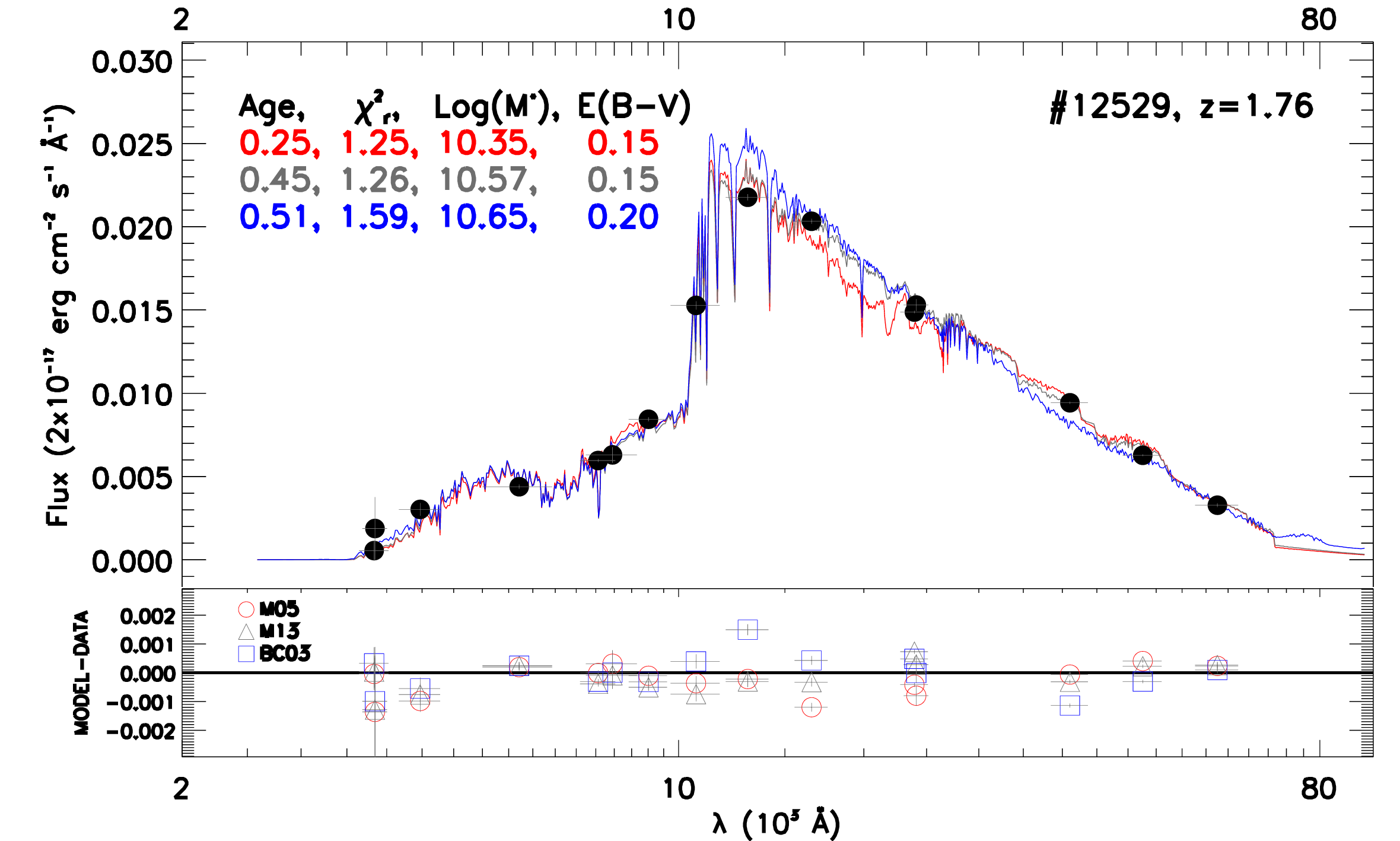}
\caption{SED fits of galaxies in the HUDF sample. Left-hand panels: no-reddening case; right-hand panels: reddened case. Observed fluxes are plotted as symbols over best-fit templates showed as lines, for M05 (red), M13 (grey) and BC03 (blue) models. Flux residuals ($MODEL-DATA$)  are plotted vs. wavelength at the bottom of each panel.}
\label{fig:Fig2}
\end{figure*}
\addtocounter{figure}{-1}
\begin{figure*}
\centering
\includegraphics[width=0.48\textwidth]{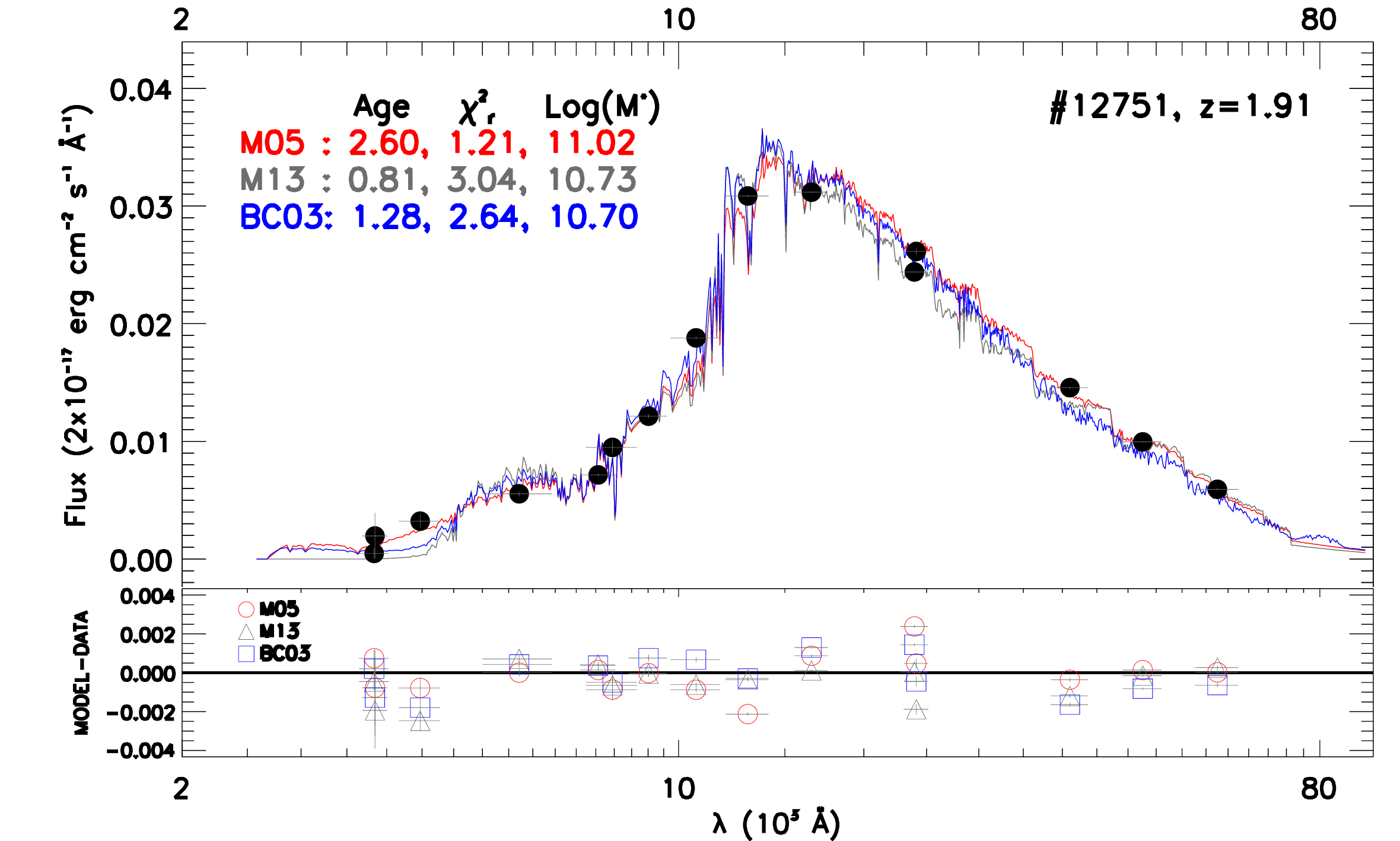}
\includegraphics[width=0.48\textwidth]{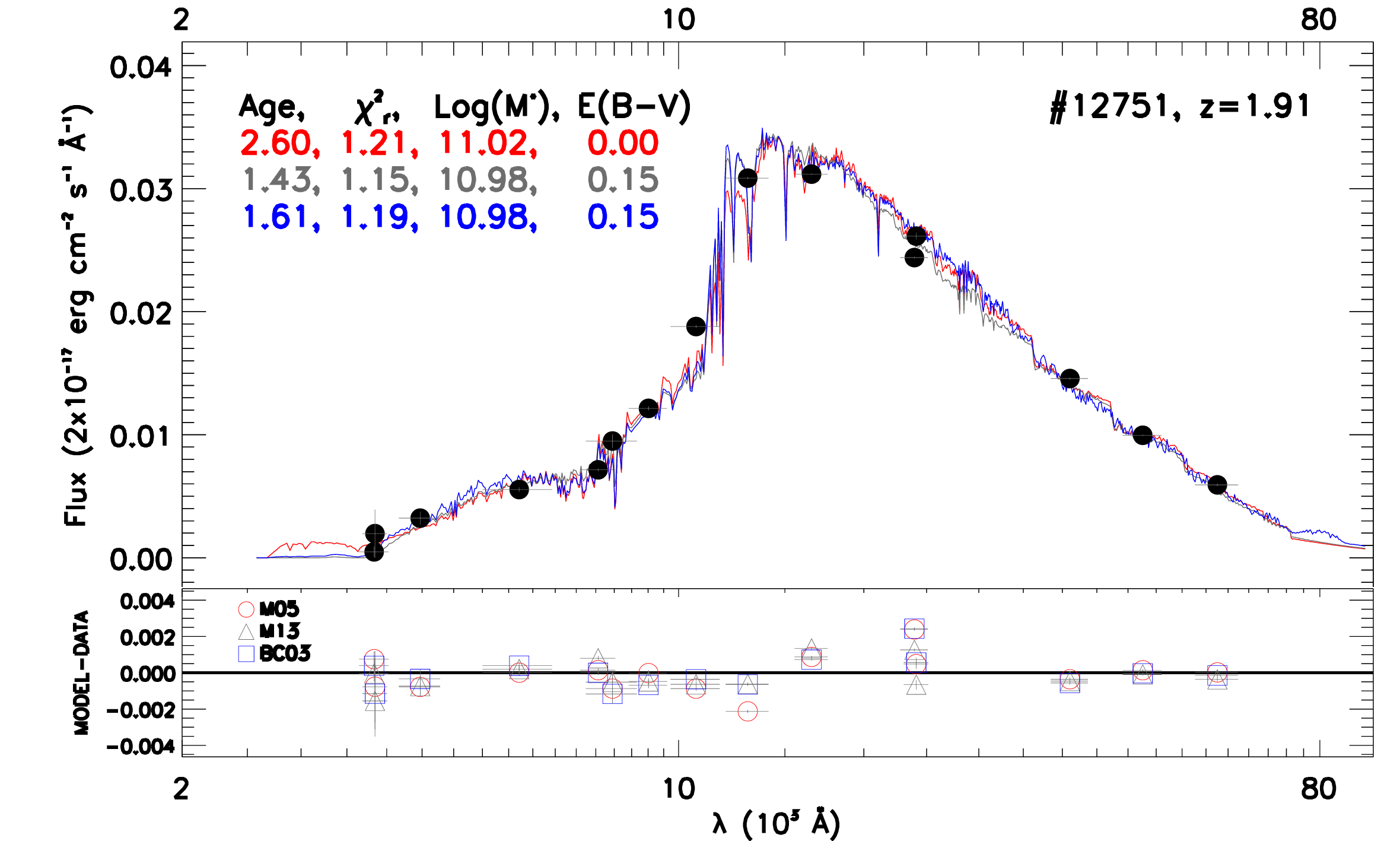}
\includegraphics[width=0.48\textwidth]{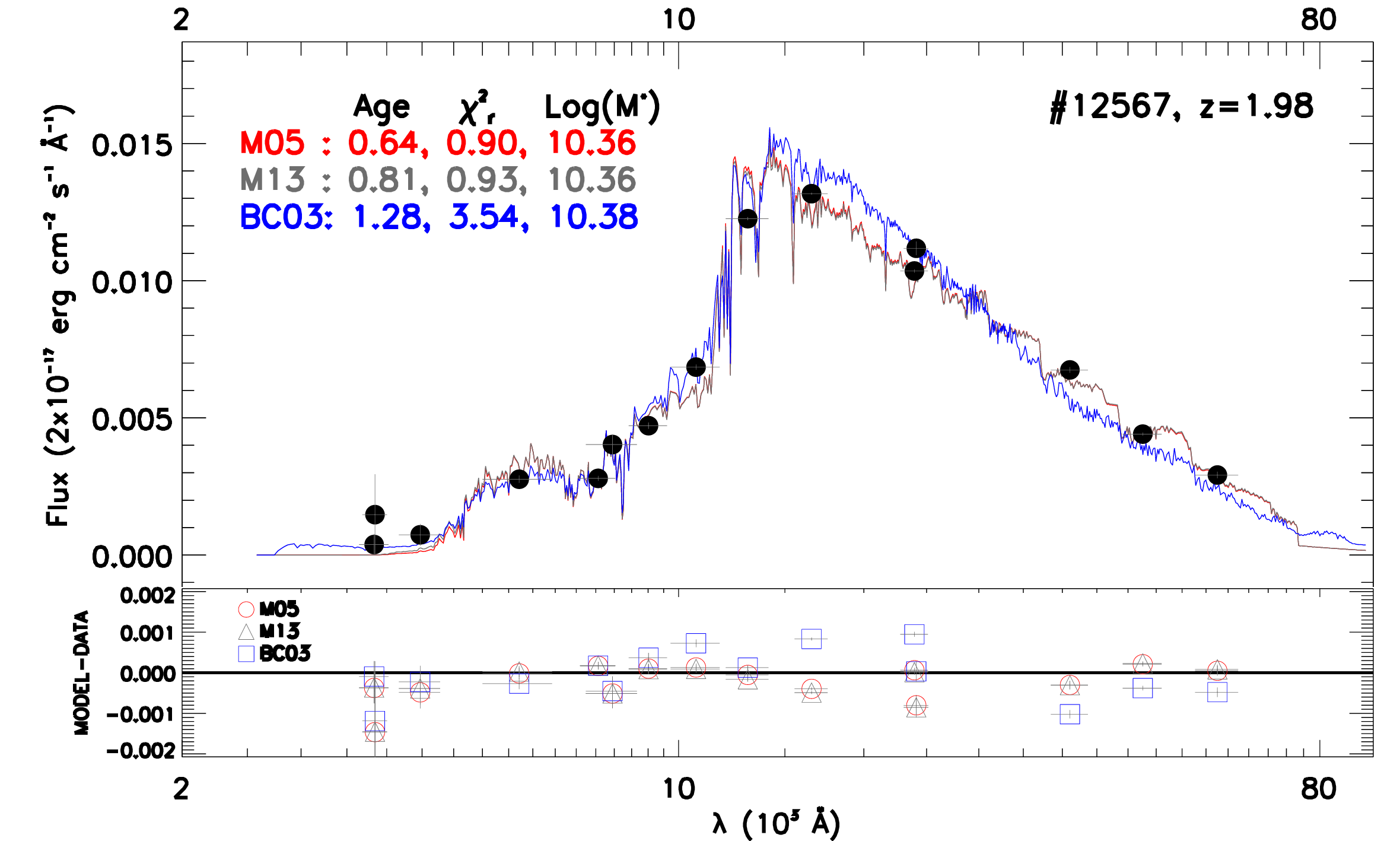}
\includegraphics[width=0.48\textwidth]{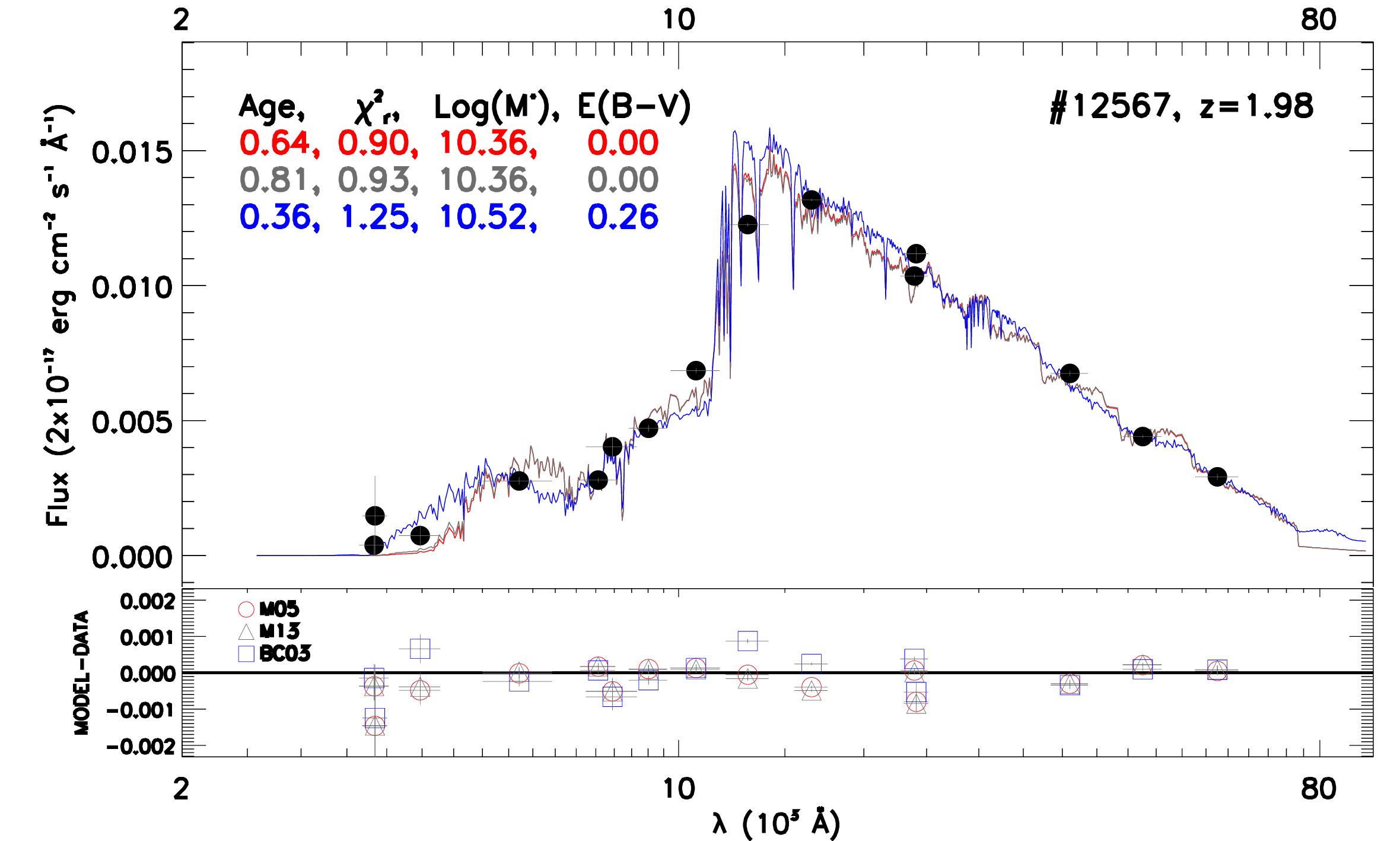}
\includegraphics[width=0.48\textwidth]{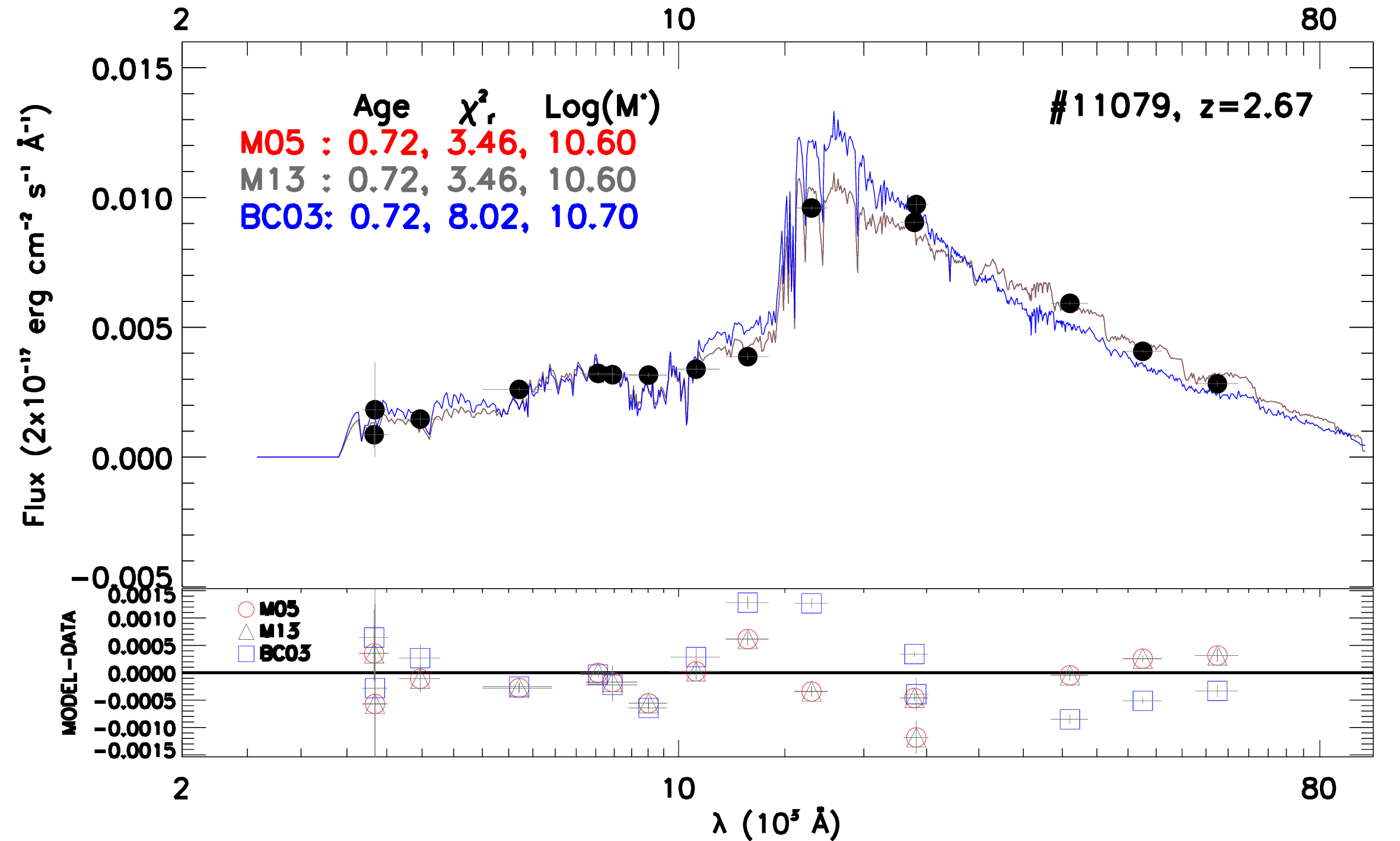}
\includegraphics[width=0.48\textwidth]{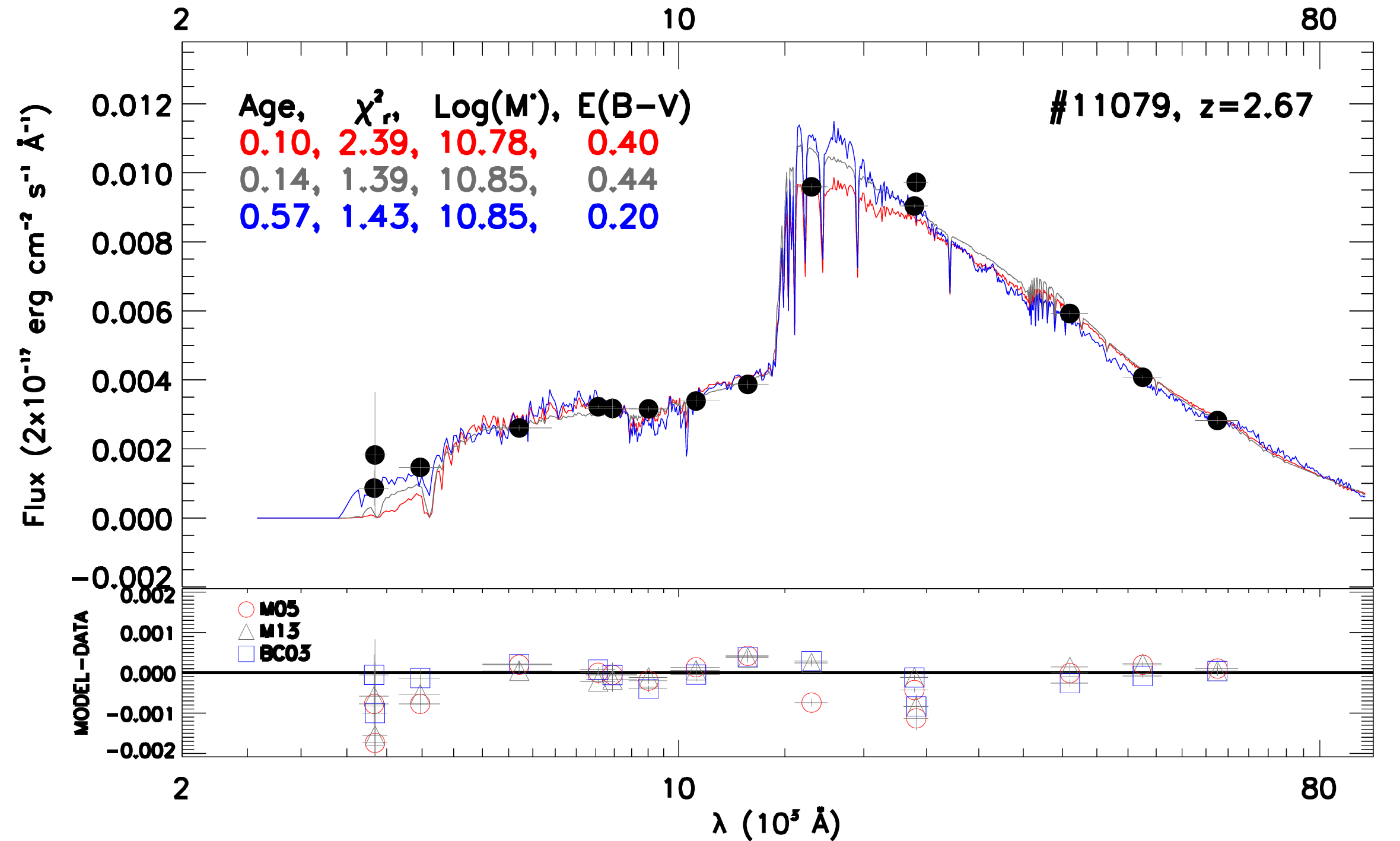}
\caption{Continued.}
\label{fig:Fig2}
\end{figure*}
\section{SED fitting results}
\label{sec:sec4}
In this section we describe the results of the SED fitting carried out on our galaxies. In particular, we also compare the results obtained with different evolutionary stellar population synthesis models, ranking them according to the $\chi^{2}_{\rm r}$ values (which are often comparable or similar) of their SED fitting solutions, but also, and more importantly, by assessing their statistical consistency.

\subsection{HUDF Sample}
\label{subsec:subsec4.1}
As in M06, we first discuss the case where reddening is not included among the free parameters of the fitting (i.e. $E(B-V)$ is set to zero). This case is important because - by removing reddening-induced degeneracies - it allows a clearer assessment of the model differences due to the underlying stellar evolution. Also, being selected as passive these galaxies  are expected to contain a low amount of cold gas and dust (see Section \ref{sec:sec5}).  
The best-fit solutions are shown in Figure \ref{fig:Fig2} (left-hand panels) and given in Table \ref{tab:Table2}.

\begin{table*}
%\begin{tiny}
\begin{center}
\caption{SED fitting results for the HUDF sample for the no-reddening case (cfr. Table 2 in M06). Values refer to the best-fit solution. Col 1: galaxy ID ; col 2: spectroscopic redshift; col 3: model; col 4: age; col 5: metallicity; col 6: star formation history; col 7: reduced $\chi^{2}$; col 8: stellar mass; col 9 : star formation rate.}
\begin{tabular}{cccclcccc}
\hline
  \multicolumn{1}{c}{\bf ID} &
  \multicolumn{1}{c}{$\mathbf{z_{\rm spec}}$} &
  \multicolumn{1}{c}{\bf Model} &
  \multicolumn{1}{c}{$\mathbf{t}$} &
  \multicolumn{1}{c}{$\mathbf{[Z/H]}$} &
  \multicolumn{1}{c}{\bf SFH} &
  \multicolumn{1}{c}{$\mathbf{\chi^{2}_{\rm r}}$} &
  \multicolumn{1}{c}{$\mathbf{M^{\ast}}$} & 
  \multicolumn{1}{c}{$\mathbf{SFR}$} \\
  \multicolumn{1}{c}{}&
  \multicolumn{1}{c}{}&
  \multicolumn{1}{c}{}&
  \multicolumn{1}{c}{\rm (Gyr)}&
  \multicolumn{1}{c}{${\rm (Z_{\odot})}$} &
  \multicolumn{1}{c}{}&
  \multicolumn{1}{c}{}&
  \multicolumn{1}{c}{$(10^{11}\ {\rm M_{\odot}})$} &
  \multicolumn{1}{c}{$({\rm M_{\odot}/ yr^{-1}})$} \\
  \multicolumn{1}{c}{(1)}&
  \multicolumn{1}{c}{(2)}&
  \multicolumn{1}{c}{(3)}&
  \multicolumn{1}{c}{(4)}&
  \multicolumn{1}{c}{(5)}&
  \multicolumn{1}{c}{(6)}&
  \multicolumn{1}{c}{(7)}&
  \multicolumn{1}{c}{(8)}&
  \multicolumn{1}{c}{(9)} \\
\hline
\hline
  \multirow{3}{*}{16273}  &      & M05  &  $1.28$ &  2   & $t_{\rm trunc}=1.0\ {\rm Gyr}$   & 2.61 & $0.18$ & $<0.1$ \\
                          & 1.39 & M13  &  $1.01$ &  0.2 & $e^{-t/0.1\ {\rm Gyr}}$ 	    & 6.93 & $0.19$ & $<0.1$ \\
			  &      & BC03 &  $2.00$ &  2   & $e^{-t/0.3\ {\rm Gyr}}$ 	    & 7.18 & $0.42$ & $0.2$  \\
\hline
  \multirow{3}{*}{13586}  &      & M05  &  $1.68$ &  0.5 & $e^{-t/0.3\ {\rm Gyr}}$          & 1.97 & $0.93$ & $1.5$  \\
                          & 1.55 & M13  &  $1.70$ &  0.2 & $e^{-t/0.3\ {\rm Gyr}}$ 	    & 6.10 & $1.23$ & $1.9$  \\
			  &      & BC03 &  $1.80$ &  2   & $e^{-t/0.3\ {\rm Gyr}}$ 	    & 6.42 & $1.66$ & $1.9$  \\
\hline
  \multirow{3}{*}{10767}  &      & M05  &  $0.51$ &  2   & $e^{-t/0.1\ {\rm Gyr}}$ 	    & 2.74 & $0.27$ & $2.1$  \\
                          & 1.73 & M13  &  $1.28$ &  0.2 & $e^{-t/0.3\ {\rm Gyr}}$ 	    & 4.85 & $0.43$ & $2.7$  \\
			  &      & BC03 &  $1.43$ &  2   & $e^{-t/0.3\ {\rm Gyr}}$ 	    & 5.60 & $0.63$ & $2.4$  \\
\hline
  \multirow{3}{*}{12529}  &      & M05  &  $0.64$ &  1   & $t_{\rm trunc}=0.1\ {\rm Gyr}$   & 2.82 & $0.22$ & $<0.1$ \\
                          & 1.76 & M13  &  $0.81$ &  0.5 & SSP 			            & 2.72 & $0.27$ & $<0.1$ \\
			  &      & BC03 &  $1.28$ &  0.2 & SSP 			            & 4.19 & $0.26$ & $<0.1$ \\
\hline
  \multirow{3}{*}{12751}  &      & M05  &  $2.60$ &  0.2 & $e^{-t/0.3\ {\rm Gyr}}$ 	    & 1.21 & $1.05$ & $0.1$  \\
                          & 1.91 & M13  &  $0.81$ &  0.5 & SSP 			            & 3.04 & $0.54$ & $<0.1$ \\
			  &      & BC03 &  $1.28$ &  0.2 & SSP 			            & 2.64 & $0.50$ & $<0.1$ \\
\hline
  \multirow{3}{*}{12567}  &      & M05  &  $0.64$ &  1   & SSP 			            & 0.90 & $0.23$ & $<0.1$ \\
                          & 1.98 & M13  &  $0.81$ &  0.2 & $t_{\rm trunc}=0.3\ {\rm Gyr}$   & 0.93 & $0.23$ & $<0.1$ \\
			  &      & BC03 &  $1.28$ &  0.2 & SSP 			            & 3.54 & $0.24$ & $<0.1$ \\
\hline
  \multirow{3}{*}{11079}  &      & M05  &  $0.72$ &  1   & $e^{-t/0.1\ {\rm Gyr}}$ 	    & 3.46 & $0.40$ & $0.4$  \\
                          & 2.67 & M13  &  $0.72$ &  0.2 & $e^{-t/0.1\ {\rm Gyr}}$ 	    & 3.46 & $0.40$ & $0.4$  \\
			  &      & BC03 &  $0.72$ &  2   & $e^{-t/0.1\ {\rm Gyr}}$ 	    & 8.02 & $0.50$ & $0.5$  \\
\hline
\hline
\end{tabular}
\label{tab:Table2}							    
\end{center}
% \end{tiny}
\end{table*}

In 6 out of 7 cases, the fits with M05 models have lower $\chi^{2}_{\rm r}$ values  compared to those obtained with the other models. This suggests  that the new, more accurate and finely sampled photometry used here did not have a major impact on the result with respect to the data used in M06.
The M13 models, which have a reduced TP-AGB contribution with respect to M05, do not perform better than M05, whereas they perform better than BC03  for 6 out of 7 galaxies. 
Hence the ranking of fits mirrors the amount of TP-AGB included in the models, which suggests that the treatment of the TP-AGB is the main driver of the goodness of fit.

It may appear surprising that, in spite of the most up-to-date calibration with MC  star clusters folded in the M13 models, the M05 ones still perform better on this sample of high-$z$~galaxies and,  as we shall see also in Section \ref{subsec:subsec4.2}, it does so also on the larger COSMOS sample. 
This suggests  that either reddening should be included or that the new age-colour calibration could be incorrect.
The former possibility would be supported by the average high  $\chi^{2}_{\rm r}$ values shown in Table \ref{tab:Table2}, though this sample consists of galaxies which should contain a low amount of dust.
We then consider the case where reddening is an additional free parameter of the fit. The results are shown in Figure \ref{fig:Fig2} (right-hand panels) and in Table \ref{tab:Table3}. As expected, with one additional free parameter the fits are generally better compared to the previous case, as reveled by the systematically lower $\chi^{2}_{\rm r}$ values.
\begin{table*}
\begin{scriptsize}
\begin{center}
\caption{As in Table 2 but now including reddening (cfr. Table 3 in M06). Reddening and reddening law are given in cols. 10 and 11}
\label{tab:Table3}
\begin{threeparttable} 
\begin{tabular}{cccclcccccl}
\hline
  \multicolumn{1}{c}{\bf ID} &
  \multicolumn{1}{c}{$\mathbf{z_{\rm spec}}$} &
  \multicolumn{1}{c}{\bf Model} &
  \multicolumn{1}{c}{$\mathbf{t}$} &
  \multicolumn{1}{c}{$\mathbf{[Z/H]}$} &
  \multicolumn{1}{c}{\bf SFH} &
  \multicolumn{1}{c}{$\mathbf{\chi^{2}_{\rm r}}$} &
  \multicolumn{1}{c}{$\mathbf{M^{\ast}}$} &
  \multicolumn{1}{c}{$\mathbf{SFR}$} &
  \multicolumn{1}{c}{$\mathbf{E(B-V)}$} &
  \multicolumn{1}{c}{\bf Reddening Law} \\
  \multicolumn{1}{c}{}&
  \multicolumn{1}{c}{}&
  \multicolumn{1}{c}{}&
  \multicolumn{1}{c}{\rm (Gyr)}&
  \multicolumn{1}{c}{${\rm (Z_{\odot})}$} &
  \multicolumn{1}{c}{}&
  \multicolumn{1}{c}{}&
  \multicolumn{1}{c}{$(10^{11}\ {\rm M_{\odot}})$} & 
  \multicolumn{1}{c}{$({\rm M_{\odot}/ yr^{-1}})$} &
  \multicolumn{1}{c}{({\rm mag})} &
  \multicolumn{1}{c}{} \\
  \multicolumn{1}{c}{(1)}&
  \multicolumn{1}{c}{(2)}&
  \multicolumn{1}{c}{(3)}&
  \multicolumn{1}{c}{(4)}&
  \multicolumn{1}{c}{(5)}&
  \multicolumn{1}{c}{(6)}&
  \multicolumn{1}{c}{(7)}&
  \multicolumn{1}{c}{(8)}&
  \multicolumn{1}{c}{(9)}&
  \multicolumn{1}{c}{(10)}&
  \multicolumn{1}{c}{(11)} \\
\hline
\hline
  \multirow{3}{*}{16273}  &      & M05  &  $0.51$ &  1   & SSP                            & 1.40 & $0.21$ & $<0.1$ & $0.15$ & \citet{Calzetti-2000}			     \\
                          & 1.39 & M13  &  $0.57$ &  0.2 & $t_{\rm trunc}=0.1\ {\rm Gyr}$ & 1.51 & $0.21$ & $<0.1$ & $0.15$ & \citet{Calzetti-2000}			     \\
			  &      & BC03 &  $2.20$ &  0.2 & $e^{-t/1.0\ {\rm Gyr}}$        & 1.31 & $0.49$ & $8.3$  & $0.44$ & \citet{Prevot-1984,Bouchet-1985}  	     \\
\hline
  \multirow{3}{*}{13586}  &      & M05  &  $0.57$ &  2   & $e^{-t/0.1\ {\rm Gyr}}$        & 1.07 & $0.69$ & $2.9$  & $0.06$ & \citet{Fitzpatrick-1986}  		   \\
                          & 1.55 & M13  &  $1.43$ &  0.2 & $e^{-t/0.3\ {\rm Gyr}}$        & 1.48 & $1.17$ & $4.4$  & $0.07$ & \citet{Prevot-1984,Bouchet-1985}  	     \\
			  &      & BC03 &  $3.75$ &  1   & $e^{-t/1.0\ {\rm Gyr}}$        & 1.48 & $2.88$ & $9.6$  & $0.15$ & \citet{Prevot-1984,Bouchet-1985}  	     \\
\hline
  \multirow{3}{*}{10767}  &      & M05  &  $1.01$ &  1   & $e^{-t/0.3\ {\rm Gyr}}$        & 0.97 & $0.39$ & $5.8$  & $0.07$ & \citet{Prevot-1984,Bouchet-1985}  	     \\
                          & 1.73 & M13  &  $1.01$ &  0.2 & $e^{-t/0.3\ {\rm Gyr}}$        & 0.97 & $0.38$ & $5.8$  & $0.07$ & \citet{Prevot-1984,Bouchet-1985}  	     \\
			  &      & BC03 &  $1.01$ &  2   & $e^{-t/0.3\ {\rm Gyr}}$        & 1.29 & $0.63$ & $9.7$  & $0.15$ & \citet{Calzetti-2000}			     \\
\hline
  \multirow{3}{*}{12529}  &      & M05  &  $0.25$ &  2   & SSP                            & 1.25 & $0.22$ & $<0.1$ & $0.15$ & \citet{Prevot-1984,Bouchet-1985}  	     \\
                          & 1.76 & M13  &  $0.45$ &  1   & SSP                            & 1.26 & $0.37$ & $<0.1$ & $0.15$ & \citet{Calzetti-2000}			     \\
			  &      & BC03 &  $0.51$ &  1   & $t_{\rm trunc}=0.1\ {\rm Gyr}$ & 1.59 & $0.45$ & $<0.1$ & $0.20$ & \citet{Calzetti-2000}			     \\
\hline
  \multirow{3}{*}{12751}  &      & M05  &  $2.60$ &  0.2 & $e^{-t/0.3\ {\rm Gyr}}$        & 1.21 & $1.05$ & $0.1$  & $0.00$ & NA\tnote{1}				     \\
                          & 1.91 & M13  &  $1.43$ &  0.2 & SSP                            & 1.15 & $0.95$ & $<0.1$ & $0.15$ & \citet{Prevot-1984,Bouchet-1985}  	     \\
			  &      & BC03 &  $1.61$ &  0.2 & $t_{\rm trunc}=1.0\ {\rm Gyr}$ & 1.19 & $0.95$ & $<0.1$ & $0.15$ & \citet{Calzetti-2000}			     \\
\hline
  \multirow{3}{*}{12567}  &      & M05  &  $0.64$ &  1   & SSP                            & 0.90 & $0.23$ & $<0.1$ & $0.00$ & NA\tnote{1}					   \\
                          & 1.98 & M13  &  $0.81$ &  0.2 & $t_{\rm trunc}=0.3\ {\rm Gyr}$ & 0.93 & $0.23$ & $<0.1$ & $0.00$ & NA\tnote{1}					   \\
			  &      & BC03 &  $0.36$ &  2   & $t_{\rm trunc}=0.1\ {\rm Gyr}$ & 1.25 & $0.33$ & $<0.1$ & $0.26$ & \citet{Fitzpatrick-1986}  		     \\
\hline
  \multirow{3}{*}{11079}  &      & M05  &  $0.10$ &  1   & SSP                            & 2.39 & $0.60$ & $<0.1$ & $0.40$ & \citet{Calzetti-2000}			     \\
                          & 2.67 & M13  &  $0.14$ &  0.2 & SSP                            & 1.39 & $0.71$ & $<0.1$ & $0.44$ & \citet{Calzetti-2000}			     \\
			  &      & BC03 &  $0.57$ &  1   & $e^{-t/0.1\ {\rm Gyr}}$        & 1.43 & $0.71$ & $3.0$  & $0.20$ & \citet{Calzetti-2000}			     \\
\hline
\hline
\end{tabular}
\begin{tablenotes}\footnotesize 
\item[1] Despite allowing for reddening, the SED of this galaxy was always best-fitted with $E(B-V)=0$, independently of the reddening law used.
\end{tablenotes}
\end{threeparttable}							    
\end{center}
\end{scriptsize}
\end{table*}

M05 perform best for 4 galaxies out of 7,  M13 and BC03 models perform best for 3  out of 7 and 1 of 7 galaxies, respectively.  The $\chi^{2}_{\rm r}$ values obtained with the three models are very similar, hence cannot strongly discriminate between them.  What varies the most are stellar ages, a well known and documented effect (see M06, \citealp{Cimatti-2008} and many papers after). 
BC03 models tend to release older and more dust-reddened fits compared to the other two models. 
As discussed in M06, this happens because a lower TP-AGB is compensated by older populations for matching the red  rest-frame optical/near-IR colors. In order to also match the optical, these old populations need to be coupled with a fraction of young stars whose optical flux is partly suppressed by reddening. 
However, disentangling the effect of reddening from that of metallicity and age is difficult. M05- and M13-based results present similar reddening values, but have significantly different metallicities, and BC03 models have both different reddening and metallicity. A consistent aspect between the models is instead the preferred reddening laws, namely the Calzetti law \citep{Calzetti-2000} and the one for the Small Magellanic Cloud  \citep{Prevot-1984, Bouchet-1985}. These results are very similar to those found in M06.

In summary, the analysis carried on the HUDF sample with the new photometry tends to confirm the  M06 results. 
We also show that decreasing the TP-AGB strength as suggested by \citet{Conroy-2010, Kriek-2010} and as adopted in our latest M13 calibration  does not lead to significant improvements. However, these results are based on a statistically very small  sample (albeit  of high quality). We then proceed by examining the results on the larger COSMOS sample perviously described.

\subsection{COSMOS Sample}
\label{subsec:subsec4.2}

SED-fitting results for the COSMOS sample are given in plots and tables identical to those for the HUDF, which we collect in Appendix \ref{sec:appendixB} (see Figure \ref{fig:Fig1_appB} for both reddening-free and reddened cases and Tables \ref{tab:Table1_appB} and \ref{tab:Table2_appB}). These results refer to the original COSMOS photometry, while the effects of applying to it zero-point offsets are discussed in Section \ref{subsec:subsec4.4}. Here we summarize our findings.

As for the HUDF sample,  in absence of reddening M05 models give solutions with lower $\chi^{2}_{\rm r}$ values for the majority of the sample (70 per cent, 31 galaxies). BC03 and M13 models are the best performing respectively for 23 (10 galaxies) and 18 per cent of the sample (8 galaxies, 5 of which have the same $\chi^{2}_{\rm r}$ value produced by M05 models). 

Again, similarly to what found for the HUDF sample, the situation changes when reddening is included. Now BC03 models perform better for 52 per cent  (23 galaxies) of the COSMOS sample. M05 and M13 models give the best fit respectively for 36 (16 galaxies) and 20 per cent of the sample (9 galaxies, 4 of which share the same $\chi^{2}_{\rm r}$ value given by M05 models). Note that for the M05 case, 19 galaxies are best-fitted with a no-reddening solution, even when  reddening is left free to vary. For these galaxies, BC03 models favour  reddened solutions. 

\begin{table*}
%\begin{tiny}
\begin{center}
\caption{Average and median reduced $\chi^{2}$ for COSMOS galaxies, for SED fits: i) using all available filters (full filter set); ii) using only filters matched in effective wavelength and bandpass to those of the HUDF sample (BB, broad-band). Col 1: info regarding inclusion or not of reddening; Col 2: model; Col 3 \& 5: average reduced $\chi^{2}$; Col 4 \& 6: median reduced $\chi^{2}$.}
\begin{tabular}{cccccc}
\hline
  \multicolumn{1}{c}{\bf Reddening law} &
  \multicolumn{1}{c}{\bf Model} &
  \multicolumn{1}{c}{$\mathbf{\langle \chi^{2}_{\rm r} \rangle}$} &
  \multicolumn{1}{c}{\bf Median $\mathbf{\chi^{2}_{\rm r}}$} &
  \multicolumn{1}{c}{$\mathbf{\langle \chi^{2}_{\rm r} \rangle}$} &
  \multicolumn{1}{c}{\bf Median $\mathbf{\chi^{2}_{\rm r}}$} \\
  \multicolumn{1}{c}{} &
  \multicolumn{1}{c}{} &
  \multicolumn{1}{c}{Full filter set} &
  \multicolumn{1}{c}{Full filter set} &
  \multicolumn{1}{c}{BB filter set} &
  \multicolumn{1}{c}{BB filter set} \\
  \multicolumn{1}{c}{(1)}&
  \multicolumn{1}{c}{(2)}&
  \multicolumn{1}{c}{(3)}&
  \multicolumn{1}{c}{(4)}&
  \multicolumn{1}{c}{(5)}&
  \multicolumn{1}{c}{(6)} \\
\hline
\hline
  \multirow{3}{*}{None}  &  M05  &  $1.43$ &  $1.28$ & $1.63$  & $1.24$  \\
                         &  M13  &  $1.79$ &  $1.50$ & $2.38$  & $1.75$  \\
			 &  BC03 &  $1.88$ &  $1.40$ & $2.72$  & $2.18$  \\
\hline
  \multirow{3}{*}{SMC}   &  M05  &  $1.30$ &  $1.17$ & $1.45$  & $1.14$  \\
                         &  M13  &  $1.35$ &  $1.20$ & $1.59$  & $1.31$  \\
			 &  BC03 &  $1.23$ &  $1.10$ & $1.54$  & $1.28$  \\
\hline
\hline
\end{tabular}
\label{tab:Table4}							    
\end{center}
% \end{tiny}
\end{table*}

\begin{figure*}
\centering
\includegraphics[width=0.85\textwidth]{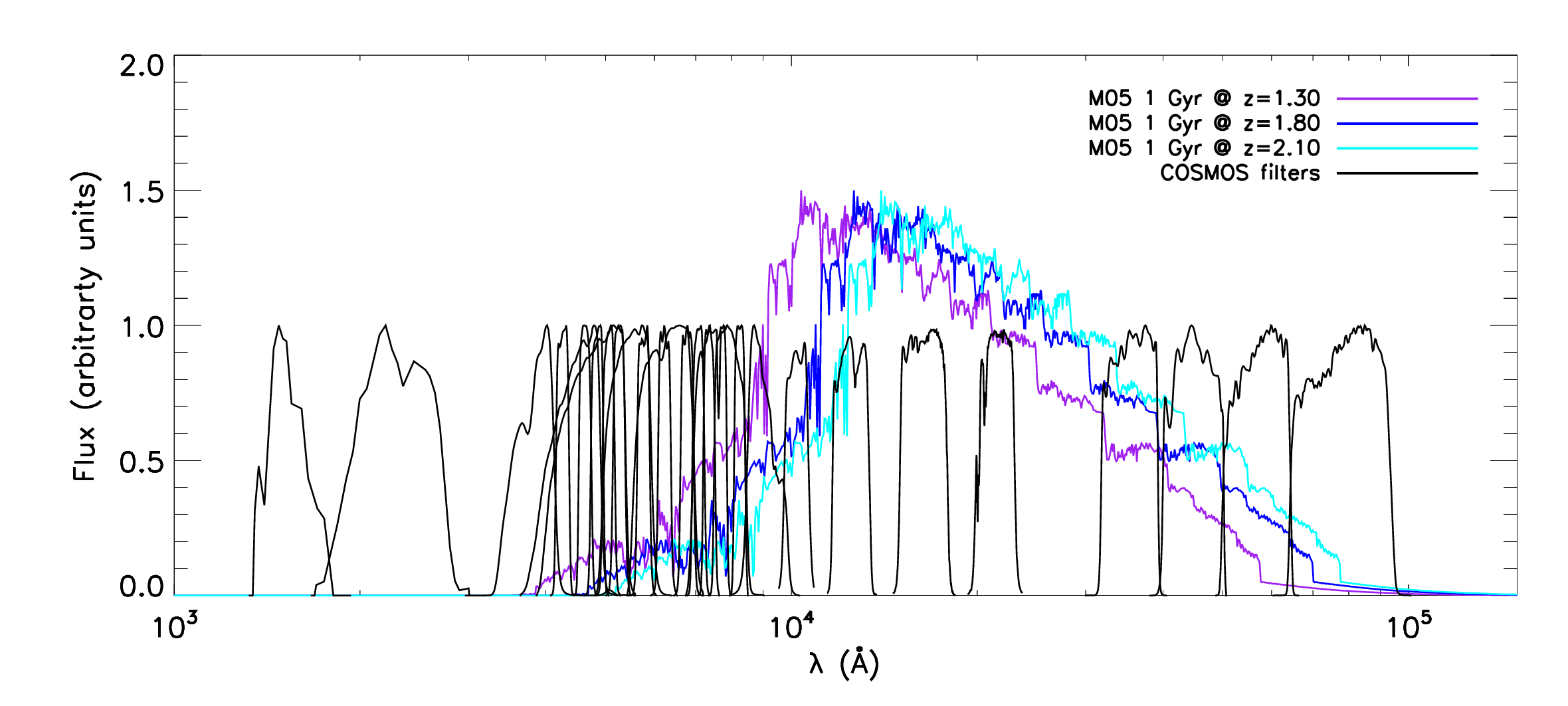}
\caption{Transmission functions (black lines) of filters available in the \citet{Muzzin-2013} sample [UV to IRAC channel 4 (MIPS excluded)]  Coloured lines represent a M05, $1\ {\rm Gyr}$-old, solar SSP model at redshifts 1.3 (purple line), 1.8 (blue line) and 2.1 (cyan line), which mimics the redshift range spanned by the COSMOS galaxies.}
\label{fig:Fig3}
\end{figure*}

 \begin{figure*}
\centering
\includegraphics[width=0.85\textwidth]{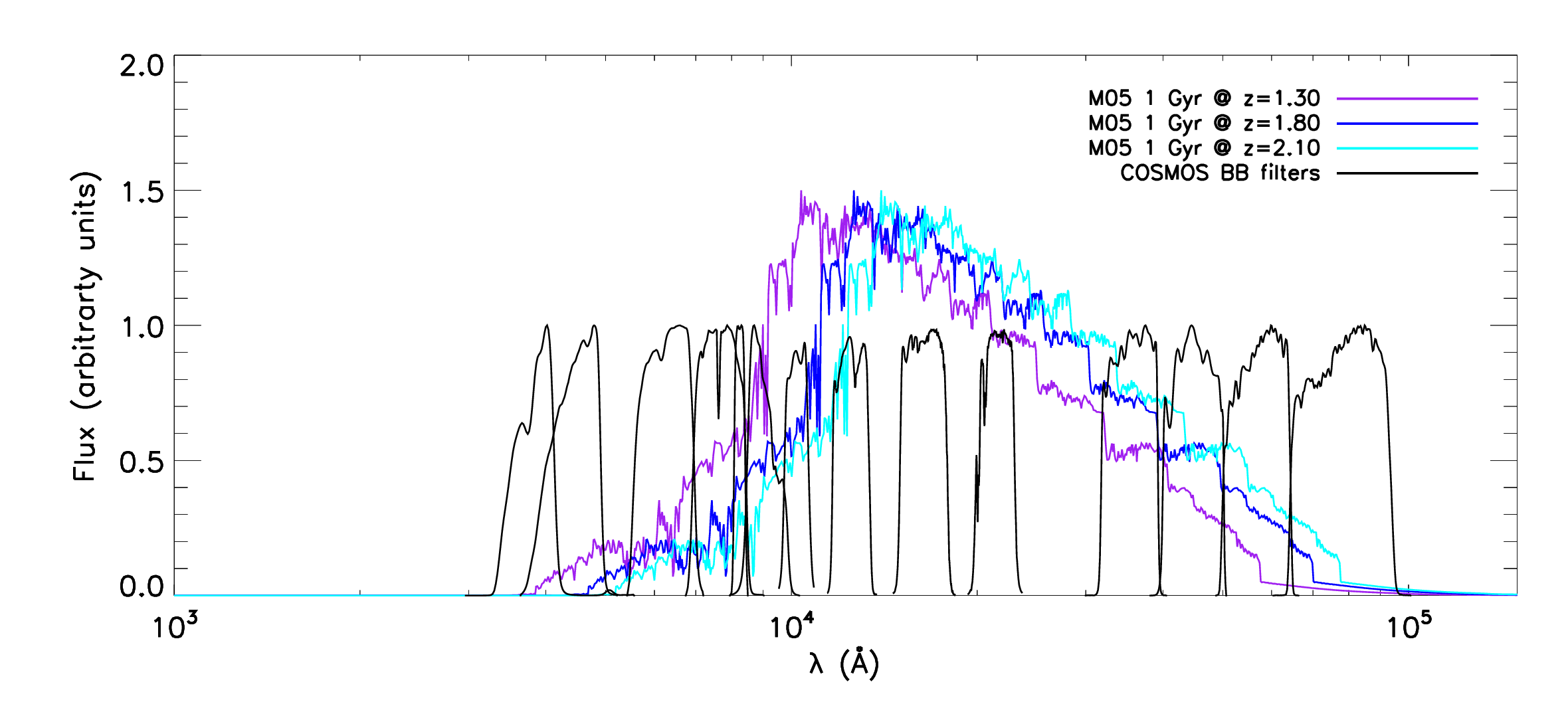}
\includegraphics[width=0.85\textwidth]{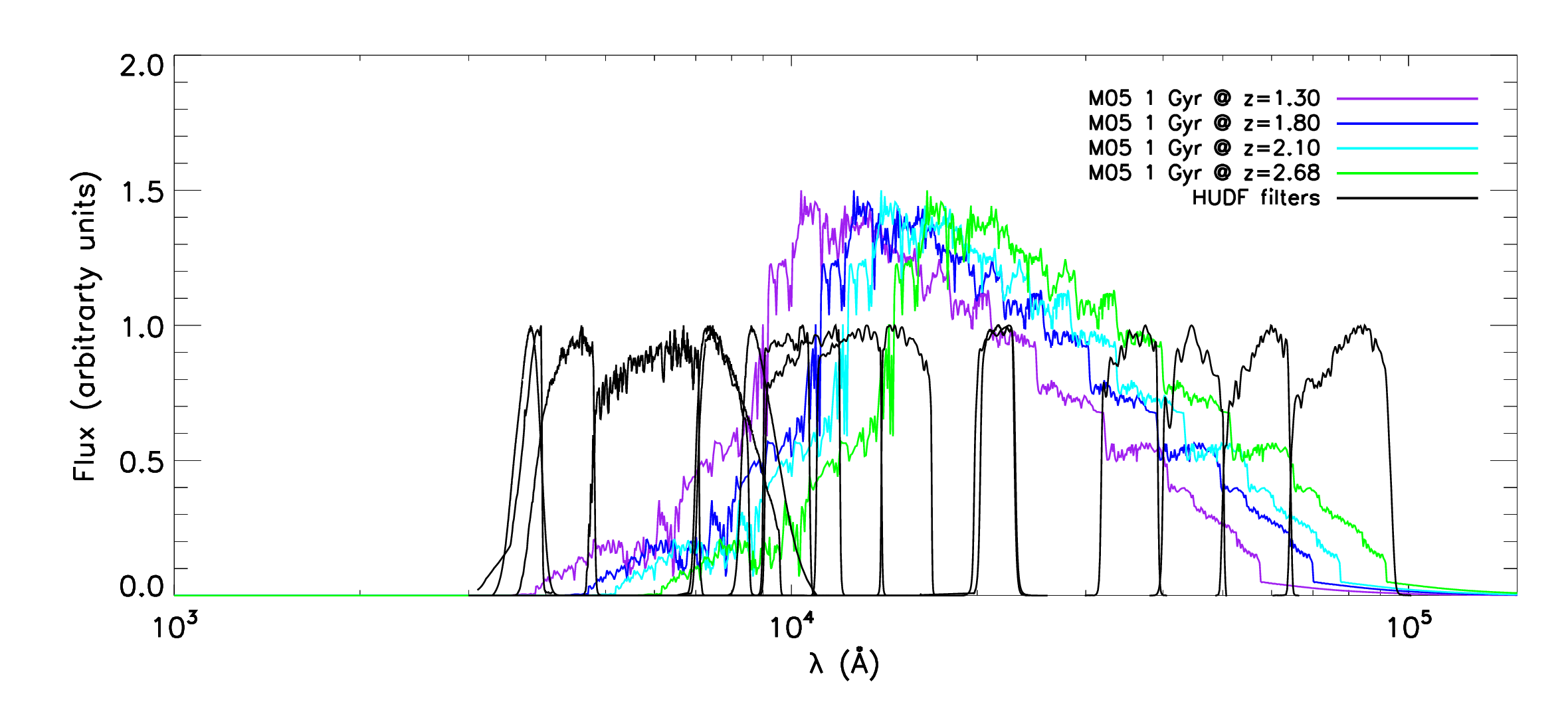}
\caption{Top panel: As in Figure \ref{fig:Fig3}, but now considering only filters matching in effective wavelength and bandpass the HUDF ones. All medium-band filters but one ({\it SuprimeCam IA827}) are discarded.
Bottom panel: As in the top panel, but for filters available in the CANDELS HUDF catalogue. One additional M05 SSP at $z=2.68$ is plotted to encompass the redshift range of the HUDF sample.}
\label{fig:Fig4}
\end{figure*}
\subsection{The effect of different SED sampling}
\label{subsec:subsec4.3}

In order to properly compare the results obtained for the HUDF and the COSMOS sample, we need to minimise differences in the analysis. To this end, we repeat the SED fitting of COSMOS galaxies using a reduced number of filters, which mimics in effective wavelength and bandpass those used for the HUDF sample. This leaves us with 14 filters (CFHT/MegaCam ${\it u^{\rm *}}$, Subaru/SuprimeCam ${\it g^{\rm +},\ r^{\rm +},\ i^{\rm +},\ B_{\rm j},\ IA827}$, UltraVISTA ${\it YJHK_{\rm s}}$ and the IRAC channels at 3.6, 4.5 5.8, 8.0 ${\rm \mu m}$, see Figure \ref{fig:Fig4}), mainly listing broad-band filters. Note that this test is also useful for inspecting the influence of medium band filters\footnote{The only such filter still used for this test is SuprimeCam {\it IA827}} on the SED fitting. 

We then repeat the analysis for both the reddening-free and reddened fitting sets (for this test SMC reddening law only). In the former case, we find that the M05, BC03 and M13 models produce better fits respectively for 86 per cent (38 galaxies), 7 per cent (3 galaxies) and 7 per cent (3 galaxies) of the sample. When reddening is allowed, the figures become  55 (24 galaxies), 25 (11 galaxies) and 20 (9 galaxies) per cent of the sample, respectively.

Differently from what obtained with the fully-sampled COSMOS SEDs, M05 models are now preferred in both cases. We note that when including reddening in the SED fitting process, M05, BC03 and M13 models chose a best fitting solution with $E(B-V)=0$ respectively for 29, 1 and 14 galaxies. Maraston's models (M05 in particular) more rarely fit with additional reddening. In general, we find that for all models, the quality of the fits is slightly  degraded, but still associated with acceptable reduced $\chi^{2}$ values (see Table \ref{tab:Table4}). 
In summary, when reddening is set to zero, M05 models do best independently of the adopted filter set. With reddening, BC03 models do best  with the full filter set and M05 models with the broad-band filter set. 

It is important to realize that using combinations of broad- and intermediate-band filters has advantages but also disadvantages. More data points should lead to a better definition of the observed SEDs, hence possibly to statistically better fits (i.e. lower $\chi^{2}_{\rm r}$). However, a variable sampling of a galaxy SED can lead to the data in the more finely sampled region of the spectrum being weighted more than the remaining ones, especially if the former have smaller uncertainties than the latter. In our case, with intermediate-band filters being confined to the optical, this may lead to underweight the near-IR part of the spectrum, which is the only one sensitive to the TP-AGB treatment.

In the specific case, when only broad-band filters  are used (see Figure \ref{fig:Fig4}), the rest-frame optical to near-IR and optical to UV are homogeneously sampled, respectively with $\sim 8$ and 6 filters, similarly to what done with HUDF SEDs. When medium-band filters (namely 13, all at $\lambda<10^4\ {\rm \AA}$) are added, the rest-frame optical to UV sampling is increased by a factor $\sim3$ while that of the rest-frame optical to near-IR remains unchanged (see Figures \ref{fig:Fig3}  and \ref{fig:Fig4}). This makes the rest-frame optical to UV sampled $\sim2.5$ times more than the rest-frame optical to near-IR, possibly leading to fits mainly driven by the rest-frame optical to UV part of the SED, hence less effective at assessing the actual contribution of TP-AGB stars.

\subsection{Effect of photometric zero-point offsets}
\label{subsec:subsec4.4}
Photometric redshifts are sensitive to errors in photometric zero points (ZP) which may well affect datasets, especially when combining data from different telescopes and instruments.
Therefore, in order to improve the accuracy of photometric redshifts, it is common practice to derive ZP offsets using galaxies with well determined spectroscopic redshift as calibrators
(e.g.,  \citealp{Ilbert-2006, Ilbert-2009, Ilbert-2013, Brammer-2011, Whitaker-2011}). As in \citet{Whitaker-2011},  \citet{Muzzin-2013} evaluated ZP offsets  by comparing photometric to spectroscopic redshifts for a sample of (mainly star-forming) galaxies from zCOSMOS \citep{Lilly-2007}, plus 19  (mainly passive) massive galaxies with $z_{\rm spec}>1.4$ \citep{van-de-Sande-2011, van-de-Sande-2013, Onodera-2012, Bezanson-2013}. Template spectra were fit to the observed SEDs using the EAZY code \citep{Brammer-2008} and fixing the redshift to the spectroscopic value. For each galaxy the residuals in each observed band between the best fit template spectrum and the observed SED were calculated and the averages over the whole sample of galaxies were derived. Such average residuals were then applied as offsets  to the photometric ZPs. 
 
As  pointed out by Muzzin et al. in the description of their online catalogue\footnote{See the Section on zero-point offsets in the README file describing \citet{Muzzin-2013} catalogue, available at http://www.strw.leidenuniv.nl/galaxyevolution/ULTRAVISTA/
README\_UVISTA\_v4.1.txt}, the derived offsets depend on the adopted spectroscopic sample and on the set of template spectra from stellar population synthesis models used in deriving the photometric redshifts.

In particular, \citet{Muzzin-2013} used as templates nine synthetic models: i) six linear combinations of PEGASE models \citep{Fioc-1999}; ii) a ``red'' template from M05 models (whose age and metallicity were not specified); iii) a $1\ {\rm Gyr}$-old BC03 SSP; iv) a slightly dust reddened model. Hence 7 of the 9 templates are either from PEGASE or BC03 models, which adopt a similar stellar evolution database with  a low TP-AGB contribution (see Figure 18 in M05).  For galaxies at high redshift the $1\ {\rm Gyr}$ template plays a critical role as this is the typical age of the stellar populations of these galaxies (e.g., \citealp{Cimatti-2008,Onodera-2015}) and is the age at which  BC03 and M05 models differ the most (see again Figure 18 in M05). 

\begin{figure}
\centering
\includegraphics[width=0.41\textwidth]{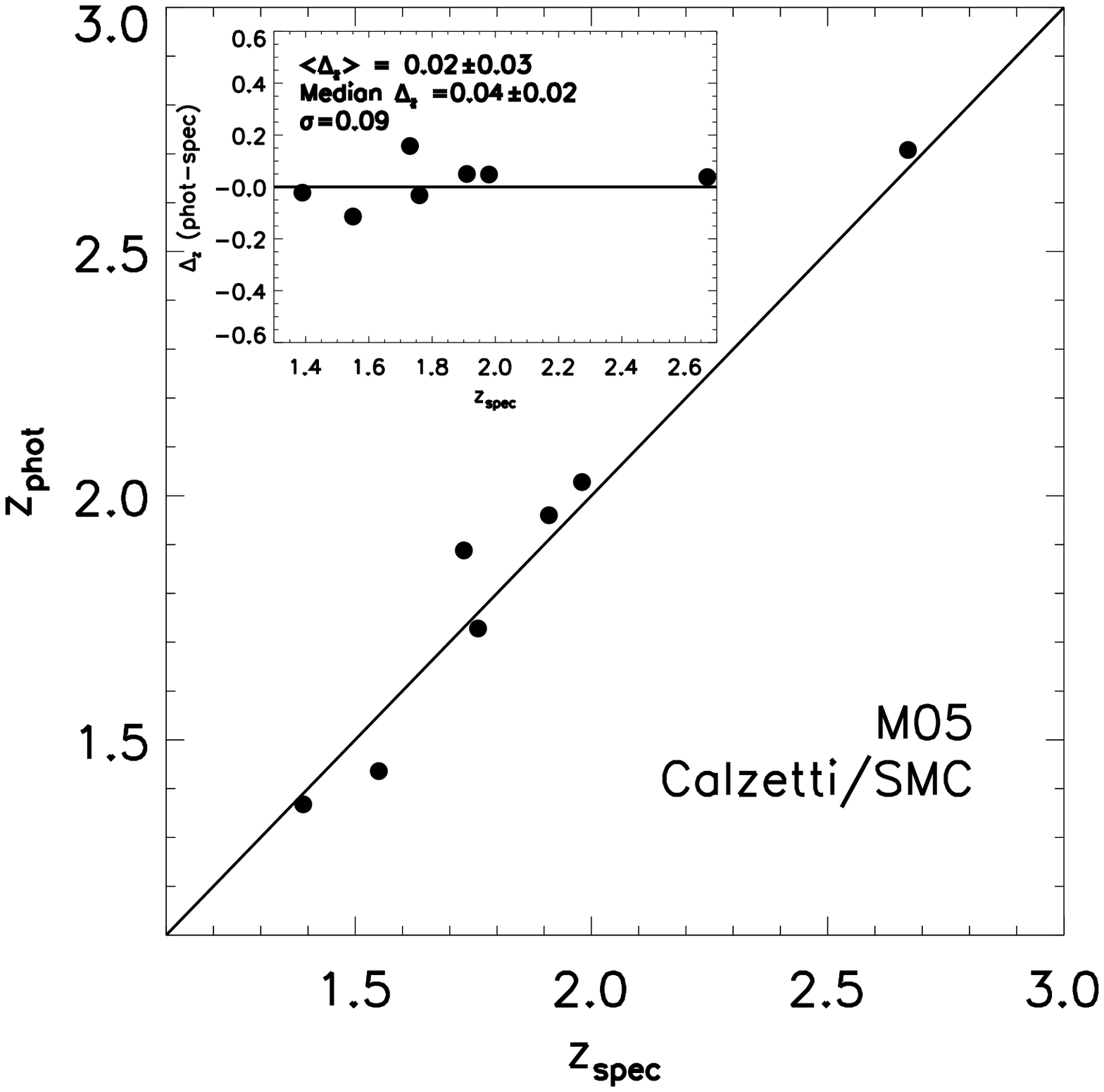}
\includegraphics[width=0.41\textwidth]{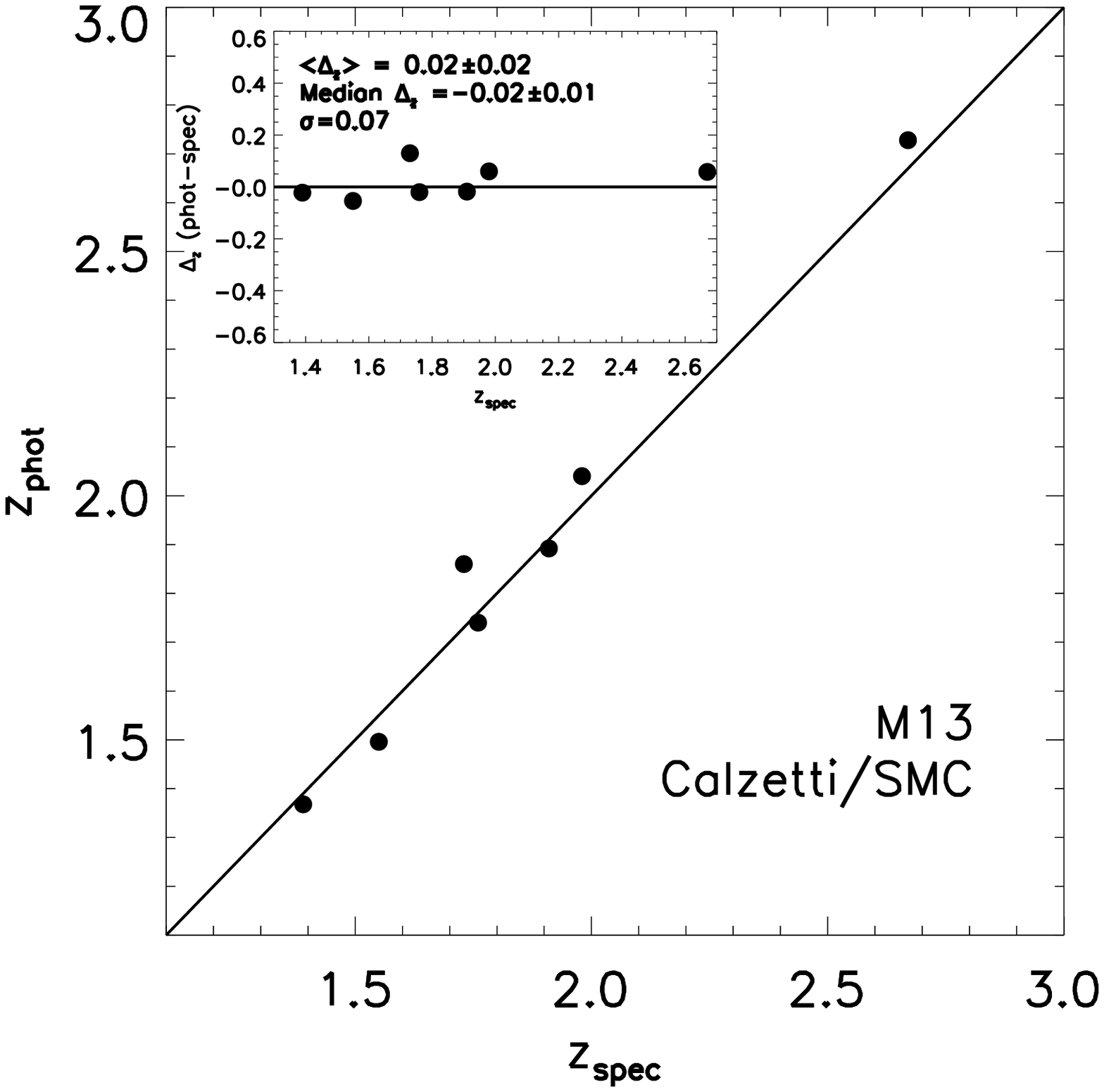}
\includegraphics[width=0.41\textwidth]{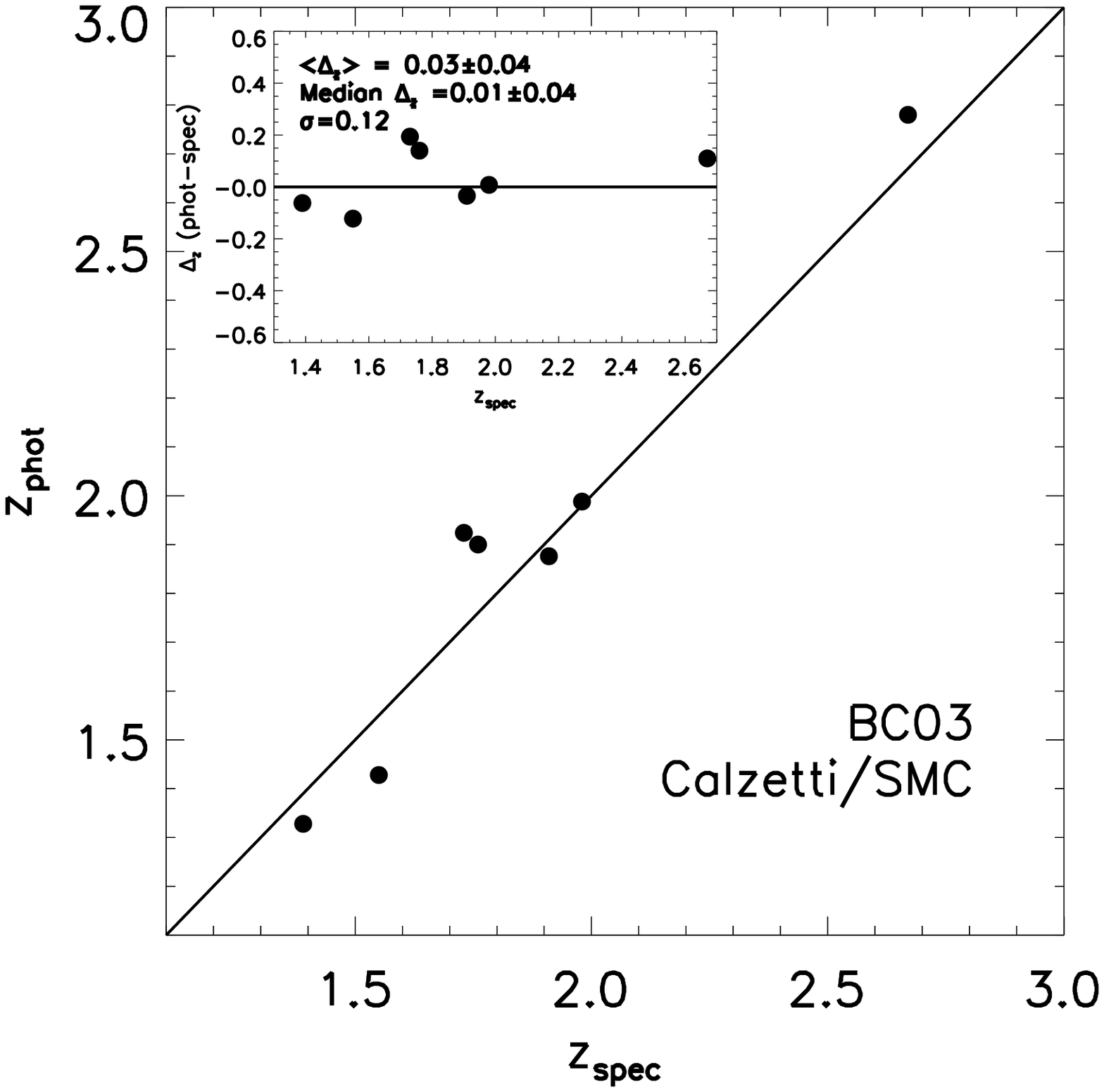}
\caption{Comparison between spectroscopic and photometric redshifts obtained with M05, M13 and and BC03 (upper, middle and bottom panel). At the top-left corners, redshift residuals ($z_{\rm phot}-z_{\rm spec}$) are plotted as a function of spectroscopic redshift.  The residual average, median and scatter are also noted.}
\label{fig:Fig5}
\end{figure}

In order to investigate the effects of photometric recalibration on our analysis, we first  assess the impact of ZP  adjustments on photometric redshifts, that we derive with HYPERZ using the same template sets described in Section \ref{sec:sec3}. Based on the results of Table \ref{tab:Table3}, we only use the Calzetti and the ``SMC'' reddening laws. 
Figure~\ref{fig:Fig5} shows the  results for the HUDF galaxies, on whose photometry no offsets were applied.
Judging from the residuals, M05 and M13 photometric redshifts are in slightly better agreement with spectroscopic redshifts compared to BC03 ones, with M13 giving the best performance. The residuals' dispersions for M05, M13 and BC03 models are indeed 0.09, 0.07 and 0.12, respectively. 
 For all models, the average and median  redshift residuals are always consistent with 0 at 1-2$\sigma$ level (see Figure \ref{fig:Fig5}). This indicates that the CANDELS photometry is free of systematics, consistently with the results of the tests by \citet{Guo-2013} aimed at assessing the need of photometric re-calibration. 

\begin{figure*}
\centering
\includegraphics[width=0.4\textwidth]{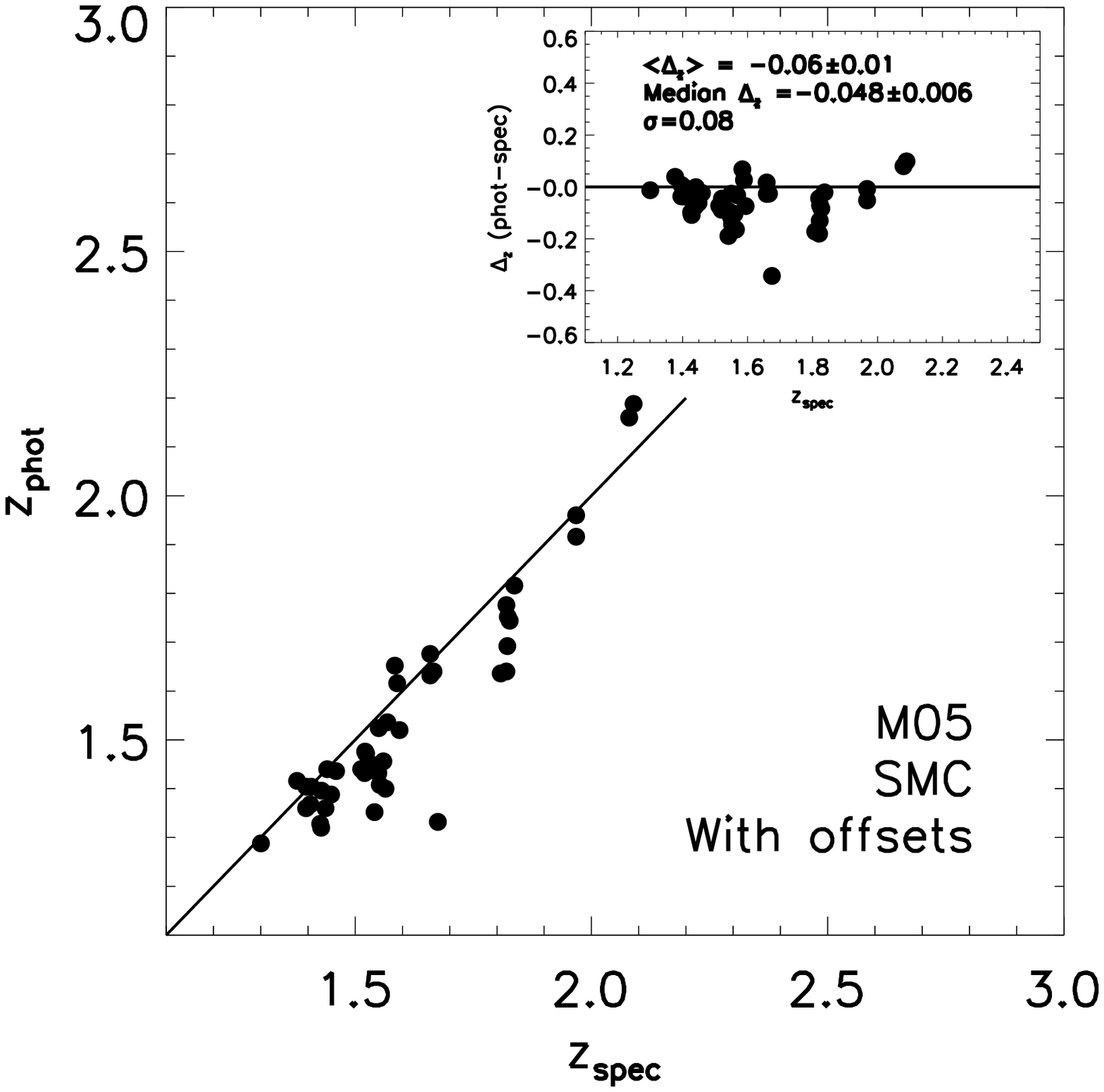}
\includegraphics[width=0.4\textwidth]{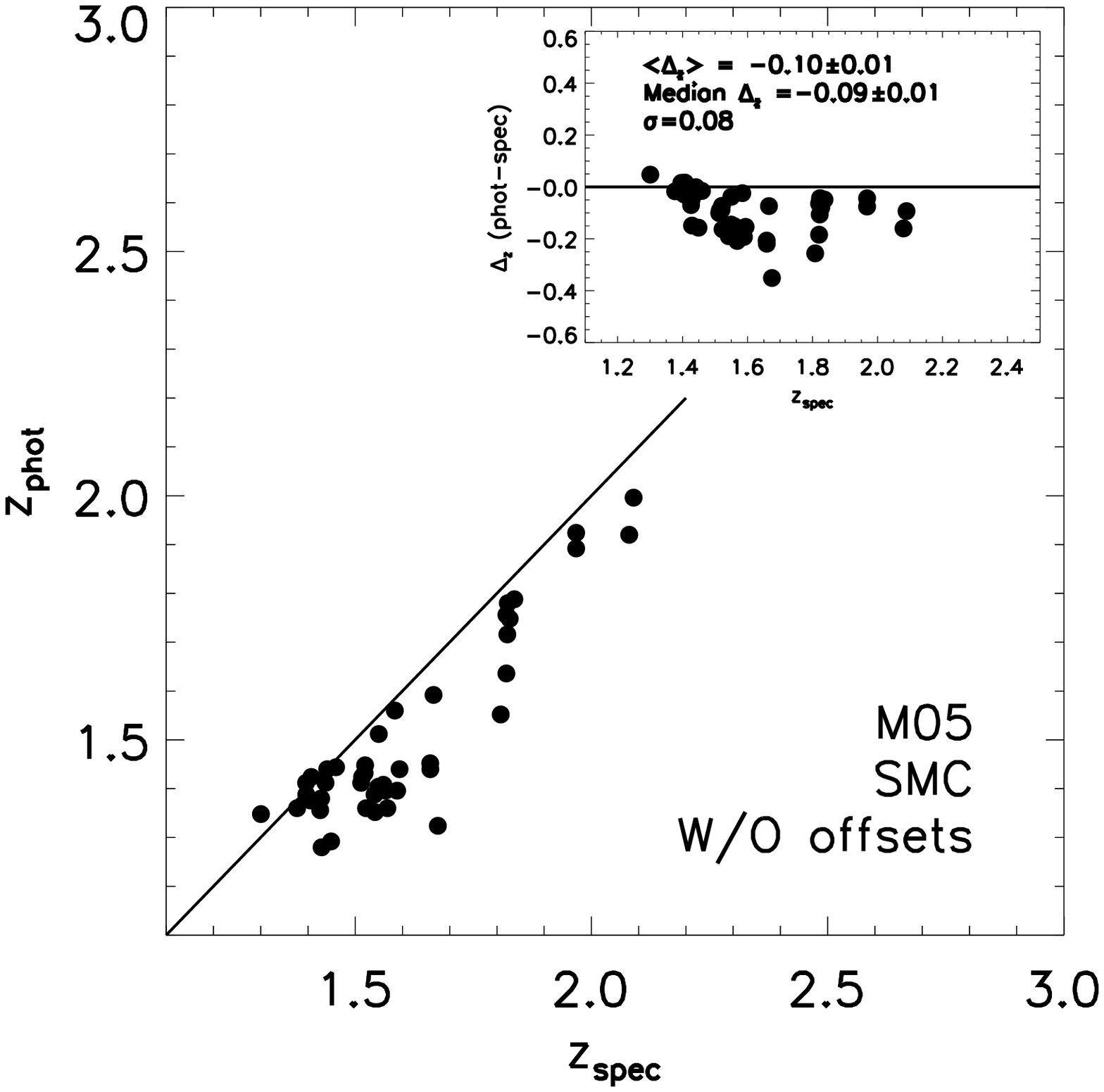}
\includegraphics[width=0.4\textwidth]{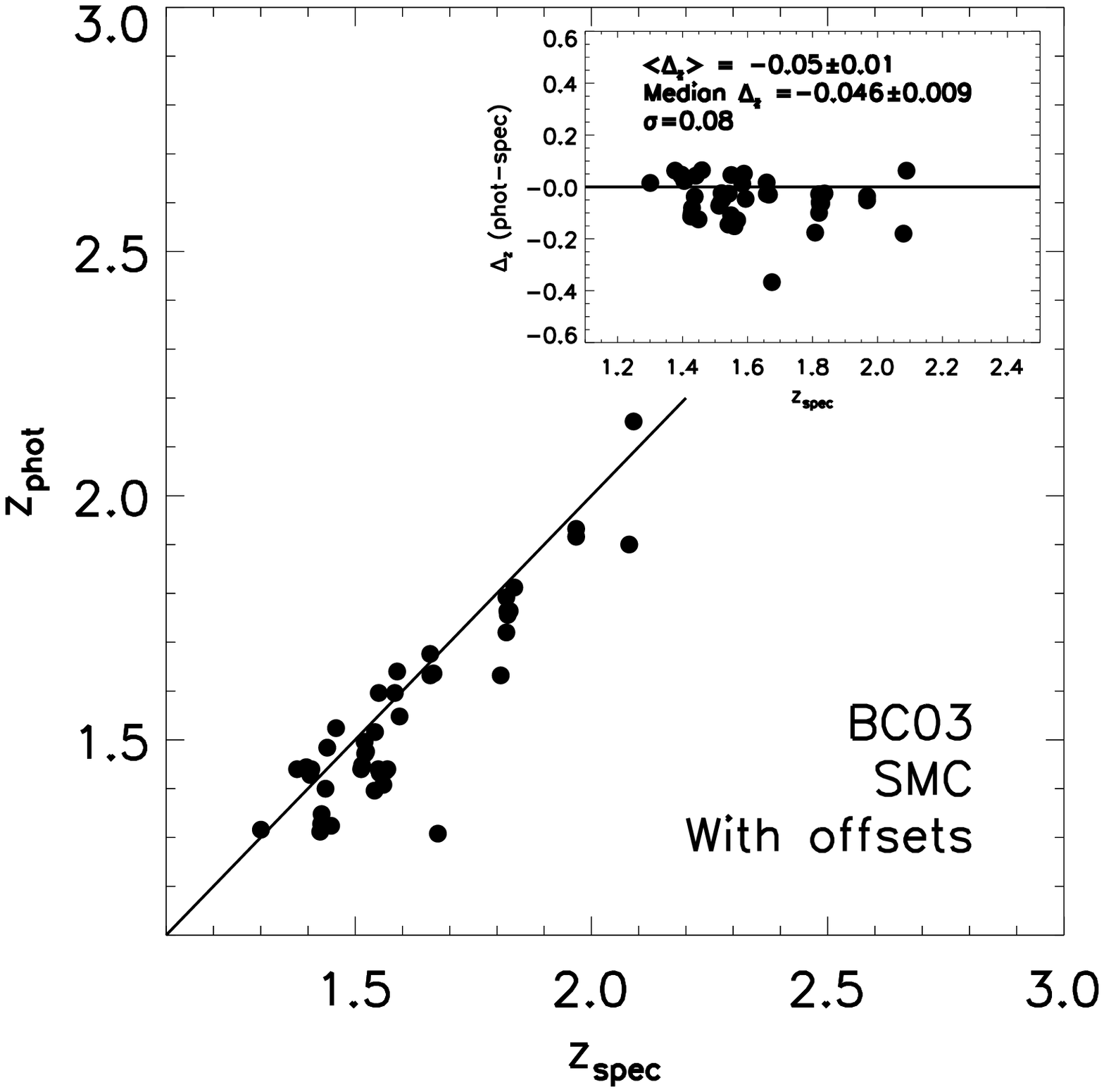}
\includegraphics[width=0.4\textwidth]{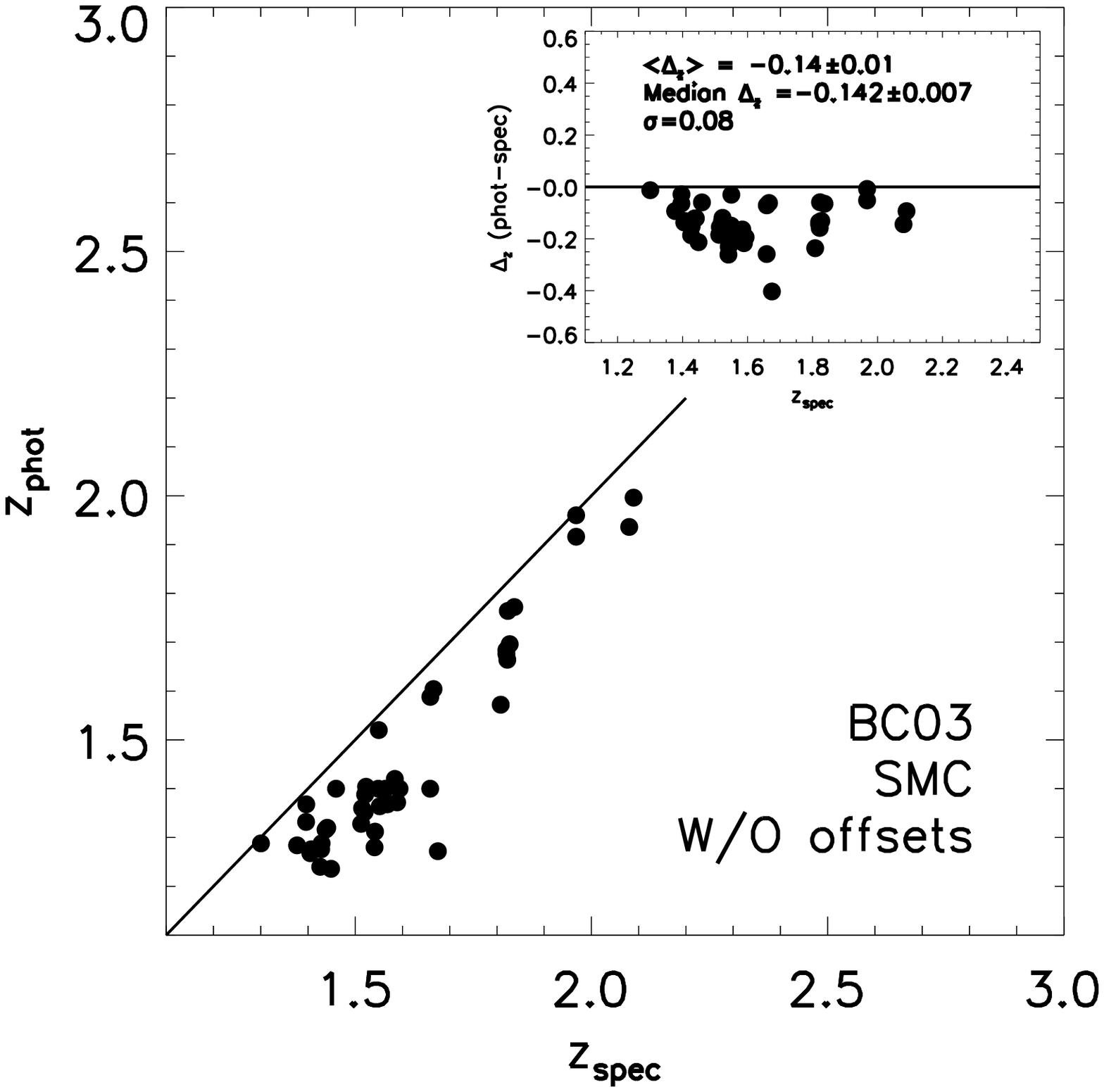}
\includegraphics[width=0.4\textwidth]{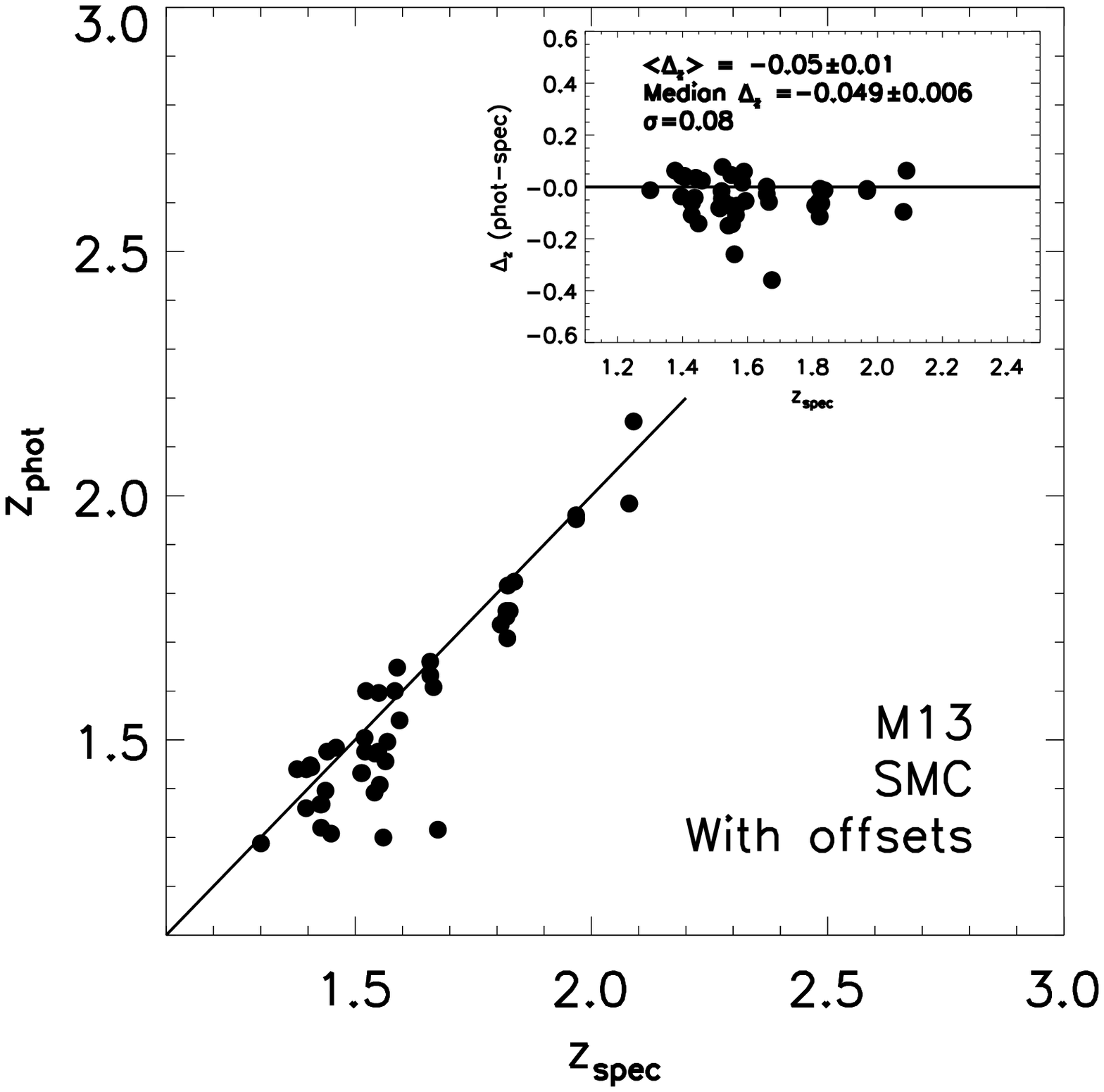}
\includegraphics[width=0.4\textwidth]{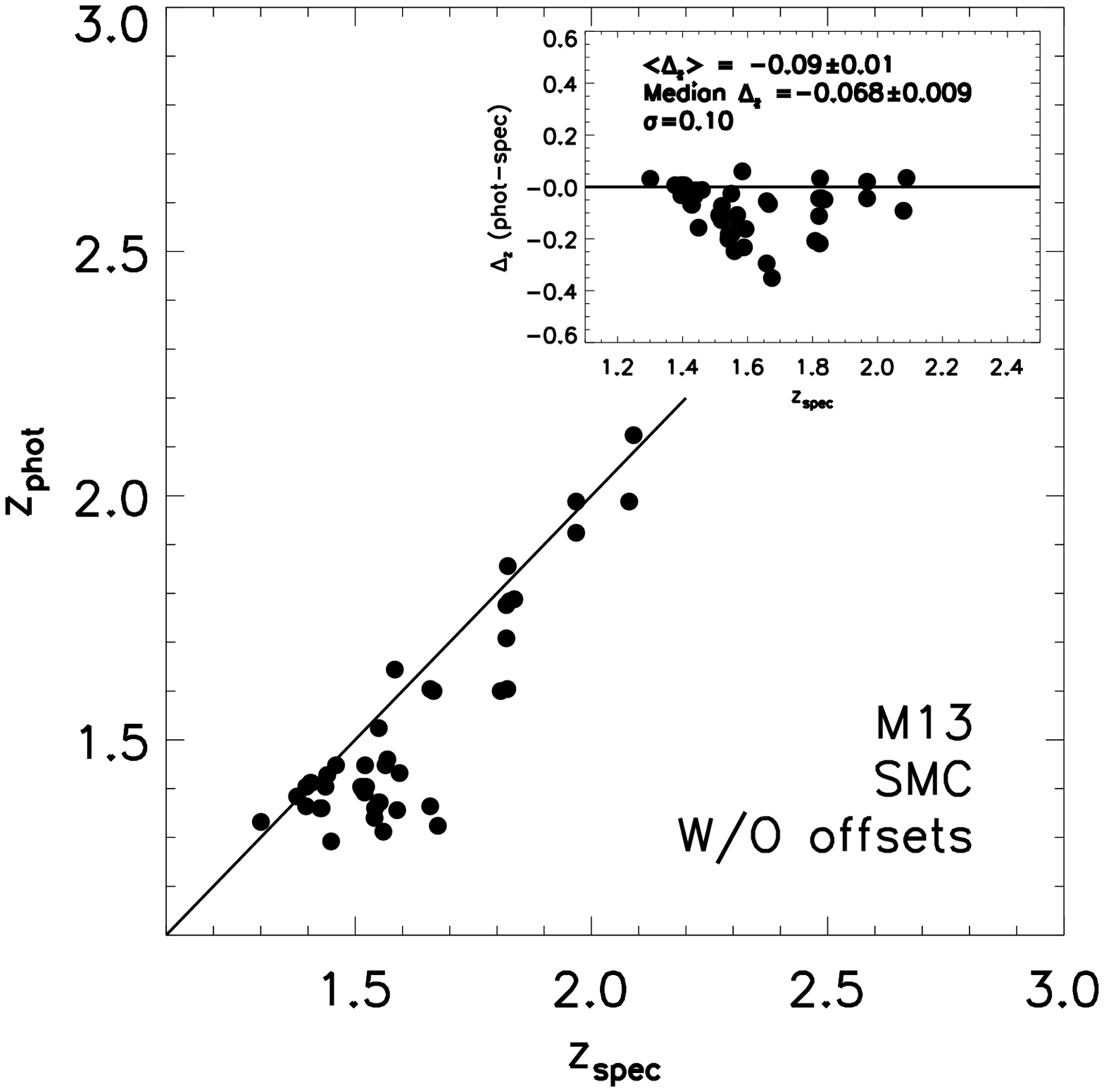}
\caption{Comparison between spectroscopic redshifts and our photometric redshifts derived with the three models (M05, BC03 and M13, from top to bottom), for the COSMOS sample. Left-hand and right-hand panels show results obtained when photometric zero-point offsets are included and excluded, respectively. At the top-right corners, redshift residuals ($z_{\rm phot}-z_{\rm spec}$) are plotted as a function of spectroscopic redshift.  Average, median and scatter of these residuals are also noted in these plots. These panels refer to the reddened case with an SMC reddening law, which is generally the most successful one for the COSMOS galaxies.}
\label{fig:Fig6}
\end{figure*}

We now repeat the same analysis  for the COSMOS galaxies, using both the original and the re-calibrated photometry. In addition, we compare the quality of the COSMOS photo-$z$'s to that of the HUDF's ones. In order to use a similar procedure, we use only the broad-band  filter set identified in the previous section, and only the SMC reddening law (based on results shown in Table \ref{tab:Table2_appB}).
The results are displayed in Figure \ref{fig:Fig6}, which shows the presence of a systematic shift when using the original photometry (right-hand plots in Figure \ref{fig:Fig6}), with  the photo-$z$'s underestimating the spectroscopic ones. This is the case for all the models (the median and mean redshift residuals are always inconsistent with 0 at $>3\sigma$ level), hence suggesting the need for a re-calibration of the photometry (as correctly pointed out by Muzzin et al.). However, although being reduced, the systematic shift is still present even after using the re-calibrated photometry (left-hand plots in Figure \ref{fig:Fig6}): the average offset between photometric and spectroscopic redshifts is reduced for all the models, but in all cases the median and mean redshift residuals remain inconsistent with 0 at $>3\sigma$ level, with  redshifts being  still systematically underestimated. The application of the ZP offsets, while reducing the systematics for all the considered models, does so significantly better for BC03 models, with the median and average residuals decreasing by a factor of $\sim 3$, compared to factors between $\sim1.4$ and $\sim 1.9$ for the other two sets of models. 

We now test to which extent the use of ZP photometric offsets from Muzzin et al. (2013) affects the results of model SED fits, hence we re-fit the COSMOS galaxies using their published, off-setted photometry, in the same way as done for the original one. For brevity we consider only the reddening-free case, also in order to avoid age-reddening degeneracy removing or blurring  any systematic effect. We find that BC03 models are now preferred for 64 per cent (28 galaxies) of the sample, against 20 (9 galaxies) and 16 (7 galaxies) per cent, for M05 and M13 models, respectively. This result is almost the opposite of what one gets without introducing the ZP offsets (cf. Section \ref{subsec:subsec4.2}).  Thus, using the original COSMOS photometry M05 models perform best whereas BC03 models are preferred when using the photometry with ZP adjustments.

One issue worth discussing a little further is whether the use of a specific set of templates may have biased this result.
From their Table 3, we see that \citet{Muzzin-2013} have {\it a priori} fixed to 0.0 the offsets for the four IRAC bands and to $-0.08$ mag the offset for the $K$ band (used as anchor for the ZP tuning), then calculated all other offsets from the average residuals as described above. The result is a negative offset of 0.1-0.2 mag for 21 out of the 22 other bands (from $u$ to $H$). In practice, given the redshift range of our galaxies (from $\sim 1.3$ to $\sim 2$) the effect of these ZP adjustments has been to increase the fluxes in all optical bands (say, shortward of $\sim 0.6-0.8\;\mu$m, depending on the redshift) relative to the near-{\it IR}, or, equivalently, to decrease the near-{\it IR} fluxes relative to the optical. The applied offsets thus mimic a reduced contribution (to the SED) of cool stars, such as TP-AGB stars. Therefore, in essence the ZP offsets introduced by \citet{Muzzin-2013} make these galaxies bluer in rest-frame optical--near-IR colours, which favours TP-AGB-light models, indeed the same kind of models used to derive the offsets themselves. Note that Maraston's and Bruzual \& Charlot's models differ more strongly for passive galaxies. Hence, one would not expect the ZP offsets derived by Muzzin et al. to show strong model dependence in the first place, as the great majority of the spectroscopic galaxies used by them for recalibration are star-forming (SF) galaxies. However, one should also consider that these models show significant differences also in the reproduction of rest-frame {\it UV}-{\it IR} colours of SF galaxies.

The application of the ZP corrections improves the photometric redshifts by 
constructions, but this does not automatically imply that it also gives a 
better match to the near-IR.  
Of course, the original ZPs are not perfect 
and   systematic differences with respect to  the ``true" ZPs are likely. 
However,  the original ZPs were derived without any use of synthetic models, hence may favour one set of 
models over another {\it by mere chance}. On the contrary, the Muzzin et al. ZPs my favour models like those which have been used to derive the ZP offsets {\it by construction}. 
We conclude that  the better performance of BC03 models when using the ZP offsets may be due to this built-in bias, whereas the original photometry is exempt from this sort of shortcoming.
For this reason, aiming at a broader exploration of the performance of the various models, we make use of the original photometry without applying ZP offsets.

\begin{table*}
%\begin{tiny}
\begin{center}
\caption{SED-fitting consistency for the total (HUDF+COSMOS) sample. Col. 1: model; col 2: inclusion of reddening; col 3: number of best fitted galaxies; col 4: number of galaxies for which the models gives a solution within $1\sigma$ from the best-fit; col 5: number of galaxies with the highest marginalised probability. Sample fractions are reported in brackets; col 6: use of $L_{\rm IR}$ limits to exclude physically unreliable solutions (for the meaning of this column see Section \ref{subsec:subsec5.2}).}
\begin{threeparttable}
\begin{tabular}{cccccc}
\hline
  \multicolumn{1}{c}{\bf Model} &
  \multicolumn{1}{c}{\bf Reddening} &  
  \multicolumn{1}{c}{$\mathbf{N_{\rm best\ fit}}$} &
  \multicolumn{1}{c}{$\mathbf{N_{\rm consistent}}$} &
  \multicolumn{1}{c}{$\mathbf{N_{\rm best\ fit}}$} &
  \multicolumn{1}{c}{\bf $\mathbf{L_{\rm IR}}$ cut}\\  
  \multicolumn{1}{c}{} &
  \multicolumn{1}{c}{} &
  \multicolumn{1}{c}{} &
  \multicolumn{1}{c}{(within $1\sigma$)} &
  \multicolumn{1}{c}{Marg. Prob.} \\
  \multicolumn{1}{c}{} &
  \multicolumn{1}{c}{} &
  \multicolumn{1}{c}{} &
  \multicolumn{1}{c}{} &
   \multicolumn{1}{c}{} & 
  \multicolumn{1}{c}{} \\
  \multicolumn{1}{c}{(1)}&
  \multicolumn{1}{c}{(2)}&
  \multicolumn{1}{c}{(3)}&
  \multicolumn{1}{c}{(4)}&
  \multicolumn{1}{c}{(5)} & 
  \multicolumn{1}{c}{(6)} \\

\hline
\hline
M05 & \multirow{3}{*}{No}	&  $37\ (72\%)$ 		   &  $43\ (84\%)$   &  $28\ (55\%)$ & \multirow{3}{*}{--}   	\\ 
M13     	&	       &  $10\ (20\%)$\tnote{1}	   &  $20\ (39\%)$   &  $16\ (31\%)$	\\ 
BC03     	&	       & 	$10\ (20 \%)$		   &  $17\ (33\%)$   &  $7\ (14\%)$	\\ 
\hline			 \hline
M05 & \multirow{3}{*}{Yes}	&  $20\ (39\%)$ 		   &  $41\ (80\%)$   &  $16\ (31\%)$	 & \multirow{3}{*}{No} \\ 
M13     &		       &  $12\ (23\%)$\tnote{2}	   &  $43\ (84\%)$   &  $11\ (22\%)$	\\  
BC03    & 		       & 	$24\ (47\%)$ 		   &  $49\ (96\%)$   &  $24\ (47\%)$	\\  
\hline			 \hline
M05 & \multirow{3}{*}{Yes}	&  $28\ (55\%)$ 		   &  $39\ (76\%)$   &  $--$	& \multirow{3}{*}{Yes}         \\ 
M13     		       & 			&  $12\ (24\%)$\tnote{3}	   &  $32\ (63\%)$   &  $--$	        \\ 
BC03     		       & 			&  $15\ (29\%)$ 		   &  $28\ (55\%)$   &  $--$	        \\ 
\hline
\hline
\end{tabular}
\begin{tablenotes}\footnotesize 
\item[1] Six galaxies have the same $\chi^{2}_{\rm r}$ values as with M05 models, see Tables \ref{tab:Table2} \& \ref{tab:Table1_appB}.
\item[2] Five galaxies have the same $\chi^{2}_{\rm r}$ values as with M05 models, see Tables \ref{tab:Table3} \& \ref{tab:Table2_appB}.
\item[3] Four galaxies have the same $\chi^{2}_{\rm r}$ values as with M05 models, see Tables \ref{tab:Table3}, \ref{tab:Table2_appB} \& \ref{tab:Table3_appB}.
\end{tablenotes}
\end{threeparttable}
\label{tab:Table5}							    
\end{center}
% \end{tiny}
\end{table*}

\begin{figure*}
\centering
\includegraphics[width=0.48\textwidth]{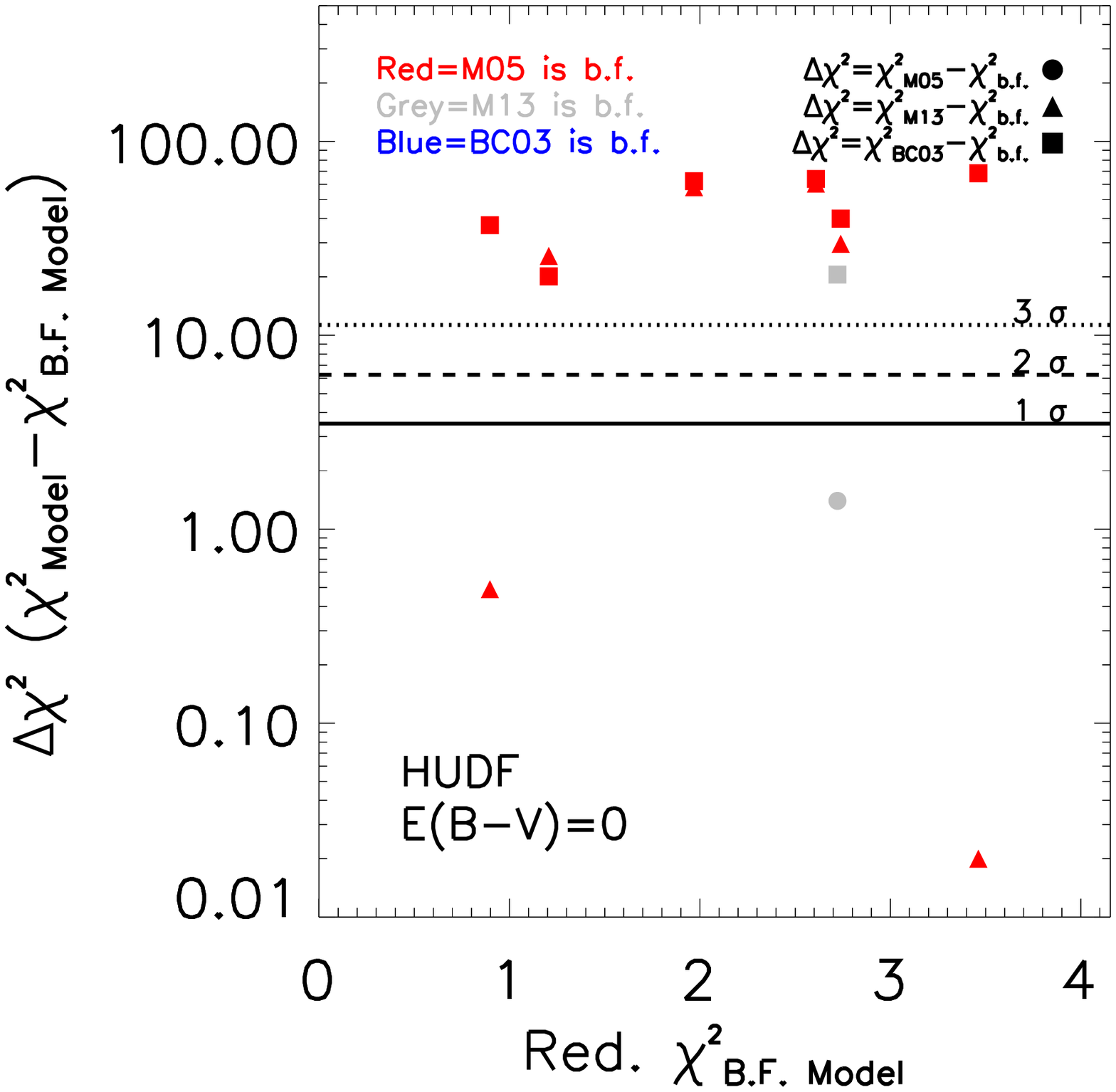}
\includegraphics[width=0.48\textwidth]{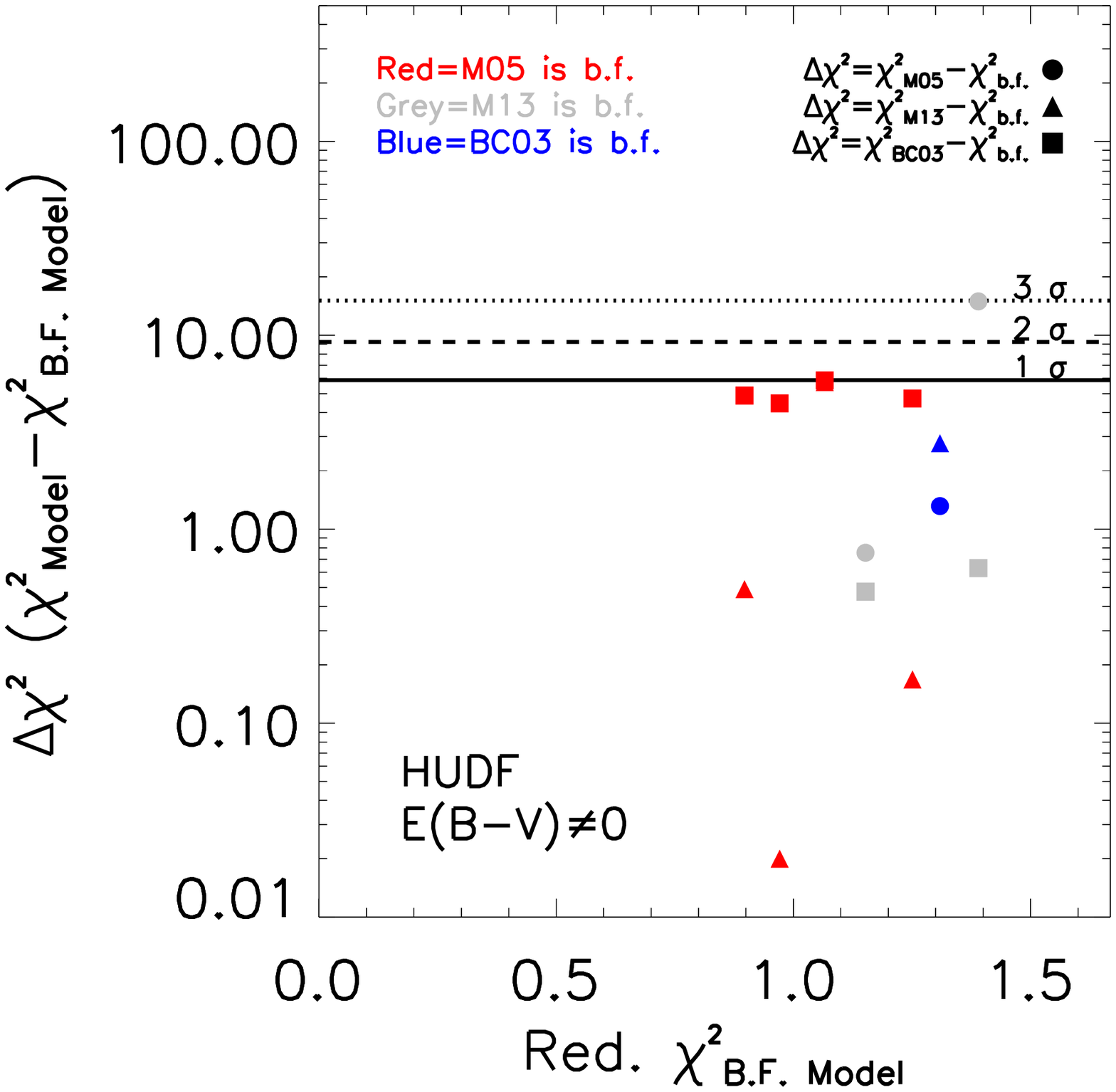}
\includegraphics[width=0.48\textwidth]{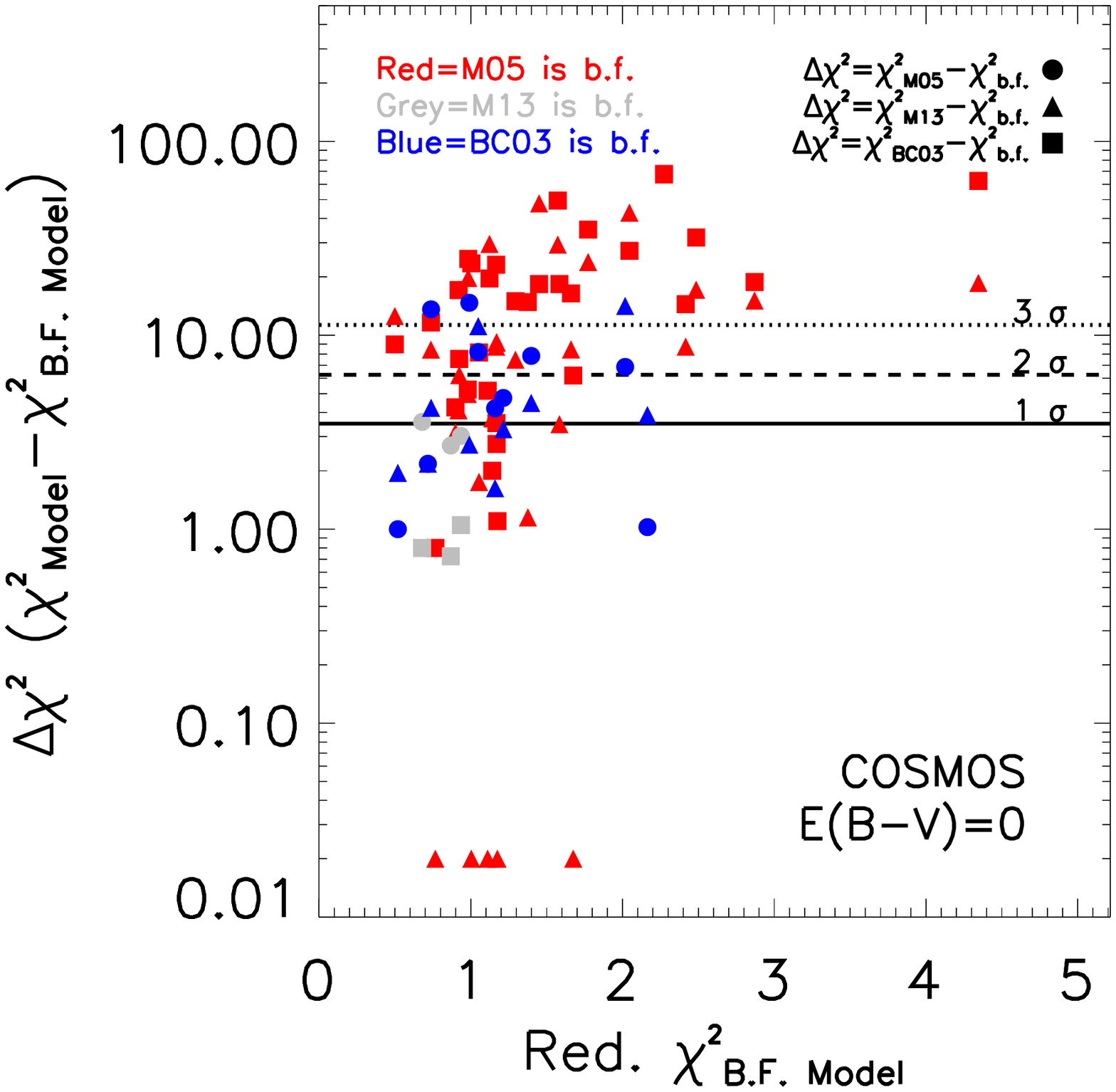}
\includegraphics[width=0.48\textwidth]{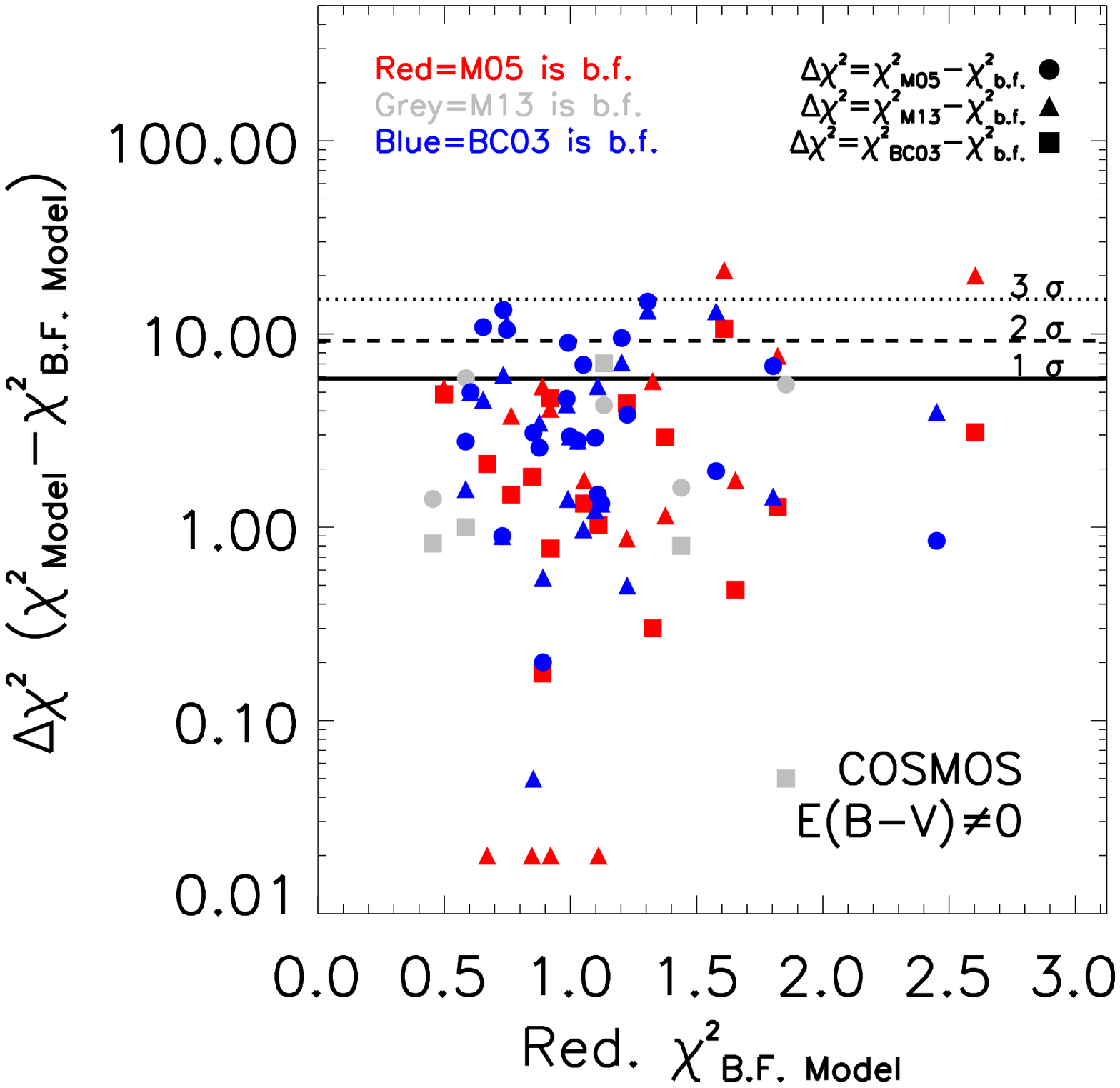}
\caption{Significance of solutions around the best-fit. For each galaxy, the lowest $\chi^{2}_{\rm r}$ value among those obtained using M05, M13 and BC03 models is plotted on the x-axis. The $\Delta \chi^{2}$ between the best fit and the other solutions are plotted on the y-axis. Points are colour coded according to the model giving the lowest $\chi^{2}_{\rm r}$ value (red, grey and blue for M05, M13 and BC03, respectively) and different symbols are used to represent the models to which the differences refer to (dots for M05, triangles for M13, squares for BC03). Horizontal lines mark the 1, 2 and 3 $\sigma$~confidence regions. Upper panels refer to the HUDF sample, lower ones to the COSMOS one. Left-hand and right-hand panels show the extinction-free and reddened cases. $\Delta \chi^{2}=0$ are plotted as  $\Delta \chi^{2}=0.02$ for visibility reasons.}
\label{fig:Fig7}
\end{figure*}

\subsection{Distinguishing among models}
\label{subsec:subsec4.5}

Based on a simple $\chi^{2}_{r}$ ranking, the results obtained so far show that M05 models are favoured in absence of reddening, while all models perform comparably when reddening is included. 

In order to statistically quantify whether a model performs significantly better than others, we study the $\Delta \chi^{2}$ among the best-fitting solutions given by the three studied models and evaluate their consistencies. For each galaxy, we identify the model (M05, M13 or BC03) giving the fit with the lowest $\chi^{2}$ and then calculate the differences between this value and the ones given by the remaining models. Following \citet{Avni-1976}, we then compare the $\Delta \chi^{2}$ with confidence intervals identified using the appropriate $\chi^{2}$ probability distributions (i.e., taking into account the number of parameters involved). 

In addition, for each galaxy we also calculate the probability of each set of models to give the best fit by marginalizing over all the parameters involved  in the fitting process (SFH, age, metallicity, $A_{\rm v}$ and attenuation law). While the analysis of the $\Delta \chi^{2}$ allows us to assess the consistency among the best-fit solutions given by the different models, the study of the marginalized probabilities for each set of models is more suited for an overall evaluation of the performance of the models by taking degeneracies into account.
Results are illustrated  in Figure \ref{fig:Fig7} and the corresponding numbers  are given in Table \ref{tab:Table5}. Note that the last row in this table refers to the figures obtained after removing too dusty unphysical solutions, as we shall explain in Section \ref{sec:sec5}.  

Both tests give a consistent picture, i.e. in absence of dust attenuation M05 models perform better for the majority of the sample galaxies and give solutions consistent within $1\sigma$ of the best fit for the majority ($84$ per cent) of the entire HUDF+COSMOS sample. The same figures for the remaining models are comprised between 30 and 40 per cent. When attenuation is included in the SED-fitting, all models perform similarly and are consistent with each other within $1\sigma$ in $>80$ per cent of the cases.

\section{Dust in passive galaxies}
\label{sec:sec5}
Our galaxies are selected for being passively evolving, i.e., sustaining star formation rates at  low level, or not at all. As such, one expects these galaxies to contain little (cold) gas and by extension little dust, in analogy to local elliptical galaxies. Moreover, their very photometric selection supports the notion of such galaxies being relatively  dust free. For example, as discussed e.g., in \cite{Labbe-2005, Wuyts-2007,Whitaker-2011, vanDokkum-2015}, the rest-frame $UVJ$ selection used by us is able to distinguish between dusty, star-forming and virtually dust-free, passively evolving galaxies. Note that the analogous NUV$rJ$ selection is also publicly available for our galaxies, as illustrated in Figure \ref{fig:Fig1_appA}.

However, even if not accreted from outside via (minor) merger events, some amount of dust is likely to be present in passively evolving galaxies. Indeed, red giant branch (RGB) and especially AGB stars are continuously loosing mass via dusty stellar winds. The wind gas is quite promptly shock-heated to the virial temperature ($\sim 10^7$ K) whereas the dust starts to be destroyed by sputtering in this hot medium. The actual amount of dust within one such galaxy will then depend on various factors, such as whether the galaxy itself supports a galactic wind or outflow and on the dust sputtering timescale, hence it is  not easily predictable. By modeling their far-IR emission, \citet{Silva-1998} estimate that local giant elliptical galaxies contain $\sim 1.5\times 10^{7}\, {\rm M_{\odot}}$ of dust. In this section we try to put constraints on the amount of dust, hence on dust reddening, in our passive galaxies at high redshift. 

\begin{figure*}
\centering
\centering
\includegraphics[width=0.48\textwidth]{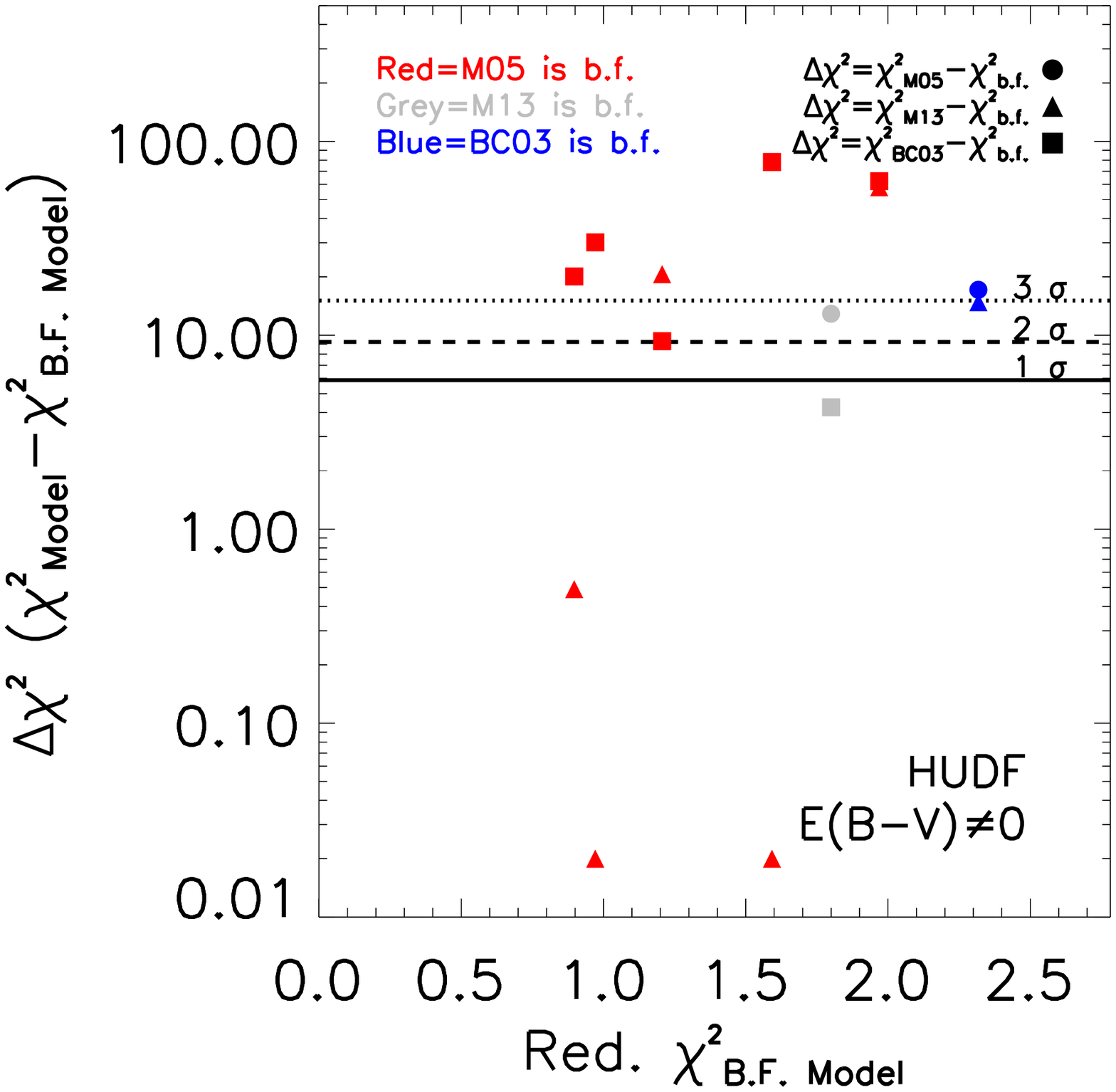}
\includegraphics[width=0.48\textwidth]{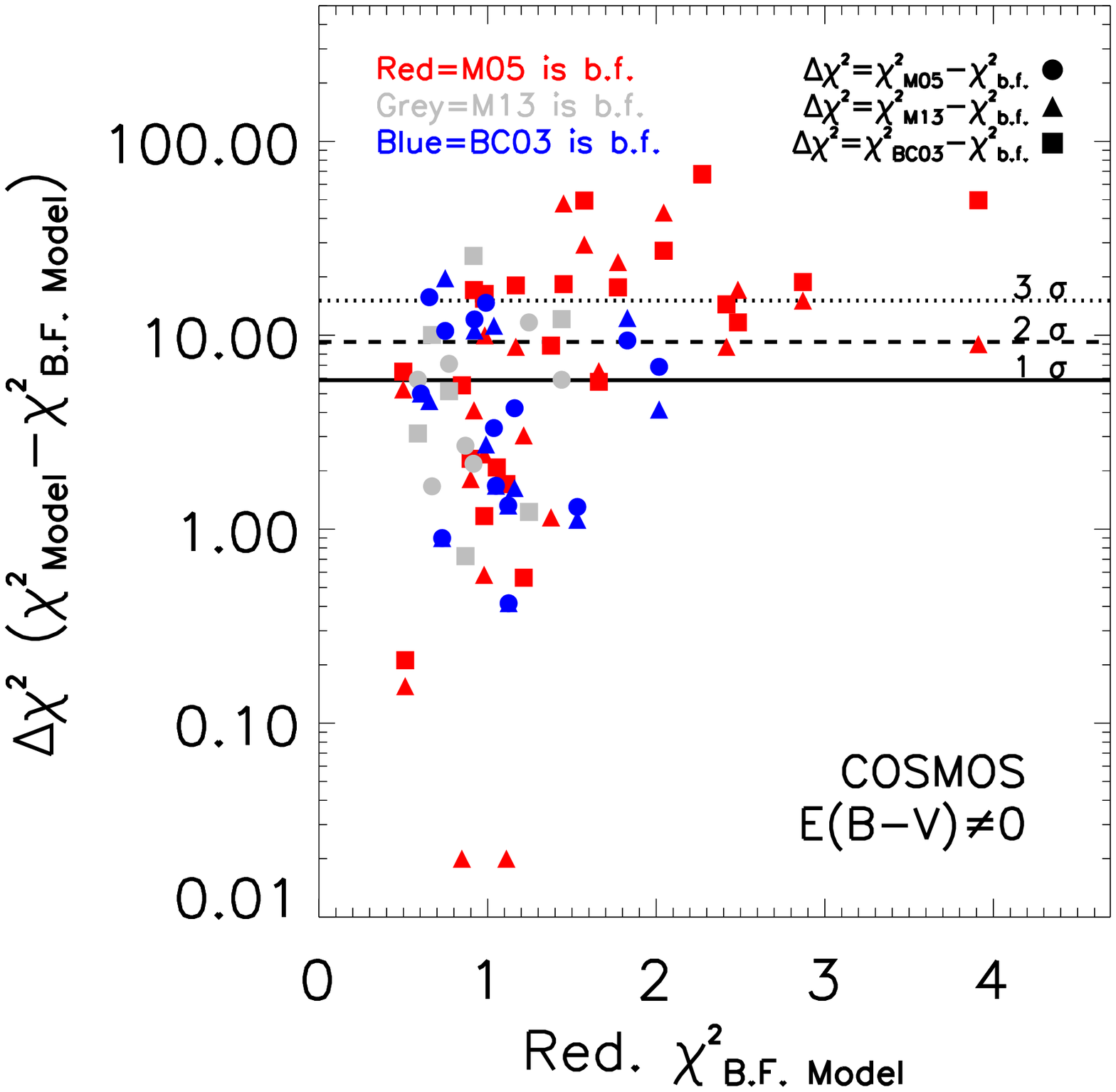}
\caption{As Figure \ref{fig:Fig7} after removal of inconsistent solutions (Section \ref{subsec:subsec5.2}).}
\label{fig:Fig8}
\end{figure*}
\subsection{Limits to dust in passive galaxies}
\label{subsec:subsec5.1}
The lack of detection of our galaxies in the MIPS $24\ {\rm \mu m}$ Spitzer channel is indicative of low/null star-formation and/or low/absence of dust (e.g., \citealp{Reddy-2006}). 
On the other hand, at the redshifts of our galaxies ($1.3\lesssim z \lesssim 2.7$), the MIPS $24\ {\rm \mu m}$ samples the rest-frame $\sim 8\ {\rm \mu m}$, which is more sensitive to hot dust, possibly coming from the circumstellar dust of AGB stars \citep{Kelson-2010}. Hence, cold dust, which emits at far-IR wavelengths, could still be present and cause reddening.
However, none of our galaxies has been individually detected at wavelengths sensitive to cold dust emission with the currently available instruments (Herschel at $100-500\ {\rm \mu m}$, AzTEC at $1.1-1.2\ {\rm mm}$, IRAM PdBI at $1.2-3.3\ {\rm mm}$, etc.), given the shallow characteristic depths of such instruments compared to the faint fluxes which may be emitted by the possibly low, though not null, dust content in our galaxies. So the next question is: what is the allowed amount of dust reddening in our galaxies held to be passive? 

To answer this question  we consider indicators that are usually adopted for normal star-forming (otherwise called Main Sequence, MS) galaxies at high-$z$, which can be used as upper limits in our case. These are the ratio between dust and stellar masses ($M_{d}/M_{*}$), the ratio between the $CO$ luminosity (directly related to the gas content) and dust mass ($L_{\rm CO}^{'}/M_{d}$) and the ratio between $CO$ luminosity and stellar mass ($L_{\rm CO}^{'}/M_{*}$) (e.g., \citealp{Rowlands-2012, di-Serego-2013, Man-2014, Sargent-2015}). 

\citet{Tan-2014} studied three sub-millimeter galaxies (SMGs) and a variety of archival data of Main Sequence and starburst galaxies and found an almost constant value from $z=0$ to $z\sim3$ of $L_{\rm CO}^{'}/M_{d}~\sim 20-30\; L_\odot/M_\odot$ .

\citet{Sargent-2015} studied the $L_{\rm CO}^{'}/M_{*}$ ratio in passive and star forming galaxies and its evolution with redshift.
They found that local early type galaxies (ETGs) from the $ATLAS^{3D}$ survey \citep{Young-2011} have $L_{\rm CO}^{'}/M_{\rm *} \sim0.0015 \ {\rm L_\odot/M_\odot}$, a factor $\sim20$ lower than local SF galaxies ($L_{\rm CO}^{'}/M_{\rm *} \sim0.03 \ {\rm L_\odot/M_\odot}$) with similar mass (see their Figure 2). They found the same factor of $\sim20$ difference when comparing the 3-$\sigma$ upper limit ($L_{\rm CO}^{'}/M_{\rm *} \le 0.01 \ {\rm L_\odot/M_\odot}$)  for p{\it BzK}-217431  (a $z\sim1.5$~passive galaxy, corresponding to  our object ID 8) with measures for z$\sim1.5$ MS galaxies ($L_{\rm CO}^{'}/M_{\rm *}\sim 0.2 \ {\rm L_\odot/M_\odot}$), suggesting that the factor $\sim 20$ difference in $L_{\rm CO}^{'}/M_{\rm *}$ between SF and passive galaxies is almost constant with $z$.

By using the $L_{\rm CO}^{'}/M_{\rm d}$ from \citet{Tan-2014} we can infer that the $M_{\rm d}/M_{\rm *}$ ratio for passive galaxies at $z\sim2$ is $\sim 3\times 10^{-4}$, while that for MS SFGs is $\sim 7\times 10^{-3}$. For a typical stellar mass $M_{\rm *}\sim5\times10^{10}\ {\rm M_{\odot}}$ we then obtain $M_{\rm d}\sim 1.7\times 10^{7}\ {\rm M_{\odot}}$ and $M_{\rm d}\sim 3\times 10^{8}\ {\rm M_{\odot}}$ for passive and SF galaxies, respectively. 

This suggests that $z\sim2$ passive galaxies have $M_{\rm d}$ values a factor $\sim20$ lower than SF galaxies at the same mass and redshift, consistent with the findings of \citet{Man-2014}.
A factor $\sim20$ difference between the dust masses of passive and SF galaxies is also consistent with \citet{Rowlands-2012}, who studied a sample of 1087 ETGs and passive spirals detected with {\it Herschel} (H-ATLAS) in comparison with an optically-selected control sample finding that the H-ATLAS detection rate of ETGs in optically selected galaxy catalogues  is 5.5 per cent (out to $r$-band apparent magnitude $\sim 20$). 

We now use this factor $\sim 20$ difference in dust content to derive an estimate of the dust reddening of our passively evolving galaxies. The extinction in the $V$ band is 
$A_{\rm V}=2.5$ log$(e)\tau$ where $\tau$ is the optical depth in the $V$ band. In turn, $\tau$ is proportional to the dust mass, hence it is $\sim 20$ times smaller in passive galaxies than in 
star-forming galaxies at the same redshift. For star-forming galaxies at an average redshift $z\sim 1.6$, \cite{Kashino-2013} found $A_{\rm H\alpha}\sim 1\ {\rm mag}$, or $E(B-V)\sim 0.4\ {\rm mag}$ (see their 
Figure 2). Hence the factor of 20 difference in optical depth implies reddening to be also a factor $\sim 20$ smaller, or $E(B-V)\sim 0.02\ {\rm mag}$, a very low reddening indeed, which should actually be regarded as an upper limit. This assumes that the distribution relative to stars and the optical properties of dust in passive galaxies to be similar to those in star-forming galaxies. Such an assumption may not be strictly valid, yet this limit on reddening is so low that we can safely conclude that in typical passively evolving galaxies at our redshifts reddening should be low, almost certainly lower than $E(B-V)\sim 0.2\ {\rm mag}$ (see Section \ref{subsec:subsec5.2}).
\begin{table*}
%\begin{tiny}
\begin{center}
\caption{Galaxy average physical properties. Col 1: sample; col 2: average spectroscopic redshift; col 3: info about whether reddening was included; col 4:  model; cols 5, 6, 7 \& 8: average values of stellar mass, age, $E(B-V)$ and reduced $\chi^{2}$. Values in brackets for the reddening case are those obtained after discarding physically unreliable solutions, Section \ref{subsec:subsec5.2}.}
\begin{tabular}{cccccccc}
\hline
  \multicolumn{1}{c}{\bf Sample} &
  \multicolumn{1}{c}{$\mathbf{\langle z \rangle}$} &
  \multicolumn{1}{c}{\bf Reddening} &
  \multicolumn{1}{c}{\bf Model} &
  \multicolumn{1}{c}{$\mathbf{\langle \log(M^{*}) \rangle}$} &
  \multicolumn{1}{c}{\bf $\mathbf{\langle t \rangle}$} &
  \multicolumn{1}{c}{$\mathbf{\langle E(B-V) \rangle}$} &
  \multicolumn{1}{c}{$\mathbf{\langle \chi^{2}_{\rm r} \rangle}$}\\
  \multicolumn{1}{c}{} &
  \multicolumn{1}{c}{} &
  \multicolumn{1}{c}{} &
  \multicolumn{1}{c}{} &
  \multicolumn{1}{c}{} &
  \multicolumn{1}{c}{\rm (Gyr)} &
  \multicolumn{1}{c}{\rm (mag)} &
  \multicolumn{1}{c}{}   \\
  \multicolumn{1}{c}{(1)}&
  \multicolumn{1}{c}{(2)}&
  \multicolumn{1}{c}{(3)}&
  \multicolumn{1}{c}{(4)}&
  \multicolumn{1}{c}{(5)}&
  \multicolumn{1}{c}{(6)}&
  \multicolumn{1}{c}{(7)}&  
  \multicolumn{1}{c}{(8)} \\
\hline
\hline
\multirow{6}{*}{COSMOS}  &	\multirow{6}{*}{$1.6\pm0.2$}    &	\multirow{3}{*}{No}    &  M05  &  $10.8\pm0.3$ &  $1.4\pm0.8$   & $--$        & $1.43$  \\
  		 	 &	  		 	        &			       &  M13  &  $10.9\pm0.2$ &  $1.3\pm0.5$   & $--$        & $1.79$  \\
		 	 &			 	        &			       &  BC03 &  $11.1\pm0.2$ &  $2.3\pm0.5$   & $--$        & $1.88$  \\
\cmidrule{3-8}
  		 	 &	  		 	        &	\multirow{3}{*}{Yes}   &  M05  &  $10.9\pm0.3\ (10.9\pm0.3)$ &  $1.4\pm0.9\ (1.6\pm0.8)$   & $0.1\pm0.2\ (0.01\pm0.03)$ & $1.25\ (1.41)$  \\
  		 	 &	  		 	        &			       &  M13  &  $11.0\pm0.3\ (10.9\pm0.3)$ &  $1.0\pm0.5\ (1.3\pm0.5)$   & $0.2\pm0.2\ (0.03\pm0.04)$ & $1.28\ (1.64)$  \\
		 	 &			 	        &			       &  BC03 &  $11.0\pm0.3\ (11.1\pm0.3)$ &  $1.5\pm0.8\ (2.0\pm0.6)$   & $0.2\pm0.1\ (0.04\pm0.04)$ & $1.17\ (1.69)$  \\
\hline				
\multirow{6}{*}{HUDF}    &	\multirow{6}{*}{$1.8\pm0.4$}    &	\multirow{3}{*}{No}    &  M05  &  $10.6\pm0.3$ &  $1.1\pm0.8$   & $--$        & $2.24$  \\
  		 	 &	  		 	 	&			       &  M13  &  $10.6\pm0.3$ &  $1.01\pm0.35$ & $--$        & $4.00$  \\
		 	 &			 	 	&			       &  BC03 &  $10.7\pm0.3$ &  $1.4\pm0.4$   & $--$        & $5.37$  \\
\cmidrule{3-8}

  		 	 &	  		 	 	&	\multirow{3}{*}{Yes}   &  M05  &  $10.6\pm0.3\ (10.6\pm0.3)$ &  $0.8\pm0.8\ (1.1\pm0.7)$   & $0.1\pm0.1\ (0.04\pm0.06)$ & $1.31\ (1.83)$  \\
  		 	 &	  		 	 	&			       &  M13  &  $10.7\pm0.3\ (10.7\pm0.3)$ &  $0.8\pm0.5\ (0.9\pm0.4)$   & $0.1\pm0.1\ (0.08\pm0.06)$ & $1.24\ (2.48)$  \\
		 	 &			 	 	&			       &  BC03 &  $10.8\pm0.3\ (10.8\pm0.2)$ &  $1\pm1\ (1.4\pm0.4)$       & $0.2\pm0.1\ (0.05\pm0.05)$ & $1.36\ (3.62)$  \\

\hline
\hline
\end{tabular}
\label{tab:Table6}							    
\end{center}
% \end{tiny}
\end{table*}
\begin{figure}
\centering
\includegraphics[width=0.5\textwidth]{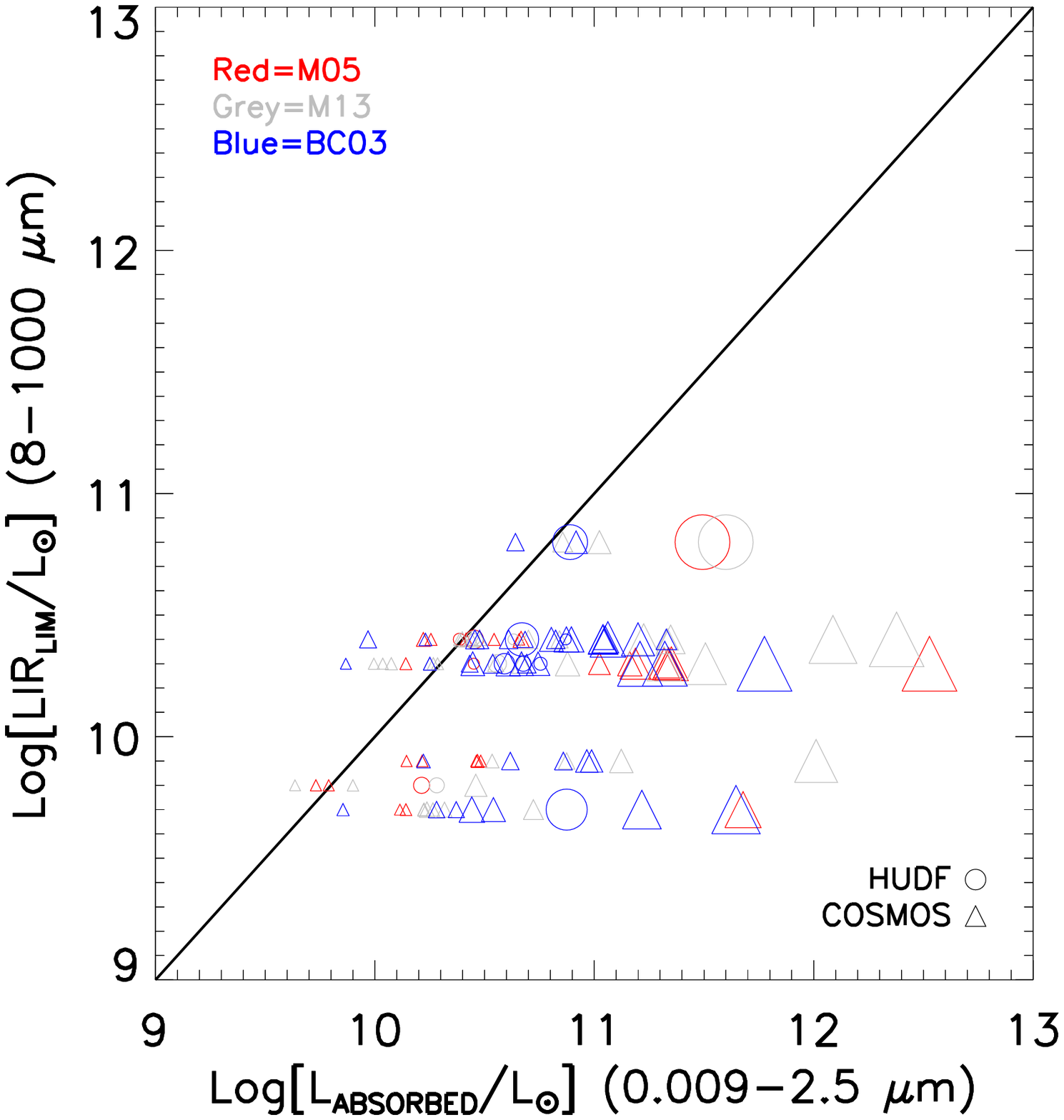}
\caption{Absorbed bolometric luminosity at $0.009-2.5\ {\rm \mu m}$ ( $L_{\rm bol}^{\rm absorbed}$) vs. $L_{\rm IR}$ ($8-1000\ {\rm \mu m}$) upper limit from \citet{Man-2014} for the best-fitting solutions obtained with M05, M13 and BC03 models (red, grey and blue symbols) for HUDF and COSMOS galaxies (circles and triangles). Here we limit ourselves to solutions listed in Tables \ref{tab:Table3} and \ref{tab:Table2_appB} with $E(B-V)>0$. A 1:1 line is plotted as reference. Data points' sizes are scaled according to $z_{\rm spec}\times E(B-V)$.}
\label{fig:Fig9}
\end{figure}
\subsection{Using far-IR observations to identify unphysical solutions}
\label{subsec:subsec5.2}
\citet{Man-2014} have selected passively evolving galaxies taken from the same UltraVISTA catalogue by \citet{McCracken-2012} and using the NUV$rJ$ selection illustrated in Figure \ref{fig:Fig1_appA} for our  44 COSMOS galaxies. 

By stacking deep {\it Spitzer} MIPS $24\mu m$ and {\it Herschel}  maps for $\sim 14,000$ quiescent galaxies (QGs) with $M_{\rm *}=10^{9.8-12.2}\ {\rm M_{\odot}}$ out to $z=3$, \citet{Man-2014} provided $L_{\rm IR}(8-1000\ {\rm \mu m})$ upper limits in bins of redshift and stellar mass, i.e., limits to the total dust emission in the mid- and far-infrared. At the typical redshift ($z\sim 1.6$) and stellar mass ($\log(M^{\star}/M_{\odot})\sim10.8$) values of our galaxy catalogue, the limit provided in their Table 2 is $\log(L_{\rm IR}^{\rm limit}/L_{\odot})\simeq10$. In particular, none of the QGs-based {\it Herschel} stacks showed significant detection (i.e., at $S/N>3$). 
These limits translate into dust-obscured SFR$<0.1-0.3\ {\rm M_{\odot}\ yr^{-1}}$ at $z\leq2$ and SFR$<6-18\ {\rm M_{\odot}\ yr^{-1}}$ at $z>2$), consistent with the low unobscured SFRs ($<0.01-1.2\ {\rm M_{\odot}\ yr^{-1}}$) inferred from modelling the ultraviolet-to-near-IR photometry.

Our COSMOS and HUDF galaxies have been selected with the {\it UVJ} selection criterion, which is very similar to the NUV$rJ$ one used for the Man et al.'s QGs. They also span a similar range of redshifts and stellar masses, hence, at least in a statistical sense, the $L_{\rm IR}(8-1000\ {\rm \mu m})$ upper limits derived by Man et al. for their 14,000 QGs should also apply to our sample of galaxies, among which 38 are actually included in their sample as satisfying their NUV$rJ$ selection criterion (see Figure \ref{fig:Fig1_appA}). We then use these upper limits to exclude SED-fitting solutions that violate them. 

\begin{figure}
\centering
\includegraphics[width=0.5\textwidth]{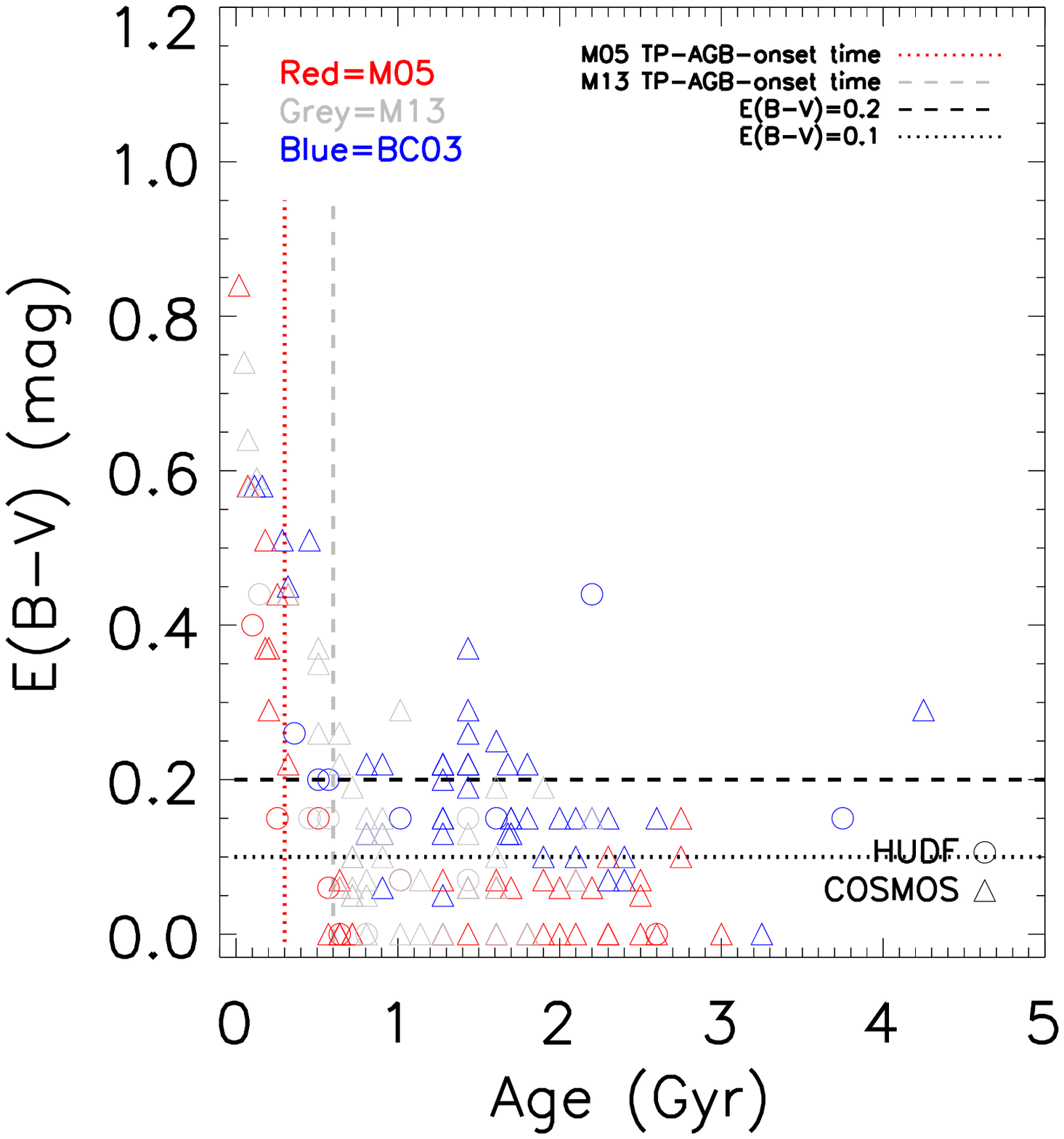}
\includegraphics[width=0.5\textwidth]{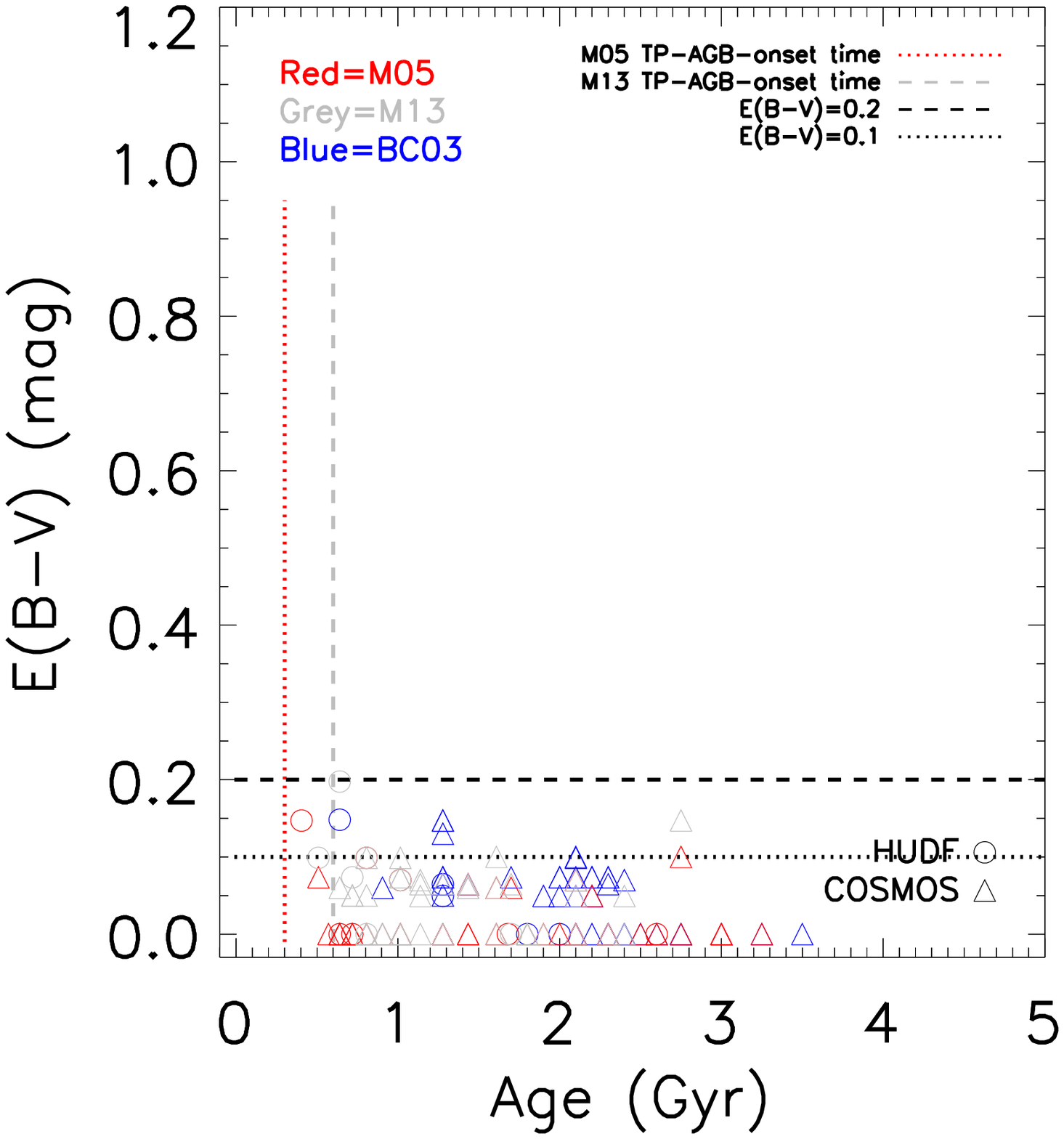}
\caption{$E(B-V)$ vs. $age$ of best-fitting solutions for M05, M13 and BC03 models (red, grey and blue symbols) for HUDF and COSMOS galaxies (circles and triangles). Upper and lower panels refer respectively to the samples prior and after discarding solutions according to Section \ref{subsec:subsec5.2}.}
\label{fig:Fig10}
\end{figure}
In order to do so, we evaluate the amount of bolometric light ($0.009<\lambda<2.5\ {\rm \mu m}$) that is supposed to be absorbed by dust for attenuated SED fitting solutions. In particular, we estimate the difference ($L_{\rm bol}^{\rm absorbed}$) between the unreddened ($L_{\rm bol}^{\rm intrinsic}$) and reddened ($L_{\rm bol}^{\rm reddened}$) bolometric luminosities for all the fitting solutions associated with our galaxies [i.e. for all combinations of  SFH, E(B-V), age, metallicity and stellar mass] . We then compare the $L_{\rm bol}^{\rm absorbed}$ values obtained with those of Man et al.'s $L_{\rm IR}$ upper limits in the appropriate redshift and stellar-mass bins (see Figure \ref{fig:Fig9}). Those solutions with reddening such that $L^{\rm bol}_{\rm absorbed}>L_{\rm IR}^{\rm limit}$ are then rejected as non self-consistent and unphysical and new best-fit solutions are identified by picking those among the remaining ones with the lowest $\chi^{2}$ values. 

This process is done under the assumption that $L_{\rm bol}^{\rm absorbed}$ is reprocessed to the {\it IR}. 
Since the limits provided by Man et al. are given over the entire mid- and far-infrared spectrum ($8<\lambda<1000\ {\rm \mu m}$), we can safely assume that all the absorbed light should be re-processed within this wavelength window, independently of the particular properties of the dust.  

Figure \ref{fig:Fig9} shows $L_{\rm bol}^{\rm absorbed}$ as from the best fitting solutions in Tables \ref{tab:Table3} and \ref{tab:Table2_appB} vs. the upper limits in \citet{Man-2014}. The number of solutions with $L^{\rm bol}_{\rm absorbed}>L_{\rm IR}^{\rm limit}$ is significant for all models. In particular, there are solutions (such COSMOS ID 12, 33 and 37) with very high [$\log(L_{\rm bol}^{\rm absorbed}/L_{\odot})>12$] absorbed luminosity. Such solutions are massive  [$\log (M^{*}/M_{\odot})\gtrsim11.2$], heavily reddened [$E(B-V)\gtrsim 0.6$] and very young ($age\lesssim 0.08 \ {\rm Gyr}$) with instantaneous $SFR<0.1$ (see Table \ref{tab:Table2_appB}), hence very likely affected by degeneracies (see also Figure \ref{fig:Fig10}). 

\begin{figure*}
\centering
\includegraphics[width=0.4\textwidth]{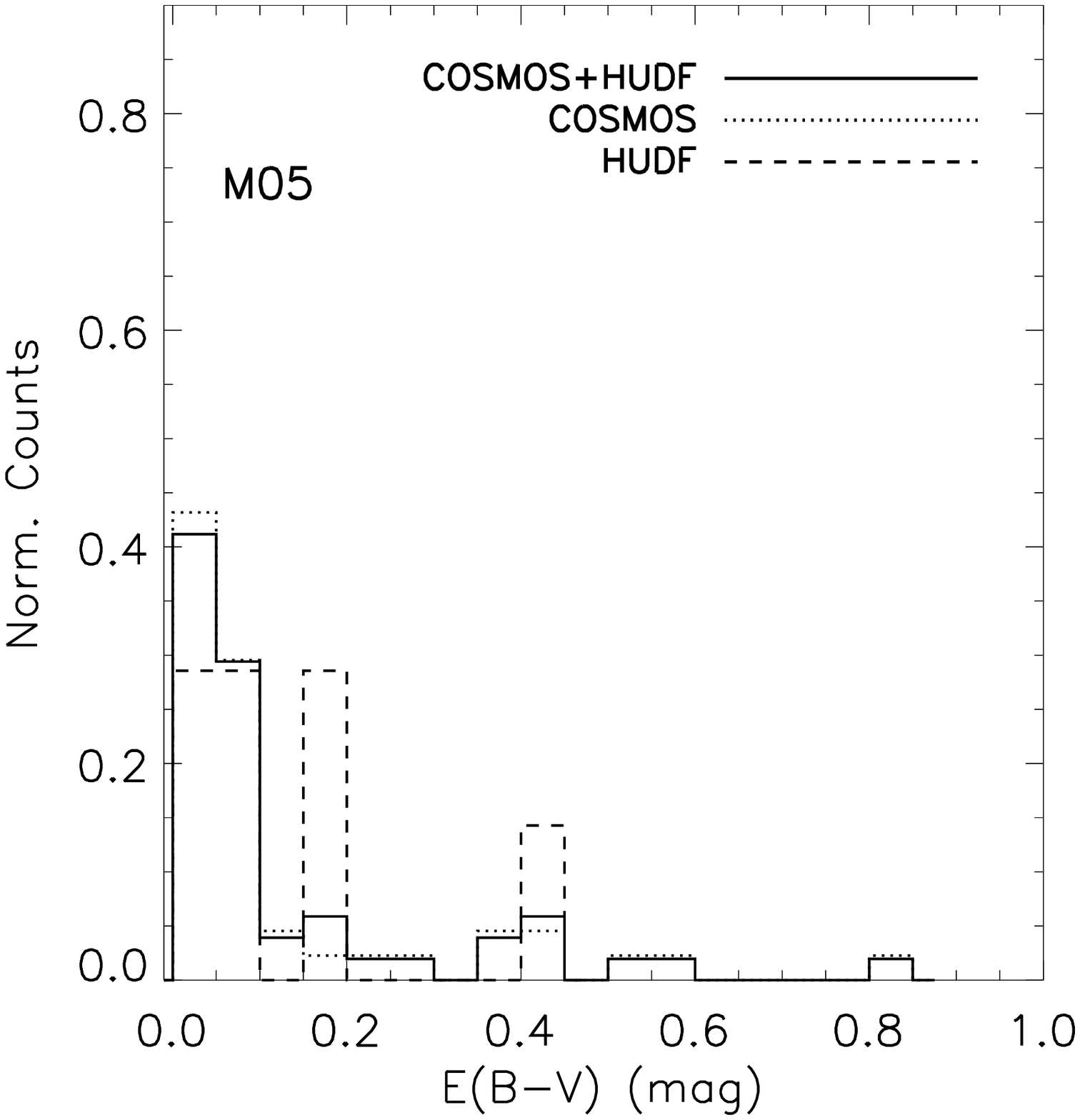}
\includegraphics[width=0.4\textwidth]{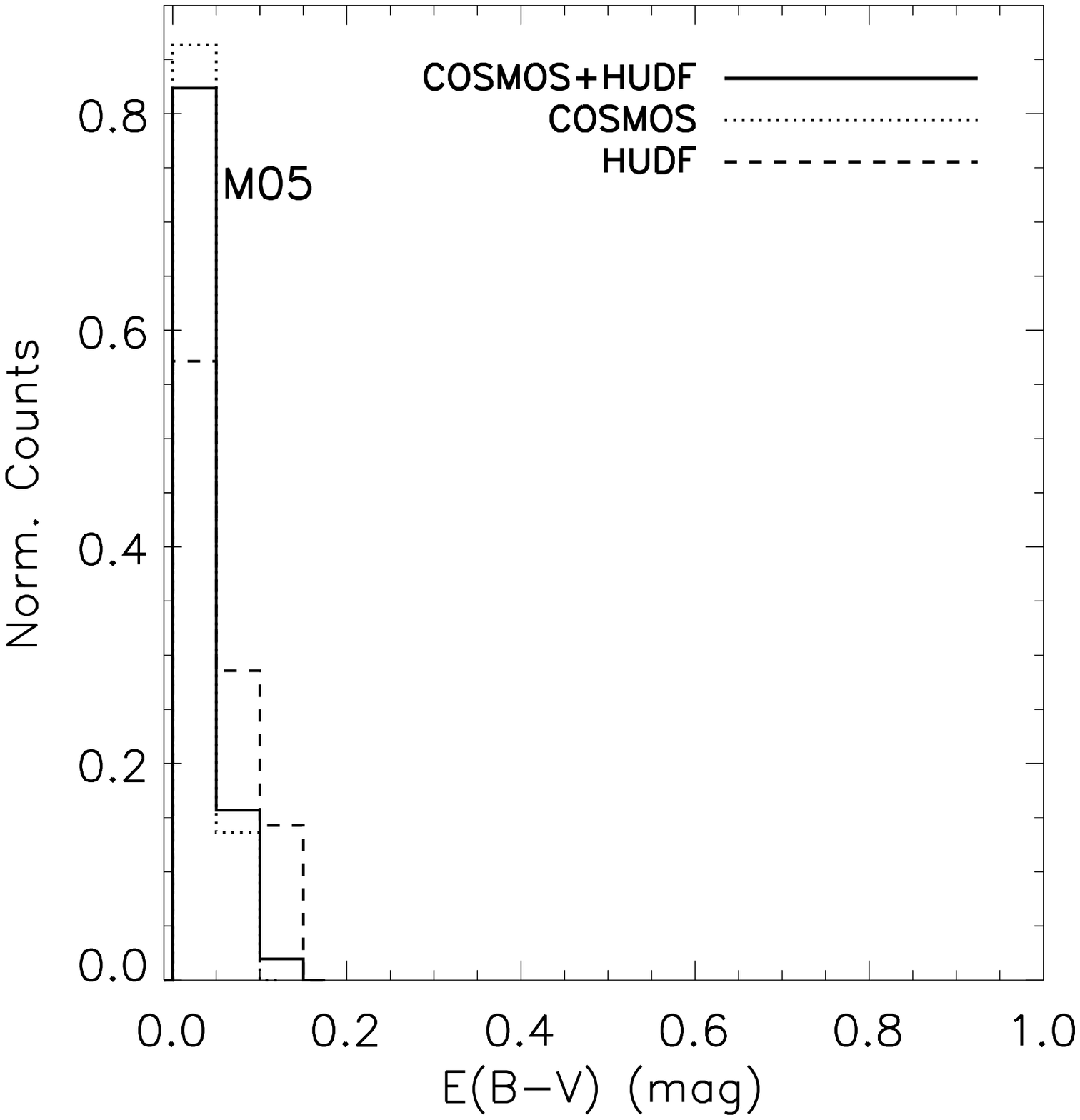}
\includegraphics[width=0.4\textwidth]{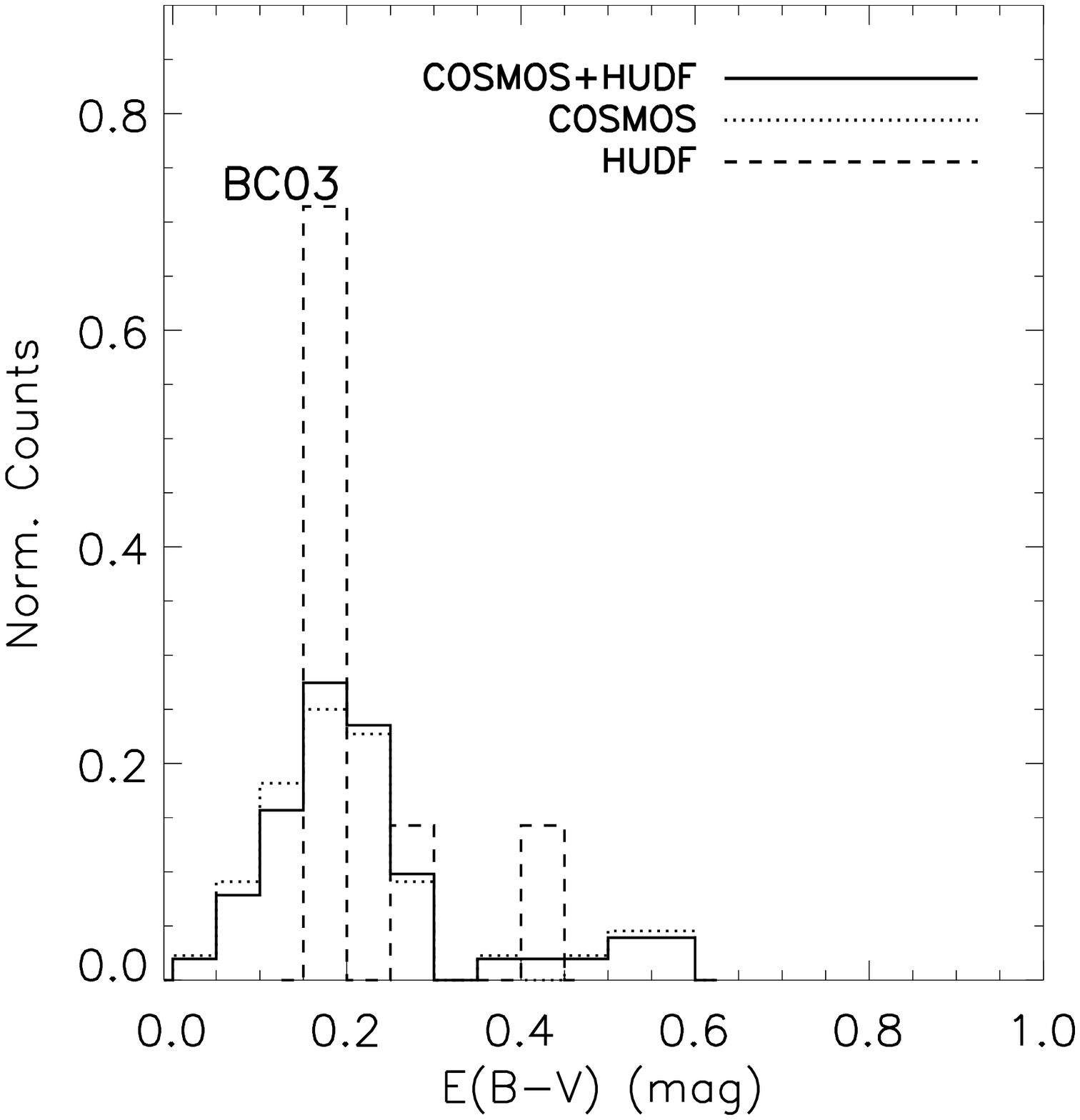}
\includegraphics[width=0.4\textwidth]{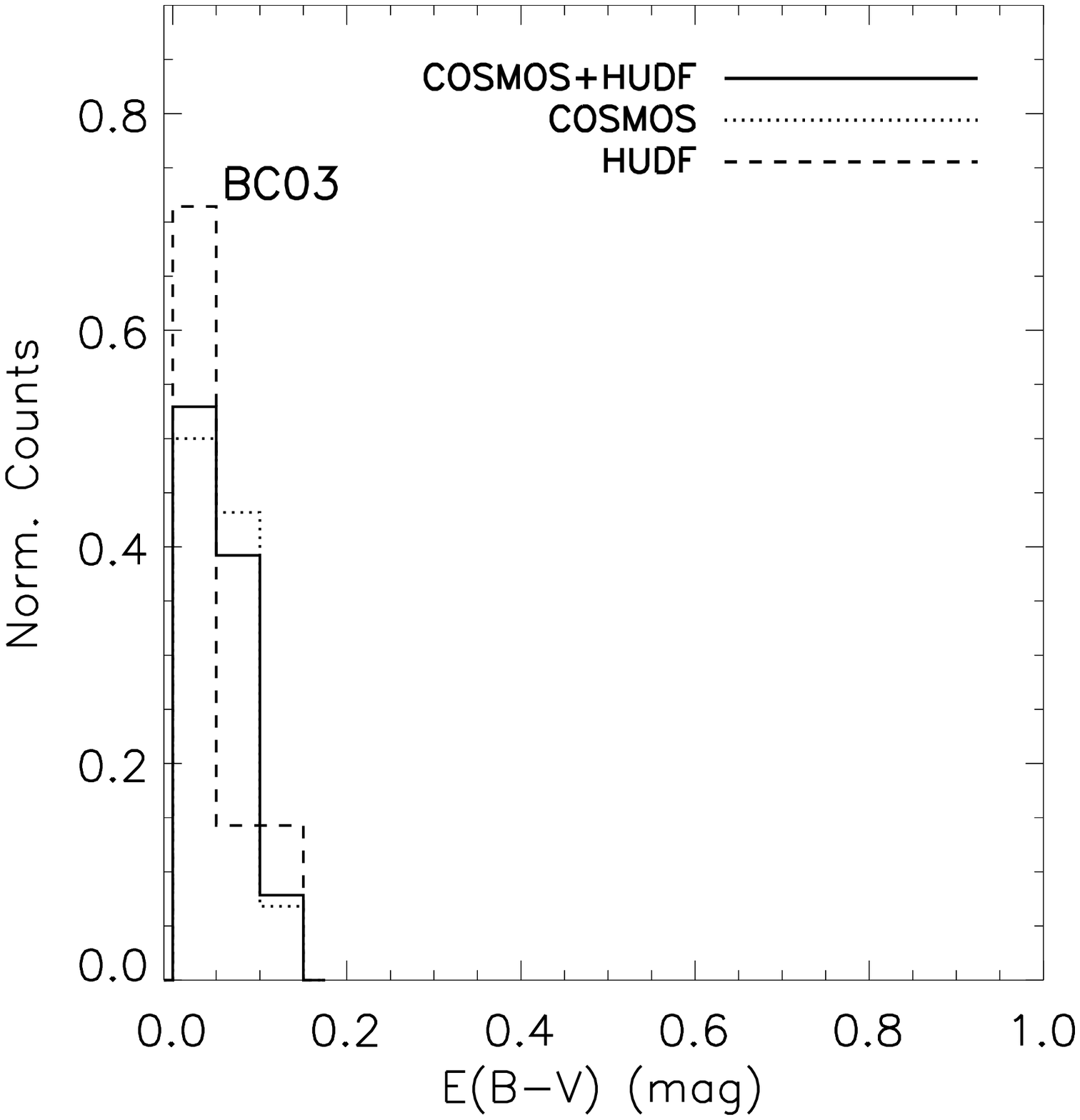}
\includegraphics[width=0.4\textwidth]{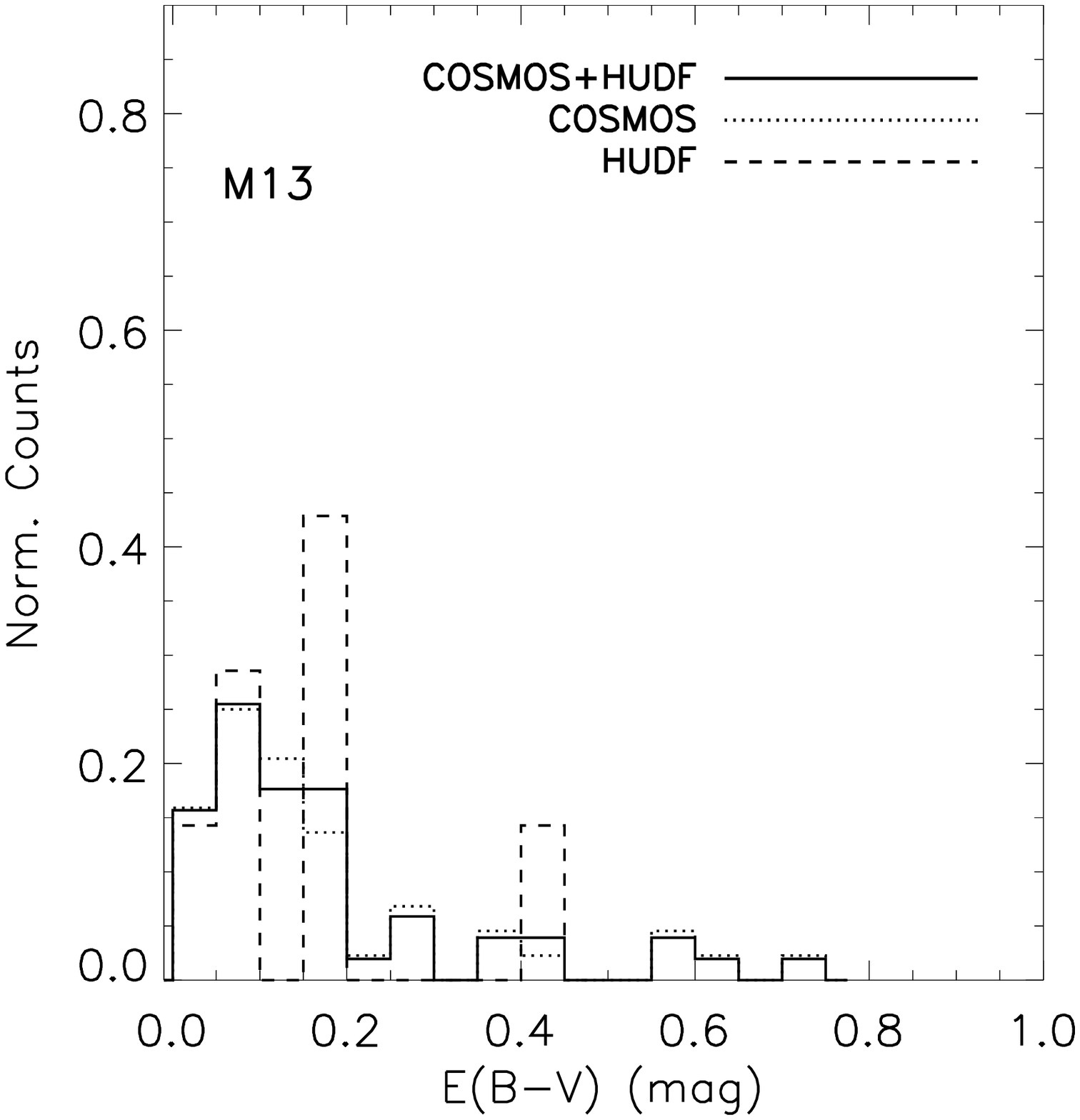}
\includegraphics[width=0.4\textwidth]{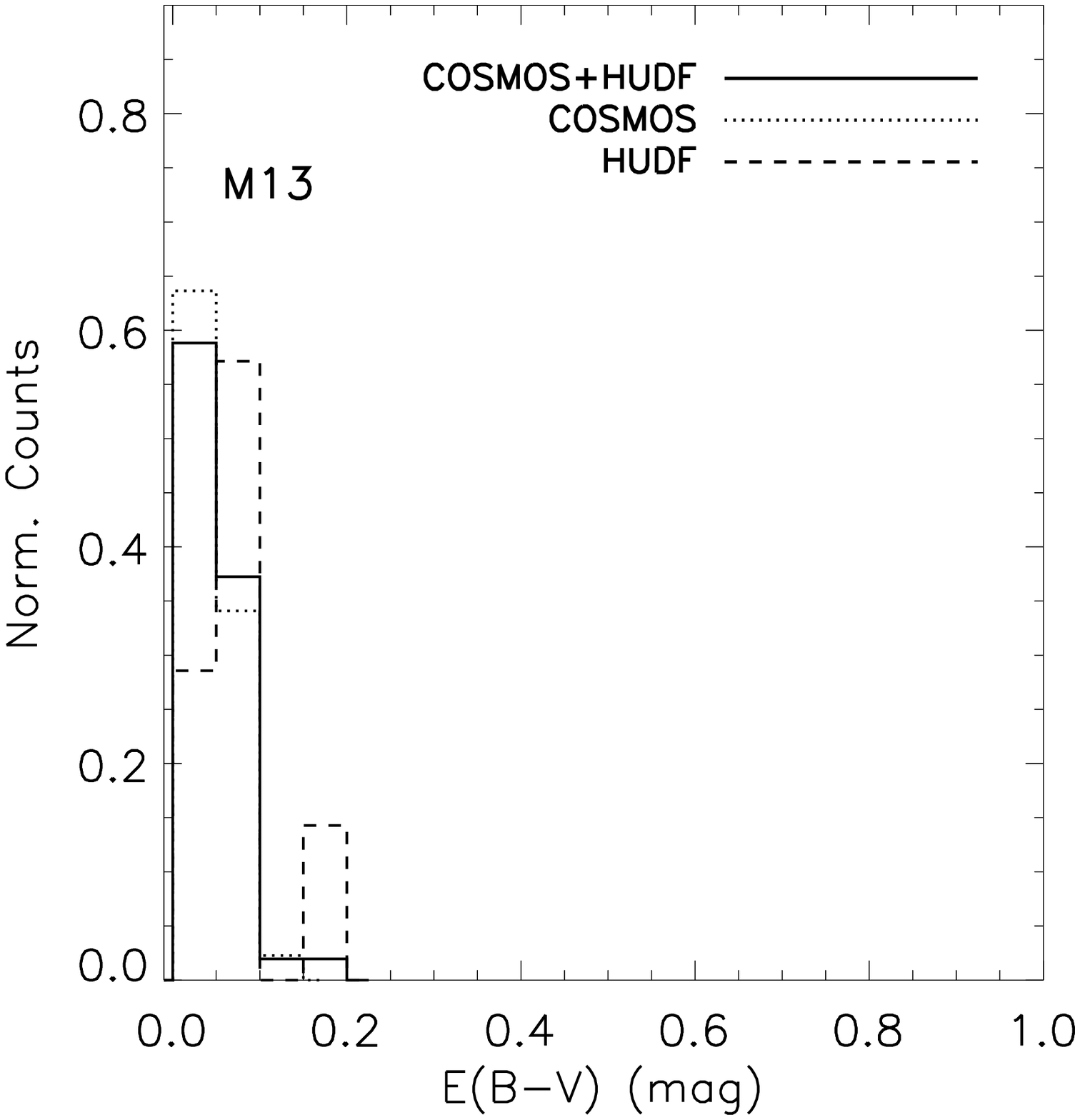}
\caption{Normalised distributions of $E(B-V)$~for HUDF, COSMOS and HUDF+COSMOS galaxies obtained with M05, BC03 and M13 models (from top to bottom). Left-hand and right-hand panels refer to the samples prior and after discarding possible unreliable solutions according to Section \ref{subsec:subsec5.2}.}
\label{fig:Fig11}
\end{figure*}

\begin{figure*}
\centering
\includegraphics[width=0.3\textwidth]{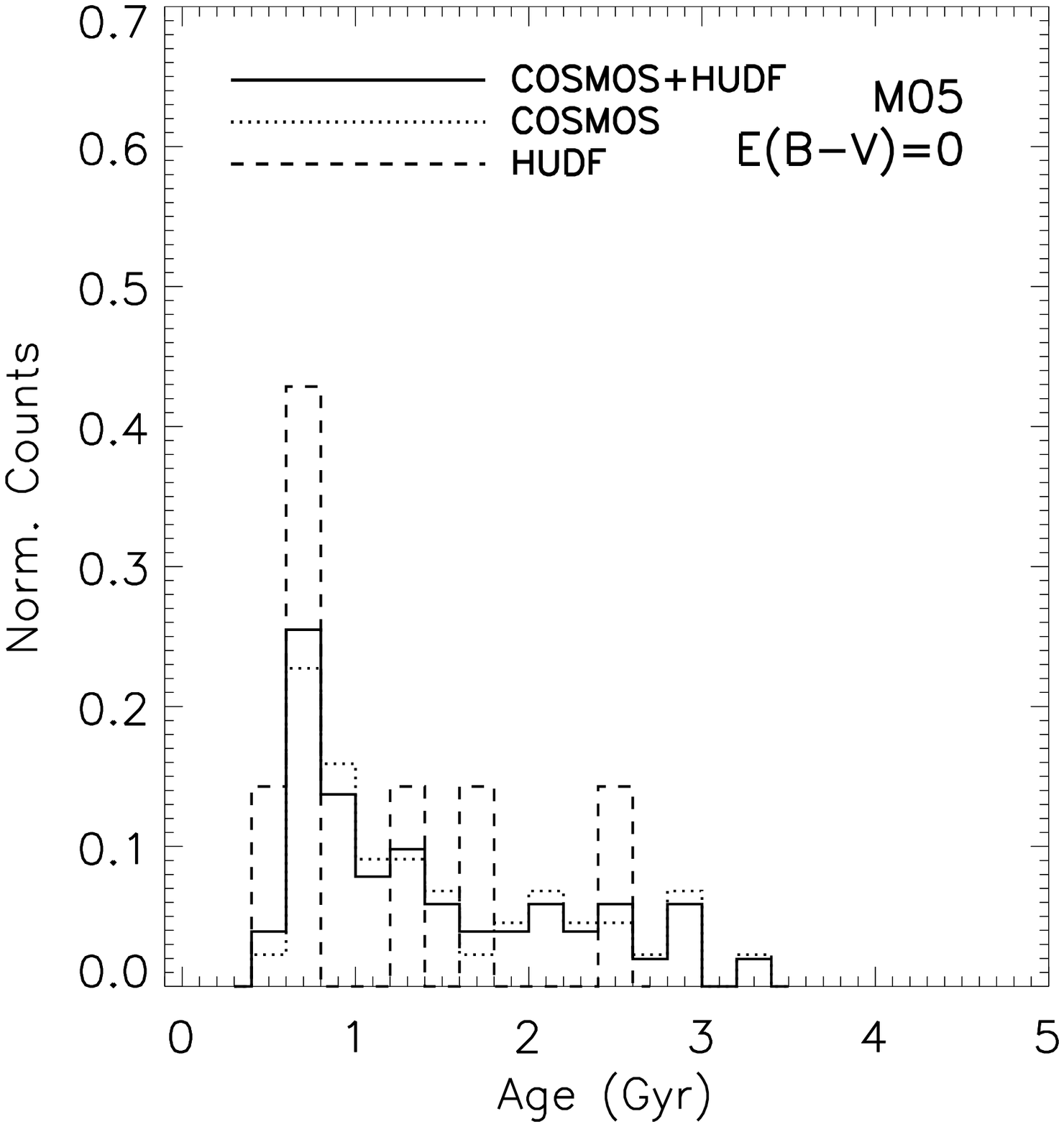}
\includegraphics[width=0.3\textwidth]{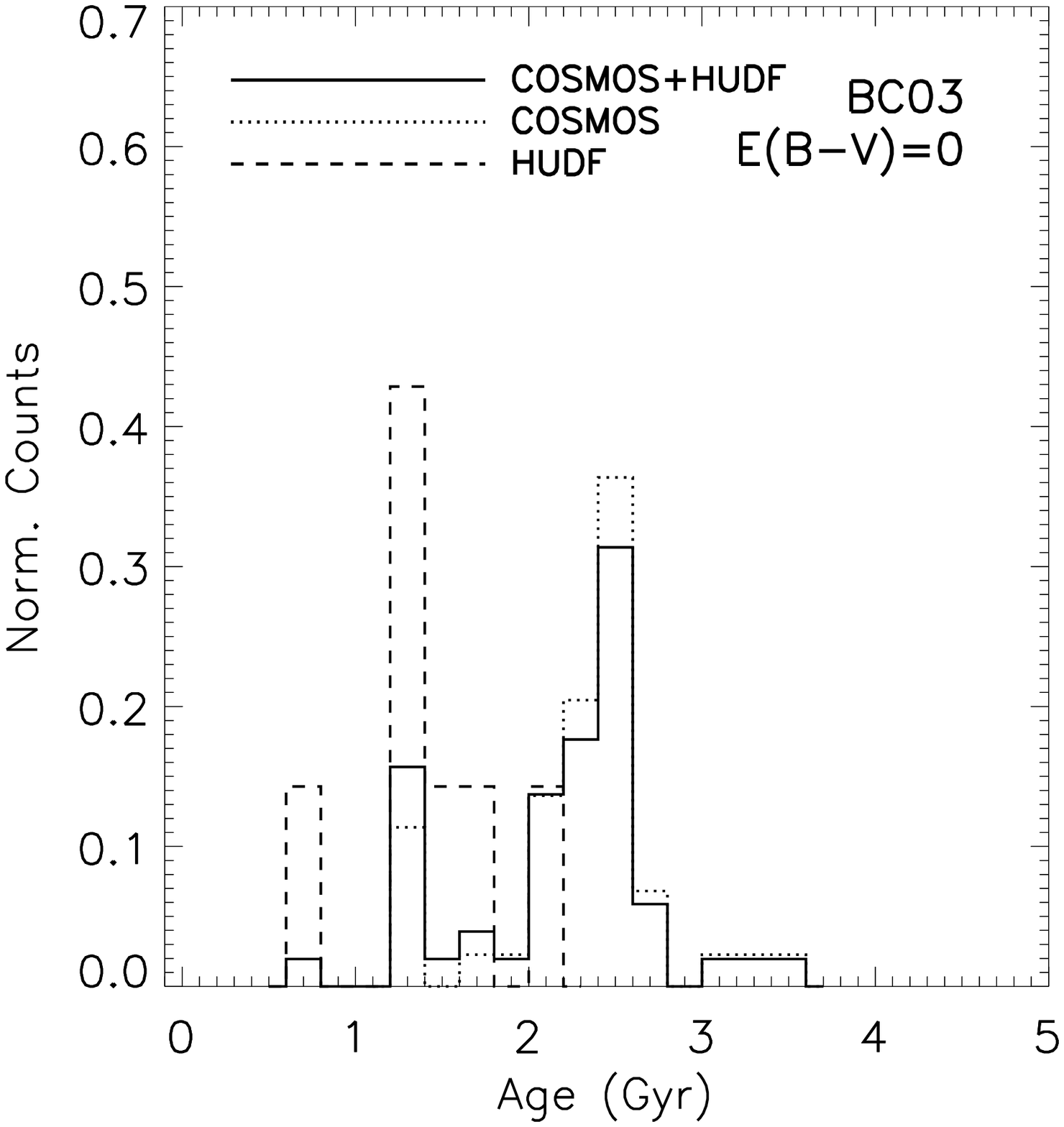}
\includegraphics[width=0.3\textwidth]{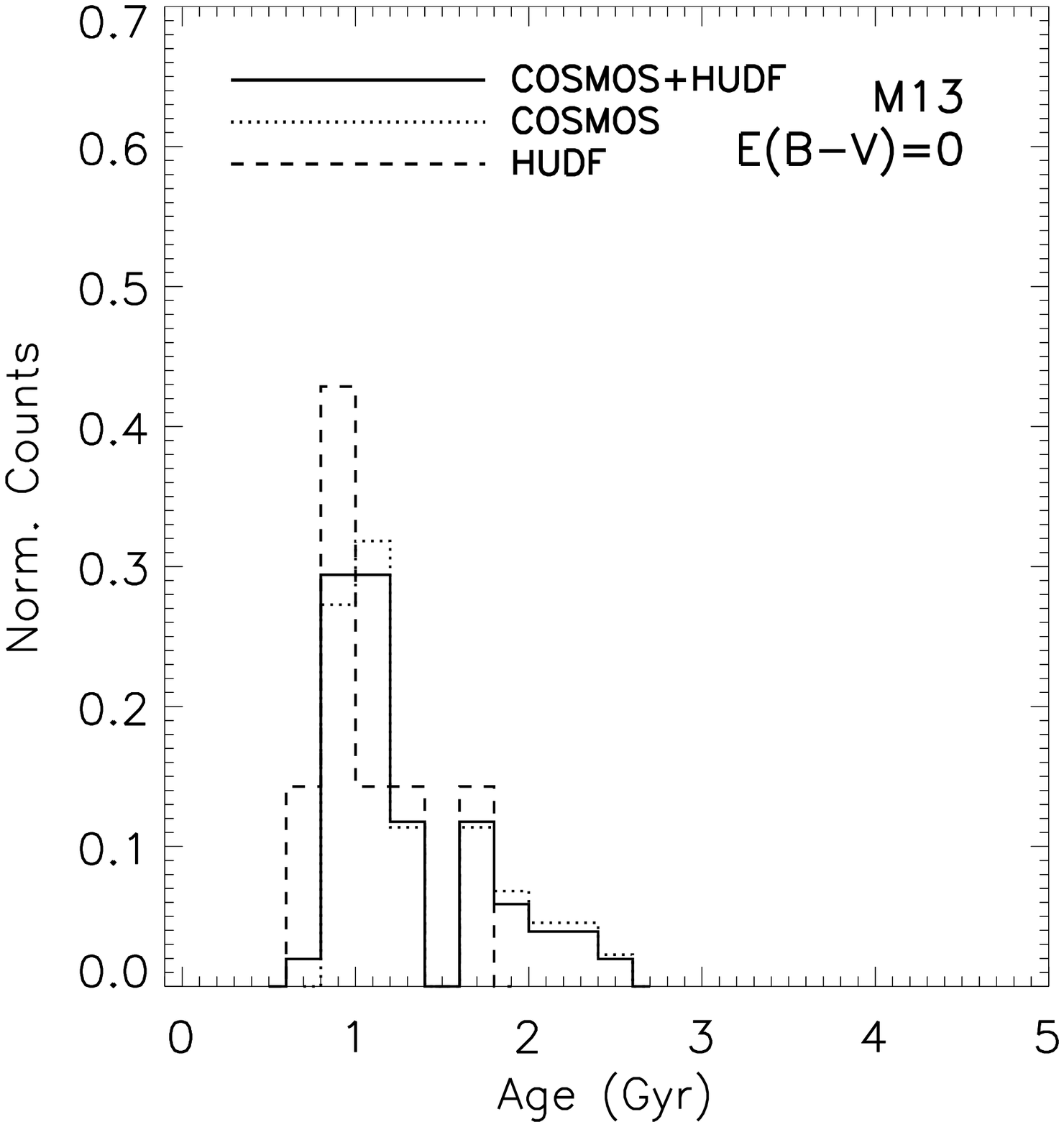}
\includegraphics[width=0.3\textwidth]{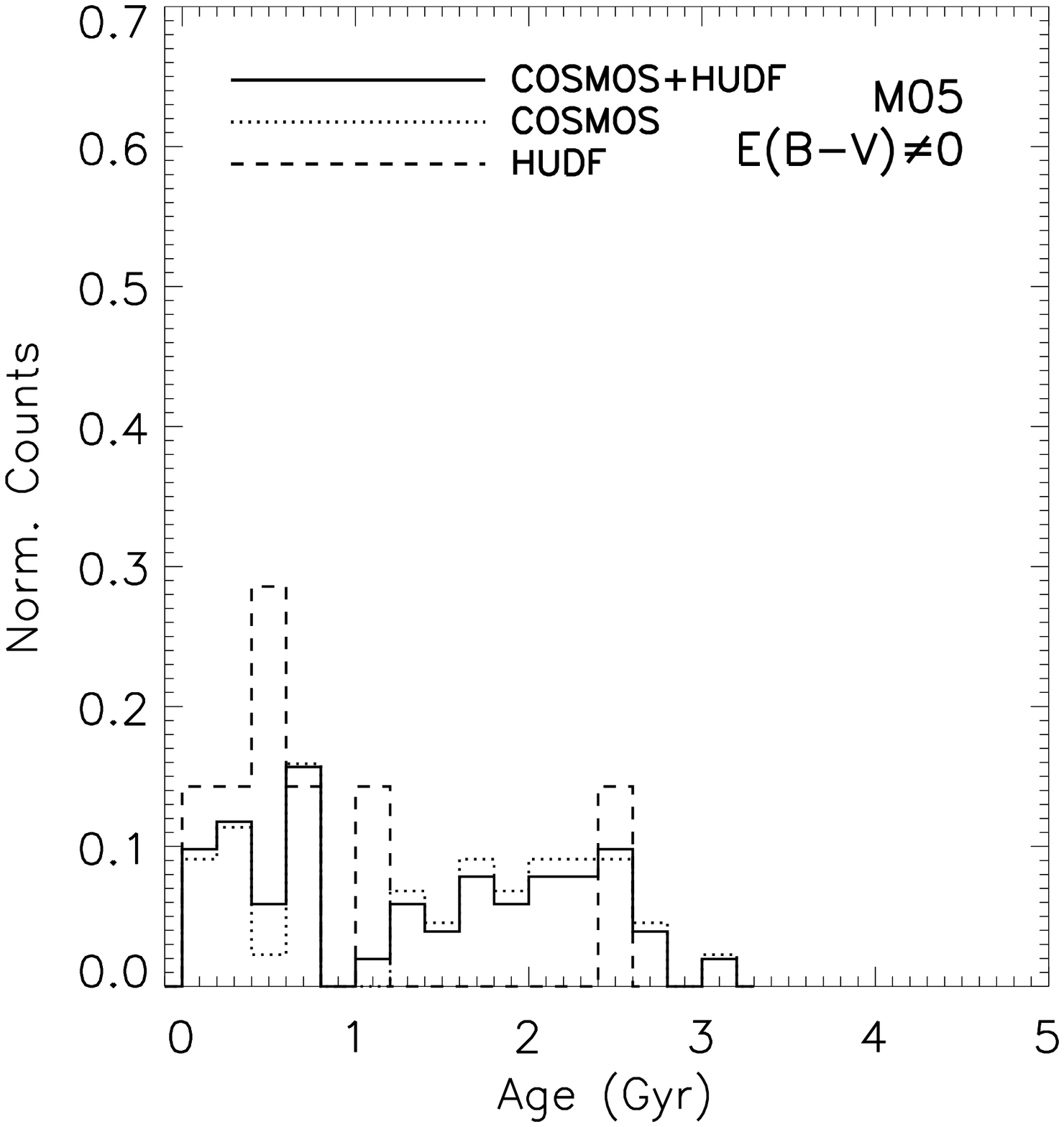}
\includegraphics[width=0.3\textwidth]{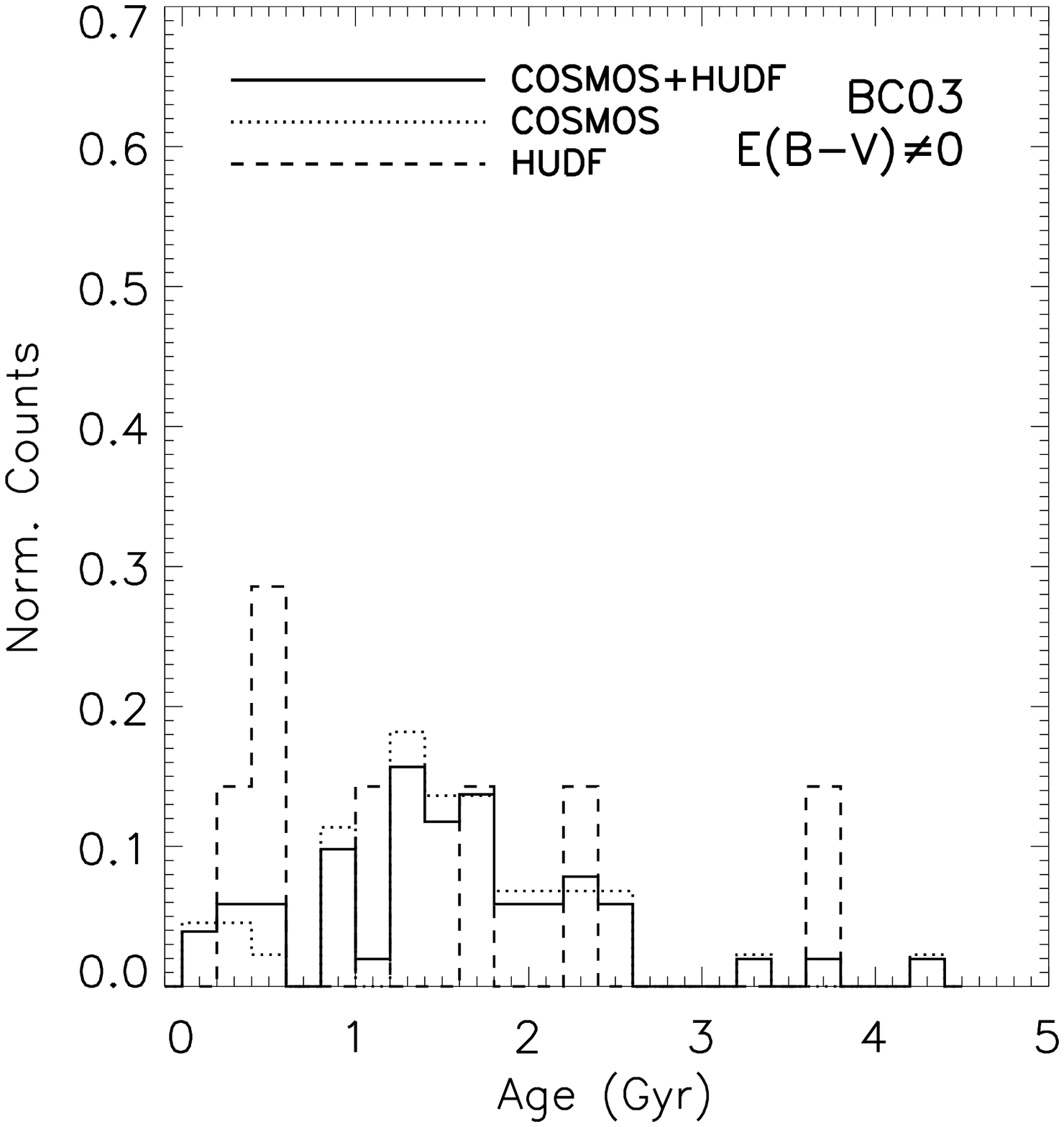}
\includegraphics[width=0.3\textwidth]{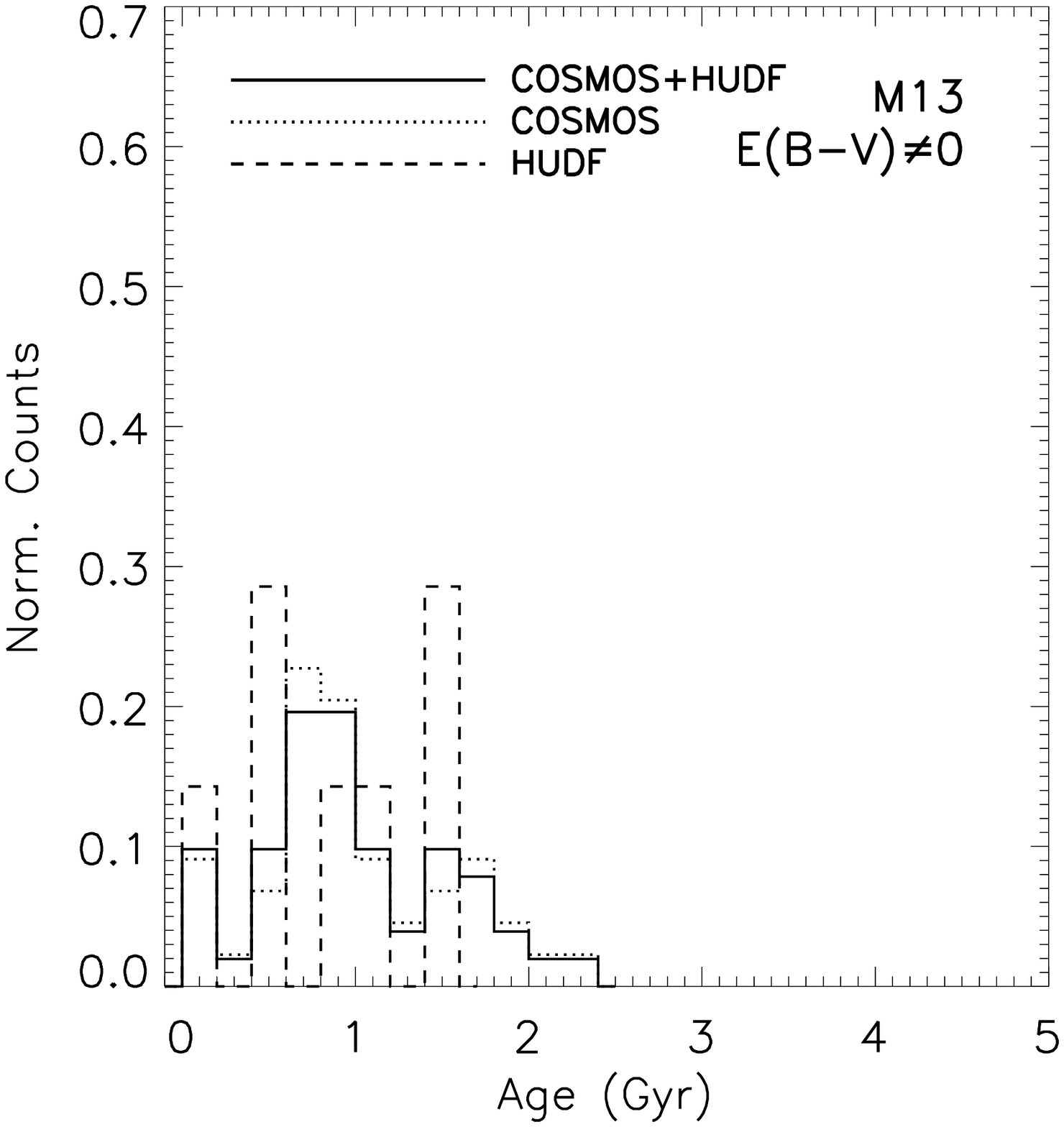}
\includegraphics[width=0.3\textwidth]{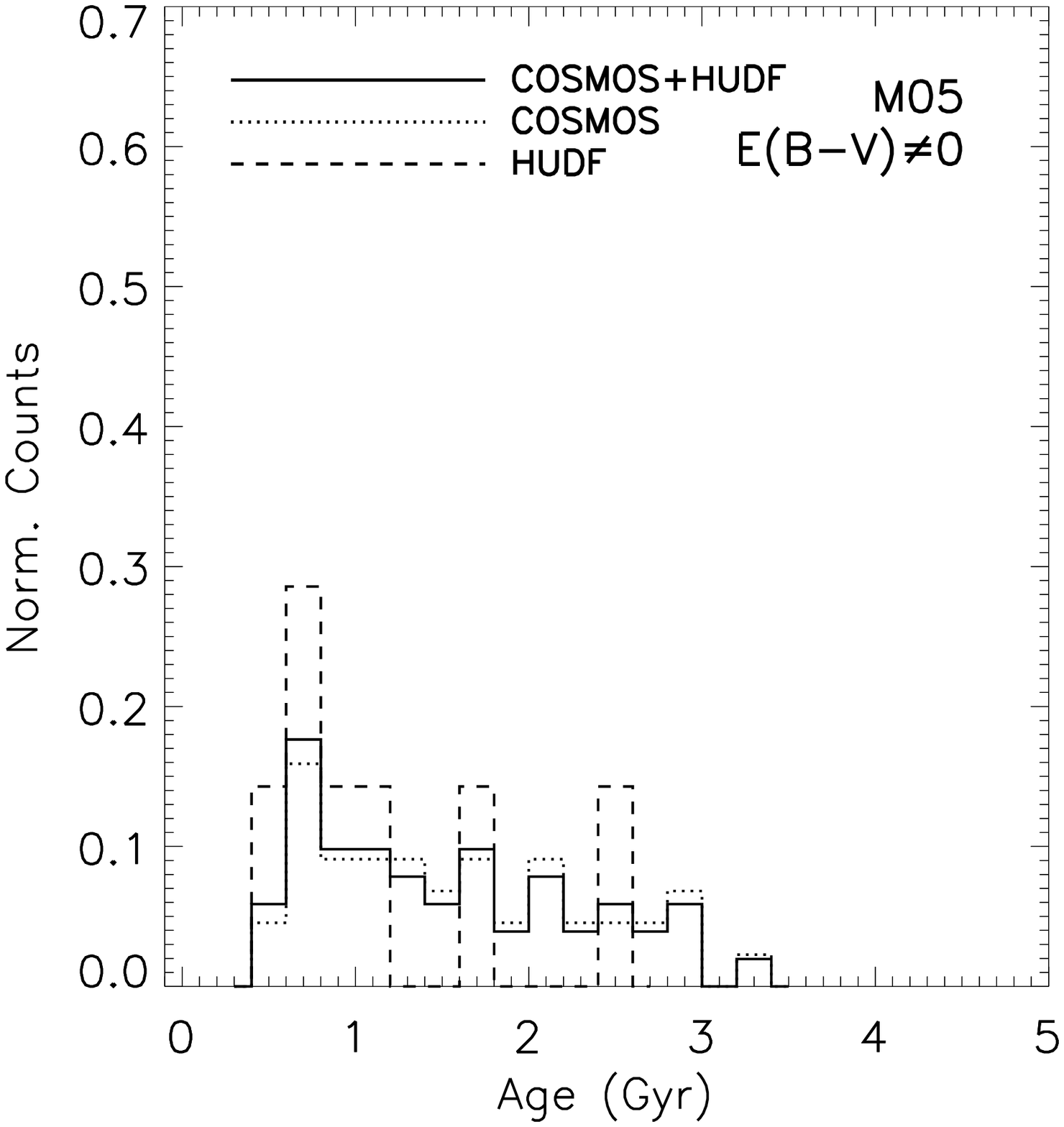}
\includegraphics[width=0.3\textwidth]{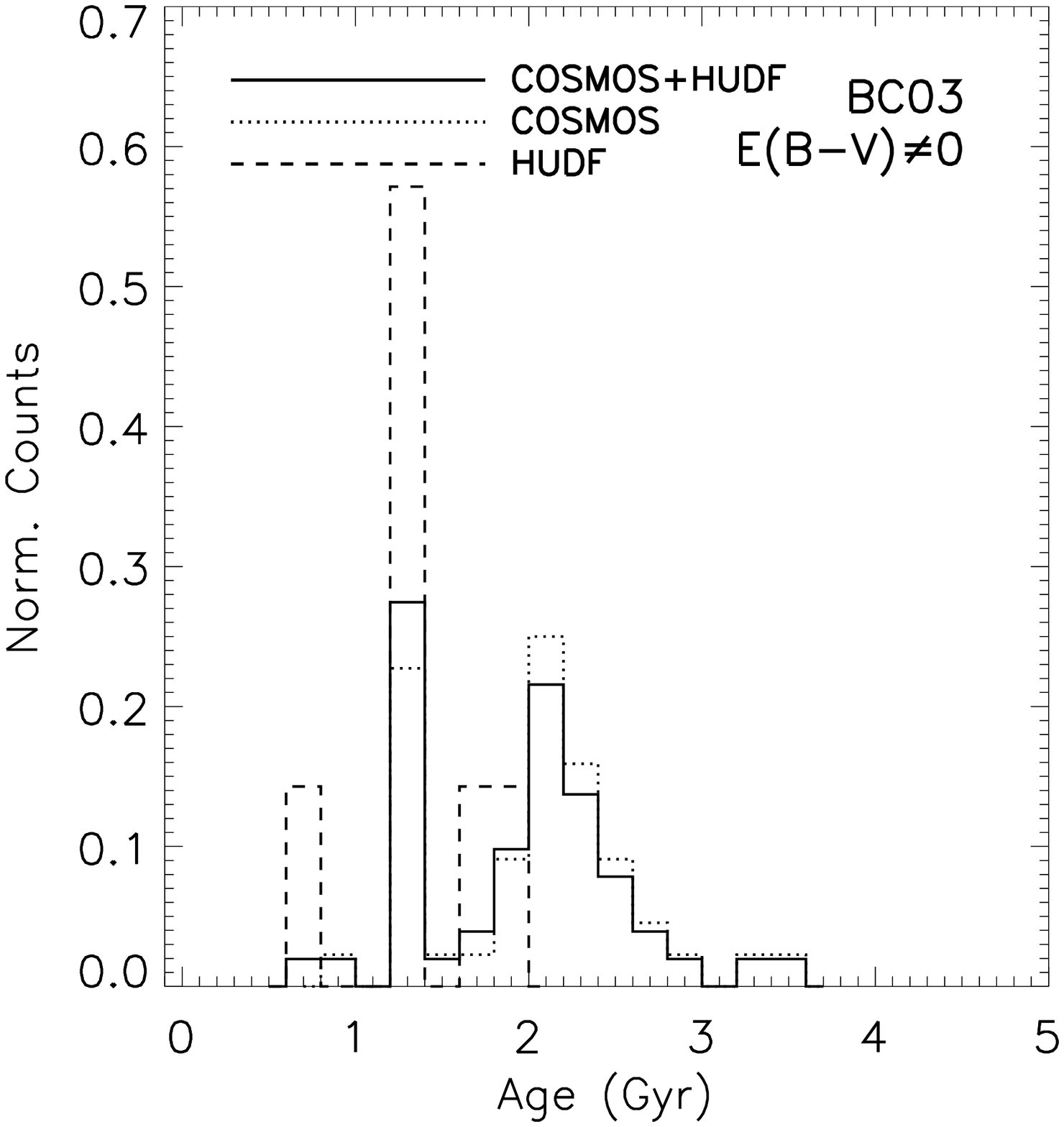}
\includegraphics[width=0.3\textwidth]{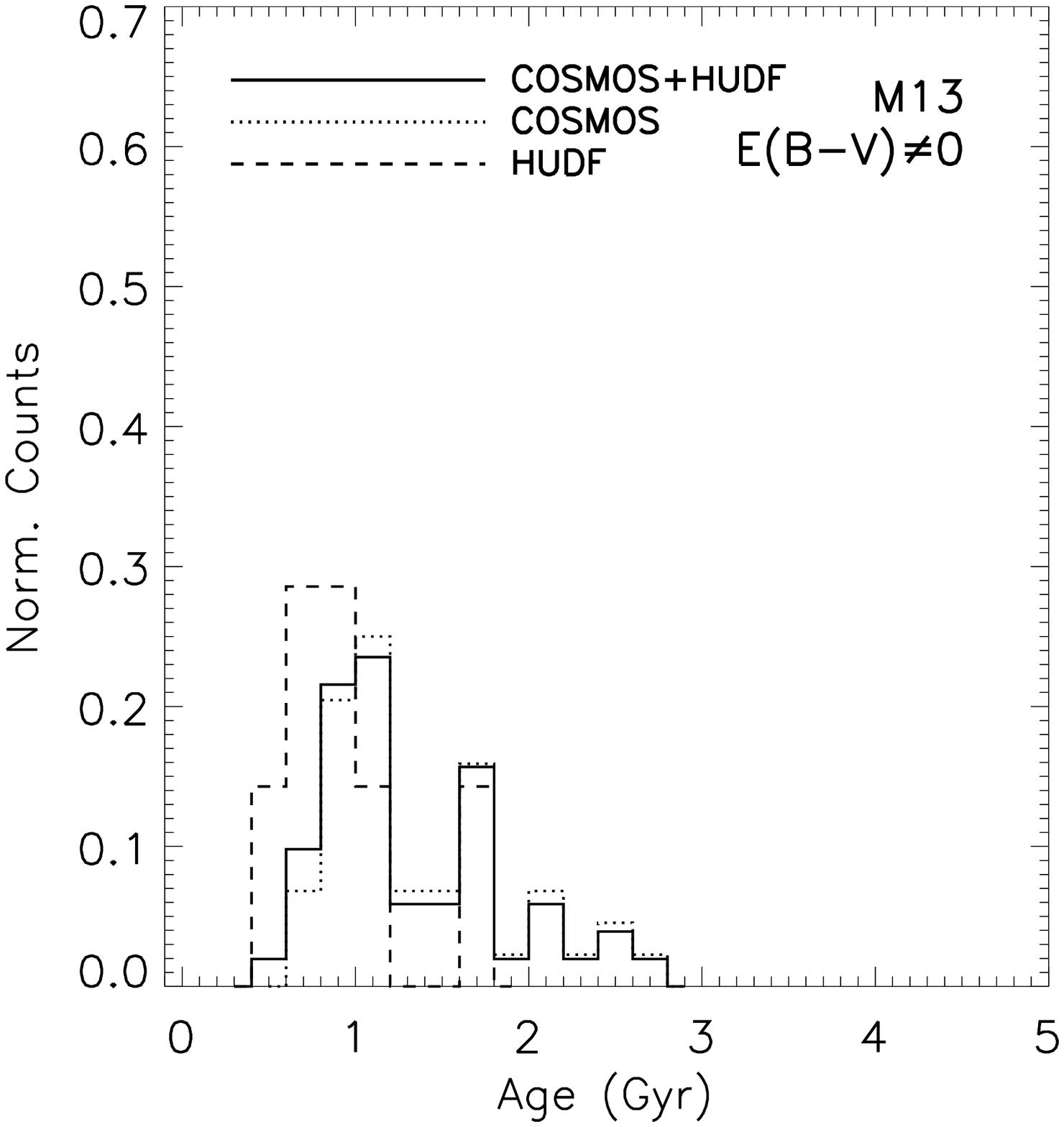}
\caption{Normalised age distributions of galaxies in HUDF, COSMOS and in the total HUDF+COSMOS sample obtained with M05, BC03 and M13 models (left, middle and right-hand panels), for no-reddening - upper plots - and reddening - middle plots. Bottom panels refer to the latter case after performing the solution cuts described in Section \ref{subsec:subsec5.2}.}
\label{fig:Fig12}
\end{figure*}

\begin{figure*}
\centering
\includegraphics[width=0.3\textwidth]{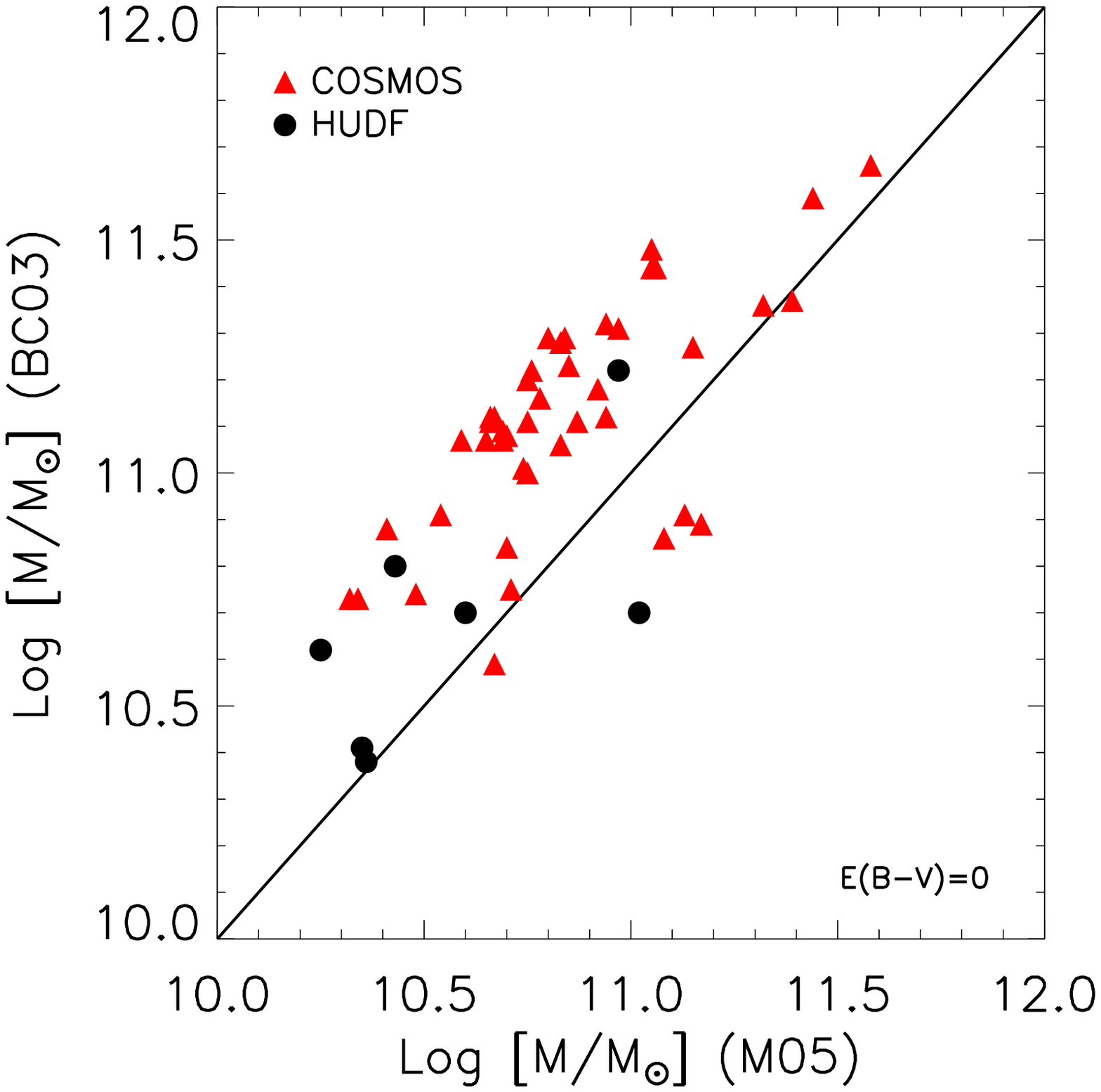}
\includegraphics[width=0.3\textwidth]{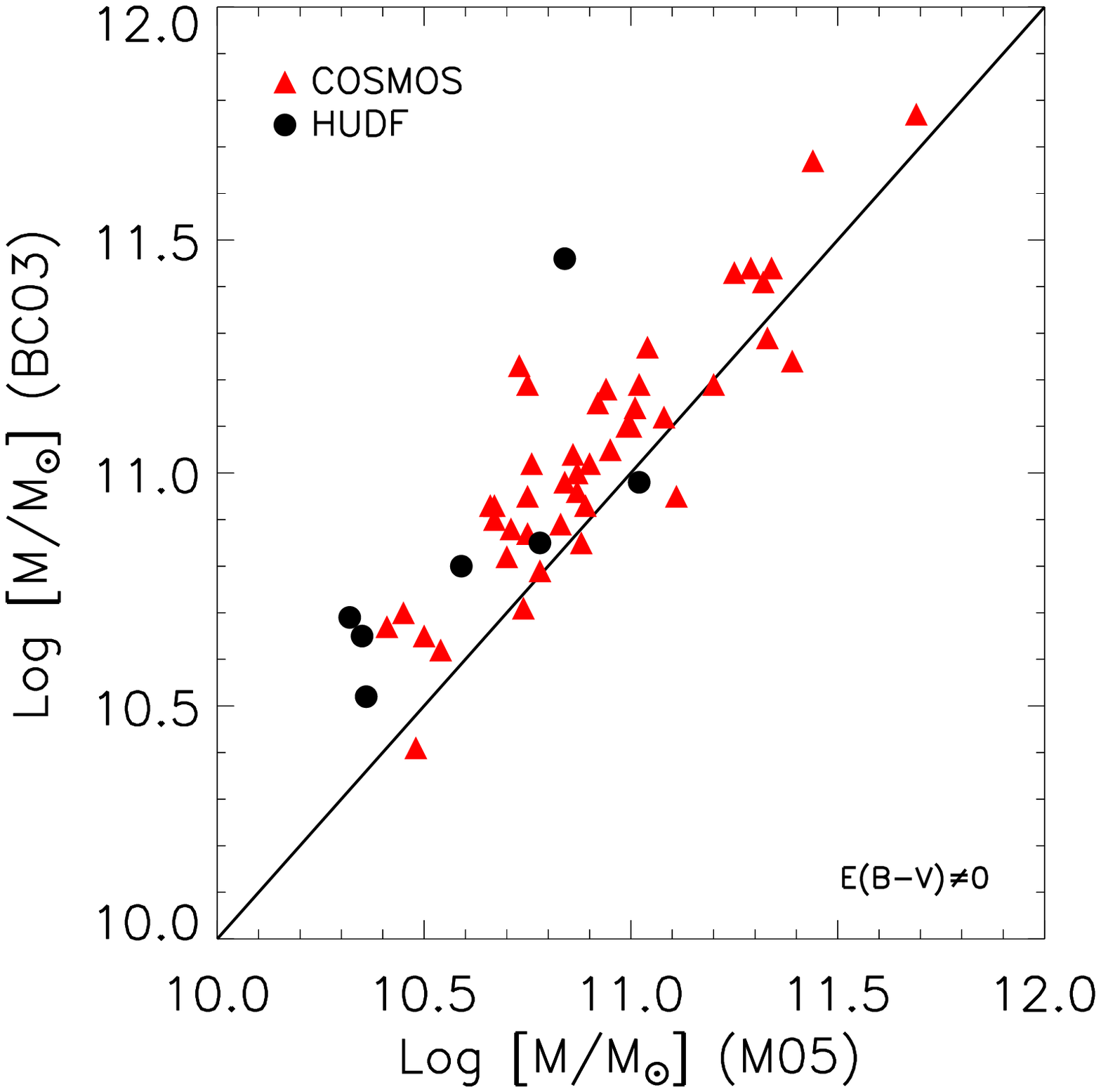}
\includegraphics[width=0.3\textwidth]{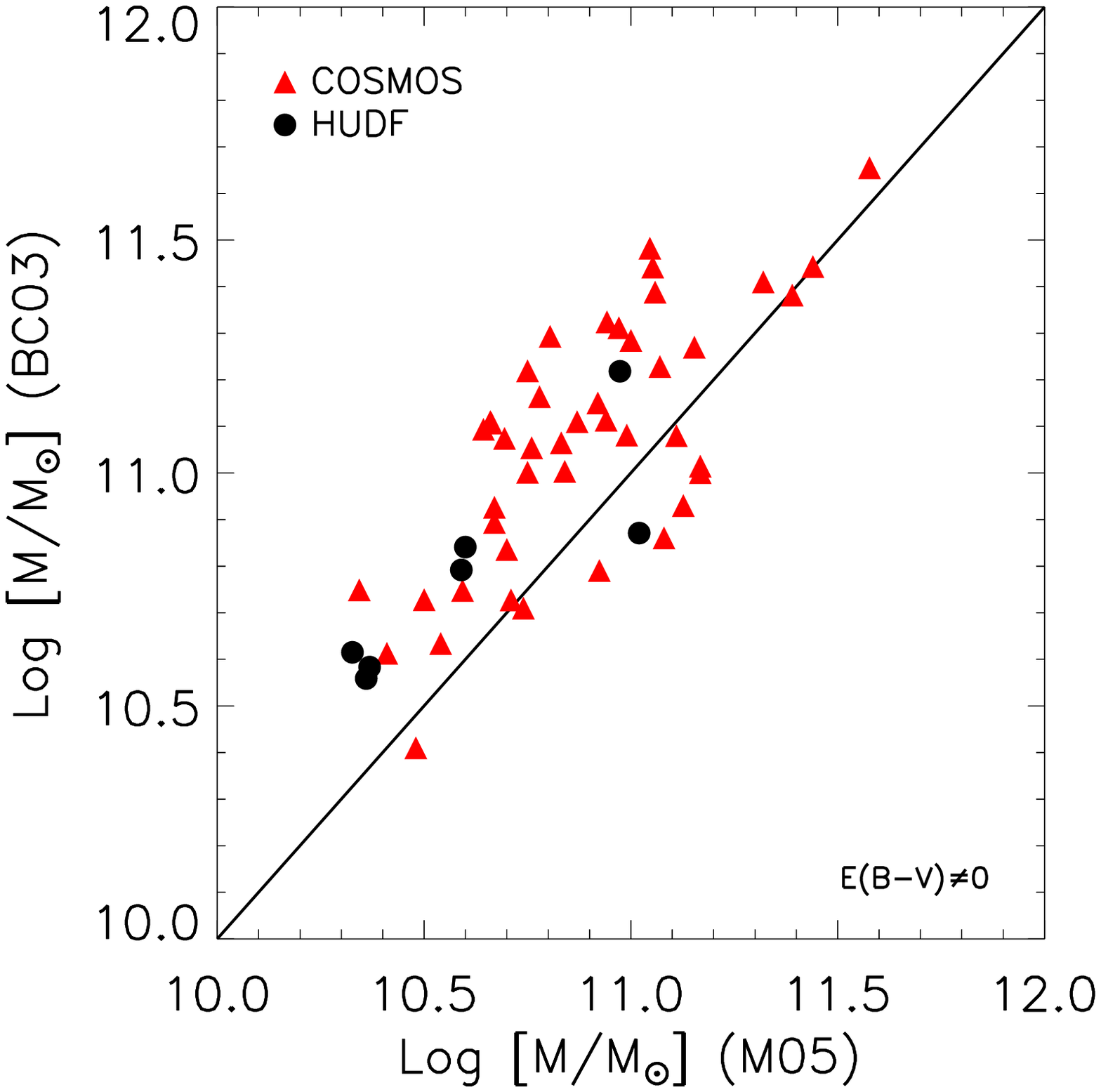}
\includegraphics[width=0.3\textwidth]{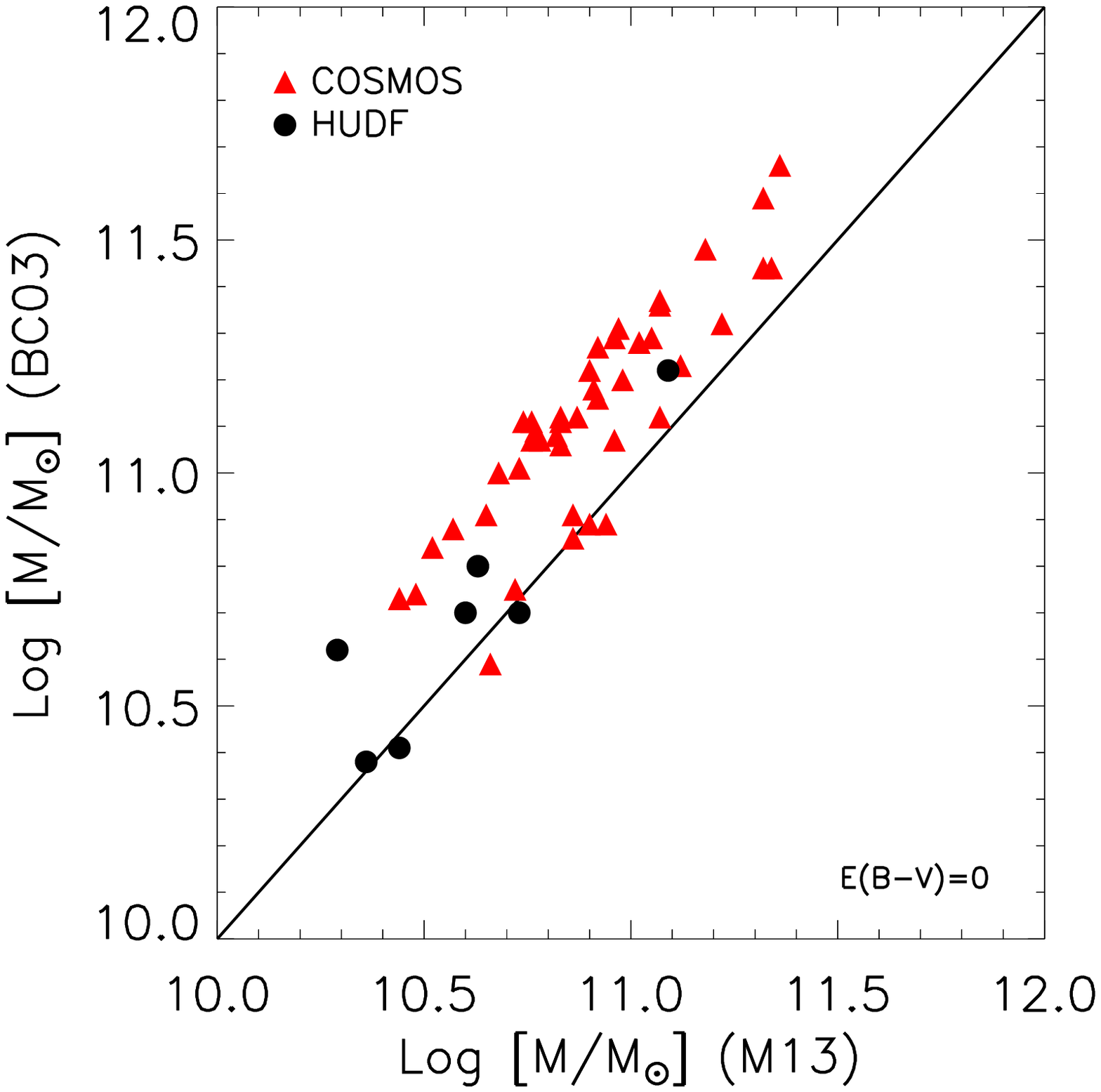}
\includegraphics[width=0.3\textwidth]{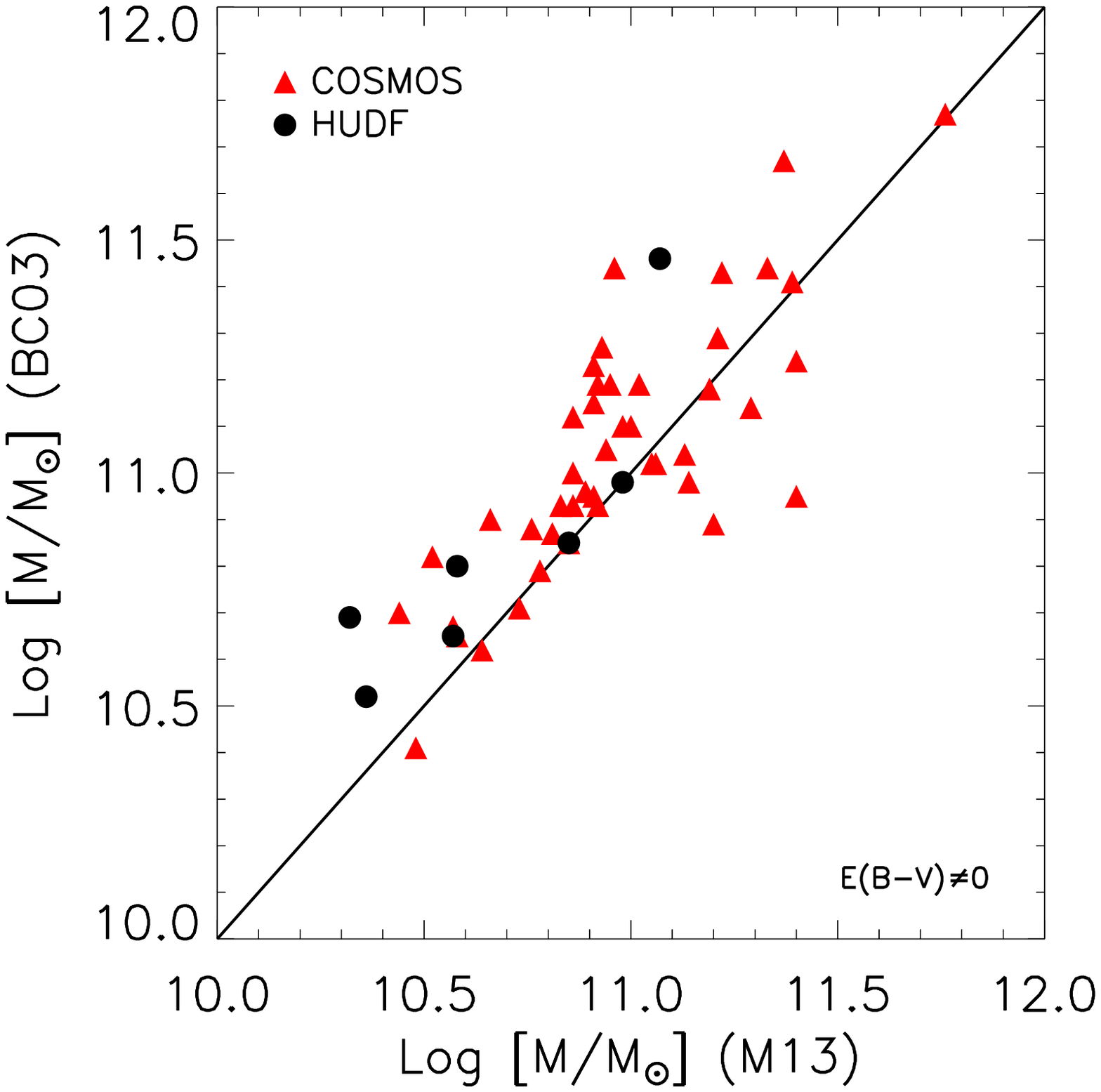}
\includegraphics[width=0.3\textwidth]{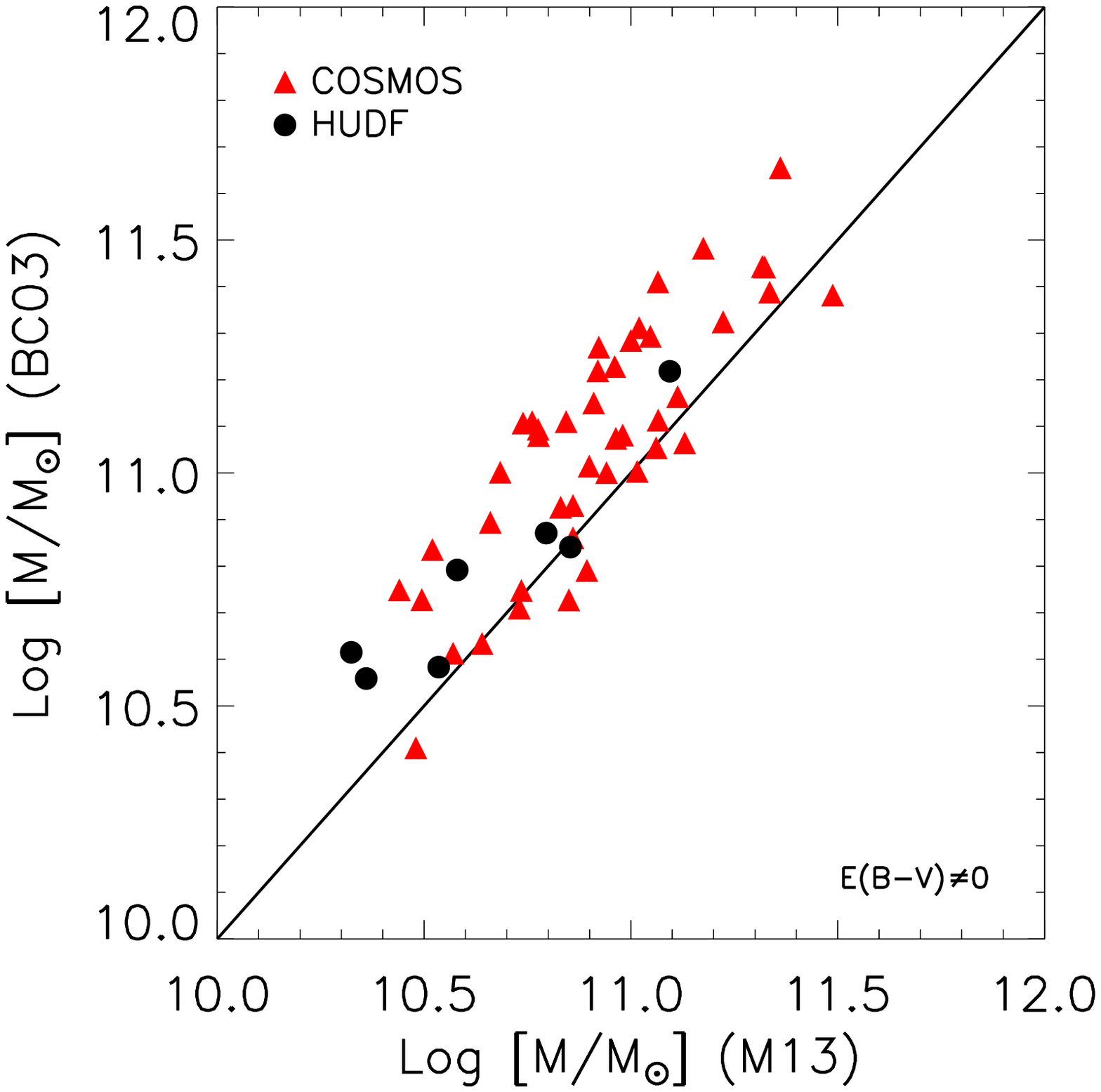}
\includegraphics[width=0.3\textwidth]{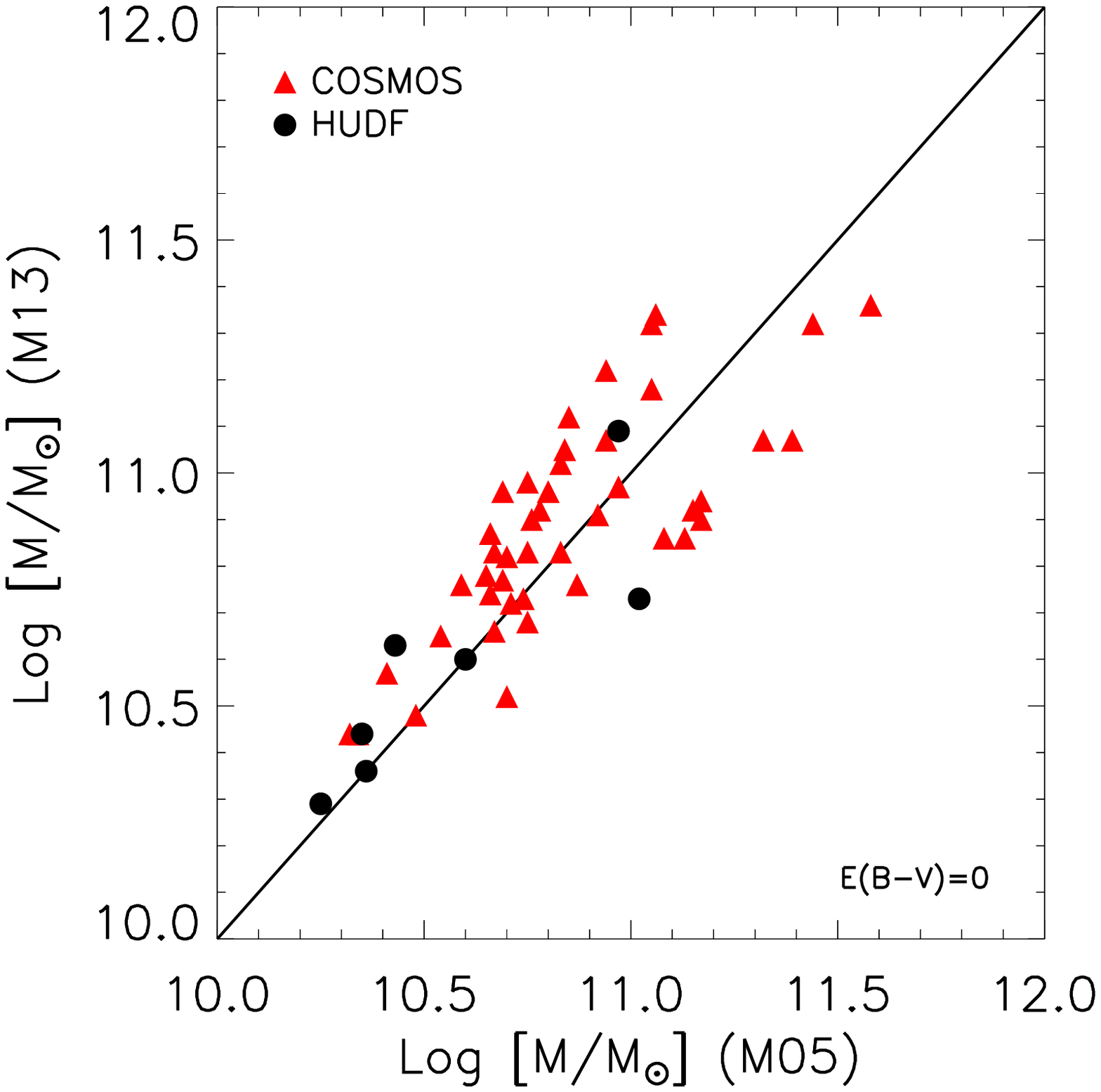}
\includegraphics[width=0.3\textwidth]{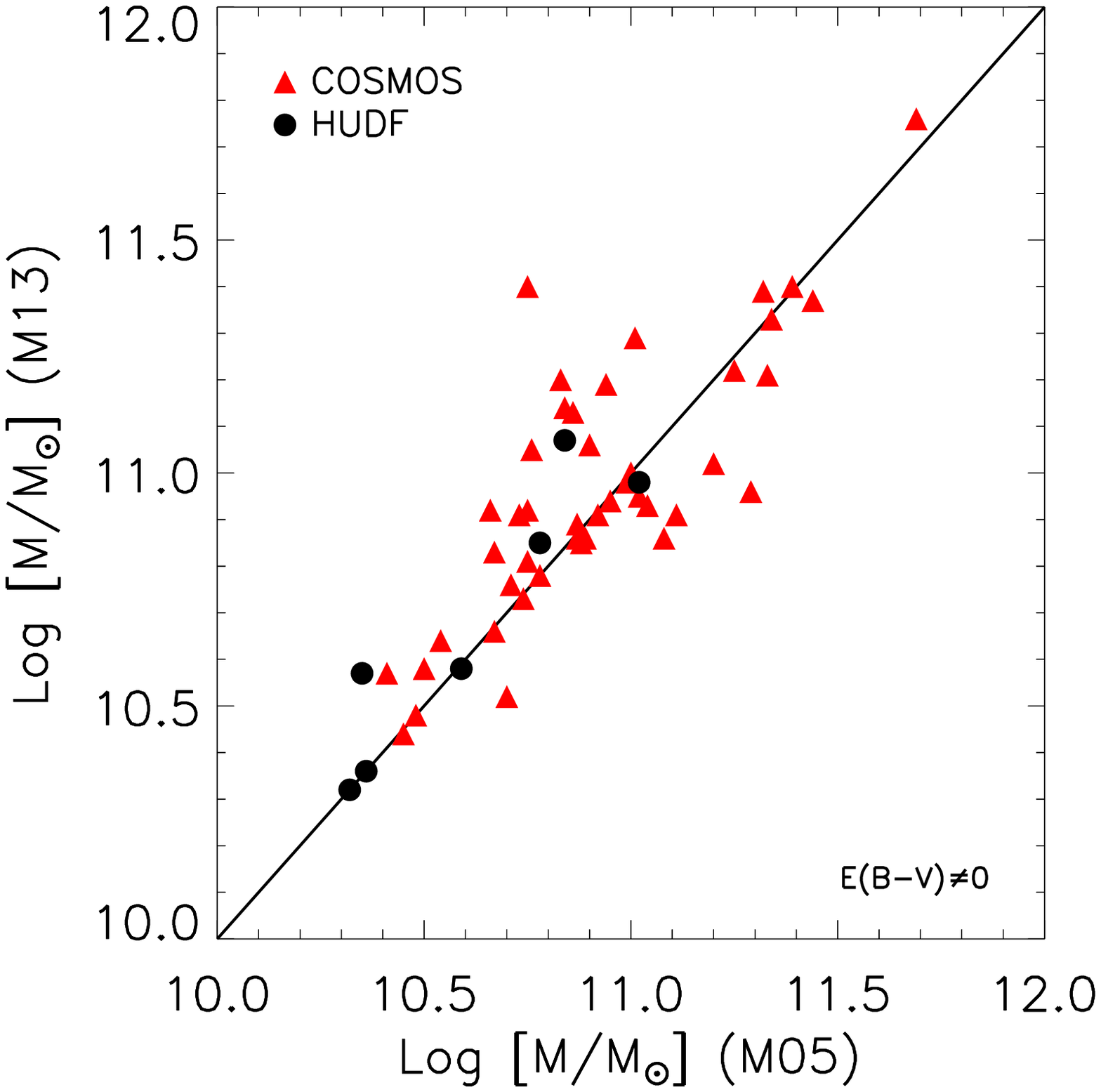}
\includegraphics[width=0.3\textwidth]{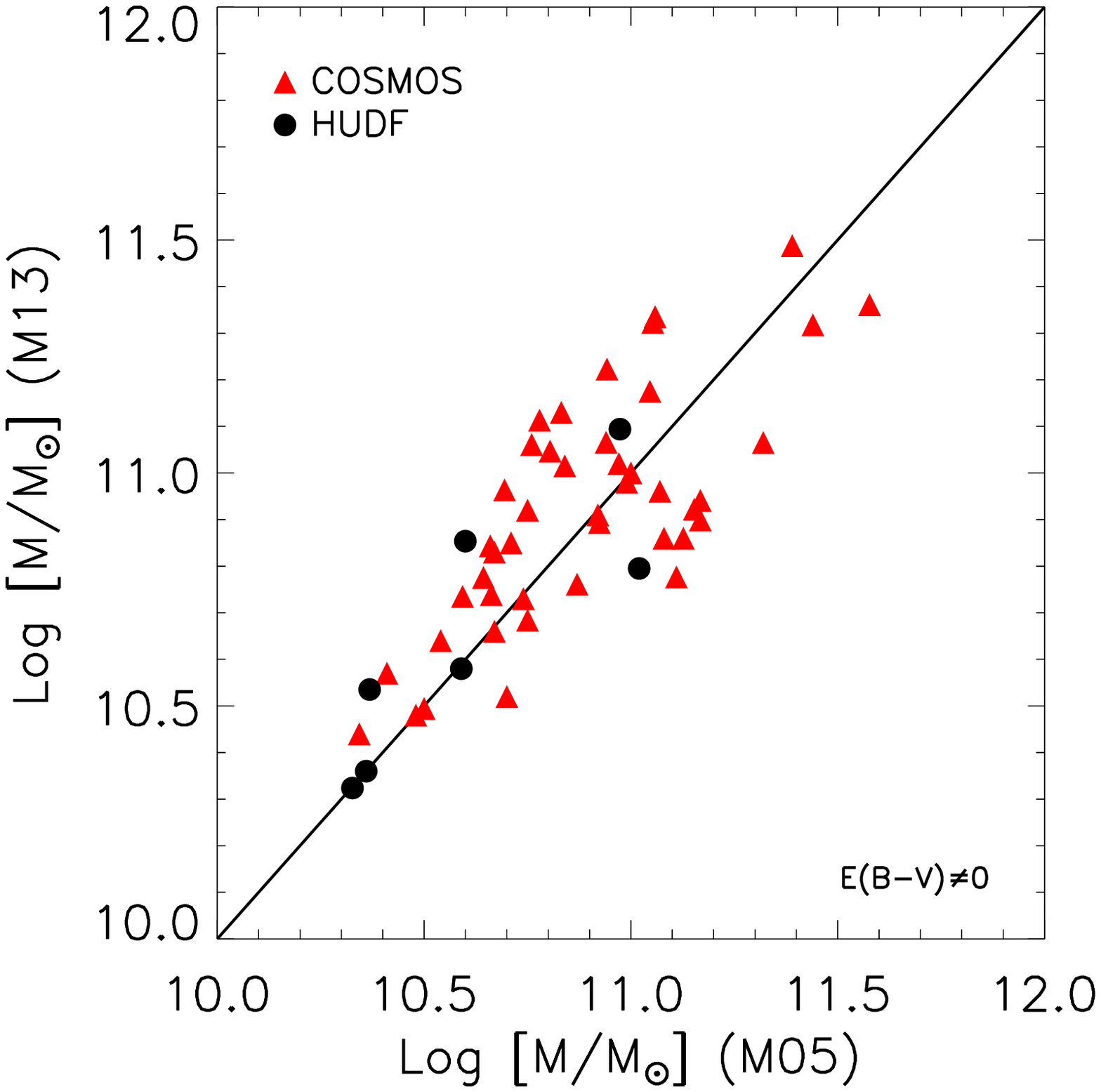}
\caption{Comparison of derived stellar masses for the COSMOS (red triangles) and the HUDF (black dots) samples and for both the reddening-free (left-hand panels) and reddening-included (middle panels) cases. Right-hand panels refer to the latter case after performing the solution cuts described in Section \ref{subsec:subsec5.2}. A one-to-one solid line is over-plotted. The COSMOS data points refer to the photometry uncorrected for zero-point offsets (see Sect. \ref{subsec:subsec4.2}).}
\label{fig:Fig13}
\end{figure*}

Having rejected non self-consistent solutions as described above, we identify 25, 45 and 32 new best-fit solutions respectively for M05, BC03 and M13 models, listed in Table \ref{tab:Table3_appB}. Plots obtained by excluding and including these newly found solutions are shown in Figures \ref{fig:Fig10}-\ref{fig:Fig13}. From Figures \ref{fig:Fig10}, \ref{fig:Fig11} and \ref{fig:Fig12} one can realise that the solutions that are removed are generally young ($age\lesssim 1\ {\rm Gyr}$) and extremely/moderately reddened ($E(B-V) \gtrsim 0.1\ {\rm mag}$). In particular for M13 and M05 models, the discarded solutions are mainly those younger than the TP-AGB phase onset time. For some galaxies (e.g., COSMOS ID 33) the fitting quality becomes quite poor (however for some of them like ID 33 itself, the fit quality was already quite poor to begin with).

The inclusion of these new solutions does not introduce major changes in our results (see Tables \ref{tab:Table5}  and \ref{tab:Table6}). The performances of all models in presence of attenuation remain comparable, with M05, BC03 and M13 finding solutions consistent with the best-fit ones within $1\sigma$ respectively 76, 63 and 55 per cent of the times (see Table \ref{tab:Table5} \& Figure \ref{fig:Fig8}). Also the sample average properties remain statistically unchanged (Table \ref{tab:Table6}). We note though, that after using the constraints from $L_{\rm IR}$, the average $E(B-V)$ of the full sample drops to $\sim0.05 \ {\rm mag}$ for all models compared to $\sim0.15\ {\rm mag}$, obtained before applying such constraints. This is comparable to the average upper limit of $E(B-V)\sim 0.02$ found in Section \ref{subsec:subsec5.1}. 
In addition, for all models the dynamic range changes from $0<E(B-V)\lesssim0.8\ {\rm mag}$ (before using the $L_{\rm IR}$ constraints) to $0<E(B-V)\lesssim0.15\ {\rm mag}$ (after using the $L_{\rm IR}$ constraints). This constitutes a significant change, evident in Figure \ref{fig:Fig10} and Table \ref{tab:Table6}. Note, however, that because of the use of statistical means, our results are valid on average but not necessarily on all individual galaxies.\\

These findings described in Sections \ref{subsec:subsec5.1} and  \ref{subsec:subsec5.2} indicate that the amount of dust in the passive galaxies studied here should be low, at least in the majority of them, and that the inclusion of unconstrained reddening in the SED fitting may lead to unphysical solutions whose lower $\chi^{2}$ values (compared to solutions fitted in absence of reddening) may be just due to the addition of another free parameter (two additional parameters if one allows more than one reddening law). 

\section{Galaxy physical properties}
\label{sec:sec6}
So far we have showed that using the reduced $\chi^2$ as a figure of merit can give some preference for one set of population models over another. However, differences are often relatively small and can also depend on other choices concerning the adopted reddening law and/or star formation history. In addition, we have also shown in Section \ref{sec:sec5} that even solutions with small $\chi^2$  may not be plausible on astrophysical grounds. In this section we discuss the differences in the galaxy properties as derived from the various stellar population models.

Table \ref{tab:Table6} lists the average physical properties of the sample galaxies and Figures \ref{fig:Fig10} to \ref{fig:Fig13} plot the main properties separately for the two samples and with and without allowing for reddening. Figures~\ref{fig:Fig9} and \ref{fig:Fig11} show  the reddening distributions for our samples, both before and after removing some solutions as illustrated in Section \ref{subsec:subsec5.2}. 
 
Figure~\ref{fig:Fig12} shows the resulting age distributions for the three sets of models, without including reddening as a free parameter (top row) and including it 
both before and after purging away unphysical solutions (middle and bottom row, respectively). When no reddening is assumed the age distributions from M05 and M13 differ markedly from those obtained with BC03 models. In the former cases the distributions peak at ages $\lesssim 1$ Gyr, with a tail extending to $\sim 2.5-3$ {\rm Gyr}. In the BC03 case, the age distribution peaks at 2.5 Gyr, with only few galaxies being assigned ages younger than $\sim 2$ Gyr.  When reddening is allowed as a free parameter these differences are all blurred.

These age differences  derive from the combined effect of an earlier TP-AGB and RGB onset in M05/M13 models with respect to the stellar evolution models used in BC03 (these effects are fully discussed in M05 and M06 to which we refer for more details). In particular,  M13 gives the largest number of young ages, BC03 that of older ages. 

In Figure \ref{fig:Fig13} we compare the values of stellar mass $M^{*}$ obtained from the various models and with various assumptions concerning the reddening. The results obtained here confirm those in  M06, \citet{Cimatti-2008} and many other works, namely that BC03 models-based stellar masses are somewhat higher (by $\sim 0.1/0.2 \ {\rm dex}$) than the M05-based ones (upper panels of Figure \ref{fig:Fig13}), especially in absence of reddening. This difference stems from the older stellar ages obtained with the BC03 models (see Figure~\ref{fig:Fig12}).The situation is roughly the same when comparing BC03-derived masses with those from M13 models (central panels of Figure \ref{fig:Fig13}). The values of $M^{*}$~obtained with M05 and M13 models (lower panels of Figure \ref{fig:Fig13}) are similar. 

\begin{figure}
\centering
\includegraphics[width=0.4\textwidth]{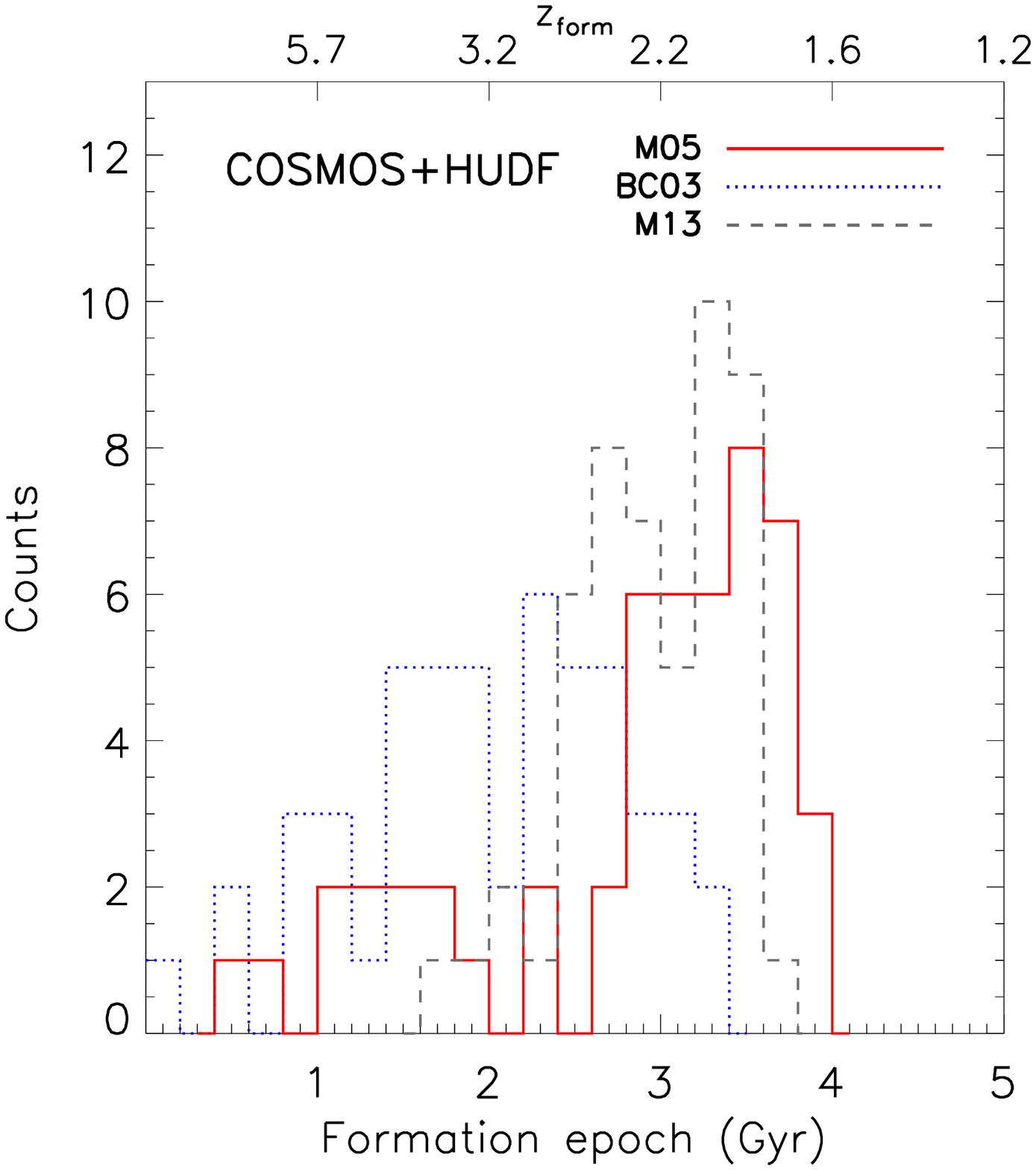} 
\caption{{The formation epoch (cosmic time at which galaxies formed) for the galaxies in our combined HUDF and COSMOS samples as predicted from the ages derived from the BC03 (blue dotted), M05 (red solid) and M13 (grey dashed) models for pure SSP fits. The upper x-axis shows the corresponding formation redshift scale.}}
\label{fig:Fig14}
\end{figure}

Finally, having argued that low/no reddening solutions are to be preferred on astrophysical grounds, we concentrate on this specific case. Indeed, the different age distributions derived from  BC03 and M05/M13 models offer an interesting opportunity to distinguish among them. Figure \ref{fig:Fig14} shows the formation epoch and redshift for our HUDF+COSMOS galaxies as inferred from their ages derived from the three sets of models. We limit ourselves to SSPs because we want to isolate age trends only. Clearly, BC03 models predict these massive, quenched galaxies to have been formed at a much earlier epoch compared to M05/M13 models. The formation redshift peaks at $z\sim 2$ for M05 and M13 models and at $z\gtrsim 3$ for BC03 models. Counts of passively evolving galaxies at these redshifts can then distinguish among these models, something that we plan to undertake in the future.

\begin{table*}
%\begin{tiny}
\begin{center}
\caption{K-S test results for COSMOS sample's galaxy properties distributions. Col 1: compared models ; col 2: inclusion of reddening; cols 3, 5 \& 7: maximum deviation between the COSMOS samples' cumulative distributions of age, logarithm of stellar mass and $E(B-V)$~obtained with 2 different models; col 4, 6 \& 8: two-sided probability of the COSMOS sample's distributions of these quantities obtained with two models of being drawn from different parent distributions. Values in brackets refer to results obtained after cutting out possible unreliable solutions, as described in Section \ref{subsec:subsec5.2}.}
\begin{tabular}{cclllllll}
\hline
  \multicolumn{1}{c}{\bf Compared models} &
  \multicolumn{1}{c}{\bf Reddening} &
  \multicolumn{1}{c}{\bf $\mathbf{D_{\rm Age}}$} &
  \multicolumn{1}{c}{$\mathbf{P_{\rm Age}}$} &
  \multicolumn{1}{c}{\bf $\mathbf{D_{\rm \log(M)}}$} &
  \multicolumn{1}{c}{$\mathbf{P_{\rm \log(M)}}$} &
  \multicolumn{1}{c}{\bf $\mathbf{D_{\rm E(B-V)}}$} &
  \multicolumn{1}{c}{$\mathbf{P_{\rm E(B-V)}}$} \\
  \multicolumn{1}{c}{} &
  \multicolumn{1}{c}{} &
  \multicolumn{1}{c}{} &
  \multicolumn{1}{c}{$(\%)$} &
  \multicolumn{1}{c}{} &
  \multicolumn{1}{c}{(\%)} &
  \multicolumn{1}{c}{} &
  \multicolumn{1}{c}{(\%)} \\\\
  \multicolumn{1}{c}{(1)}&
  \multicolumn{1}{c}{(2)}&
  \multicolumn{1}{c}{(3)}&
  \multicolumn{1}{c}{(4)}&
  \multicolumn{1}{c}{(5)} &
  \multicolumn{1}{c}{(6)} &
  \multicolumn{1}{c}{(7)} &
  \multicolumn{1}{c}{(8)} \\
\hline
\hline
M05 vs BC03       &  \multirow{3}{*}{No}             &  $0.60$            &  $>99$         &  $0.48$         &  $>99$       &  $--$           &  $--$ \\
M05 vs M13        &	        		     &  $0.25$            &  $89 $         &  $0.20$         &  $72 $       &  $--$           &  $--$ \\
BC03 vs M13       &	        		     &  $0.75$            &  $>99$         &  $0.45$         &  $>99$       &  $--$           &  $--$ \\
%\cline{2-9}
\cmidrule{1-8}
M05 vs BC03       &  \multirow{3}{*}{Yes}            &  $0.27\ (0.39)$    &  $94\ (>99) $  &  $0.29\ (0.31)$ &  $97\ (>99)$ &  $0.61\ (0.43)$ &  $>99\ (>99)$ \\
M05 vs M13        & 		                     &  $0.34\ (0.18)$    &  $>99\ (63) $  &  $0.18\ (0.16)$ &  $60\ (48) $ &  $0.32\ (0.33)$ &  $98\ (>99) $ \\
BC03 vs M13       & 		                     &  $0.48\ (0.53)$    &  $>99\ (>99)$  &  $0.20\ (0.29)$ &  $88\ (98) $ &  $0.34\ (0.14)$ &  $>99\ (31)$  \\
\hline
\hline
\end{tabular}
\label{tab:Table7}							    
\end{center}
% \end{tiny}
\end{table*}
\section{Discussion and Conclusions}
\label{sec:sec7}
With now 51 galaxies, the sample analyzed here is to date the largest sample of high-$z$ passively-evolving  galaxies with spectroscopic redshifts which are used  to study evolutionary synthesis models and their effect on galaxy properties. Our analysis also aims at understanding discrepant results in the recent literature (e.g., \citealp{Kriek-2010} \& \citealp{Perez-Gonzalez-2013}).

\subsection{Comparison of the HUDF and COSMOS samples}
\label{subsec:subsec7.1}

In this section we compare with each other the results obtained for the two galaxy samples studied here, even if the HUDF sample is certainly affected by small number statistics.

With regards to this sample, we confirm the results by Maraston et al. (2006) that M05 models perform better than other models when reddening is not included in the SED fitting 
while all models have comparable $\chi^{2}_{\rm r}$ values when reddening is included. 

The results we obtained on the COSMOS sample are similar to those for the HUDF one, namely M05 models perform better (70 per cent of the sample) than the other models for the no-reddening case.  However, when reddening is included in the fitting procedure, the BC03 models are preferred for about half (52 per cent) the sample (however this figure drops to 32 per cent after applying the $L_{\rm IR}$-based cuts described in Section \ref{subsec:subsec5.2}), while for the HUDF sample M05 was preferred for 60 per cent (71 per cent after applying the same $L_{\rm IR}$-based cuts) of the galaxies. Despite this, even for the COSMOS sample the $\chi^{2}_{\rm r}$ values obtained with the three models in the reddening case are similar.

The average values of $E(B-V)$ and age distributions for the two samples seem consistent with each other at  $\sim1\sigma$ level (see Table \ref{tab:Table6} and Figures \ref{fig:Fig11} and  \ref{fig:Fig12}).  This consistency, however, is likely the result of sparsely sampled distributions (HUDF sample in particular), which lead to relatively large dispersions. In fact, when analysing the mass and age distributions of the two samples  via K-S tests\footnote{So measuring the maximum deviation of the parameters' cumulative distributions (D) and its significance (P).}, for the no-reddening case one obtains: i) $D_{\rm \log Mass}=0.74$ and $D_{\rm Age}=0.84$, both with $P>99$ per cent in the BC03 case; ii) $D_{\rm \log Mass}=0.58$ with $P>97$ per cent for the M05 case; iii) $D_{\rm \log Mass}=0.63$ with $P>99$ per cent for the M13 case. This means that the mass and the age (the latter only for BC03 models) distributions of the HUDF and the COSMOS samples are statistically different (i.e. drawn from different parent distributions) in the no-reddening case. However, given that the HUDF sample only counts 7 galaxies, the significance of the K-S test is certainly not stringent. When analyzing the cases where reddening is included, no statistically significant differences are found between the two galaxy samples. Significant differences between the age (for M13 and BC03 models) and mass (for M05 and BC03 models) distributions of the two catalogues emerge only after excluding non self-consistent, unphysical solutions, as described in Section \ref{subsec:subsec5.2}.

\subsection{Comparison of Stellar Populations Models}
\label{subsec:subsec7.2} 

For both samples,  BC03 models are generally associated with older ages or with higher reddening values compared with M05- and M13-based results. This offers an opportunity to distinguish between such models as they predict significantly different number densities of passively evolving galaxies beyond redshift $\sim 2$ or 3,
especially in the case of no reddening. With regards to stellar masses, while M05 and M13 models-based values are very similar, those based on BC03 models are higher, as well known in the literature. Photometric redshifts derived with the three models find M13 giving the best recovery of spectroscopic redshifts, with an accuracy $\sim1.3$ and $\sim1.6$ times better than those derived using M05 and BC03 models. 

With the COSMOS sample, we can explore the impact of different models on the statistical properties of our galaxy sample. Table \ref{tab:Table7} reports the results of  K-S tests performed on the mass, age and reddening distributions obtained with different models. One notices again that when fits are performed without reddening, there are significant differences in the age and mass distributions obtained with BC03 models compared to those yielded by M05 or M13 models, which are instead quite similar to each other. This is so because the TP-AGB fuel consumption in M13 is  just a reduced version of that adopted in M05, all the rest being the same. Thus,  without reddening as a free parameter these two models behave similarly. It is only when reddening is included that statistically significant differences appear between M05- and M13-based galaxy properties. This is so because, depending on the galaxy SED and the accessible age range set by its redshift, M13 models need a combination of AGB flux and reddening to best adjust to the observed galaxy SEDs. M05 models instead need to resort more rarely to reddening,  due to their higher TP-AGB contribution. In fact, note that when reddening is included in the analysis, 43 per cent (19 galaxies) of the COSMOS sample is associated with $E(B-V)=0$ when M05 models are used. This figure is 16 (7 galaxies) and $\sim 2$ (1 galaxy) per cent, respectively for M13 and BC03 models.  When comparing M05 and M13 models with BC03 ones, the differences seen when excluding reddening are still statistically significant apart from the M05 and BC03 age distributions and the M13 and BC03 stellar mass distributions (after purging for unphysical solutions, differences between the latter distributions appear). The reddening distributions result instead statistically different in the majority of the analysed cases (apart from the comparison between BC03 and M13 models after applying the $L_{\rm IR}$-based cuts as in Section \ref{subsec:subsec5.2}). According to the results in Table \ref{tab:Table7}, statistical differences are mainly seen between M05 and BC03.

The M13 models, which have roughly 50 per cent less fuel in TP-AGB than M05, do not fit the data substantially better than the old M05 models. This suggests that star clusters and galaxies may speak different languages, maybe due to the fact that clusters have too low a mass to properly sample short evolutionary phases such as the TP-AGB.  Note also that the age calibration of \citet{Noel-2013} used to fix the M13 models is not derived from fully homogeneously-measured star-cluster ages, even if it is the most homogenous one that could be found in the literature. 

\subsection{Comparison with Some Previous Results}
\label{subsec:subsec7.3}
\citet{Kriek-2010} selected allegedly ``post-starburst" galaxies for being bound within a narrow range of  rest-frame $UVB$ colours, namely $0.98\lesssim (U-B)\lesssim 1.02;\; 0.50\lesssim (B-V)\lesssim 0.58$, with galaxies being spread over a wide range of (photometric) redshifts between $\sim 0.7$ and $\sim 2$. Data were taken from  a photometric galaxy catalogue based on the NOAO Extremely Wide-Field Infrared Imager (NEWFIRM) medium-band survey (NMBS, \citealp{vanDokkum-2009}). SED fits over the optical part of the composite spectrum indicated an age of $\sim 1$ Gyr when using either the BC03 and M05 models, hence coinciding with the expected maximum contribution of TP-AGB stars in M05 models. When extending the fit to include the rest-frame near-IR the M05 models grossly overestimate the near-IR flux whereas the BC03 models provide an excellent fit (with $\chi_{\rm r}^2=0.73$ vs. 5.08 with M05 models). Thus, not much contribution from TP-AGB stars seems to be present in these galaxies.

Kriek et al. used photometric data which include model-dependent zero-point offsets \citep{Whitaker-2011}, in much the same fashion as in \cite{Muzzin-2013}, and therefore may have introduced a similar bias (in favour of BC03 models). Photometric redshifts are especially uncertain in the range $1.4\lesssim z\lesssim 2$, so erroneous redshifts may dilute somewhat the TP-AGB signal, but this is unlikely the reason for the strong discrepancy between M05 and BC03 results.
More problematical is the choice of the star formation history of these starburst galaxies, for which Kriek et al. adopt an exponentially declining SFR. 
Post-starbursts at different redshifts may contain a largely different mass  in the youngest (burst)  component relative to  the much older bulk of the  galaxy, especially at lower redshifts. For example,  a post-starburst galaxy may just have 10 per cent of mass in a post-starburst phase, which makes the contribution of the TP-AGB barely detectable. Note that at high-redshift instead, when galaxies are young, the concept of post-starburst coincides with what is classified as `passive' where a major fraction of the galaxy mass is made by $\sim 1$ Gyr old stars, hence their TP-AGB contribution should be maximal.  For these reasons we opted for $z\gtrsim1.5$ passive galaxies for assessing the contribution of the TP-AGB. 
Given the discrepancy between our and the Kriek et al.  results, one interesting exercise would consist in fitting with our method their six galaxies with spectroscopic redshifts, especially exploring different star formation histories. Unfortunately, these galaxies cannot be selected from the publicly available NMBS catalogue\footnote{\url{http://www.astro.yale.edu/nmbs/Data_Products.html}}, because the synthetic colours used for the post-starburst selection are not available in the database. 
Even if our SED-fitting analysis inclusive of reddening shows some differences between M05 and BC03 models, we do not see the large discrepancies between models quoted by \citet{Kriek-2010} for their post-starburst galaxy sample. 

\citet{Zibetti-2013} explored the possibility of substantial old stellar populations ``hiding'' behind the post-starburst population in 16 SDSS galaxies at low-$z$ and found that it was possible  to reconcile the optical-{\it NIR} colours with the BC03 models but not with the M05 ones. As emphasized several times, the two sets of models differ substantially (for their TP-AGB treatment) only for ages in the range $\sim 0.5-2$ Gyr. Hence synthetic composite stellar populations mimicking post star-burst galaxies and made of stars younger of $\sim 0.5$ Gyr (the burst component)  and older than $\sim 2$ Gyr should be nearly identical using either the BC03 or the M05 models. So, the issue reduces to the precise  distribution of stellar ages in the 16 galaxies studied by Zibetti et al., which may still be outside the small number of combinations they have explored.

\citet{Perez-Gonzalez-2013} studied a sample of 27 passive galaxies (as indicated by D4000 and Mg$_{\rm UV}$ features) in the GOODS-N field with spectroscopic redshifts in the range $1.0<z_{\rm spec}\lesssim1.4$, by using photometry taken from a set of 24 SHARDS (Survey for High-$z$ Absorption Red and Dead Sources) medium-band filters in the UV/optical part of spectrum ($5000<\lambda<9500\ {\rm \AA}$) coupled with available ground- and space-based broad-band photometry. They explored several evolutionary population synthesis models, including M05 and BC03 [in addition to PEGASE and unpublished Charlot \& Bruzual, 2009 (hereafter CB09)] and different IMFs (i.e., \citealp{Chabrier-2003} and \citealp{Kroupa-2001}).

They found that BC03 models with a Chabrier IMF provided the best fits for ~95 per cent of their sample, while second best-solutions are given by  M05 (with a Kroupa IMF) for 93 per cent of their galaxies. Also in their case the reduced $\chi^{2}$ of the various models (in particular of BC03 and M05) were not statistically different for individual galaxies.

Our results for the COSMOS sample are qualitatively consistent with those of \citet{Perez-Gonzalez-2013}, in presence of reddening. However, we find a substantially lower fraction of BC03 best fitting models, i.e. $\sim 50$ per cent compared to 95 per cent in P\'{e}rez-Gonz\'{a}lez et al.. 
This may be due to the fact that they  fixed both the SFH and the reddening law, respectively to an exponentially declining model and to a \citet{Calzetti-2000} law, whereas  we allow for four different SFHs and 5 different reddening laws (see Section \ref{sec:sec3}).  It is worth noting that \citet{Perez-Gonzalez-2013} found  that BC03 models lead to systematically higher extinctions than all other models, in agreement with our findings.
Also in this case, it would be interesting to study their galaxy sample with  our method, which also includes the option of  no reddening at all. 

Despite the general agreement, one result stands at odd, namely that BC03 models (with Chabrier IMF) are found to give less massive and younger galaxies than all the other models, including M05 ones. This is in contrast with our and many other findings in the literature, which report  BC03 models yielding on average more massive and older galaxies than M05 models as shown in Figure \ref{fig:Fig13}. The use of different IMFs  (Chabrier for BC03 and Kroupa for M05) should not be responsible for such difference, according to the simulations of \citet{Pforr-2012} (see their Table B1). Another possibility is the non-uniform SED sampling that results from using both medium- and broad-band filters, as  in P\'{e}rez-Gonz\'{a}lez et al. medium-band filters  are all at $\lambda<10^{4}\ {\rm \AA}$. The inclusion of many additional bands only in the rest-frame UV/optical  is likely to dilute the effect of the TP-AGB, as discussed in Section \ref{subsec:subsec4.3}. 

Similarly to what done in this paper, \citet{Perez-Gonzalez-2013} tested the effect of the medium-band filters on galaxy SED fitting by removing them and re-fitting their galaxies with broad-band filters only.  They found that M05 models performed best for 20-30 per cent of their sample and that TP-AGB light models achieved the best results for $\sim50-70$ per cent of the same sample\footnote{ BC03 with Chabrier IMF, CB09 with Kroupa and Chabrier IMFs and PEGASE, whose best-fit fractions are respectively 15, 20-30, 20-30 and 8 per cent.}. This confirms a lower performance of TP-AGB light models when medium-band filters are removed. 

\citet{Perez-Gonzalez-2013}  used one TP-AGB-heavy model set (M05) and 4 TP-AGB-light model sets,whereas we only use one representative model set for each class of models (TP-AGB heavy, mild and light, i.e. M05, M13 and BC03). Restricting the  comparison to only  BC03 and M05 models, one finds that in  P\'{e}rez-Gonz\'{a}lez et al. BC03 models perform better than M05 for $100$ per cent of these galaxies when all filters are used, whereas their performance reduces to $\sim 33-45$ per cent when only broad-band filters are considered, compared to $\sim 55-67$ per cent achieved by M05 models, once more confirming that M05 models perform somewhat better when optical medium passbands are not used.

\citet{Perez-Gonzalez-2013} interpreted their result as evidence that all models (except PEGASE) almost equally fit SEDs based on broad-band photometry, while BC03 ones better fit the UV/optical (in particular the rest-frame range $2000\lesssim \lambda \lesssim 4000\  {\rm \AA}$) when medium-band photometry is included as well. So, it is not clear whether in this case the better performance of BC03 models is due to a better rendition of the optical SED or of the (TP-AGB sensitive) near-IR SED.

\section{Summary }
\label{sec:sec8}
This study revisits the debated contribution of Thermally-Pulsating AGB stars to galaxy SEDs, for calibration of evolutionary synthesis models. In particular, we investigate models by \citet{Bruzual-2003}, \citet{Maraston-2005} and a new set of Maraston's models with reduced TP-AGB (M13).
 We have re-analyzed the high-$z$ ($1.39<z_{\rm spec}<2.67$) galaxies previously studied  in M06 (HUDF sample) with the new, more accurate and extended photometry available in the CANDELS Multi-Wavelength Catalogue. Furthermore, we have enlarged the sample with additional 44 spectroscopically confirmed, high-$z$ ($1.3<z_{\rm spec}<2.1$) galaxies taken from the COSMOS \citep{Scoville-2007} field and with multi-band photometry from the COSMOS/UltraVISTA catalogue \citep{Muzzin-2013}.  We also aim at understanding some conflicting results in recent studies of high-$z$ galaxies in the literature (e.g., \citealp{Kriek-2010}, \citealp{Perez-Gonzalez-2013}).

The two samples studied here  count 51 (7+44) galaxies with spectroscopic redshifts in the range $1.3\lesssim z\lesssim 2.67$. The passive nature of these galaxies is confirmed by spectroscopic evidence (\citealp{Daddi-2005}, Gobat et al., in prep.) based on the D4000 break and/or the Mg$_{\rm UV}$ absorption features and by their location in the zone of 'passive' galaxies in two-colour plots such as the {\it BzK} and rest-frame {\it UVJ} plots. However, the wide range of template star formation histories allowed by the SED fits, still leaves the possibility to prefer a star-forming template if their SEDs still indicates ongoing star formation, but this choice was made only in a few cases.  We consider three different sets of stellar population models which encompass the widest range of TP-AGB contribution  presently available. The Maraston (2005; M05) models which have been referred to as  ``TP-AGB-heavy'', the Bruzual \& Charlot (2003, BC03) models, which are  ``TP-AGB-light'', and a new set of models with a TP-AGB contribution which is intermediate between the two (M13 models). 

We investigate effects that can influence the results, encompassing inclusion and exclusion of reddening, choice of reddening law, inclusion/exclusion of intermediate-band filters in the optical domain and a variety  of assumption concerning the star formation history. We  also considered the possible bias induced by heterogeneous SED sampling and model-dependent zero-point photometric corrections applied to the raw data. 

Our main results can be summarised as follows:

\begin{itemize}

\item[(i)]  In absence of attenuation as a free parameter in the SED fit,  M05 models, with their strong TP-AGB, best reproduce the majority ($72$ per cent) of the observed galaxy SEDs in our sample, with solutions which are $\gtrsim 80$ per cent of the times within $1\sigma$ from the best-fit ones.\\

\item[(ii)] 
SED-fits including reddening as an additional free parameter, find all sets of models performing almost equally, with a marginal preference for BC03. The SED fits converge to systematically different ages, reddening and masses for different models, with BC03 giving typically older ages, higher masses and reddening than M05 models.  This result is robust against the choice of attenuation law, in the sense that even when focusing only on a steep reddening law like the SMC one we do not see significant changes in these results.  Our results are also robust against degeneracies, whose influence has been studied by comparing probabilities marginalized over the involved fitting parameters and by assessing the  physical reliability of solutions in comparison with observation-based limits.\\

\item[(iii)] 
We provide  plausibility arguments  for this class of {\it passive} galaxies being indeed affected by low reddening/extinction, as their  flux from the mid- ({\it Spitzer/MIPS}) to the  far-IR ({\it Herschel}/{\it PACS}) is very low.\\

\item[(iv)]
The results for M13 models sit in between those for M05 and BC03. In particular they do not provide better fits than those of  M05 models, despite  M13 ones having been calibrated using the most recent age determinations for the onset of the TP-AGB stellar phase in Magellanic Cloud star clusters and a new estimate of their average $V-K$ colours (as in Conroy \& Gunn~2010 and No\"el et al. 2013).\\

\item[(v)] Our analysis of the COSMOS sample shows that model-dependent recalibration aiming to correct errors in photometric zero-points is not able to completely remove systematics in the photometric-redshift determination of our galaxies. Moreover, we argue that, being model dependent, it biases the results by favouring the same kind of models used to evaluate the zero-point offsets themselves. Specifically, we show that when TP-AGB-light models are used as templates, the resulting zero-point offsets make our passive galaxies similarly TP-AGB-light. We argue that when assessing the performances of different sets of models, model-dependent photometric re-calibrations should not be applied.\\  

\item[(vi)] We test the influence on SED fitting of the use of medium-band filters finely sampling the SEDs preferentially at $\lambda<10^{4}\ {\rm \AA}$. We find that the SED-fitting process is sensitive to this. In particular, such non-uniform SED sampling gives less weight to  the TP-AGB-sensitive  near-IR, hence dilutes the AGB signal reducing the effectiveness of the tests. \\

\item[(vii)]  In any event, the differences in performance among the three sets of models (as measured by their reduced $\chi^2$ values) are rather small after all and no compelling evidence in favour of one or another was reached by these comparisons. Nevertheless, solutions with nearly identical $\chi^2$ values may significantly differ in other, more astrophysical respects. For example, when limiting to SSPs, BC03 models give systematically older ages for these passive galaxies, compared to M05/M13 models, hence they predict a substantially higher number of passive galaxies at redshifts beyond $\sim 2-3$. We suggest that to distinguish between stellar population models, this systematic difference can be exploited in the future using number density counts of passive galaxies.
\end{itemize}

\subsection*{Acknowledgments}
The authors thank the referee for a report which significantly improved the paper.
They are also thankful to Daniel Thomas and Rob Crittenden for helpful discussions on the statistical analysis, to Daniela Calzetti and Karl Gordon for their valuable comments on the dust content of galaxies, to Maarten Baes for important comments on radiative transfer models, to Margherita Talia and Giulia Rodighiero for their advice with regards to galaxy IR light estimation methods and to Pablo G. P\'{e}rez-Gonz\'{a}lez for helpful discussions about his and his collaborators' results.

\bibliographystyle{mn2e}

\bibliography{Reference}
\appendix
\clearpage
\section{Published {\it NUV-r} vs {\it r-J} classification \& photo-$z$ for our COSMOS sample}
\label{sec:appendixA}
\begin{figure}
\centering
\hspace*{-1.6cm}\includegraphics[width=0.6\textwidth]{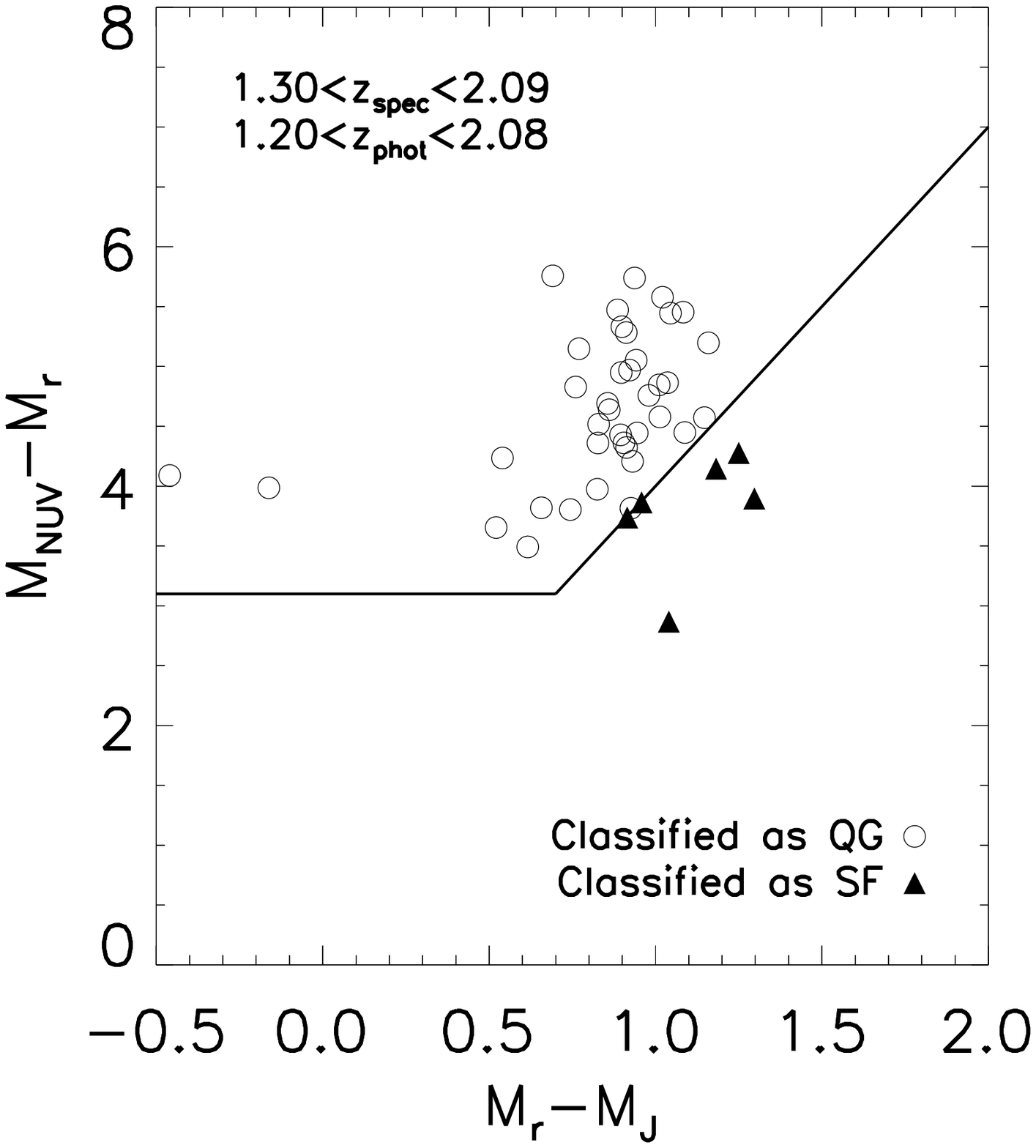}
\includegraphics[width=0.5\textwidth]{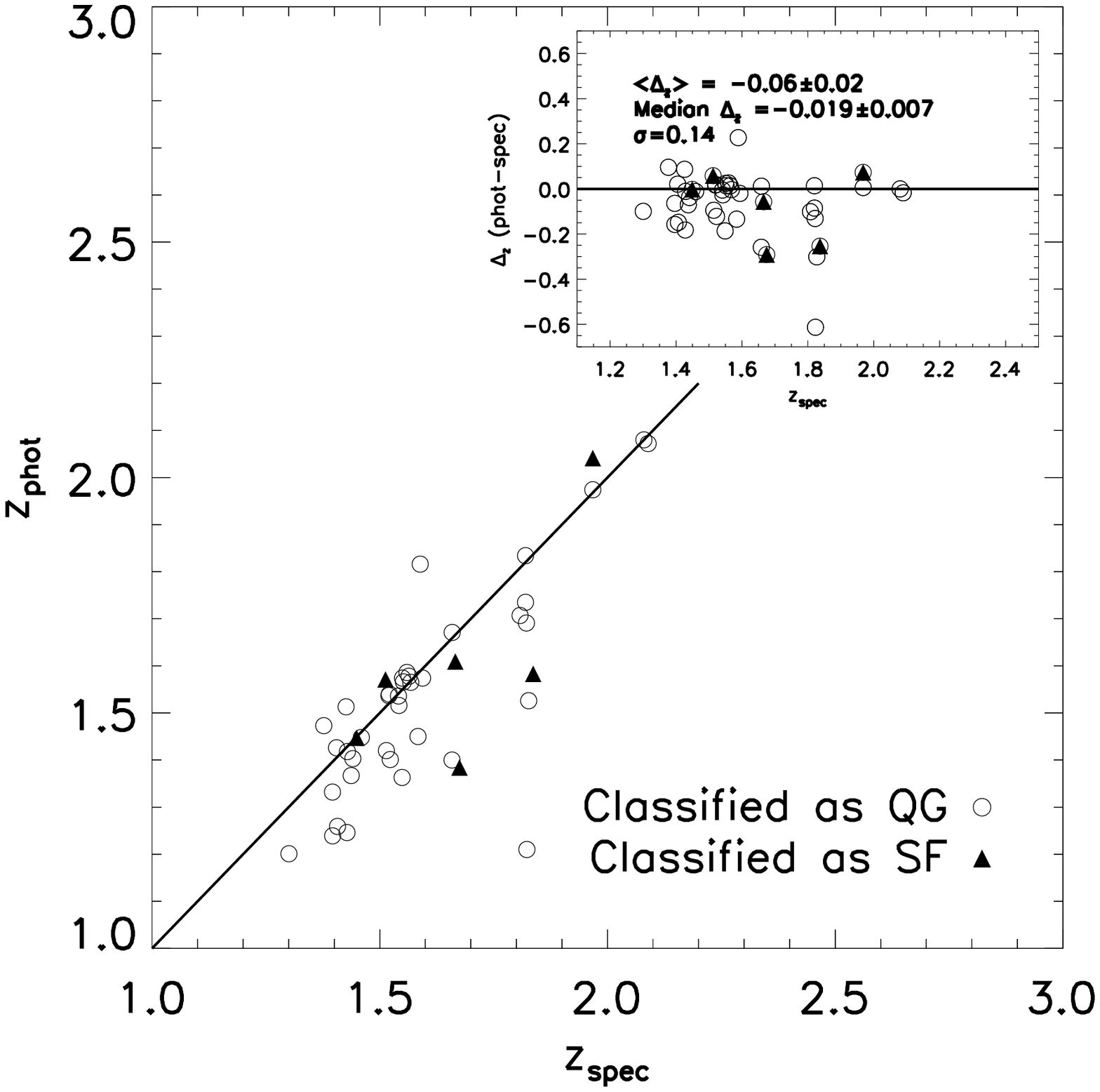}
\caption{Upper panel: {\it NUV-r} vs. {\it r-J} colour-colour plot showing the classification by \citet{Ilbert-2013} of our COSMOS galaxies as quiescent (Q, circles) or star forming (SF, triangles). Lower panel: comparison between spectroscopic and photometric redshifts for our COSMOS galaxies, as in Figures \ref{fig:Fig5} \&  \ref{fig:Fig6}. The photometric redshifts are now those estimated by  \citet{Ilbert-2013}. We point out that despite 6 galaxies are classified as SF by using the  classification by \citet{Ilbert-2013}, all 44 COSMOS galaxies studied by us are classified as passive according to both {\it BzK} and {\it UVJ} selection criteria based on spectroscopic redshifts.}
\label{fig:Fig1_appA}
\end{figure}

Upper panel of Figure \ref{fig:Fig1_appA} shows the classification scheme into Quiescent (Q, open circles) and Star Forming (SF, triangles) galaxies, using rest-frame {\it NUV-r} and {\it r-J} two-colour criterion, similar to the $UVJ$  criterion \citep{Williams-2009} used by us. The rest-frame colours used in this plot were taken from the publicly available galaxy properties from \citet{Ilbert-2013} based on BC03 models. The vast majority of galaxies lie in the 'quiescent' zone of the diagram, or  close to the separation line. This plot aims only at showing that 38 of our 44 COSMOS galaxies are indeed included in the catalogue of quiescent galaxies used by \citet{Man-2014} for {\it Herschel} stacks construction.
As the classification between quiescent and star forming galaxies we show in this figure was obtained by \citet{Ilbert-2013} using photometric redshifts, we can check their quality for our COSMOS galaxy sample by comparing them to our spectroscopic redshifts. Results are shown in lower panel of Figure \ref{fig:Fig1_appA}. We find median and mean residuals consistent with zero within $3\sigma$, corroborating on average the reliability of their Q/SF classification. However, since this Q/SF classification was based on photometric redshifts, it can lead to misclassification. In fact, it is worth pointing out that all the 44 COSMOS galaxies plotted here are classified as passive via both {\it BzK} and {\it UVJ} selection criteria based on spectroscopic redshifts. As an example, the most offset galaxy (ID 40) from the quiescent region in the Q/SF classification plot in Figure \ref{fig:Fig1_appA} presents also a large $\Delta_{\rm z}=z_{\rm phot}- z_{\rm spec}=-0.25$, which may be responsible for its classification as SF. In fact, we do classify it as passive.

\section{Additional tables and figures}
\label{sec:appendixB}
Here we show plots and tables summarising the SED fitting results for the COSMOS sample in both the reddening-free (see Table \ref {tab:Table1_appB} \& left-hand panels of Figure \ref{fig:Fig1_appB}) and reddening (see Table \ref {tab:Table2_appB} \& right-hand panels of Figure \ref{fig:Fig1_appB}) cases, when using the full filter set and the original photometry. 

In Figure \ref{fig:Fig2_appB}, we show the SED fits for three COSMOS galaxies (as examples) obtained for the full-filter-set case by including (left-hand panels) and excluding (right-hand panels) model dependent zero-point offset corrections. This figure shows the sensitivity of the SED fitting process to such corrections. 

Finally, in Table \ref{tab:Table3_appB} we show new best-fitting solutions obtained in presence of attenuation for our entire galaxy sample (HUDF+COSMOS), after excluding non self-consistent solutions as described in Section \ref{subsec:subsec5.2}.\\

\clearpage
\begin{table*}
%\begin{tiny}
\begin{center}
\caption{SED fitting results for the COSMOS sample using the full-filter-set and the original photometry: no-reddening case. Values refer to the best-fit solution. Col 1: galaxy ID ; col 2: spectroscopic redshift; col 3: model; col 4: age; col 5: metallicity; col 6: star formation history; col 7: reduced $\chi^{2}$; col 8: stellar mass; col 9: star formation rate.}
\begin{tabular}{cccclcccc}
\hline
  \multicolumn{1}{c}{\bf ID} &
  \multicolumn{1}{c}{$\mathbf{z_{\rm spec}}$} &
  \multicolumn{1}{c}{\bf Model} &
  \multicolumn{1}{c}{$\mathbf{t}$} &
  \multicolumn{1}{c}{$\mathbf{[Z/H]}$} &
  \multicolumn{1}{c}{\bf SFH} &
  \multicolumn{1}{c}{$\mathbf{\chi^{2}_{\rm r}}$} &
  \multicolumn{1}{c}{$\mathbf{M^{\ast}}$} &
  \multicolumn{1}{c}{$\mathbf{SFR}$} \\
  \multicolumn{1}{c}{}&
  \multicolumn{1}{c}{}&
  \multicolumn{1}{c}{}&
  \multicolumn{1}{c}{\rm (Gyr)}&
  \multicolumn{1}{c}{${\rm (Z_{\odot})}$} &
  \multicolumn{1}{c}{}&
  \multicolumn{1}{c}{}&
  \multicolumn{1}{c}{$(10^{11}\ {\rm M_{\odot}})$} &
  \multicolumn{1}{c}{$({\rm M_{\odot}/ yr^{-1}})$} \\
  \multicolumn{1}{c}{(1)}&
  \multicolumn{1}{c}{(2)}&
  \multicolumn{1}{c}{(3)}&
  \multicolumn{1}{c}{(4)}&
  \multicolumn{1}{c}{(5)}&
  \multicolumn{1}{c}{(6)}&
  \multicolumn{1}{c}{(7)}&
  \multicolumn{1}{c}{(8)}&
  \multicolumn{1}{c}{(9)} \\
\hline
\hline    
  \multirow{3}{*}{1}	  &	 & M05  &  $1.02$ & 2.0  & $e^{-t/0.1\ {\rm Gyr}}$		 &2.30 & $0.60$ &  $<0.1$ \\
                 	  &1.3005& M13  &  $1.14$ & 2.0  & SSP  				 &2.61 & $0.83$ &  $<0.1$ \\
		 	  &	 & BC03 &  $2.50$ & 2.0  & $e^{-t/0.3\ {\rm Gyr}}$		 &2.02 & $1.45$ &  $0.2$  \\
\hline
  \multirow{3}{*}{2}	  &	 & M05  &  $1.28$ & 2.0  & $t_{\rm trunc}=1\ {\rm Gyr}$ 	 &0.74 & $0.22$ &  $<0.1$ \\
                 	  &1.3770& M13  &  $1.14$ & 0.2  & $e^{-t/0.1\ {\rm Gyr}}$		 &1.07 & $0.28$ &  $<0.1$ \\
		 	  &	 & BC03 &  $2.10$ & 2.0  & $e^{-t/0.3\ {\rm Gyr}}$		 &1.20 & $0.54$ &  $0.2$  \\
\hline
  \multirow{3}{*}{3}	  &	 & M05  &  $1.02$ & 2.0  & $t_{\rm trunc}=0.3\ {\rm Gyr}$	 &1.06 & $0.63$ &  $<0.1$ \\
                 	  &1.3961& M13  &  $1.14$ & 2.0  & SSP  				 &0.94 & $0.91$ &  $<0.1$ \\
		 	  &	 & BC03 &  $3.50$ & 1.0  & $e^{-t/0.3\ {\rm Gyr}}$		 &0.98 & $1.95$ &  $<0.1$ \\
\hline
  \multirow{3}{*}{4}	  &	 & M05  &  $2.75$ & 1.0  & $e^{-t/0.3\ {\rm Gyr}}$		 &1.58 & $1.41$ &  $0.1$  \\
                 	  &1.3965& M13  &  $1.02$ & 2.0  & SSP  				 &1.10 & $0.83$ &  $<0.1$ \\
		 	  &	 & BC03 &  $3.25$ & 1.0  & $e^{-t/0.3\ {\rm Gyr}}$		 &0.99 & $1.86$ &  $<0.1$ \\
\hline
  \multirow{3}{*}{5}	  &	 & M05  &  $0.81$ & 2.0  & $e^{-t/0.1\ {\rm Gyr}}$		 &1.57 & $0.87$ &  $0.4$  \\
                 	  &1.4050& M13  &  $2.10$ & 0.2  & $e^{-t/0.3\ {\rm Gyr}}$		 &2.75 & $1.66$ &  $0.7$  \\
		 	  &	 & BC03 &  $2.10$ & 2.0  & $e^{-t/0.3\ {\rm Gyr}}$		 &3.55 & $2.09$ &  $0.9$  \\
\hline
  \multirow{3}{*}{6}	  &	 & M05  &  $2.00$ & 1.0  & $t_{\rm trunc}=1\ {\rm Gyr}$ 	 &1.33 & $0.74$ &  $<0.1$ \\
                 	  &1.4072& M13  &  $0.81$ & 2.0  & SSP  				 &1.23 & $0.58$ &  $<0.1$ \\
		 	  &	 & BC03 &  $2.75$ & 1.0  & $e^{-t/0.3\ {\rm Gyr}}$		 &1.16 & $1.29$ &  $0.1$  \\
\hline
  \multirow{3}{*}{7}	  &	 & M05  &  $0.90$ & 2.0  & $e^{-t/0.1\ {\rm Gyr}}$		 &1.45 & $0.49$ &  $0.1$  \\
                 	  &1.4260& M13  &  $2.30$ & 0.2  & $e^{-t/0.3\ {\rm Gyr}}$		 &3.36 & $0.91$ &  $0.2$  \\
		 	  &	 & BC03 &  $2.30$ & 2.0  & $e^{-t/0.3\ {\rm Gyr}}$		 &2.18 & $1.17$ &  $0.3$  \\
\hline
  \multirow{3}{*}{8}	  &	 & M05  &  $1.43$ & 2.0  & $t_{\rm trunc}=1\ {\rm Gyr}$ 	 &2.05 & $1.12$ &  $<0.1$ \\
                 	  &1.4280& M13  &  $2.40$ & 0.2  & $e^{-t/0.3\ {\rm Gyr}}$		 &3.76 & $2.09$ &  $0.3$  \\
		 	  &	 & BC03 &  $2.40$ & 2.0  & $e^{-t/0.3\ {\rm Gyr}}$		 &3.13 & $2.75$ &  $0.4$  \\
\hline
  \multirow{3}{*}{9}	  &	 & M05  &  $0.81$ & 2.0  & SSP  				 &2.42 & $1.12$ &  $<0.1$ \\
                 	  &1.4290& M13  &  $1.02$ & 2.0  & SSP  				 &2.77 & $1.51$ &  $<0.1$ \\
		 	  &	 & BC03 &  $3.00$ & 1.0  & $e^{-t/0.3\ {\rm Gyr}}$		 &2.99 & $3.02$ &  $0.1$  \\
\hline
  \multirow{3}{*}{10}	  &	 & M05  &  $0.72$ & 2.0  & SSP  				 &0.56 & $0.21$ &  $<0.1$ \\
                 	  &1.4372& M13  &  $0.90$ & 2.0  & SSP  				 &0.60 & $0.28$ &  $<0.1$ \\
		 	  &	 & BC03 &  $2.30$ & 2.0  & $e^{-t/0.3\ {\rm Gyr}}$		 &0.52 & $0.54$ &  $0.1$  \\
\hline
  \multirow{3}{*}{11}	  &	 & M05  &  $2.30$ & 1.0  & $e^{-t/0.3\ {\rm Gyr}}$		 &0.82 & $0.50$ &  $0.1$  \\
                 	  &1.4408& M13  &  $0.90$ & 2.0  & SSP  				 &0.68 & $0.33$ &  $<0.1$ \\
		 	  &	 & BC03 &  $2.75$ & 1.0  & $e^{-t/0.3\ {\rm Gyr}}$		 &0.71 & $0.69$ &  $<0.1$ \\
\hline
  \multirow{3}{*}{12}	  &	 & M05  &  $1.02$ & 0.5  & $t_{\rm trunc}=0.1\ {\rm Gyr}$	 &2.27 & $0.46$ &  $<0.1$ \\
                 	  &1.4490& M13  &  $1.02$ & 0.2  & $e^{-t/0.1\ {\rm Gyr}}$		 &4.97 & $0.55$ &  $<0.1$ \\
		 	  &	 & BC03 &  $2.10$ & 2.0  & $e^{-t/0.3\ {\rm Gyr}}$		 &4.99 & $1.29$ &  $0.5$  \\
\hline
  \multirow{3}{*}{13}	  &	 & M05  &  $1.90$ & 1.0  & $t_{\rm trunc}=1\ {\rm Gyr}$ 	 &0.98 & $0.56$ &  $<0.1$ \\
                 	  &1.4595& M13  &  $0.81$ & 2.0  & SSP  				 &0.87 & $0.48$ &  $<0.1$ \\
		 	  &	 & BC03 &  $2.60$ & 1.0  & $e^{-t/0.3\ {\rm Gyr}}$		 &0.90 & $1.00$ &  $0.1$  \\
\hline
  \multirow{3}{*}{14}	  &	 & M05  &  $0.64$ & 2.0  & $t_{\rm trunc}=0.1\ {\rm Gyr}$	 &1.66 & $0.39$ &  $<0.1$ \\
                 	  &1.5126& M13  &  $1.28$ & 0.2  & $t_{\rm trunc}=0.3\ {\rm Gyr}$	 &2.00 & $0.58$ &  $<0.1$ \\
		 	  &	 & BC03 &  $2.50$ & 1.0  & $e^{-t/0.3\ {\rm Gyr}}$		 &2.32 & $1.17$ &  $0.1$  \\
\hline
  \multirow{3}{*}{15}	  &	 & M05  &  $0.57$ & 2.0  & $t_{\rm trunc}=0.1\ {\rm Gyr}$	 &1.05 & $0.26$ &  $<0.1$ \\
                 	  &1.5145& M13  &  $1.02$ & 0.2  & $t_{\rm trunc}=0.1\ {\rm Gyr}$	 &1.12 & $0.37$ &  $<0.1$ \\
		 	  &	 & BC03 &  $2.30$ & 1.0  & $e^{-t/0.3\ {\rm Gyr}}$		 &1.38 & $0.76$ &  $0.2$  \\
\hline
  \multirow{3}{*}{16}	  &	 & M05  &  $1.28$ & 1.0  & $t_{\rm trunc}=0.3\ {\rm Gyr}$	 &1.17 & $0.55$ &  $<0.1$ \\
                 	  &1.5199& M13  &  $1.28$ & 0.2  & $t_{\rm trunc}=0.3\ {\rm Gyr}$	 &1.17 & $0.54$ &  $<0.1$ \\
		 	  &	 & BC03 &  $2.40$ & 1.0  & $e^{-t/0.3\ {\rm Gyr}}$		 &1.22 & $1.02$ &  $0.2$  \\
\hline
\hline
\end{tabular}
\label{tab:Table1_appB}							    
\end{center}
% \end{tiny}
\end{table*}

\addtocounter{table}{-1}

\begin{table*}
%\begin{tiny}
\begin{center}
\caption{Continued.}
\begin{tabular}{cccclcccc}
\hline
  \multicolumn{1}{c}{\bf ID} &
  \multicolumn{1}{c}{$\mathbf{z_{\rm spec}}$} &
  \multicolumn{1}{c}{\bf Model} &
  \multicolumn{1}{c}{$\mathbf{t}$} &
  \multicolumn{1}{c}{$\mathbf{[Z/H]}$} &
  \multicolumn{1}{c}{\bf SFH} &
  \multicolumn{1}{c}{$\mathbf{\chi^{2}_{\rm r}}$} &
  \multicolumn{1}{c}{$\mathbf{M^{\ast}}$} &
  \multicolumn{1}{c}{$\mathbf{SFR}$} \\
  \multicolumn{1}{c}{}&
  \multicolumn{1}{c}{}&
  \multicolumn{1}{c}{}&
  \multicolumn{1}{c}{\rm (Gyr)}&
  \multicolumn{1}{c}{${\rm (Z_{\odot})}$} &
  \multicolumn{1}{c}{}&
  \multicolumn{1}{c}{}&
  \multicolumn{1}{c}{$(10^{11}\ {\rm M_{\odot}})$} &
  \multicolumn{1}{c}{$({\rm M_{\odot}/ yr^{-1}})$} \\
  \multicolumn{1}{c}{(1)}&
  \multicolumn{1}{c}{(2)}&
  \multicolumn{1}{c}{(3)}&
  \multicolumn{1}{c}{(4)}&
  \multicolumn{1}{c}{(5)}&
  \multicolumn{1}{c}{(6)}&
  \multicolumn{1}{c}{(7)}&
  \multicolumn{1}{c}{(8)}&
  \multicolumn{1}{c}{(9)} \\
\hline
\hline
  \multirow{3}{*}{17}	  &	 & M05  &  $0.72$ & 2.0  & $t_{\rm trunc}=0.3\ {\rm Gyr}$	 &1.38 & $0.47$ & $<0.1$  \\
                 	  &1.5210& M13  &  $1.14$ & 0.2  & $e^{-t/0.1\ {\rm Gyr}}$		 &1.42 & $0.68$ & $<0.1$  \\
		 	  &	 & BC03 &  $2.30$ & 1.0  & $e^{-t/0.3\ {\rm Gyr}}$		 &1.97 & $1.32$ & $0.3$  \\
\hline
  \multirow{3}{*}{18}	  &	 & M05  &  $1.43$ & 2.0  & $t_{\rm trunc}=1\ {\rm Gyr}$ 	 &1.38 & $0.50$ & $<0.1$  \\
                 	  &1.5230& M13  &  $1.02$ & 2.0  & SSP  				 &1.49 & $0.66$ & $<0.1$  \\
		 	  &	 & BC03 &  $2.40$ & 2.0  & $e^{-t/0.3\ {\rm Gyr}}$		 &1.05 & $1.20$ & $0.2$   \\
\hline
  \multirow{3}{*}{19}	  &	 & M05  &  $1.28$ & 1.0  & $e^{-t/0.1\ {\rm Gyr}}$		 &0.77 & $0.30$ & $<0.1$  \\
                 	  &1.5410& M13  &  $1.28$ & 0.2  & $e^{-t/0.1\ {\rm Gyr}}$		 &0.77 & $0.30$ & $<0.1$  \\
		 	  &	 & BC03 &  $2.40$ & 1.0  & $e^{-t/0.3\ {\rm Gyr}}$		 &0.80 & $0.55$ & $0.1$   \\
\hline    
  \multirow{3}{*}{20}	  &	 & M05  &  $1.43$ & 2.0  & $t_{\rm trunc}=1\ {\rm Gyr}$ 	 &0.50 & $0.35$ & $<0.1$  \\
                 	  &1.5420& M13  &  $1.02$ & 2.0  & SSP  				 &1.00 & $0.45$ & $<0.1$  \\
		 	  &	 & BC03 &  $2.40$ & 2.0  & $e^{-t/0.3\ {\rm Gyr}}$		 &0.86 & $0.81$ & $0.1$   \\
\hline
  \multirow{3}{*}{21}	  &	 & M05  &  $0.64$ & 2.0  & SSP  				 &0.98 & $0.46$ & $<0.1$  \\
                 	  &1.5490& M13  &  $1.90$ & 0.2  & $t_{\rm trunc}=1\ {\rm Gyr}$ 	 &1.18 & $0.74$ & $<0.1$  \\
		 	  &	 & BC03 &  $2.60$ & 1.0  & $e^{-t/0.3\ {\rm Gyr}}$		 &1.19 & $1.32$ & $0.1$   \\
\hline
  \multirow{3}{*}{22}	  &	 & M05  &  $1.80$ & 1.0  & $t_{\rm trunc}=1\ {\rm Gyr}$ 	 &0.80 & $0.83$ & $<0.1$  \\
                 	  &1.5499& M13  &  $1.80$ & 0.2  & $t_{\rm trunc}=1\ {\rm Gyr}$ 	 &0.80 & $0.81$ & $<0.1$  \\
		 	  &	 & BC03 &  $2.50$ & 1.0  & $e^{-t/0.3\ {\rm Gyr}}$		 &0.72 & $1.51$ & $0.2$   \\
\hline
  \multirow{3}{*}{23}	  &	 & M05  &  $0.72$ & 2.0  & $t_{\rm trunc}=0.1\ {\rm Gyr}$	 &1.14 & $0.56$ & $<0.1$  \\
                 	  &1.5519& M13  &  $2.00$ & 0.2  & $t_{\rm trunc}=1\ {\rm Gyr}$ 	 &1.29 & $0.95$ & $<0.1$  \\
		 	  &	 & BC03 &  $2.60$ & 1.0  & $e^{-t/0.3\ {\rm Gyr}}$		 &1.22 & $1.58$ & $0.1$   \\
\hline
  \multirow{3}{*}{24}	  &	 & M05  &  $2.10$ & 0.2  & $t_{\rm trunc}=1\ {\rm Gyr}$ 	 &2.48 & $0.87$ & $<0.1$  \\
                 	  &1.5596& M13  &  $1.80$ & 0.2  & $e^{-t/0.3\ {\rm Gyr}}$		 &3.17 & $1.17$ & $1.3$   \\
		 	  &	 & BC03 &  $1.90$ & 1.0  & $e^{-t/0.3\ {\rm Gyr}}$		 &3.76 & $1.32$ & $1.1$   \\
\hline
  \multirow{3}{*}{25}	  &	 & M05  &  $0.64$ & 2.0  & $t_{\rm trunc}=0.1\ {\rm Gyr}$	 &1.58 & $0.69$ & $<0.1$  \\
                 	  &1.5642& M13  &  $1.80$ & 0.2  & $t_{\rm trunc}=1\ {\rm Gyr}$ 	 &1.72 & $1.12$ & $<0.1$  \\
		 	  &	 & BC03 &  $2.40$ & 1.0  & $e^{-t/0.3\ {\rm Gyr}}$		 &2.32 & $1.95$ & $0.3$   \\
\hline
  \multirow{3}{*}{26}	  &	 & M05  &  $0.81$ & 2.0  & $t_{\rm trunc}=0.3\ {\rm Gyr}$	 &0.98 & $0.45$ & $<0.1$  \\
                 	  &1.5681& M13  &  $1.28$ & 0.2  & $t_{\rm trunc}=0.3\ {\rm Gyr}$	 &1.77 & $0.60$ & $<0.1$  \\
		 	  &	 & BC03 &  $2.20$ & 2.0  & $e^{-t/0.3\ {\rm Gyr}}$		 &1.97 & $1.17$ & $0.4$   \\
\hline
  \multirow{3}{*}{27}	  &	 & M05  &  $1.61$ & 1.0  & $t_{\rm trunc}=1\ {\rm Gyr}$ 	 &1.11 & $0.47$ & $<0.1$  \\
                 	  &1.5839& M13  &  $1.61$ & 0.2  & $t_{\rm trunc}=1\ {\rm Gyr}$ 	 &1.11 & $0.46$ & $<0.1$  \\
		 	  &	 & BC03 &  $1.28$ & 0.5  & SSP  				 &1.32 & $0.39$ & $<0.1$  \\
\hline
  \multirow{3}{*}{28}	  &	 & M05  &  $0.90$ & 2.0  & $e^{-t/0.1\ {\rm Gyr}}$		 &1.12 & $0.71$ & $0.1$   \\
                 	  &1.5888& M13  &  $2.30$ & 0.2  & $e^{-t/0.3\ {\rm Gyr}}$		 &2.31 & $1.32$ & $0.3$   \\
		 	  &	 & BC03 &  $2.30$ & 2.0  & $e^{-t/0.3\ {\rm Gyr}}$		 &1.91 & $1.70$ & $0.4$   \\
\hline
  \multirow{3}{*}{29}	  &	 & M05  &  $0.81$ & 1.0  & SSP  				 &1.71 & $0.51$ & $<0.1$  \\
                 	  &1.5939& M13  &  $1.02$ & 0.5  & SSP  				 &1.58 & $0.52$ & $<0.1$  \\
		 	  &	 & BC03 &  $1.28$ & 0.5  & SSP  				 &1.40 & $0.56$ & $<0.1$  \\
\hline
  \multirow{3}{*}{30}	  &	 & M05  &  $0.81$ & 2.0  & $e^{-t/0.1\ {\rm Gyr}}$		 &1.29 & $0.49$ & $0.2$   \\
                 	  &1.6587& M13  &  $1.14$ & 0.2  & $e^{-t/0.1\ {\rm Gyr}}$		 &1.59 & $0.59$ & $<0.1$  \\
		 	  &	 & BC03 &  $2.40$ & 1.0  & $e^{-t/0.3\ {\rm Gyr}}$		 &1.89 & $1.23$ & $0.2$   \\
\hline
  \multirow{3}{*}{31}	  &	 & M05  &  $3.25$ & 0.2  & $t_{\rm trunc}=1\ {\rm Gyr}$ 	 &1.40 & $1.48$ & $<0.1$  \\
                 	  &1.6590& M13  &  $0.90$ & 1.0  & SSP  				 &1.35 & $0.87$ & $<0.1$  \\
		 	  &	 & BC03 &  $1.28$ & 0.5  & $t_{\rm trunc}=0.1\ {\rm Gyr}$	 &1.21 & $0.78$ & $<0.1$  \\
\hline
  \multirow{3}{*}{32}	  &	 & M05  &  $3.00$ & 0.2  & $e^{-t/0.3\ {\rm Gyr}}$		 &2.20 & $1.48$ & $<0.1$  \\
                 	  &1.6658& M13  &  $0.81$ & 1.0  & SSP  				 &2.32 & $0.79$ & $<0.1$  \\
		 	  &	 & BC03 &  $1.28$ & 0.5  & $e^{-t/0.1\ {\rm Gyr}}$		 &2.16 & $0.78$ & $<0.1$  \\
\hline
\hline
\end{tabular}
\label{tab:Table1_appB}							    
\end{center}
% \end{tiny}
\end{table*}

\addtocounter{table}{-1}

\begin{table*}
%\begin{tiny}
\begin{center}
\caption{Continued.}
\begin{tabular}{cccclcccc}
\hline
  \multicolumn{1}{c}{\bf ID} &
  \multicolumn{1}{c}{$\mathbf{z_{\rm spec}}$} &
  \multicolumn{1}{c}{\bf Model} &
  \multicolumn{1}{c}{$\mathbf{t}$} &
  \multicolumn{1}{c}{$\mathbf{[Z/H]}$} &
  \multicolumn{1}{c}{\bf SFH} &
  \multicolumn{1}{c}{$\mathbf{\chi^{2}_{\rm r}}$} &
  \multicolumn{1}{c}{$\mathbf{M^{\ast}}$} &
  \multicolumn{1}{c}{$\mathbf{SFR}$} \\
  \multicolumn{1}{c}{}&
  \multicolumn{1}{c}{}&
  \multicolumn{1}{c}{}&
  \multicolumn{1}{c}{\rm (Gyr)}&
  \multicolumn{1}{c}{${\rm (Z_{\odot})}$} &
  \multicolumn{1}{c}{}&
  \multicolumn{1}{c}{}&
  \multicolumn{1}{c}{$(10^{11}\ {\rm M_{\odot}})$} &
  \multicolumn{1}{c}{$({\rm M_{\odot}/ yr^{-1}})$} \\
  \multicolumn{1}{c}{(1)}&
  \multicolumn{1}{c}{(2)}&
  \multicolumn{1}{c}{(3)}&
  \multicolumn{1}{c}{(4)}&
  \multicolumn{1}{c}{(5)}&
  \multicolumn{1}{c}{(6)}&
  \multicolumn{1}{c}{(7)}&
  \multicolumn{1}{c}{(8)}&
  \multicolumn{1}{c}{(9)} \\
\hline
\hline
  \multirow{3}{*}{33}	  &	 & M05  &  $0.72$ & 2.0  & $e^{-t/0.1\ {\rm Gyr}}$		 &4.34 & $0.56$ & $0.5$   \\
                 	  &1.6750& M13  &  $1.02$ & 0.2  & $e^{-t/0.1\ {\rm Gyr}}$		 &5.09 & $0.68$ & $<0.1$  \\
		 	  &	 & BC03 &  $2.10$ & 1.0  & $e^{-t/0.3\ {\rm Gyr}}$		 &6.84 & $1.29$ & $0.5$   \\
\hline
  \multirow{3}{*}{34}	  &	 & M05  &  $0.72$ & 2.0  & $t_{\rm trunc}=0.1\ {\rm Gyr}$	 &0.92 & $0.68$ & $<0.1$  \\
                 	  &1.8080& M13  &  $1.90$ & 0.2  & $t_{\rm trunc}=1\ {\rm Gyr}$ 	 &1.17 & $1.05$ & $<0.1$  \\
		 	  &	 & BC03 &  $2.60$ & 1.0  & $e^{-t/0.3\ {\rm Gyr}}$		 &1.23 & $1.91$ & $0.1$   \\
\hline
  \multirow{3}{*}{35}	  &	 & M05  &  $2.50$ & 0.5  & $e^{-t/0.3\ {\rm Gyr}}$		 &1.17 & $2.75$ & $0.3$   \\
                 	  &1.8200& M13  &  $1.80$ & 0.2  & $t_{\rm trunc}=1\ {\rm Gyr}$ 	 &1.52 & $2.09$ & $<0.1$  \\
		 	  &	 & BC03 &  $2.50$ & 1.0  & $e^{-t/0.3\ {\rm Gyr}}$		 &2.09 & $3.89$ & $0.4$   \\
\hline
  \multirow{3}{*}{36}	  &	 & M05  &  $0.64$ & 2.0  & SSP  				 &0.90 & $0.58$ & $<0.1$  \\
                 	  &1.8200& M13  &  $0.81$ & 2.0  & SSP  				 &1.02 & $0.79$ & $<0.1$  \\
		 	  &	 & BC03 &  $2.60$ & 1.0  & $e^{-t/0.3\ {\rm Gyr}}$		 &1.07 & $1.66$ & $0.1$   \\
\hline
  \multirow{3}{*}{37}	  &	 & M05  &  $0.72$ & 2.0  & $e^{-t/0.1\ {\rm Gyr}}$		 &1.77 & $1.15$ & $1.1$   \\
                 	  &1.8220& M13  &  $1.90$ & 0.2  & $e^{-t/0.3\ {\rm Gyr}}$		 &2.73 & $2.19$ & $1.7$   \\
		 	  &	 & BC03 &  $2.10$ & 1.0  & $e^{-t/0.3\ {\rm Gyr}}$		 &3.18 & $2.75$ & $1.1$   \\
\hline
  \multirow{3}{*}{38}	  &	 & M05  &  $2.10$ & 0.2  & SSP  				 &2.87 & $3.80$ & $<0.1$  \\
                 	  &1.8230& M13  &  $0.90$ & 0.2  & $e^{-t/0.1\ {\rm Gyr}}$		 &3.47 & $2.29$ & $0.4$   \\
		 	  &	 & BC03 &  $2.20$ & 0.5  & $e^{-t/0.3\ {\rm Gyr}}$		 &3.62 & $4.57$ & $1.4$   \\
\hline
  \multirow{3}{*}{39}	  &	 & M05  &  $1.02$ & 1.0  & $e^{-t/0.1\ {\rm Gyr}}$		 &1.00 & $0.93$ & $<0.1$  \\
                 	  &1.8270& M13  &  $1.02$ & 0.2  & $e^{-t/0.1\ {\rm Gyr}}$		 &1.00 & $0.93$ & $<0.1$  \\
		 	  &	 & BC03 &  $2.30$ & 1.0  & $e^{-t/0.3\ {\rm Gyr}}$		 &1.94 & $2.04$ & $0.4$   \\
\hline    
  \multirow{3}{*}{40}	  &	 & M05  &  $1.28$ & 1.0  & $t_{\rm trunc}=1\ {\rm Gyr}$ 	 &1.67 & $0.68$ & $<0.1$  \\
                 	  &1.8368& M13  &  $1.28$ & 0.2  & $t_{\rm trunc}=1\ {\rm Gyr}$ 	 &1.67 & $0.68$ & $<0.1$  \\
		 	  &	 & BC03 &  $2.10$ & 0.2  & $e^{-t/0.3\ {\rm Gyr}}$		 &1.92 & $1.15$ & $0.5$   \\
\hline
  \multirow{3}{*}{41}	  &	 & M05  &  $2.30$ & 0.2  & $t_{\rm trunc}=1\ {\rm Gyr}$ 	 &0.92 & $1.20$ & $<0.1$  \\
                 	  &1.9677& M13  &  $0.81$ & 0.2  & $e^{-t/0.1\ {\rm Gyr}}$		 &1.08 & $0.72$ & $0.3$   \\
		 	  &	 & BC03 &  $1.28$ & 0.2  & SSP  				 &1.60 & $0.72$ & $<0.1$  \\
\hline
  \multirow{3}{*}{42}	  &	 & M05  &  $3.00$ & 0.2  & $e^{-t/0.3\ {\rm Gyr}}$		 &1.28 & $1.35$ & $<0.1$  \\
                 	  &1.9677& M13  &  $0.81$ & 1.0  & SSP  				 &0.91 & $0.72$ & $<0.1$  \\
		 	  &	 & BC03 &  $1.80$ & 0.5  & $t_{\rm trunc}=1\ {\rm Gyr}$ 	 &0.74 & $0.81$ & $<0.1$  \\
\hline
  \multirow{3}{*}{43}	  &	 & M05  &  $2.60$ & 0.2  & $e^{-t/0.3\ {\rm Gyr}}$		 &1.17 & $2.09$ & $0.2$   \\
                 	  &2.0799& M13  &  $0.90$ & 0.2  & $e^{-t/0.1\ {\rm Gyr}}$		 &1.53 & $1.17$ & $0.2$   \\
		 	  &	 & BC03 &  $2.20$ & 0.5  & $e^{-t/0.3\ {\rm Gyr}}$		 &1.31 & $2.29$ & $0.7$   \\
\hline
  \multirow{3}{*}{44}	  &	 & M05  &  $3.00$ & 0.2  & $e^{-t/0.3\ {\rm Gyr}}$		 &1.17 & $2.45$ & $0.1$   \\
                 	  &2.0892& M13  &  $0.90$ & 0.5  & SSP  				 &1.54 & $1.17$ & $<0.1$  \\
		 	  &	 & BC03 &  $2.75$ & 0.2  & $e^{-t/0.3\ {\rm Gyr}}$		 &1.28 & $2.34$ & $0.1$   \\
\hline
\hline
\end{tabular}
\label{tab:Table1_appB}							    
\end{center}
% \end{tiny}
\end{table*}

\clearpage

\begin{table*}
\begin{scriptsize}
\begin{center}
\caption{Same as Table B1 for the reddening case. Cols. 10 and 11 list reddening and the reddening law.}
\label{tab:Table2_appB}
\begin{threeparttable} 
\begin{tabular}{cccclcccccl}
\hline
  \multicolumn{1}{c}{\bf ID} &
  \multicolumn{1}{c}{$\mathbf{z_{\rm spec}}$} &
  \multicolumn{1}{c}{\bf Model} &
  \multicolumn{1}{c}{$\mathbf{t}$} &
  \multicolumn{1}{c}{$\mathbf{[Z/H]}$} &
  \multicolumn{1}{c}{\bf SFH} &
  \multicolumn{1}{c}{$\mathbf{\chi^{2}_{\rm r}}$} &
  \multicolumn{1}{c}{$\mathbf{M^{\ast}}$} &
  \multicolumn{1}{c}{$\mathbf{SFR}$} & 
  \multicolumn{1}{c}{$\mathbf{E(B-V)}$} &
  \multicolumn{1}{c}{\bf Reddening Law} \\
  \multicolumn{1}{c}{}&
  \multicolumn{1}{c}{}&
  \multicolumn{1}{c}{}&
  \multicolumn{1}{c}{\rm (Gyr)}&
  \multicolumn{1}{c}{${\rm (Z_{\odot})}$} &
  \multicolumn{1}{c}{}&
  \multicolumn{1}{c}{}&
  \multicolumn{1}{c}{$(10^{11}\ {\rm M_{\odot}})$} &
  \multicolumn{1}{c}{$({\rm M_{\odot}/ yr^{-1}})$} &
  \multicolumn{1}{c}{({\rm mag})}&
  \multicolumn{1}{c}{} \\
  \multicolumn{1}{c}{(1)}&
  \multicolumn{1}{c}{(2)}&
  \multicolumn{1}{c}{(3)}&
  \multicolumn{1}{c}{(4)}&
  \multicolumn{1}{c}{(5)}&
  \multicolumn{1}{c}{(6)}&
  \multicolumn{1}{c}{(7)}&
  \multicolumn{1}{c}{(8)}&
  \multicolumn{1}{c}{(9)}&
  \multicolumn{1}{c}{(10)}&
  \multicolumn{1}{c}{(11)} \\
\hline
\hline
  \multirow{3}{*}{1}      &	 & M05  &  $0.64$ & 2.0 & SSP				 & 2.09 & $0.54$ & $<0.1$ & $0.06$ & \citet{Allen-1976}  		     \\
                          &1.3005& M13  &  $0.72$ & 2.0 & SSP				 & 1.86 & $0.81$ & $<0.1$ & $0.10$ & \citet{Calzetti-2000}		     \\
			  &	 & BC03 &  $2.40$ & 1.0 & $e^{-t/0.3\ {\rm Gyr}}$	 & 1.80 & $1.70$ & $0.3$ & $0.10$ & \citet{Calzetti-2000}		     \\
\hline
  \multirow{3}{*}{2}      &	 & M05  &  $1.43$ & 1.0 & $t_{\rm trunc}=1\ {\rm Gyr}$	 & 0.67 & $0.28$ & $<0.1$ & $0.06$ & \citet{Allen-1976}  		     \\
                          &1.3770& M13  &  $1.43$ & 0.2 & $t_{\rm trunc}=1\ {\rm Gyr}$	 & 0.67 & $0.28$ & $<0.1$ & $0.06$ & \citet{Allen-1976}  		     \\
			  &	 & BC03 &  $1.61$ & 0.2 & $t_{\rm trunc}=1\ {\rm Gyr}$	 & 0.75 & $0.50$ & $<0.1$ & $0.25$ & \citet{Calzetti-2000}		     \\
\hline
  \multirow{3}{*}{3}      &	 & M05  &  $2.00$ & 1.0 & $t_{\rm trunc}=1\ {\rm Gyr}$	 & 0.70 & $1.05$ & $<0.1$ & $0.06$ & \citet{Allen-1976}  		     \\
                          &1.3961& M13  &  $0.72$ & 2.0 & SSP				 & 0.65 & $0.89$ & $<0.1$ & $0.10$ & \citet{Calzetti-2000}		     \\
			  &	 & BC03 &  $2.30$ & 0.5 & $t_{\rm trunc}=1\ {\rm Gyr}$	 & 0.58 & $1.55$ & $<0.1$ & $0.15$ & \citet{Calzetti-2000}		     \\
\hline
  \multirow{3}{*}{4}      &	 & M05  &  $1.61$ & 1.0 & $t_{\rm trunc}=0.3\ {\rm Gyr}$ & 1.35 & $1.10$ & $<0.1$ & $0.07$ & \citet{Prevot-1984,Bouchet-1985}	     \\
                          &1.3965& M13  &  $0.81$ & 2.0 & SSP				 & 1.05 & $0.85$ & $<0.1$ & $0.07$ & \citet{Prevot-1984,Bouchet-1985}	     \\
			  &	 & BC03 &  $3.25$ & 1.0 & $e^{-t/0.3\ {\rm Gyr}}$	 & 0.99 & $1.86$ & $<0.1$ & $0.00$ & NA\tnote{1} 			     \\
\hline
  \multirow{3}{*}{5}      &	 & M05  &  $2.30$ & 0.2 & $t_{\rm trunc}=0.1\ {\rm Gyr}$ & 1.33 & $2.14$ & $<0.1$ & $0.10$ & \citet{Calzetti-2000}		     \\
                          &1.4050& M13  &  $1.02$ & 0.2 & SSP				 & 1.55 & $1.62$ & $<0.1$ & $0.29$ & \citet{Prevot-1984,Bouchet-1985}	     \\
			  &	 & BC03 &  $1.68$ & 0.5 & $e^{-t/0.3\ {\rm Gyr}}$	 & 1.34 & $1.95$ & $3.2$ & $0.22$ & \citet{Prevot-1984,Bouchet-1985}	     \\
\hline
  \multirow{3}{*}{6}      &	 & M05  &  $2.00$ & 1.0 & $t_{\rm trunc}=1\ {\rm Gyr}$	 & 1.33 & $0.74$ & $<0.1$ & $0.00$ & NA\tnote{1} 			     \\
                          &1.4072& M13  &  $0.90$ & 1.0 & SSP				 & 1.09 & $0.78$ & $<0.1$ & $0.13$ & \citet{Seaton-1979} 		     \\
			  &	 & BC03 &  $1.70$ & 1.0 & $t_{\rm trunc}=1\ {\rm Gyr}$	 & 1.05 & $0.91$ & $<0.1$ & $0.13$ & \citet{Seaton-1979} 		     \\
\hline
  \multirow{3}{*}{7}      &	 & M05  &  $1.28$ & 2.0 & $t_{\rm trunc}=1\ {\rm Gyr}$	 & 1.22 & $0.51$ & $<0.1$ & $0.07$ & \citet{Prevot-1984,Bouchet-1985}	     \\
                          &1.4260& M13  &  $0.64$ & 0.2 & $t_{\rm trunc}=0.1\ {\rm Gyr}$ & 1.26 & $0.58$ & $<0.1$ & $0.22$ & \citet{Prevot-1984,Bouchet-1985}	     \\
			  &	 & BC03 &  $0.32$ & 2.0 & $t_{\rm trunc}=0.1\ {\rm Gyr}$ & 1.40 & $0.76$ & $<0.1$ & $0.45$ & \citet{Allen-1976}  		     \\
\hline
  \multirow{3}{*}{8}      &	 & M05  &  $0.64$ & 2.0 & $t_{\rm trunc}=0.1\ {\rm Gyr}$ & 1.82 & $1.02$ & $<0.1$ & $0.07$ & \citet{Prevot-1984,Bouchet-1985}	     \\
                          &1.4280& M13  &  $0.90$ & 1.0 & SSP				 & 2.13 & $1.95$ & $<0.1$ & $0.15$ & \citet{Calzetti-2000}		     \\
			  &	 & BC03 &  $1.28$ & 0.2 & SSP				 & 1.87 & $1.38$ & $<0.1$ & $0.22$ & \citet{Prevot-1984,Bouchet-1985}	     \\
\hline
  \multirow{3}{*}{9}      &	 & M05  &  $1.90$ & 1.0 & $t_{\rm trunc}=1\ {\rm Gyr}$	 & 2.07 & $1.78$ & $<0.1$ & $0.07$ & \citet{Prevot-1984,Bouchet-1985}	     \\
                          &1.4290& M13  &  $0.72$ & 2.0 & SSP				 & 1.85 & $1.66$ & $<0.1$ & $0.10$ & \citet{Calzetti-2000}		     \\
			  &	 & BC03 &  $1.70$ & 0.5 & $t_{\rm trunc}=0.1\ {\rm Gyr}$ & 1.86 & $2.69$ & $<0.1$ & $0.15$ & \citet{Calzetti-2000}		     \\
\hline
  \multirow{3}{*}{10}     &	 & M05  &  $1.70$ & 1.0 & $t_{\rm trunc}=1\ {\rm Gyr}$	 & 0.51 & $0.32$ & $<0.1$ & $0.06$ & \citet{Allen-1976}  		     \\
                          &1.4372& M13  &  $0.64$ & 1.0 & SSP				 & 0.46 & $0.38$ & $<0.1$ & $0.26$ & \citet{Seaton-1979} 		     \\
			  &	 & BC03 &  $1.43$ & 1.0 & $t_{\rm trunc}=1\ {\rm Gyr}$	 & 0.49 & $0.45$ & $<0.1$ & $0.26$ & \citet{Seaton-1979} 		     \\
\hline
  \multirow{3}{*}{11}     &	 & M05  &  $2.30$ & 1.0 & $e^{-t/0.3\ {\rm Gyr}}$	 & 0.82 & $0.50$ & $0.1$ & $0.00$ & NA\tnote{1} 			     \\
                          &1.4408& M13  &  $0.72$ & 2.0 & SSP				 & 0.59 & $0.33$ & $<0.1$ & $0.05$ & \citet{Calzetti-2000}		     \\
			  &	 & BC03 &  $2.30$ & 1.0 & $e^{-t/0.3\ {\rm Gyr}}$	 & 0.63 & $0.66$ & $0.1$ & $0.07$ & \citet{Prevot-1984,Bouchet-1985}	     \\
\hline
  \multirow{3}{*}{12}     &	 & M05  &  $0.07$ & 2.0 & SSP				 & 1.30 & $0.68$ & $<0.1$ & $0.58$ & \citet{Allen-1976}  		     \\
                          &1.4490& M13  &  $0.07$ & 2.0 & SSP				 & 1.13 & $1.58$ & $<0.1$ & $0.64$ & \citet{Calzetti-2000}		     \\
			  &	 & BC03 &  $0.11$ & 2.0 & SSP				 & 1.41 & $0.78$ & $<0.1$ & $0.58$ & \citet{Allen-1976}  		     \\
\hline
  \multirow{3}{*}{13}     &	 & M05  &  $1.90$ & 1.0 & $t_{\rm trunc}=1\ {\rm Gyr}$	 & 0.98 & $0.56$ & $<0.1$ & $0.00$ & NA\tnote{1} 			     \\
                          &1.4595& M13  &  $0.90$ & 1.0 & SSP				 & 0.86 & $0.65$ & $<0.1$ & $0.13$ & \citet{Seaton-1979} 		     \\
			  &	 & BC03 &  $1.68$ & 1.0 & $t_{\rm trunc}=1\ {\rm Gyr}$	 & 0.85 & $0.74$ & $<0.1$ & $0.13$ & \citet{Seaton-1979} 		     \\
\hline
  \multirow{3}{*}{14}     &	 & M05  &  $0.18$ & 0.5 & SSP				 & 1.65 & $0.60$ & $<0.1$ & $0.51$ & \citet{Prevot-1984,Bouchet-1985}	     \\
                          &1.5126& M13  &  $0.13$ & 0.5 & SSP				 & 1.72 & $0.60$ & $<0.1$ & $0.59$ & \citet{Prevot-1984,Bouchet-1985}	     \\
			  &	 & BC03 &  $0.29$ & 0.2 & SSP				 & 1.67 & $0.62$ & $<0.1$ & $0.51$ & \citet{Prevot-1984,Bouchet-1985}	     \\
\hline
  \multirow{3}{*}{15}     &	 & M05  &  $0.57$ & 2.0 & $t_{\rm trunc}=0.1\ {\rm Gyr}$ & 1.05 & $0.26$ & $<0.1$ & $0.00$ & NA\tnote{1} 			     \\
                          &1.5145& M13  &  $1.02$ & 0.2 & $t_{\rm trunc}=0.1\ {\rm Gyr}$ & 1.12 & $0.37$ & $<0.1$ & $0.00$ & NA\tnote{1} 			     \\
			  &	 & BC03 &  $0.45$ & 0.2 & $t_{\rm trunc}=0.3\ {\rm Gyr}$ & 1.11 & $0.47$ & $<0.1$ & $0.51$ & \citet{Prevot-1984,Bouchet-1985}	     \\
\hline
  \multirow{3}{*}{16}     &	 & M05  &  $1.28$ & 1.0 & $t_{\rm trunc}=0.3\ {\rm Gyr}$ & 1.17 & $0.55$ & $<0.1$ & $0.00$ & NA\tnote{1} 			     \\
                          &1.5199& M13  &  $1.28$ & 0.2 & $t_{\rm trunc}=0.3\ {\rm Gyr}$ & 1.17 & $0.54$ & $<0.1$ & $0.00$ & NA\tnote{1} 			     \\
			  &	 & BC03 &  $1.28$ & 0.5 & SSP				 & 1.12 & $0.51$ & $<0.1$ & $0.05$ & \citet{Calzetti-2000}		     \\
\hline
  \multirow{3}{*}{17}     &	 & M05  &  $0.72$ & 2.0 & $t_{\rm trunc}=0.3\ {\rm Gyr}$ & 1.38 & $0.47$ & $<0.1$ & $0.00$ & NA\tnote{1} 			     \\
                          &1.5210& M13  &  $1.14$ & 0.2 & $e^{-t/0.1\ {\rm Gyr}}$	 & 1.42 & $0.68$ & $<0.1$ & $0.00$ & NA\tnote{1} 			     \\
			  &	 & BC03 &  $0.81$ & 2.0 & $e^{-t/0.1\ {\rm Gyr}}$	 & 1.49 & $0.85$ & $0.3$ & $0.13$ & \citet{Seaton-1979} 		     \\
\hline
  \multirow{3}{*}{18}     &	 & M05  &  $2.10$ & 1.0 & $e^{-t/0.3\ {\rm Gyr}}$	 & 1.12 & $0.98$ & $0.4$ & $0.07$ & \citet{Prevot-1984,Bouchet-1985}	     \\
                          &1.5230& M13  &  $2.10$ & 0.2 & $e^{-t/0.3\ {\rm Gyr}}$	 & 1.12 & $0.95$ & $0.4$ & $0.07$ & \citet{Prevot-1984,Bouchet-1985}	     \\
			  &	 & BC03 &  $2.10$ & 1.0 & $e^{-t/0.3\ {\rm Gyr}}$	 & 1.00 & $1.26$ & $0.5$ & $0.15$ & \citet{Prevot-1984,Bouchet-1985}	     \\
\hline
\hline
\end{tabular}
\begin{tablenotes}\footnotesize 
\item[1] The SED of this galaxy was best-fitted with $E(B-V)=0$ also when reddening was allowed.
\end{tablenotes}
\end{threeparttable}							    
\end{center}
\end{scriptsize}
\end{table*}

\addtocounter{table}{-1}

\begin{table*}
\begin{scriptsize}
\begin{center}
\caption{Continued.}
\label{tab:Table2_appB}
\begin{threeparttable} 
\begin{tabular}{cccclcccccl}
\hline
  \multicolumn{1}{c}{\bf ID} &
  \multicolumn{1}{c}{$\mathbf{z_{\rm spec}}$} &
  \multicolumn{1}{c}{\bf Model} &
  \multicolumn{1}{c}{$\mathbf{t}$} &
  \multicolumn{1}{c}{$\mathbf{[Z/H]}$} &
  \multicolumn{1}{c}{\bf SFH} &
  \multicolumn{1}{c}{$\mathbf{\chi^{2}_{\rm r}}$} &
  \multicolumn{1}{c}{$\mathbf{M^{\ast}}$} &
  \multicolumn{1}{c}{$\mathbf{SFR}$} &  
  \multicolumn{1}{c}{$\mathbf{E(B-V)}$} &
  \multicolumn{1}{c}{\bf Reddening Law} \\
  \multicolumn{1}{c}{}&
  \multicolumn{1}{c}{}&
  \multicolumn{1}{c}{}&
  \multicolumn{1}{c}{\rm (Gyr)}&
  \multicolumn{1}{c}{${\rm (Z_{\odot})}$} &
  \multicolumn{1}{c}{}&
  \multicolumn{1}{c}{}&
  \multicolumn{1}{c}{$(10^{11}\ {\rm M_{\odot}})$} & 
  \multicolumn{1}{c}{$({\rm M_{\odot}/ yr^{-1}})$} &
  \multicolumn{1}{c}{({\rm mag})}&
  \multicolumn{1}{c}{} \\
  \multicolumn{1}{c}{(1)}&
  \multicolumn{1}{c}{(2)}&
  \multicolumn{1}{c}{(3)}&
  \multicolumn{1}{c}{(4)}&
  \multicolumn{1}{c}{(5)}&
  \multicolumn{1}{c}{(6)}&
  \multicolumn{1}{c}{(7)}&
  \multicolumn{1}{c}{(8)}&
  \multicolumn{1}{c}{(9)}&
  \multicolumn{1}{c}{(10)}&
  \multicolumn{1}{c}{(11)} \\
\hline
\hline
  \multirow{3}{*}{19}     &	 & M05  &  $1.28$ & 1.0 & $e^{-t/0.1\ {\rm Gyr}}$	 & 0.77 & $0.30$ & $<0.1$ & $0.00$ & NA\tnote{1} 			     \\
                          &1.5410& M13  &  $1.28$ & 0.2 & $e^{-t/0.1\ {\rm Gyr}}$	 & 0.77 & $0.30$ & $<0.1$ & $0.00$ & NA\tnote{1} 			     \\
			  &	 & BC03 &  $1.28$ & 0.2 & SSP				 & 0.73 & $0.26$ & $<0.1$ & $0.13$ & \citet{Seaton-1979} 		     \\
\hline
  \multirow{3}{*}{20}     &	 & M05  &  $1.43$ & 2.0 & $t_{\rm trunc}=1\ {\rm Gyr}$	 & 0.50 & $0.35$ & $<0.1$ & $0.00$ & NA\tnote{1} 			     \\
                          &1.5420& M13  &  $1.14$ & 0.2 & $t_{\rm trunc}=0.1\ {\rm Gyr}$ & 0.71 & $0.44$ & $<0.1$ & $0.07$ & \citet{Prevot-1984,Bouchet-1985}	     \\
			  &	 & BC03 &  $1.28$ & 0.2 & SSP				 & 0.69 & $0.42$ & $<0.1$ & $0.22$ & \citet{Prevot-1984,Bouchet-1985}	     \\
\hline
  \multirow{3}{*}{21}     &	 & M05  &  $0.64$ & 2.0 & SSP				 & 0.98 & $0.46$ & $<0.1$ & $0.00$ & NA\tnote{1} 			     \\
                          &1.5490& M13  &  $0.72$ & 1.0 & SSP				 & 1.02 & $0.83$ & $<0.1$ & $0.19$ & \citet{Seaton-1979} 		     \\
			  &	 & BC03 &  $1.43$ & 0.2 & $e^{-t/0.1\ {\rm Gyr}}$	 & 0.88 & $0.85$ & $<0.1$ & $0.22$ & \citet{Prevot-1984,Bouchet-1985}	     \\
\hline
  \multirow{3}{*}{22}     &	 & M05  &  $1.80$ & 1.0 &  $t_{\rm trunc}=1\ {\rm Gyr}$  & 0.80 & $0.83$ & $<0.1$ & $0.00$ & NA\tnote{1} 			     \\
                          &1.5499& M13  &  $1.80$ & 0.2 &  $t_{\rm trunc}=1\ {\rm Gyr}$  & 0.80 & $0.81$ & $<0.1$ & $0.00$ & NA\tnote{1} 			     \\
			  &	 & BC03 &  $2.40$ & 0.5 & $e^{-t/0.3\ {\rm Gyr}}$	 & 0.60 & $1.41$ & $0.2$  & $0.07$ & \citet{Prevot-1984,Bouchet-1985}	     \\
\hline
  \multirow{3}{*}{23}     &	 & M05  &  $0.72$ & 2.0 &  $t_{\rm trunc}=0.1\ {\rm Gyr}$& 1.14 & $0.56$ & $<0.1$ & $0.00$ & NA\tnote{1} 			     \\
                          &1.5519& M13  &  $0.72$ & 2.0 & SSP				 & 1.14 & $0.83$ & $<0.1$ & $0.05$ & \citet{Calzetti-2000}		     \\
			  &	 & BC03 &  $2.20$ & 0.5 & $e^{-t/0.3\ {\rm Gyr}}$	 & 1.03 & $1.55$ & $0.5$  & $0.15$ & \citet{Prevot-1984,Bouchet-1985}	     \\
\hline
  \multirow{3}{*}{24}     &	 & M05  &  $2.10$ & 0.2 & $t_{\rm trunc}=1\ {\rm Gyr}$	 & 2.48 & $0.87$ & $<0.1$ & $0.00$ & NA\tnote{1} 			     \\
                          &1.5596& M13  &  $0.51$ & 0.2 & SSP				 & 2.61 & $1.55$ & $<0.1$ & $0.35$ & \citet{Calzetti-2000}		     \\
			  &	 & BC03 &  $1.80$ & 0.2 & $e^{-t/0.3\ {\rm Gyr}}$	 & 2.45 & $1.51$ & $1.7$  & $0.15$ & \citet{Calzetti-2000}		     \\
\hline
  \multirow{3}{*}{25}     &	 & M05  &  $0.64$ & 2.0 & $t_{\rm trunc}=0.1\ {\rm Gyr}$ & 1.58 & $0.69$ & $<0.1$ & $0.00$ & NA\tnote{1} 			     \\
                          &1.5642& M13  &  $0.90$ & 1.0 & SSP				 & 1.49 & $1.38$ & $<0.1$ & $0.10$ & \citet{Calzetti-2000}		     \\
			  &	 & BC03 &  $1.28$ & 0.2 & SSP				 & 1.20 & $0.95$ & $<0.1$ & $0.15$ & \citet{Prevot-1984,Bouchet-1985}	     \\
\hline
  \multirow{3}{*}{26}     &	 & M05  &  $0.32$ & 1.0 & $t_{\rm trunc}=0.1\ {\rm Gyr}$ & 0.92 & $0.74$ & $<0.1$ & $0.44$ & \citet{Prevot-1984,Bouchet-1985}	     \\
                          &1.5681& M13  &  $0.32$ & 0.2 & $t_{\rm trunc}=0.1\ {\rm Gyr}$ & 0.92 & $0.72$ & $<0.1$ & $0.44$ & \citet{Prevot-1984,Bouchet-1985}	     \\
			  &	 & BC03 &  $1.43$ & 0.2 & $e^{-t/0.3\ {\rm Gyr}}$	 & 0.95 & $1.00$ & $3.8$  & $0.37$ & \citet{Prevot-1984,Bouchet-1985}	     \\
\hline
  \multirow{3}{*}{27}     &	 & M05  &  $1.61$ & 1.0 & $t_{\rm trunc}=1\ {\rm Gyr}$	 & 1.11 & $0.47$ & $<0.1$ & $0.00$ & NA\tnote{1} 			     \\
                          &1.5839& M13  &  $1.61$ & 0.2 & $t_{\rm trunc}=1\ {\rm Gyr}$	 & 1.11 & $0.46$ & $<0.1$ & $0.00$ & NA\tnote{1} 			     \\
			  &	 & BC03 &  $1.80$ & 0.2 & $e^{-t/0.3\ {\rm Gyr}}$	 & 1.15 & $0.79$ & $0.9$  & $0.22$ & \citet{Prevot-1984,Bouchet-1985}	     \\
\hline
  \multirow{3}{*}{28}     &	 & M05  &  $2.75$ & 0.2 & $e^{-t/0.3\ {\rm Gyr}}$	 & 0.89 & $1.95$ & $0.1$  & $0.15$ & \citet{Calzetti-2000}		     \\
                          &1.5888& M13  &  $1.43$ & 0.2 & $t_{\rm trunc}=1\ {\rm Gyr}$	 & 1.10 & $0.91$ & $<0.1$ & $0.13$ & \citet{Allen-1976}  		     \\
			  &	 & BC03 &  $4.25$ & 0.2 & $e^{-t/1\ {\rm Gyr}}$ 	 & 0.90 & $2.75$ & $5.7$  & $0.29$ & \citet{Prevot-1984,Bouchet-1985}	     \\
\hline
  \multirow{3}{*}{29}     &	 & M05  &  $0.25$ & 1.0 & $t_{\rm trunc}=0.1\ {\rm Gyr}$ & 1.38 & $0.76$ & $<0.1$ & $0.44$ & \citet{Prevot-1984,Bouchet-1985}	     \\
                          &1.5939& M13  &  $0.81$ & 1.0 & SSP				 & 1.25 & $0.71$ & $<0.1$ & $0.05$ & \citet{Calzetti-2000}		     \\
			  &	 & BC03 &  $1.28$ & 0.2 & $e^{-t/0.1\ {\rm Gyr}}$	 & 1.23 & $0.71$ & $<0.1$ & $0.15$ & \citet{Prevot-1984,Bouchet-1985}	     \\
\hline
  \multirow{3}{*}{30}     &	 & M05  &  $2.75$ & 0.2 & $e^{-t/0.3\ {\rm Gyr}}$	 & 1.21 & $1.29$ & $0.1$  & $0.10$ & \citet{Calzetti-2000}		     \\
                          &1.6587& M13  &  $0.51$ & 1.0 & SSP				 & 1.15 & $0.81$ & $<0.1$ & $0.26$ & \citet{Allen-1976}  		     \\
			  &	 & BC03 &  $1.28$ & 0.2 & $e^{-t/0.1\ {\rm Gyr}}$	 & 1.10 & $0.89$ & $<0.1$ & $0.20$ & \citet{Calzetti-2000}		     \\
\hline
  \multirow{3}{*}{31}     &	 & M05  &  $0.18$ & 2.0 & SSP				 & 1.27 & $0.89$ & $<0.1$ & $0.37$ & \citet{Prevot-1984,Bouchet-1985}	     \\
                          &1.6590& M13  &  $0.72$ & 1.0 & SSP				 & 0.98 & $0.87$ & $<0.1$ & $0.06$ & \citet{Allen-1976}  		     \\
			  &	 & BC03 &  $0.90$ & 0.2 & SSP				 & 0.73 & $1.12$ & $<0.1$ & $0.22$ & \citet{Prevot-1984,Bouchet-1985}	     \\
\hline
  \multirow{3}{*}{32}     &	 & M05  &  $0.20$ & 2.0 & SSP				 & 1.89 & $0.79$ & $<0.1$ & $0.29$ & \citet{Prevot-1984,Bouchet-1985}	     \\
                          &1.6658& M13  &  $0.51$ & 0.2 & SSP				 & 1.83 & $1.15$ & $<0.1$ & $0.37$ & \citet{Prevot-1984,Bouchet-1985}	     \\
			  &	 & BC03 &  $0.81$ & 0.2 & SSP				 & 1.31 & $1.05$ & $<0.1$ & $0.22$ & \citet{Prevot-1984,Bouchet-1985}	     \\
\hline
  \multirow{3}{*}{33}     &	 & M05  &  $0.02$ & 0.5 & SSP				 & 2.60 & $0.56$ & $<0.1$ & $0.84$ & \citet{Allen-1976}  		     \\
                          &1.6750& M13  &  $0.05$ & 2.0 & SSP				 & 3.41 & $2.51$ & $<0.1$ & $0.74$ & \citet{Calzetti-2000}		     \\
			  &	 & BC03 &  $0.16$ & 2.0 & $t_{\rm trunc}=0.1\ {\rm Gyr}$ & 2.73 & $0.89$ & $<0.1$ & $0.58$ & \citet{Allen-1976}  		     \\
\hline
  \multirow{3}{*}{34}     &	 & M05  &  $1.61$ & 1.0 & $t_{\rm trunc}=1\ {\rm Gyr}$	 & 0.85 & $1.00$ & $<0.1$ & $0.06$ & \citet{Seaton-1979} 		     \\
                          &1.8080& M13  &  $1.61$ & 0.2 & $t_{\rm trunc}=1\ {\rm Gyr}$	 & 0.85 & $1.00$ & $<0.1$ & $0.06$ & \citet{Seaton-1979} 		     \\
			  &	 & BC03 &  $1.43$ & 0.2 & $t_{\rm trunc}=0.1\ {\rm Gyr}$ & 0.92 & $1.26$ & $<0.1$ & $0.19$ & \citet{Seaton-1979} 		     \\
\hline
  \multirow{3}{*}{35}     &	 & M05  &  $2.50$ & 0.5 & $e^{-t/0.3\ {\rm Gyr}}$	 & 1.17 & $2.75$ & $0.3$  & $0.00$ & NA\tnote{1} 			     \\
                          &1.8200& M13  &  $0.81$ & 1.0 & SSP				 & 1.32 & $2.34$ & $<0.1$ & $0.13$ & \citet{Allen-1976}  		     \\
			  &	 & BC03 &  $2.60$ & 0.2 & $e^{-t/0.3\ {\rm Gyr}}$	 & 1.11 & $4.68$ & $0.4$  & $0.15$ & \citet{Calzetti-2000}		     \\
\hline
  \multirow{3}{*}{36}     &	 & M05  &  $0.64$ & 2.0 & SSP				 & 0.90 & $0.58$ & $<0.1$ & $0.00$ & NA\tnote{1} 			     \\
                          &1.8200& M13  &  $0.81$ & 1.0 & SSP				 & 0.91 & $1.12$ & $<0.1$ & $0.15$ & \citet{Calzetti-2000}		     \\
			  &	 & BC03 &  $0.90$ & 2.0 & $e^{-t/0.1\ {\rm Gyr}}$	 & 0.89 & $1.05$ & $0.2$  & $0.13$ & \citet{Fitzpatrick-1986}		     \\
\hline
  \multirow{3}{*}{37}     &	 & M05  &  $2.20$ & 0.2 & $e^{-t/0.3\ {\rm Gyr}}$	 & 1.61 & $2.19$ & $0.6$  & $0.06$ & \citet{Seaton-1979} 		     \\
                          &1.8220& M13  &  $0.08$ & 2.0 & SSP				 & 2.46 & $2.14$ & $<0.1$ & $0.58$ & \citet{Allen-1976}  		     \\
			  &	 & BC03 &  $1.90$ & 0.5 & $e^{-t/0.3\ {\rm Gyr}}$	 & 2.03 & $2.75$ & $2.2$  & $0.10$ & \citet{Calzetti-2000}		     \\
\hline
\hline
\end{tabular}
\begin{tablenotes}\footnotesize 
\item[1] The SED of this galaxy was best-fitted with $E(B-V)=0$ also when reddening was allowed.
\end{tablenotes}
\end{threeparttable}							    
\end{center}
\end{scriptsize}
\end{table*}

\addtocounter{table}{-1}

\begin{table*}
\begin{scriptsize}
\begin{center}
\caption{Continued.}
\label{tab:Table2_appB}
\begin{threeparttable} 
\begin{tabular}{cccclcccccl}
\hline
  \multicolumn{1}{c}{\bf ID} &
  \multicolumn{1}{c}{$\mathbf{z_{\rm spec}}$} &
  \multicolumn{1}{c}{\bf Model} &
  \multicolumn{1}{c}{$\mathbf{t}$} &
  \multicolumn{1}{c}{$\mathbf{[Z/H]}$} &
  \multicolumn{1}{c}{\bf SFH} &
  \multicolumn{1}{c}{$\mathbf{\chi^{2}_{\rm r}}$} &
  \multicolumn{1}{c}{$\mathbf{M^{\ast}}$} &
  \multicolumn{1}{c}{$\mathbf{SFR}$} & 
  \multicolumn{1}{c}{$\mathbf{E(B-V)}$} &
  \multicolumn{1}{c}{\bf Reddening Law} \\
  \multicolumn{1}{c}{}&
  \multicolumn{1}{c}{}&
  \multicolumn{1}{c}{}&
  \multicolumn{1}{c}{\rm (Gyr)}&
  \multicolumn{1}{c}{${\rm (Z_{\odot})}$} &
  \multicolumn{1}{c}{}&
  \multicolumn{1}{c}{}&
  \multicolumn{1}{c}{$(10^{11}\ {\rm M_{\odot}})$} & 
  \multicolumn{1}{c}{$({\rm M_{\odot}/ yr^{-1}})$} &
  \multicolumn{1}{c}{({\rm mag})}&
  \multicolumn{1}{c}{} \\
  \multicolumn{1}{c}{(1)}&
  \multicolumn{1}{c}{(2)}&
  \multicolumn{1}{c}{(3)}&
  \multicolumn{1}{c}{(4)}&
  \multicolumn{1}{c}{(5)}&
  \multicolumn{1}{c}{(6)}&
  \multicolumn{1}{c}{(7)}&
  \multicolumn{1}{c}{(8)}&
  \multicolumn{1}{c}{(9)}&
  \multicolumn{1}{c}{(10)}&
  \multicolumn{1}{c}{(11)} \\
\hline
\hline
  \multirow{3}{*}{38}     &	 & M05  &  $2.50$ & 0.2 & $e^{-t/0.3\ {\rm Gyr}}$	 & 1.65 & $4.90$ & $0.5$  & $0.05$ & \citet{Calzetti-2000}		     \\
                          &1.8230& M13  &  $1.90$ & 0.2 & SSP				 & 2.10 & $5.75$ & $<0.1$ & $0.19$ & \citet{Allen-1976}  		     \\
			  &	 & BC03 &  $2.00$ & 0.2 & $e^{-t/0.3\ {\rm Gyr}}$	 & 1.58 & $5.89$ & $3.5$  & $0.15$ & \citet{Calzetti-2000}		     \\
\hline
  \multirow{3}{*}{39}     &	 & M05  &  $2.50$ & 0.2 & $t_{\rm trunc}=1\ {\rm Gyr}$	 & 0.76 & $1.58$ & $<0.1$ & $0.07$ & \citet{Prevot-1984,Bouchet-1985}	     \\
                          &1.8270& M13  &  $1.43$ & 0.2 & $t_{\rm trunc}=1\ {\rm Gyr}$	 & 0.92 & $1.05$ & $<0.1$ & $0.06$ & \citet{Fitzpatrick-1986}		     \\
			  &	 & BC03 &  $1.43$ & 0.2 & $e^{-t/0.3\ {\rm Gyr}}$	 & 0.82 & $1.55$ & $5.9$  & $0.29$ & \citet{Prevot-1984,Bouchet-1985}	     \\
\hline
  \multirow{3}{*}{40}     &	 & M05  &  $0.32$ & 1.0 & $t_{\rm trunc}=0.1\ {\rm Gyr}$ & 1.50 & $0.72$ & $<0.1$ & $0.22$ & \citet{Prevot-1984,Bouchet-1985}	     \\
                          &1.8368& M13  &  $1.61$ & 0.2 & SSP				 & 1.44 & $1.35$ & $<0.1$ & $0.10$ & \citet{Calzetti-2000}		     \\
			  &	 & BC03 &  $1.28$ & 0.2 & $e^{-t/0.3\ {\rm Gyr}}$	 & 1.47 & $1.10$ & $6.9$  & $0.22$ & \citet{Prevot-1984,Bouchet-1985}	     \\
\hline
  \multirow{3}{*}{41}     &	 & M05  &  $2.30$ & 0.2 & $t_{\rm trunc}=1\ {\rm Gyr}$	 & 0.92 & $1.20$ & $<0.1$ & $0.00$ & NA\tnote{1} 			     \\
                          &1.9677& M13  &  $0.81$ & 0.2 & $e^{-t/0.1\ {\rm Gyr}}$	 & 1.08 & $0.72$ & $0.3$  & $0.00$ & NA\tnote{1} 			     \\
			  &	 & BC03 &  $1.43$ & 0.2 & $e^{-t/0.3\ {\rm Gyr}}$	 & 1.11 & $1.32$ & $5.0$  & $0.22$ & \citet{Prevot-1984,Bouchet-1985}	     \\
\hline
  \multirow{3}{*}{42}     &	 & M05  &  $0.20$ & 1.0 & SSP				 & 1.09 & $0.78$ & $<0.1$ & $0.37$ & \citet{Prevot-1984,Bouchet-1985}	     \\
                          &1.9677& M13  &  $0.64$ & 1.0 & SSP				 & 0.84 & $0.72$ & $<0.1$ & $0.06$ & \citet{Seaton-1979} 		     \\
			  &	 & BC03 &  $0.90$ & 1.0 & $e^{-t/0.1\ {\rm Gyr}}$	 & 0.65 & $0.85$ & $0.1$  & $0.06$ & \citet{Fitzpatrick-1986}		     \\
\hline
  \multirow{3}{*}{43}     &	 & M05  &  $2.60$ & 0.2 & $e^{-t/0.3\ {\rm Gyr}}$	 & 1.17 & $2.09$ & $0.2$  & $0.00$ & NA\tnote{1} 			     \\
                          &2.0799& M13  &  $1.61$ & 0.2 & SSP				 & 1.20 & $2.45$ & $<0.1$ & $0.19$ & \citet{Fitzpatrick-1986}		     \\
			  &	 & BC03 &  $2.10$ & 0.2 & $e^{-t/0.3\ {\rm Gyr}}$	 & 0.75 & $2.57$ & $1.1$  & $0.10$ & \citet{Calzetti-2000}		     \\
\hline
  \multirow{3}{*}{44}     &	 & M05  &  $3.00$ & 0.2 & $e^{-t/0.3\ {\rm Gyr}}$	 & 1.17 & $2.45$ & $0.1$  & $0.00$ & NA\tnote{1} 			     \\
                          &2.0892& M13  &  $2.20$ & 0.2 & SSP				 & 1.16 & $2.51$ & $<0.1$ & $0.15$ & \citet{Prevot-1984,Bouchet-1985}	     \\
			  &	 & BC03 &  $1.70$ & 0.2 & $t_{\rm trunc}=1\ {\rm Gyr}$	 & 0.99 & $1.74$ & $<0.1$ & $0.15$ & \citet{Prevot-1984,Bouchet-1985}	     \\
\hline
\hline
\end{tabular}
\begin{tablenotes}\footnotesize 
\item[1] The SED of this galaxy was best-fitted with $E(B-V)=0$, also when reddening was allowed.
\end{tablenotes}
\end{threeparttable}							    
\end{center}
\end{scriptsize}
\end{table*}

\clearpage

\begin{figure*}
\centering
\includegraphics[width=0.48\textwidth]{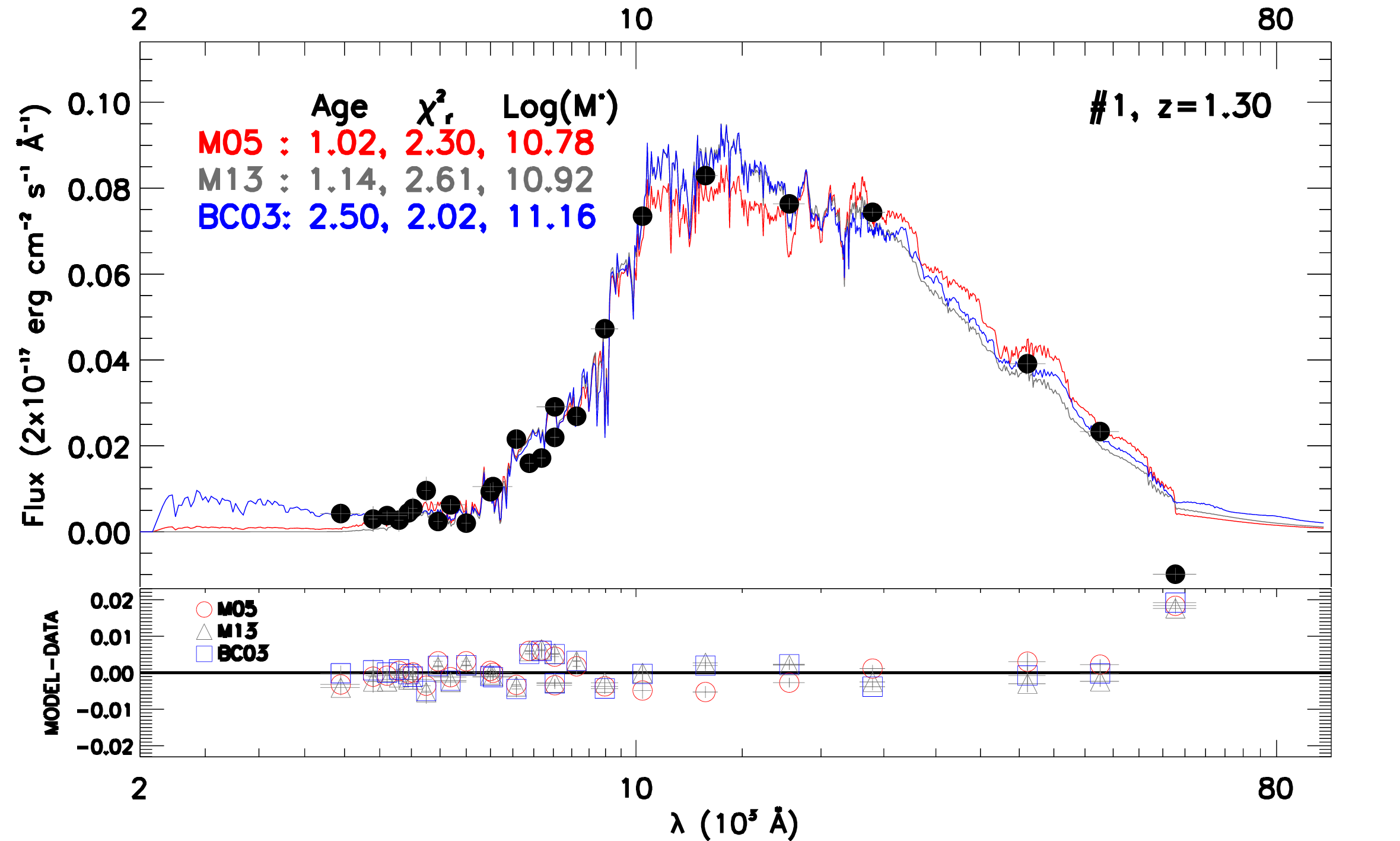}
\includegraphics[width=0.48\textwidth]{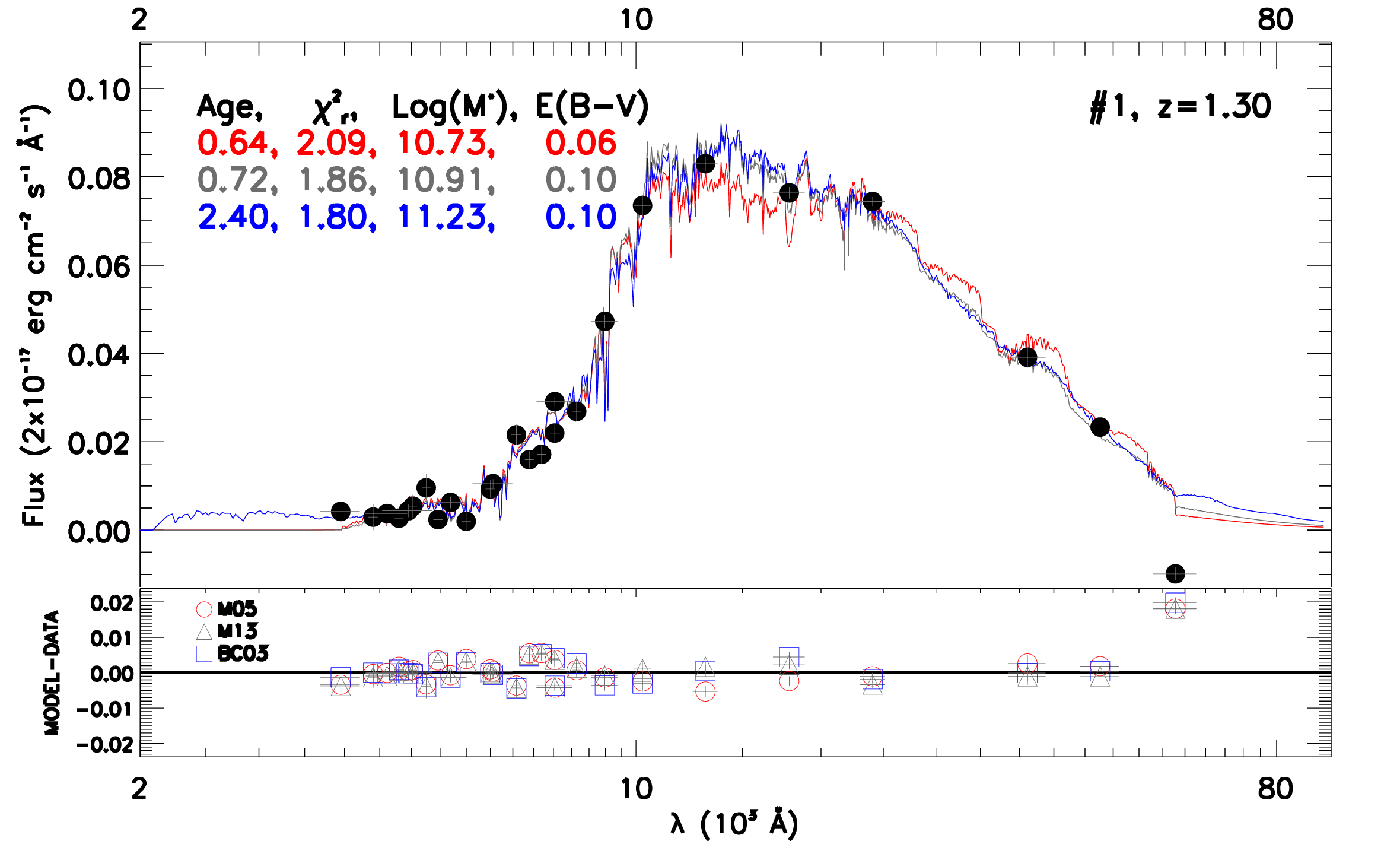}
\includegraphics[width=0.48\textwidth]{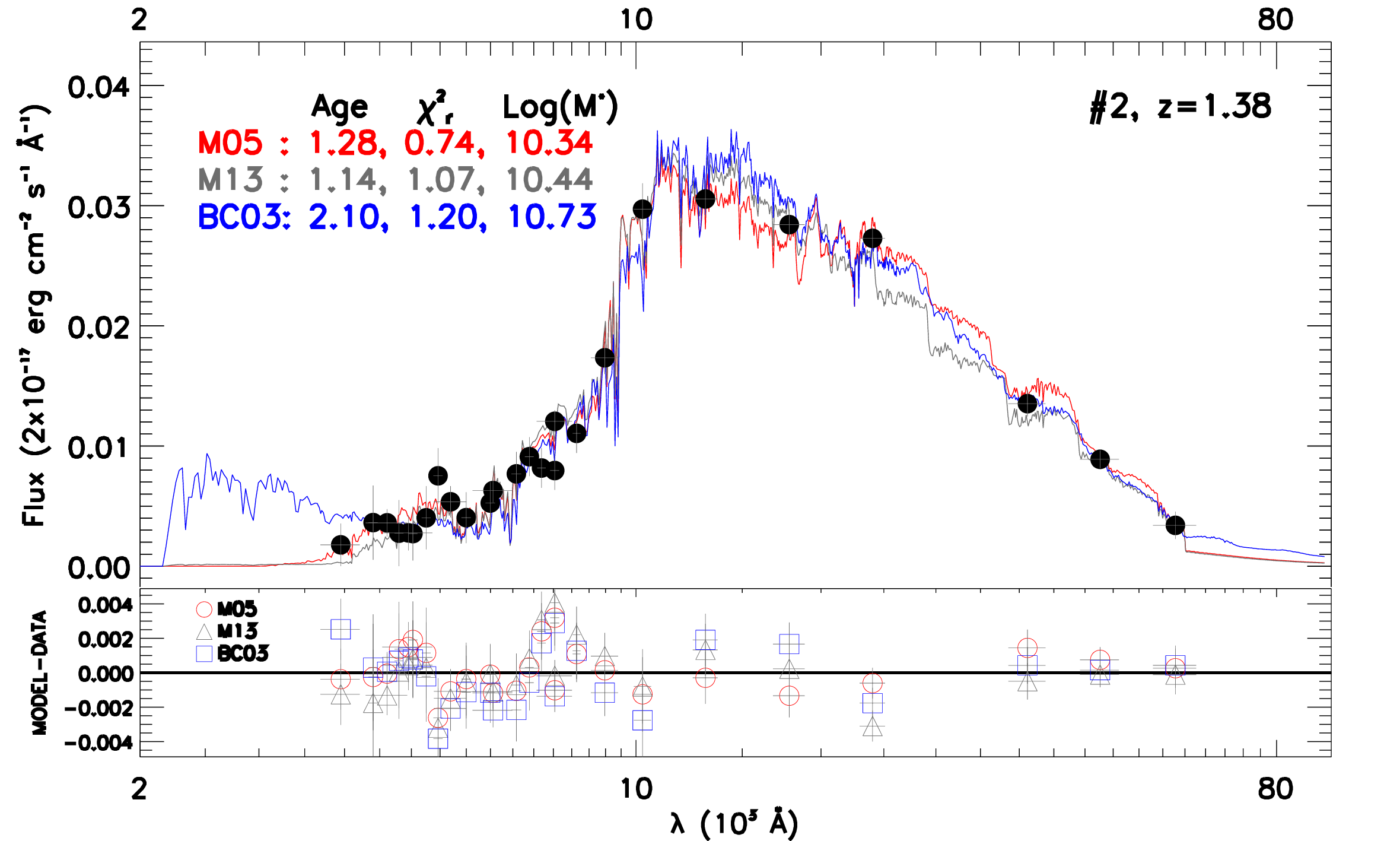}
\includegraphics[width=0.48\textwidth]{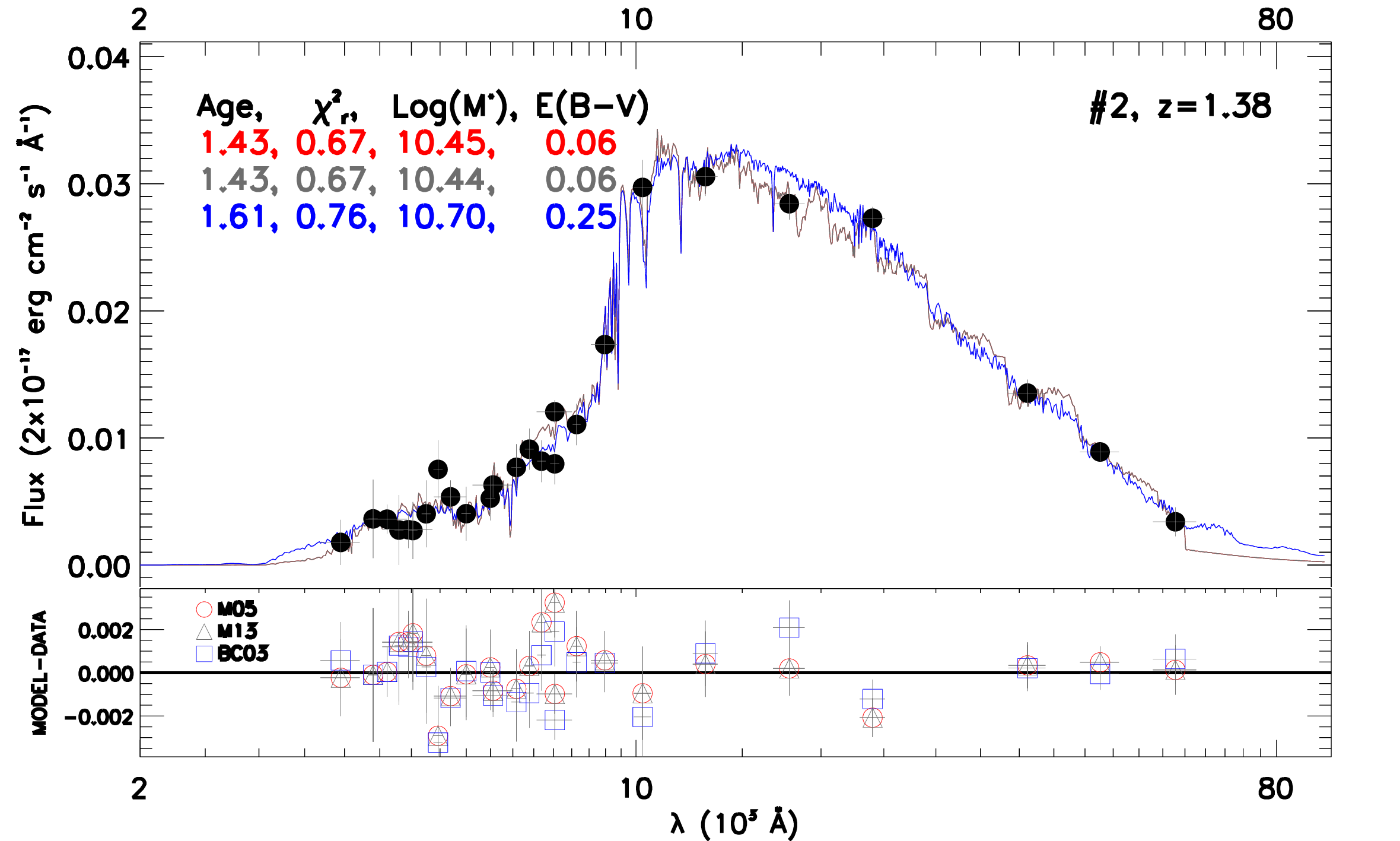}
\includegraphics[width=0.48\textwidth]{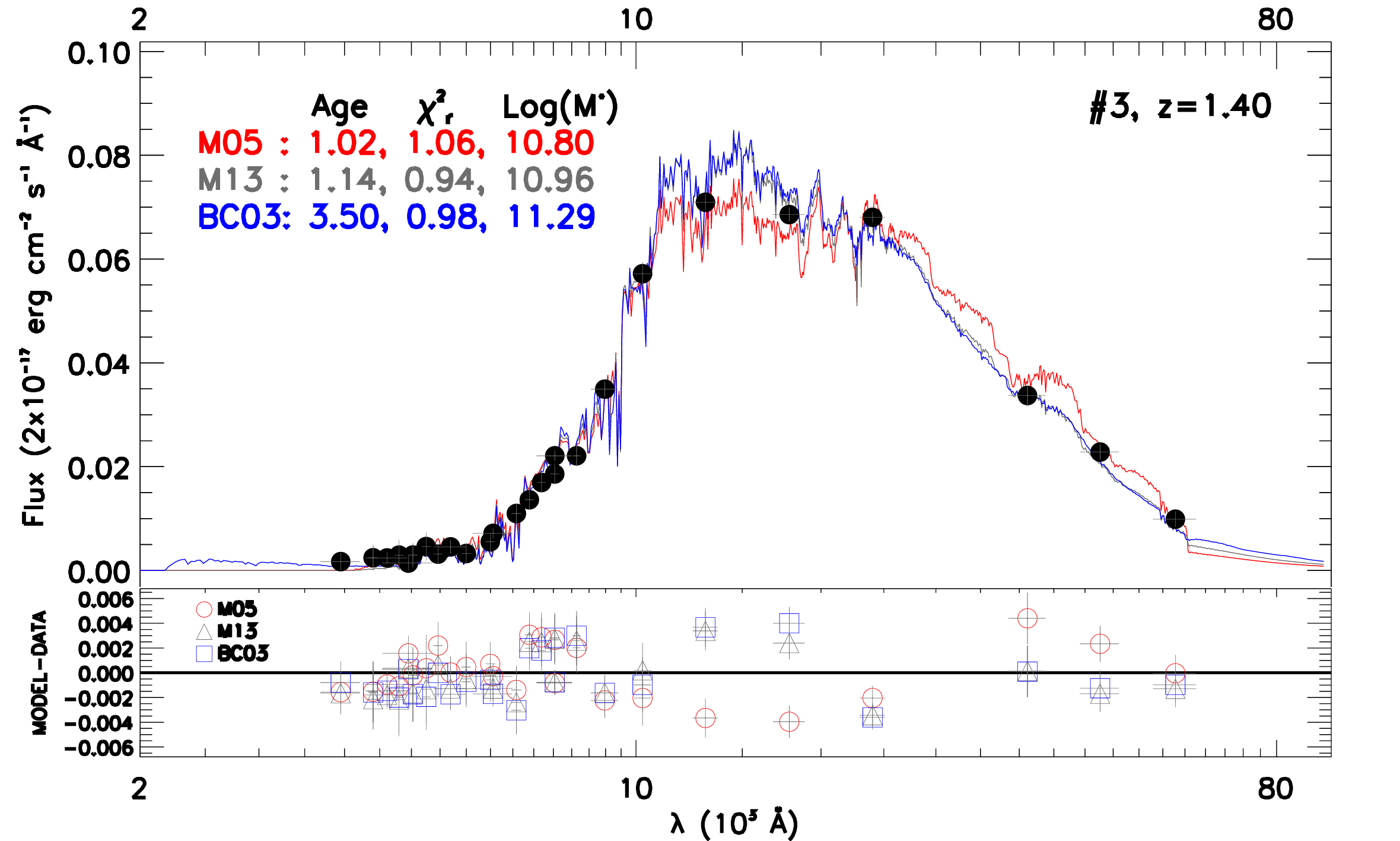}
\includegraphics[width=0.48\textwidth]{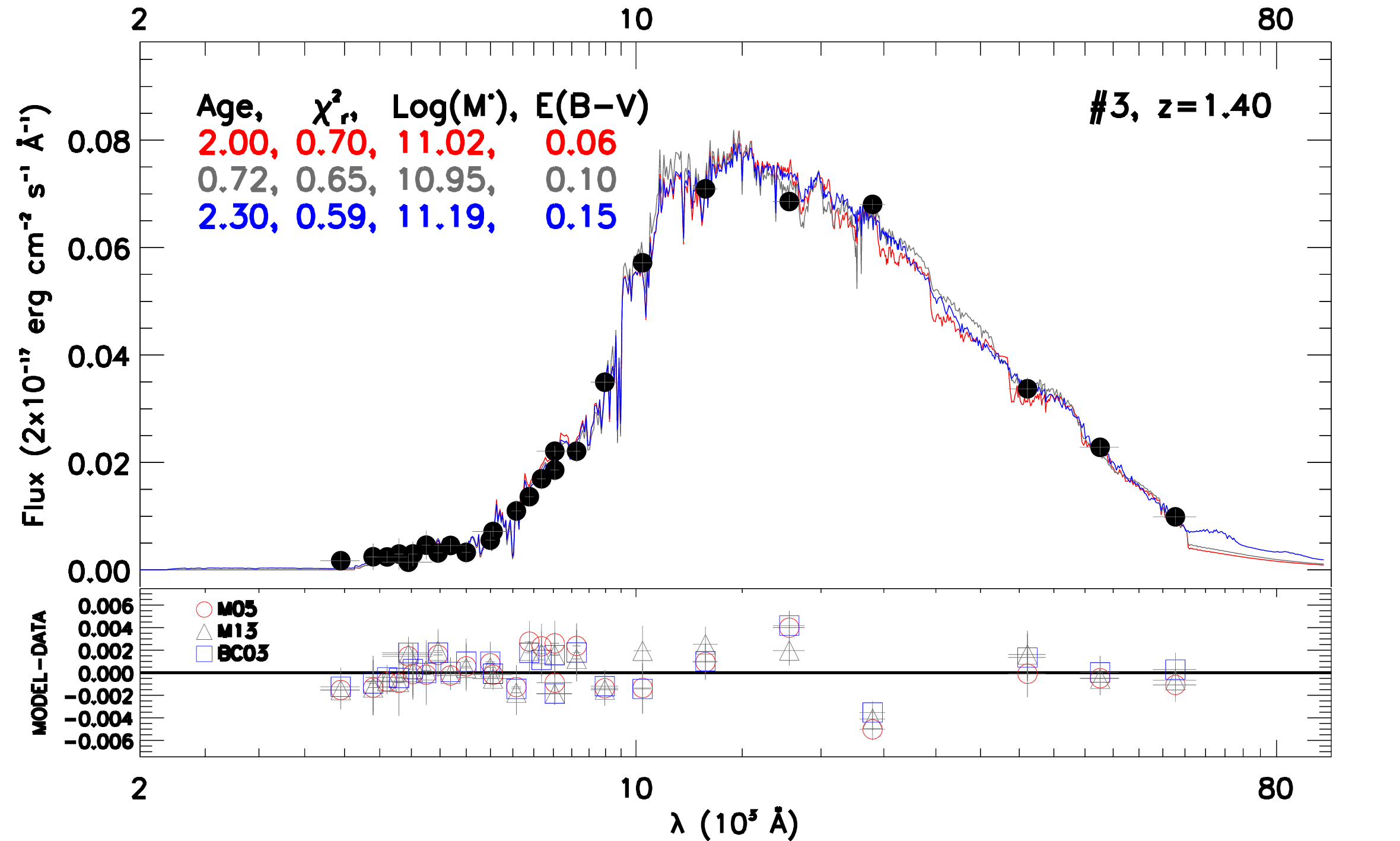}
\includegraphics[width=0.48\textwidth]{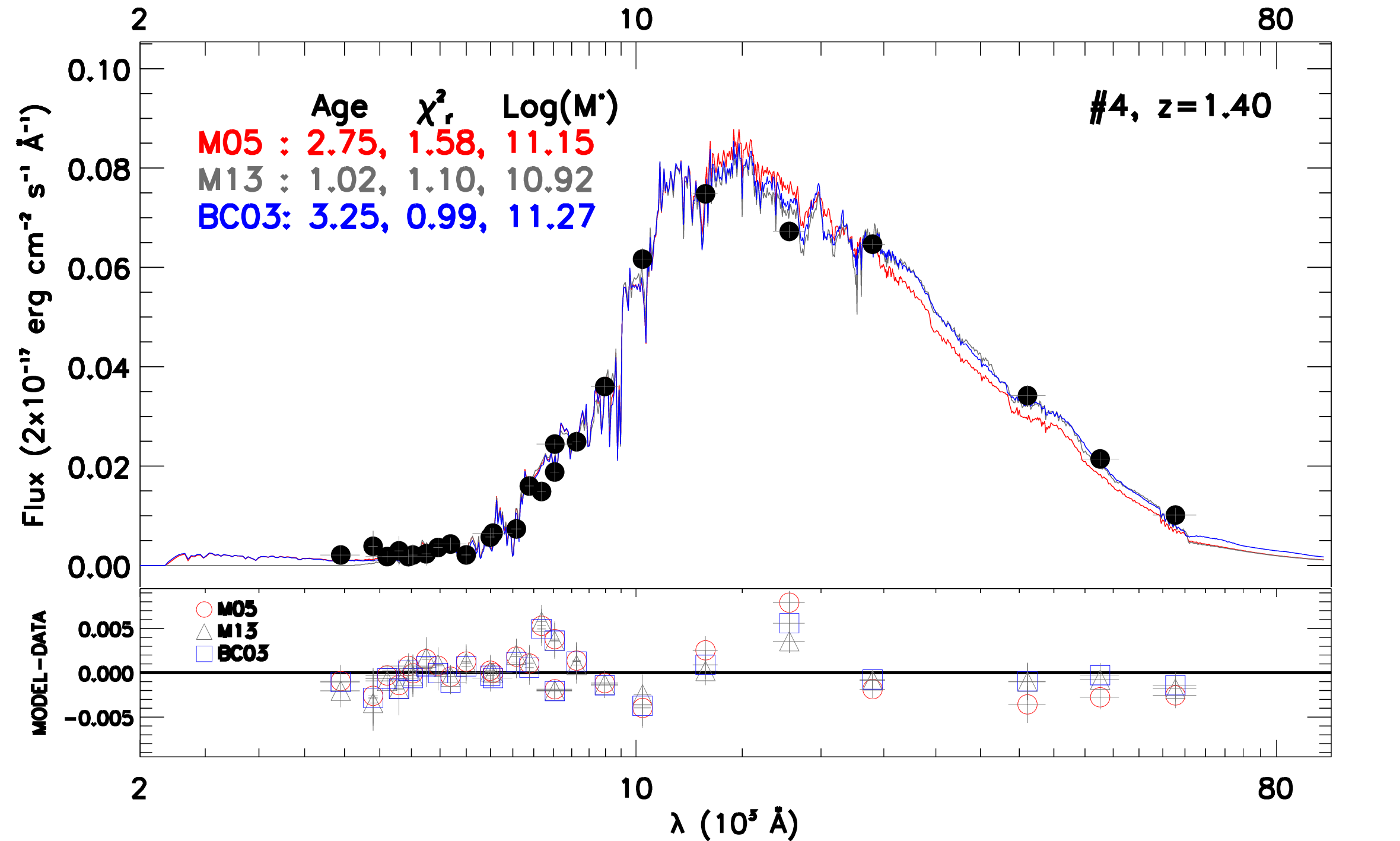}
\includegraphics[width=0.48\textwidth]{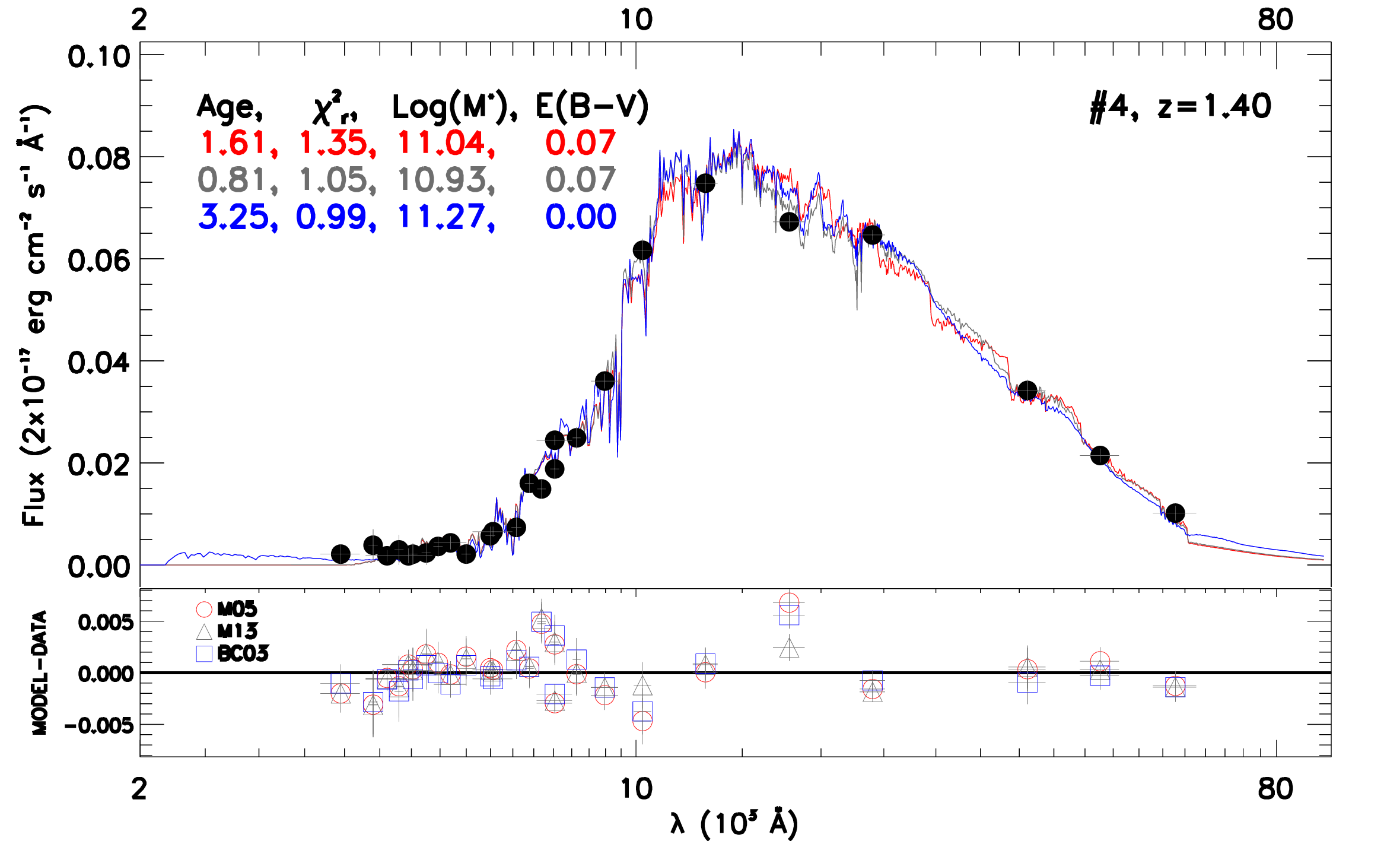}
\caption{SED fits of galaxies in the COSMOS sample. Left-hand panels: no-reddening case; right-hand panels: reddened case. Observed fluxes are plotted as symbols over best-fit templates showed as lines, for M05 (red), M13 (grey) and BC03 (blue) models. Flux residuals ($MODEL-DATA$) are plotted vs. wavelength at the bottom of each panel.}
\label{fig:Fig1_appB}
\end{figure*}

\addtocounter{figure}{-1}
\begin{figure*}
\centering
\includegraphics[width=0.48\textwidth]{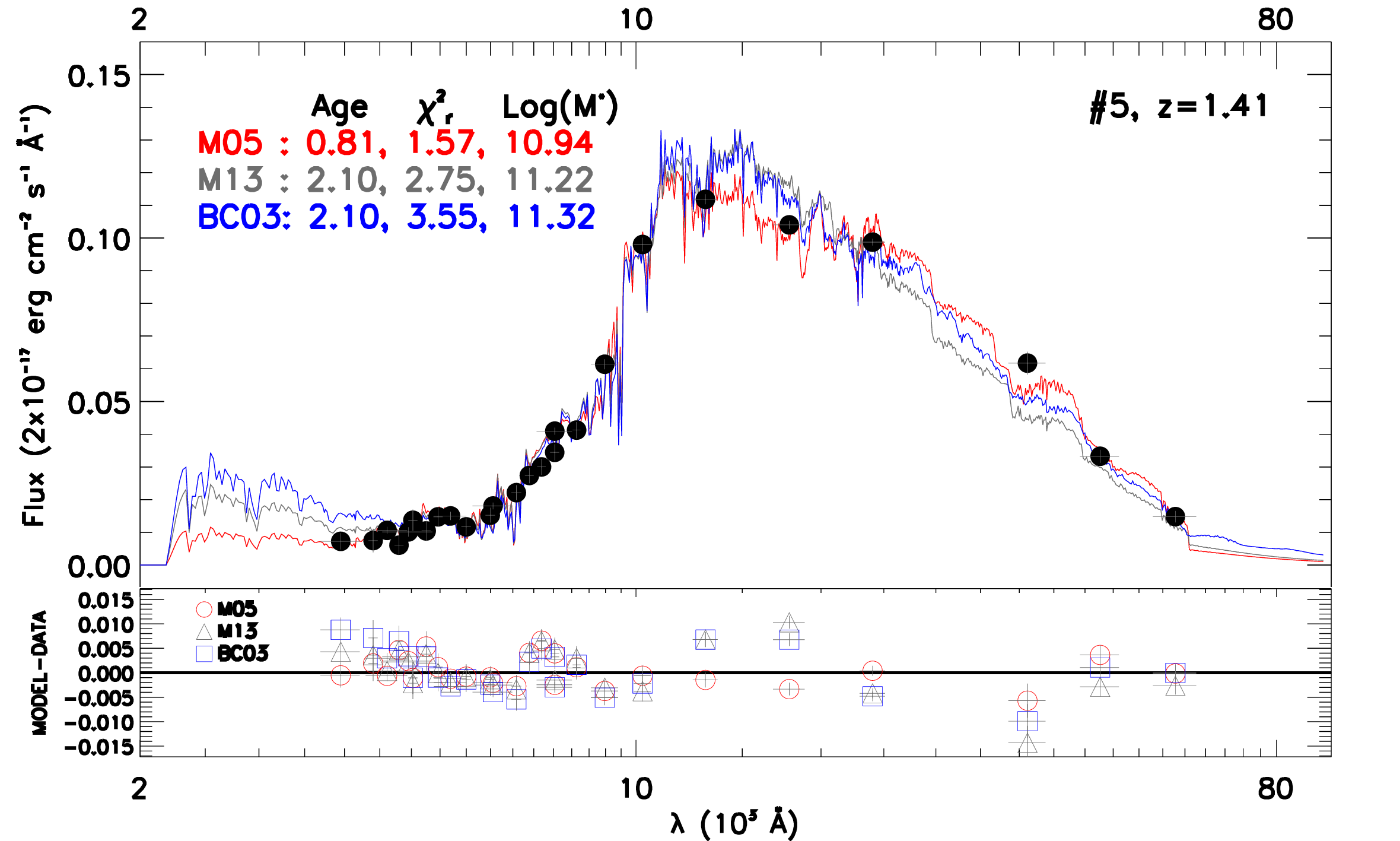}
\includegraphics[width=0.48\textwidth]{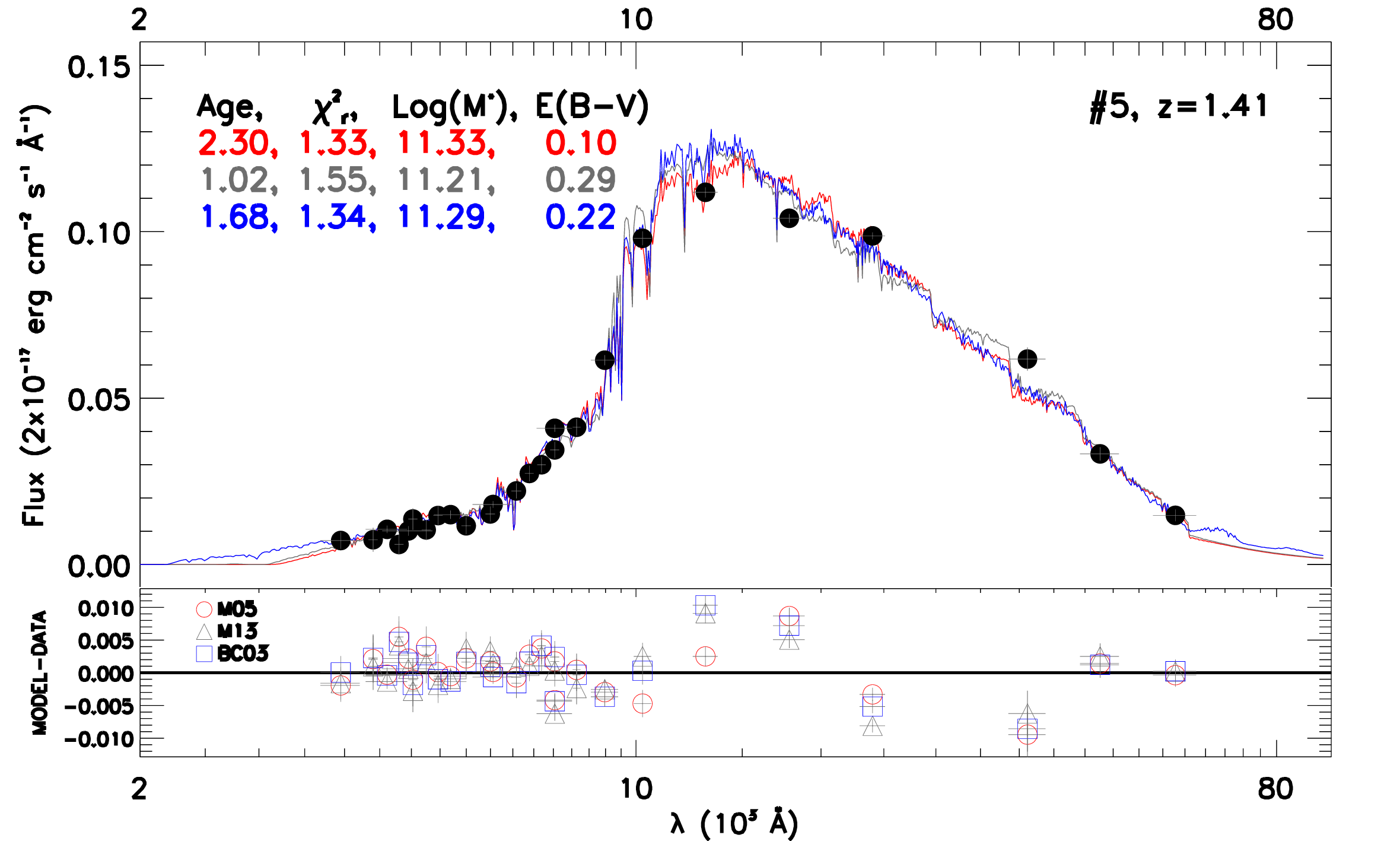}
\includegraphics[width=0.48\textwidth]{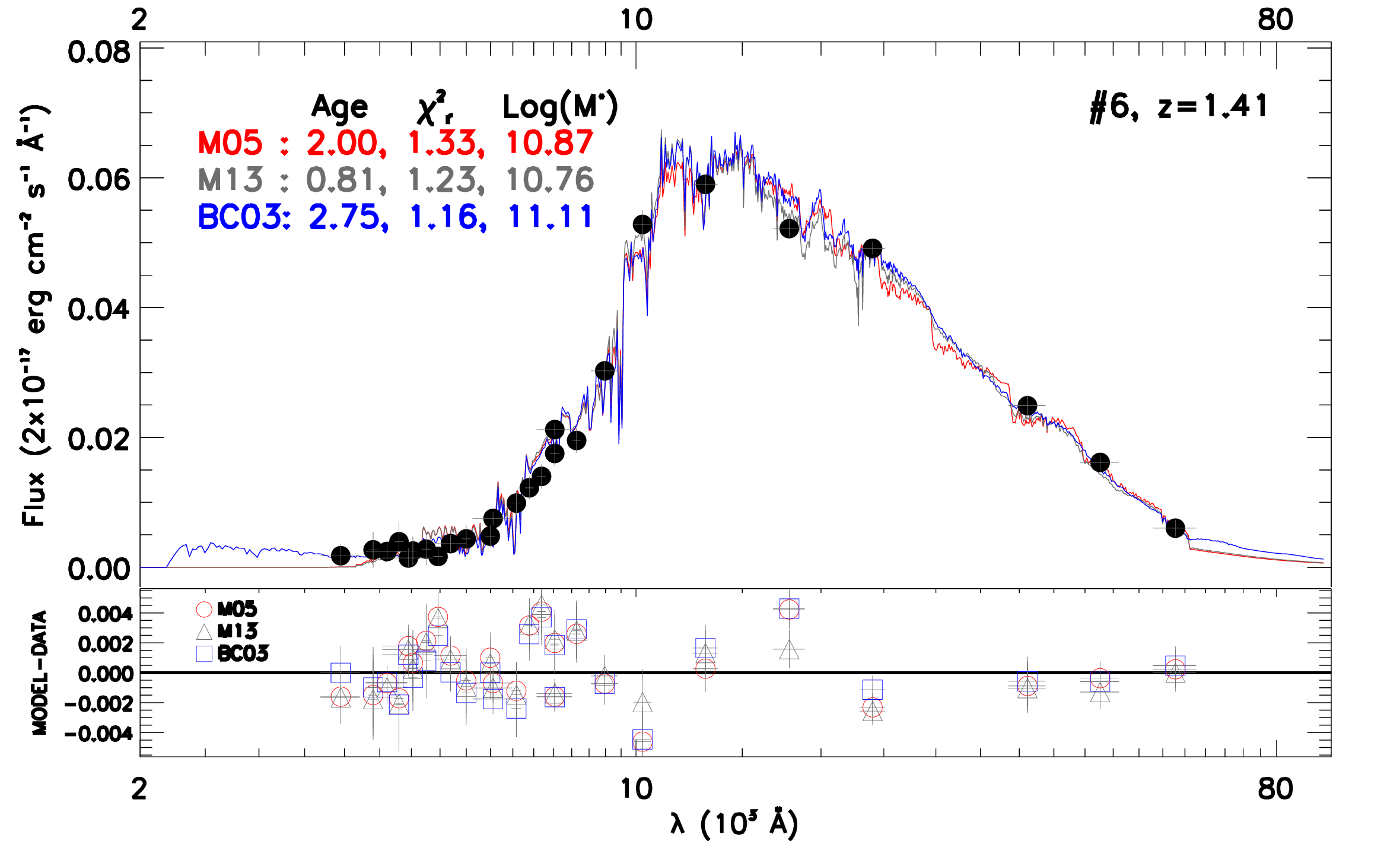}
\includegraphics[width=0.48\textwidth]{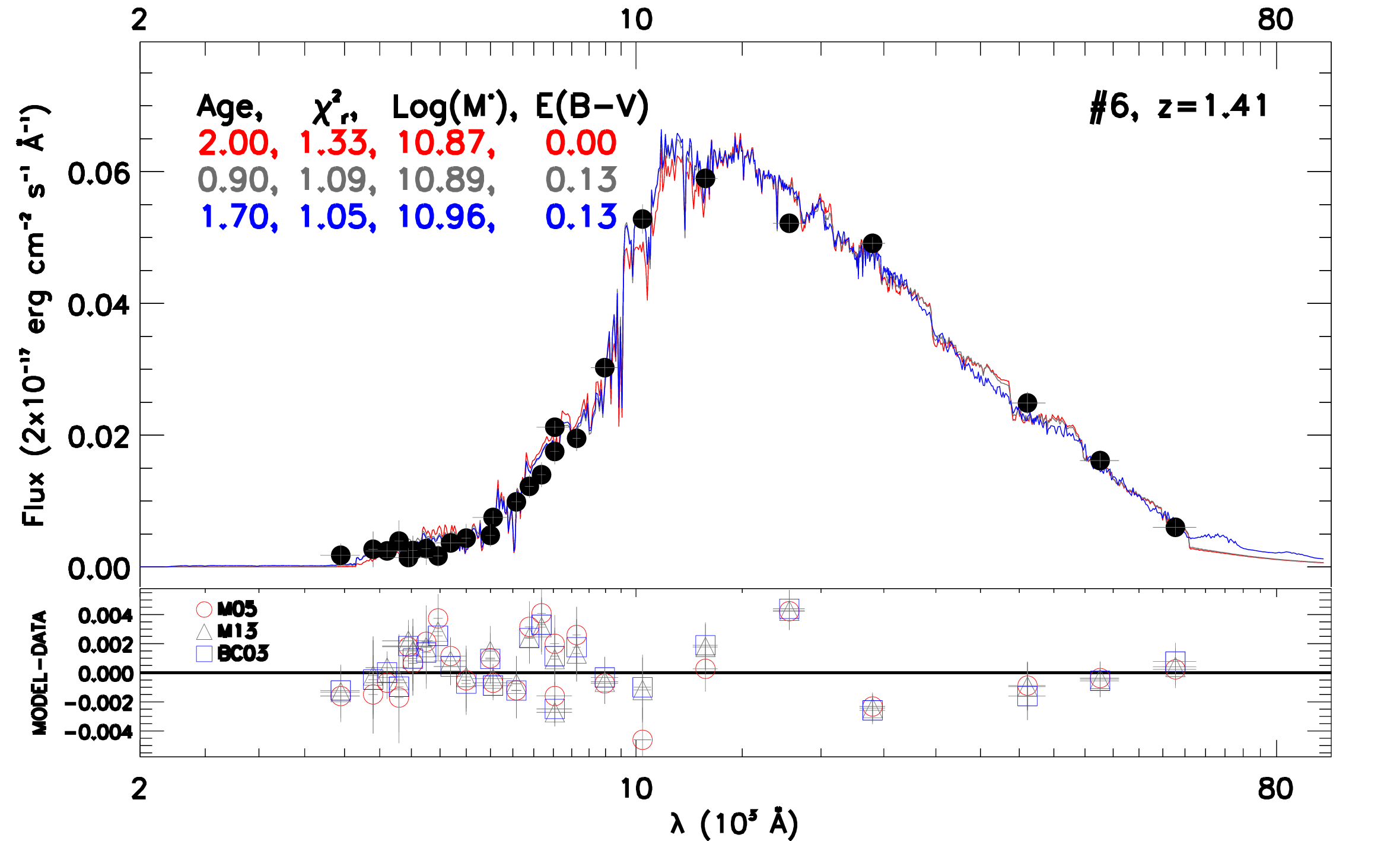}
\includegraphics[width=0.48\textwidth]{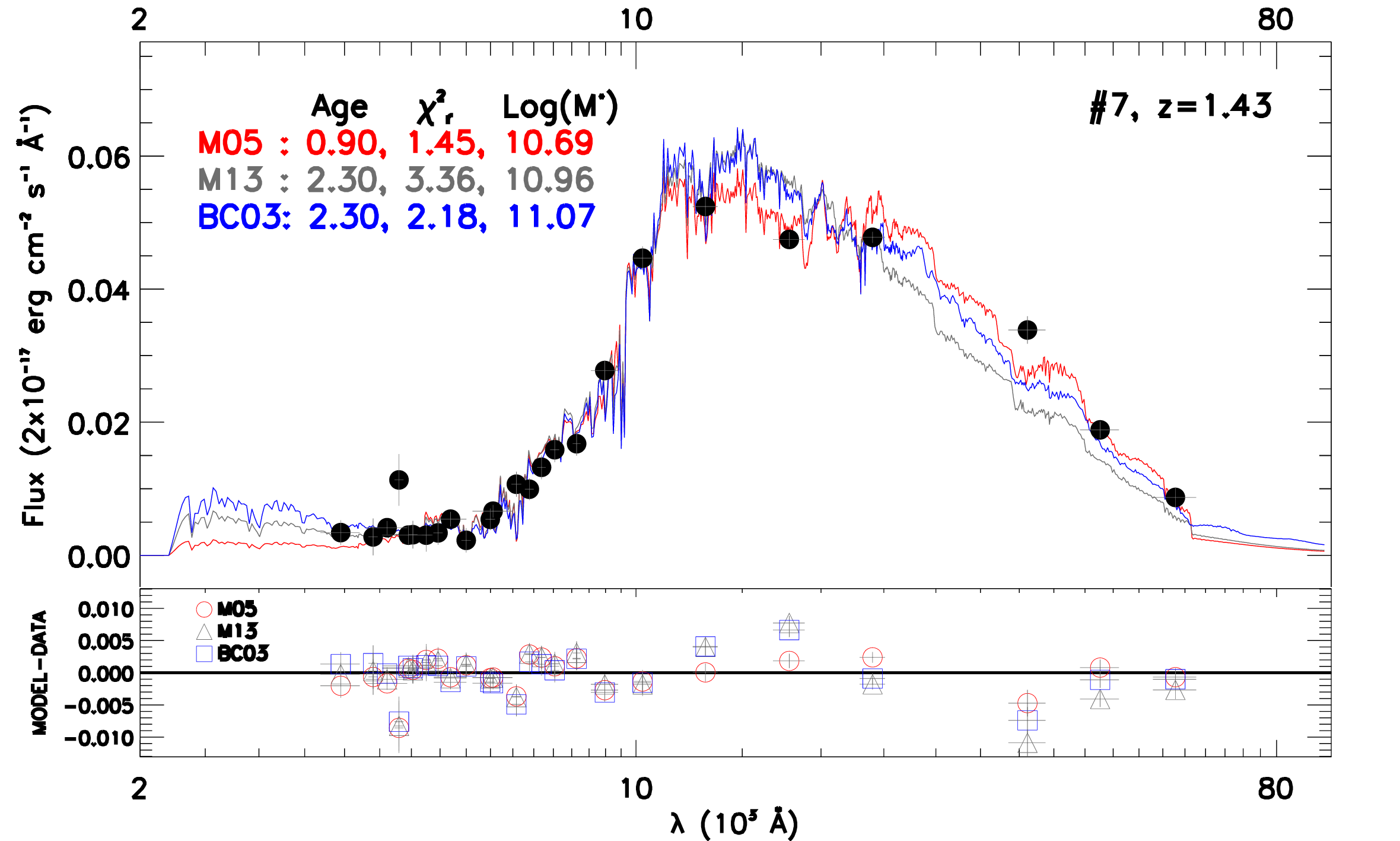}
\includegraphics[width=0.48\textwidth]{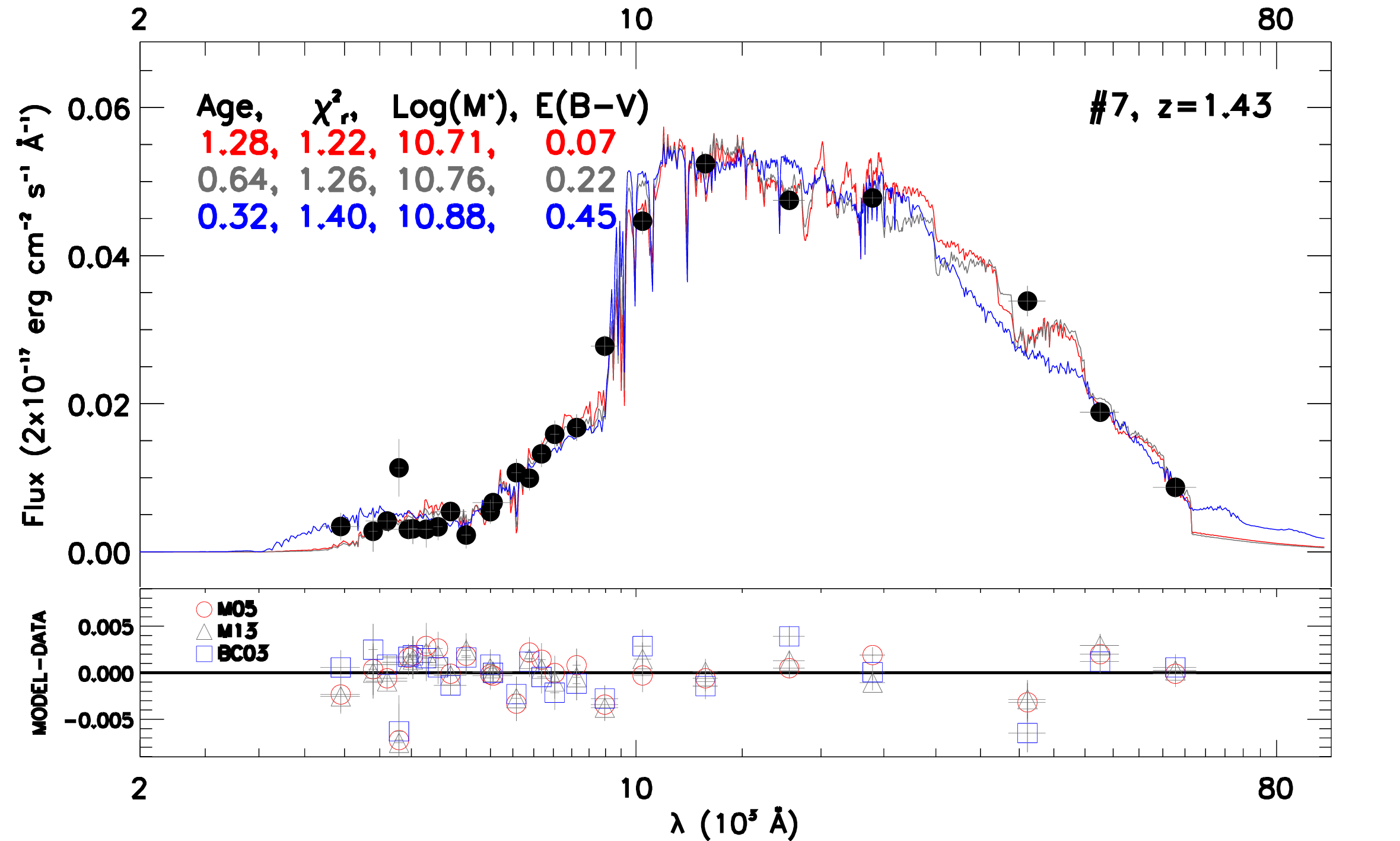}
\includegraphics[width=0.48\textwidth]{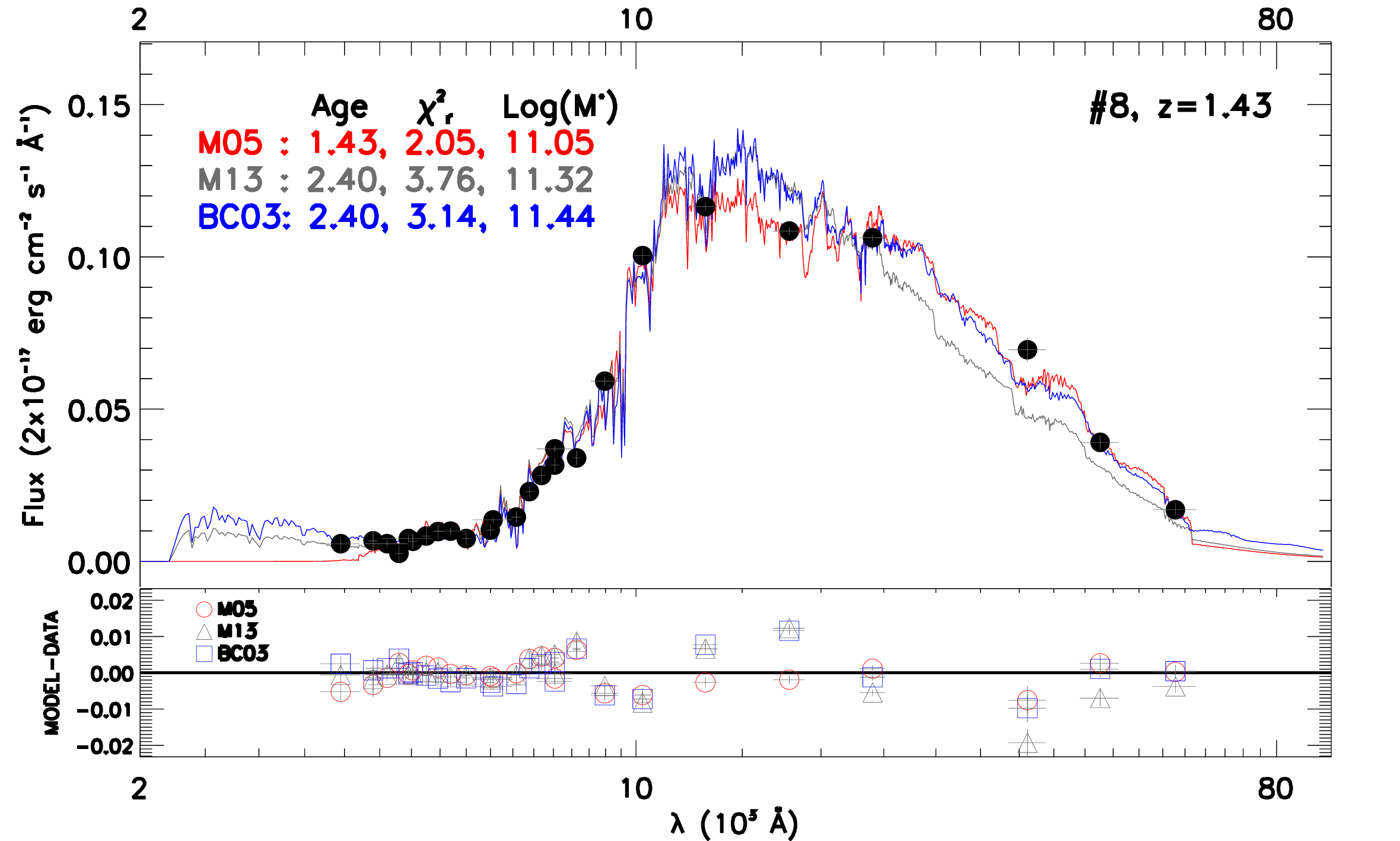}
\includegraphics[width=0.48\textwidth]{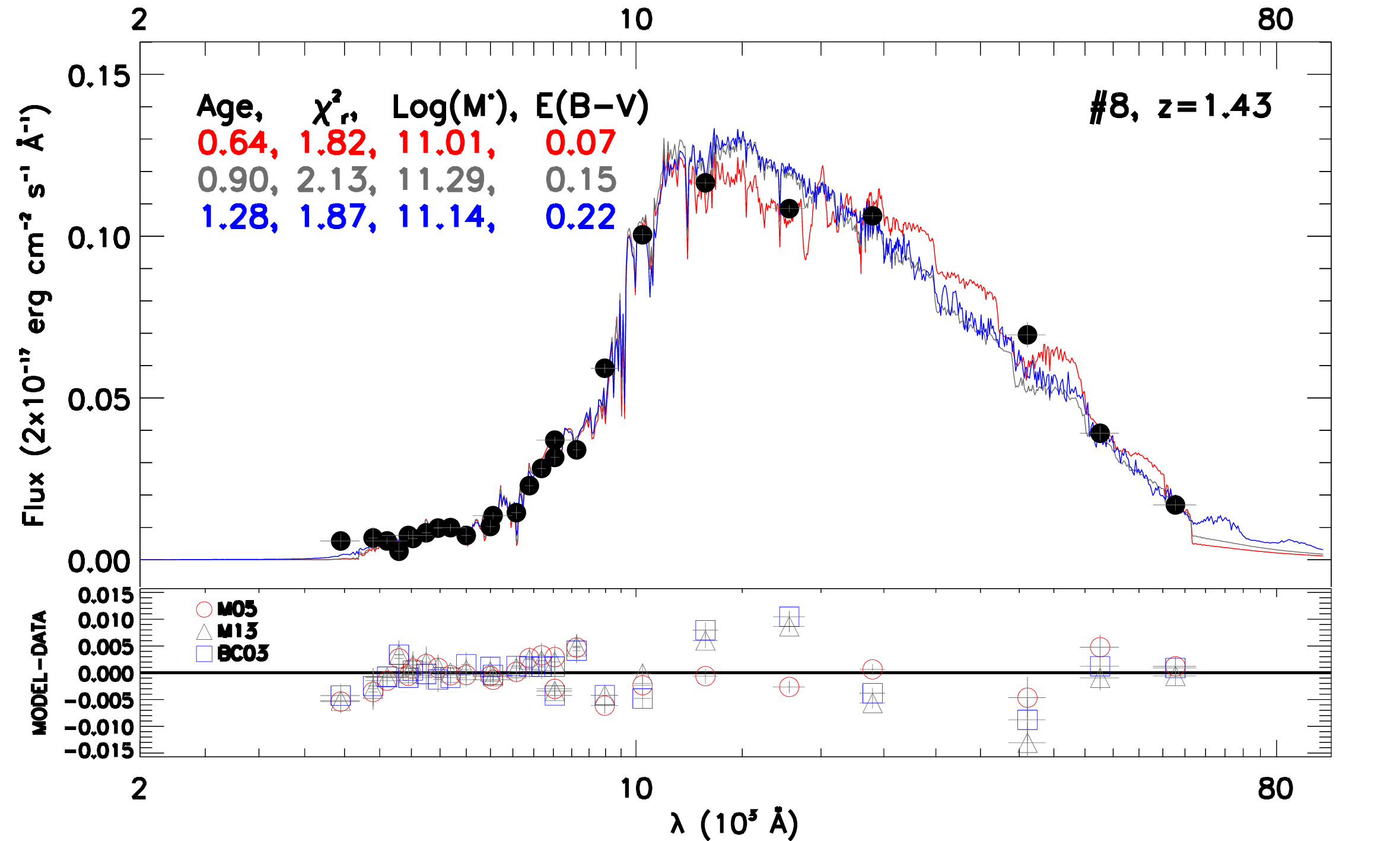}
\caption{Continued.}
\label{fig:Fig1_appB}
\end{figure*}

\addtocounter{figure}{-1}
\begin{figure*}
\centering
\includegraphics[width=0.48\textwidth]{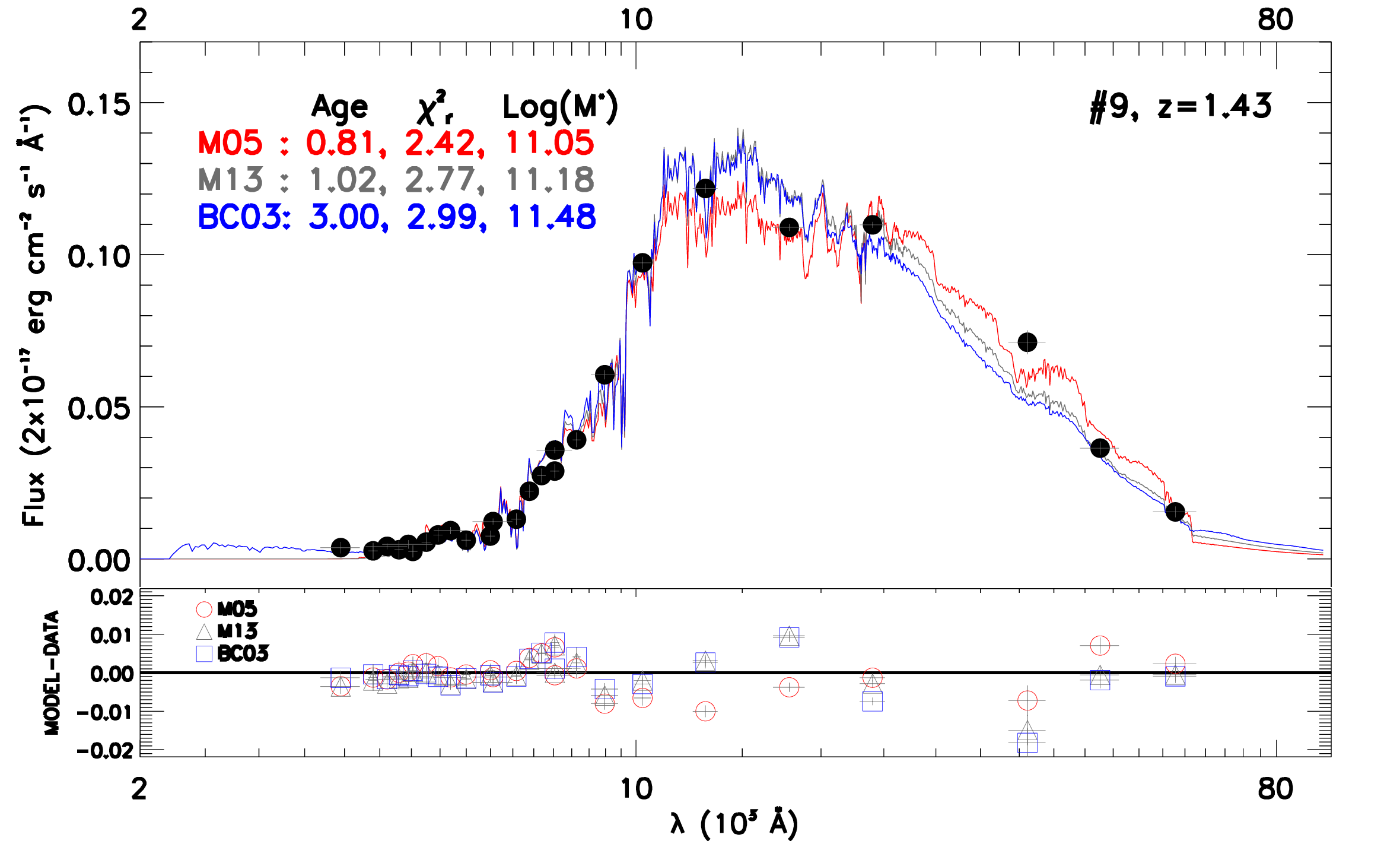}
\includegraphics[width=0.48\textwidth]{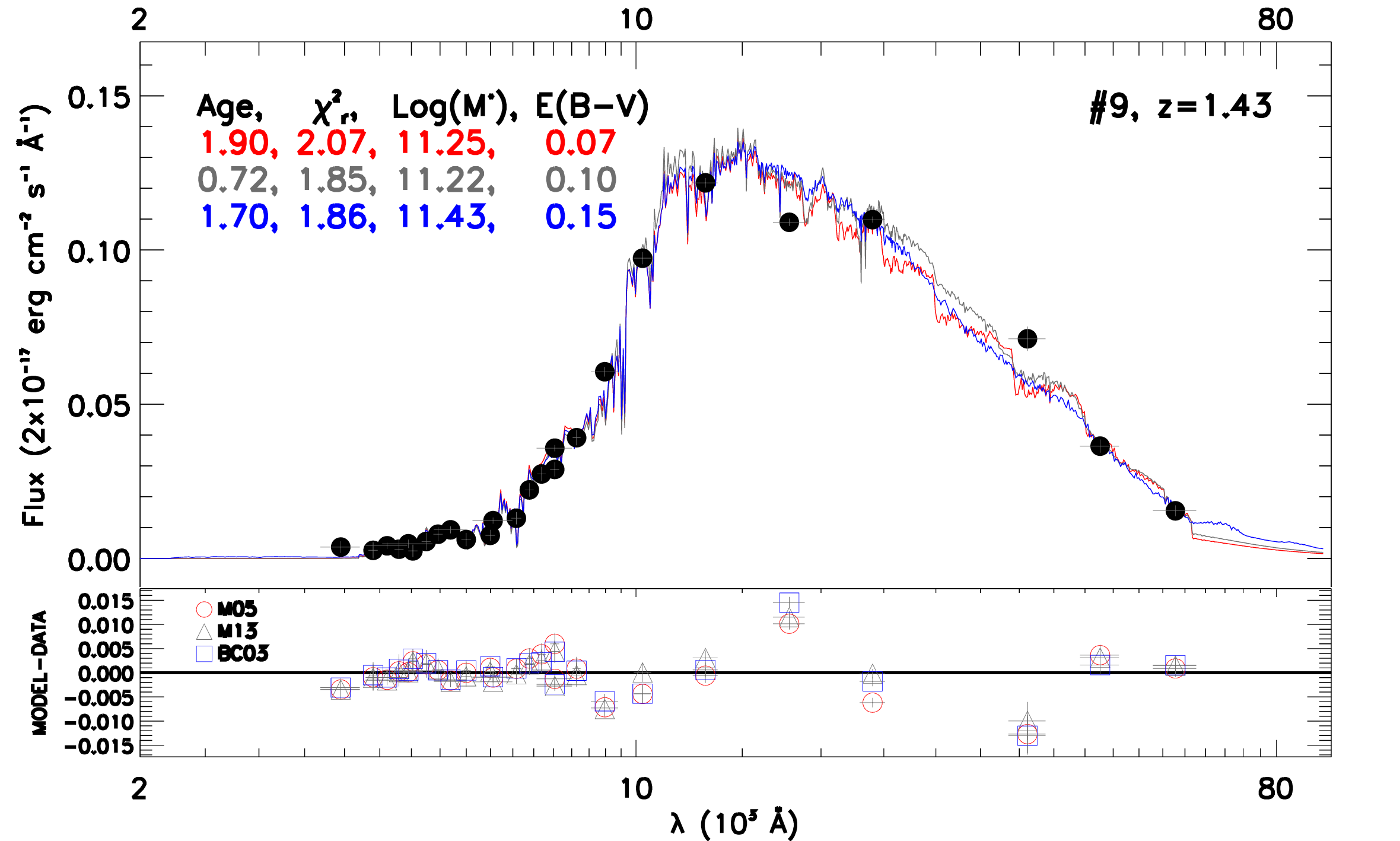}
\includegraphics[width=0.48\textwidth]{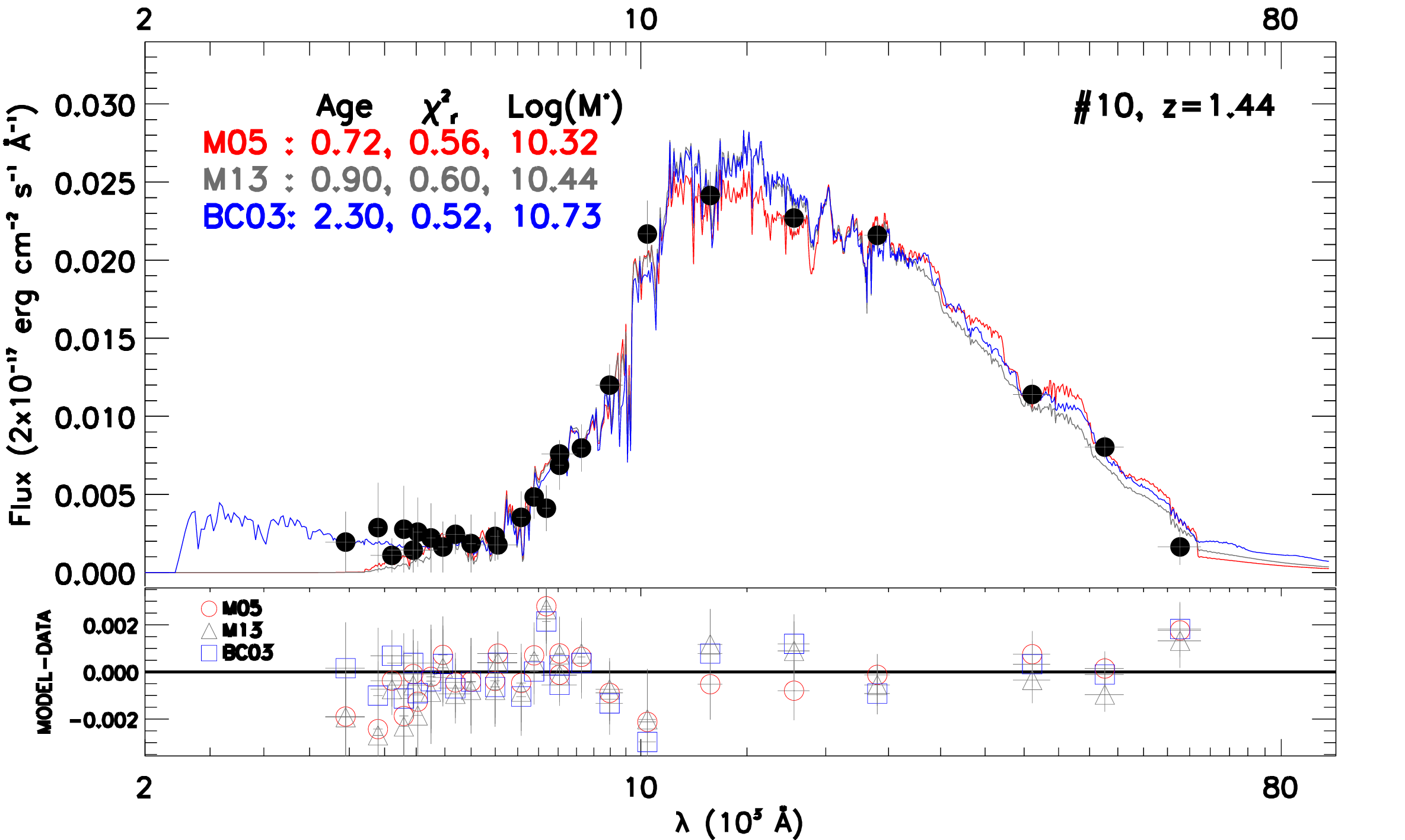}
\includegraphics[width=0.48\textwidth]{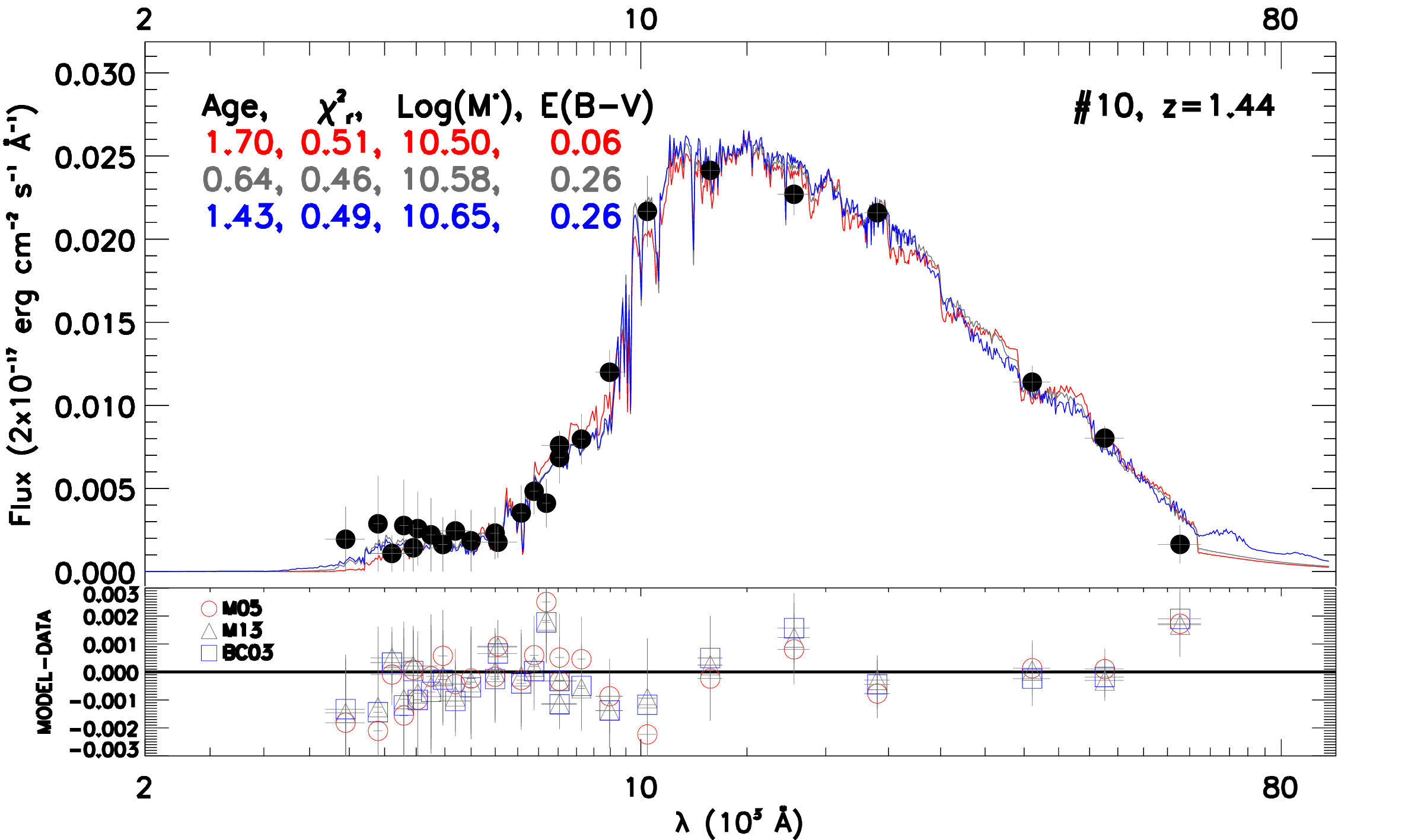}
\includegraphics[width=0.48\textwidth]{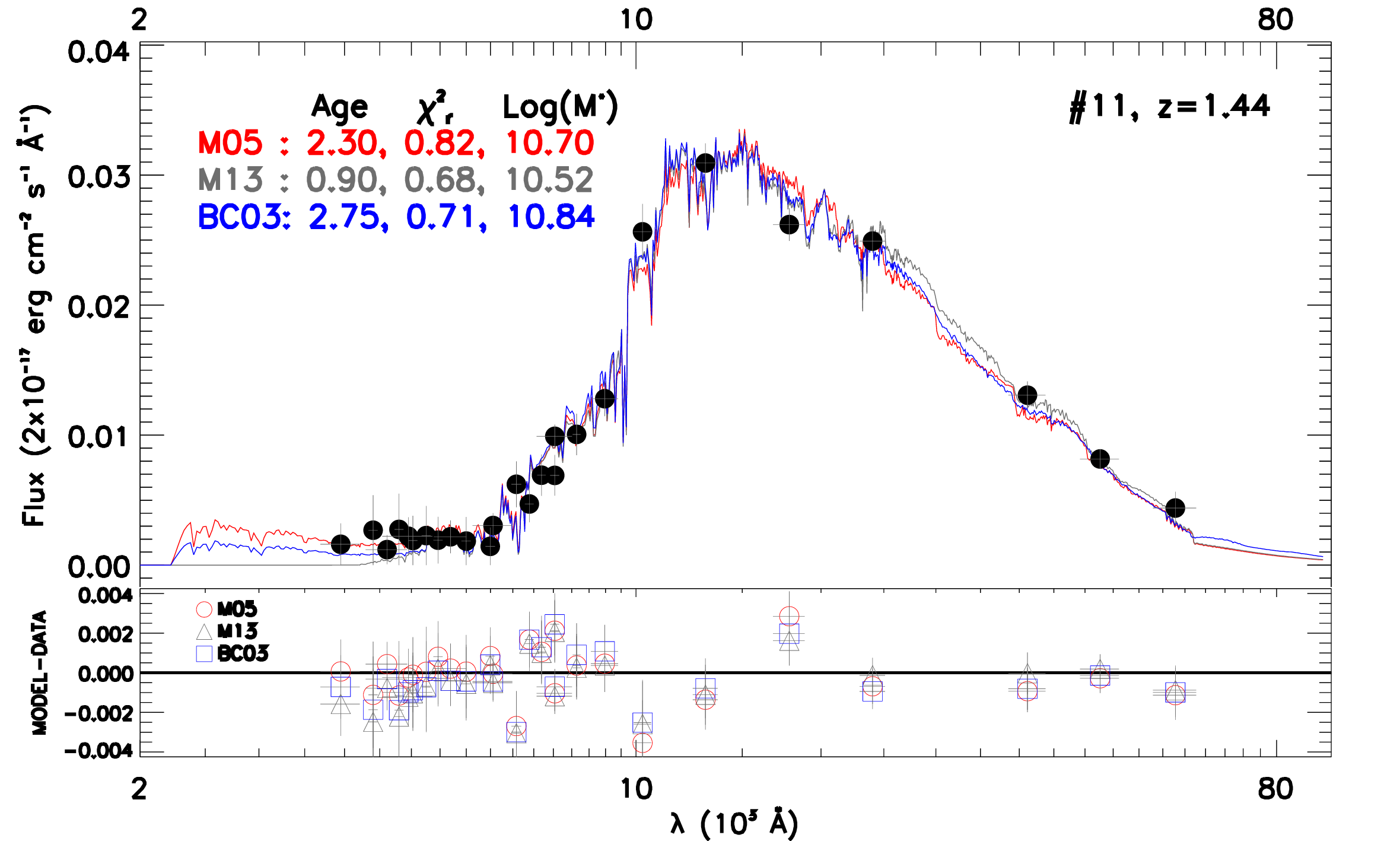}
\includegraphics[width=0.48\textwidth]{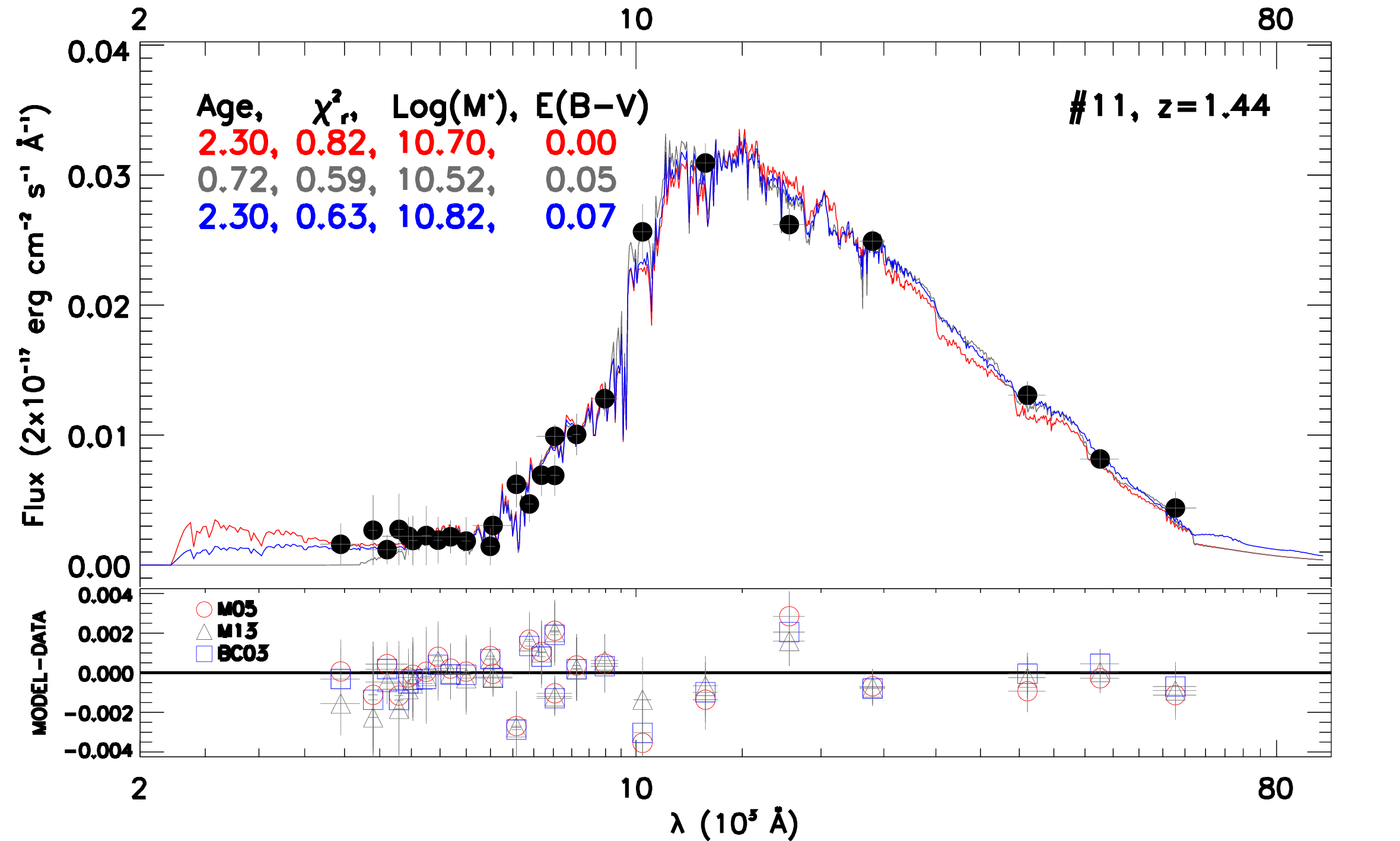}
\includegraphics[width=0.48\textwidth]{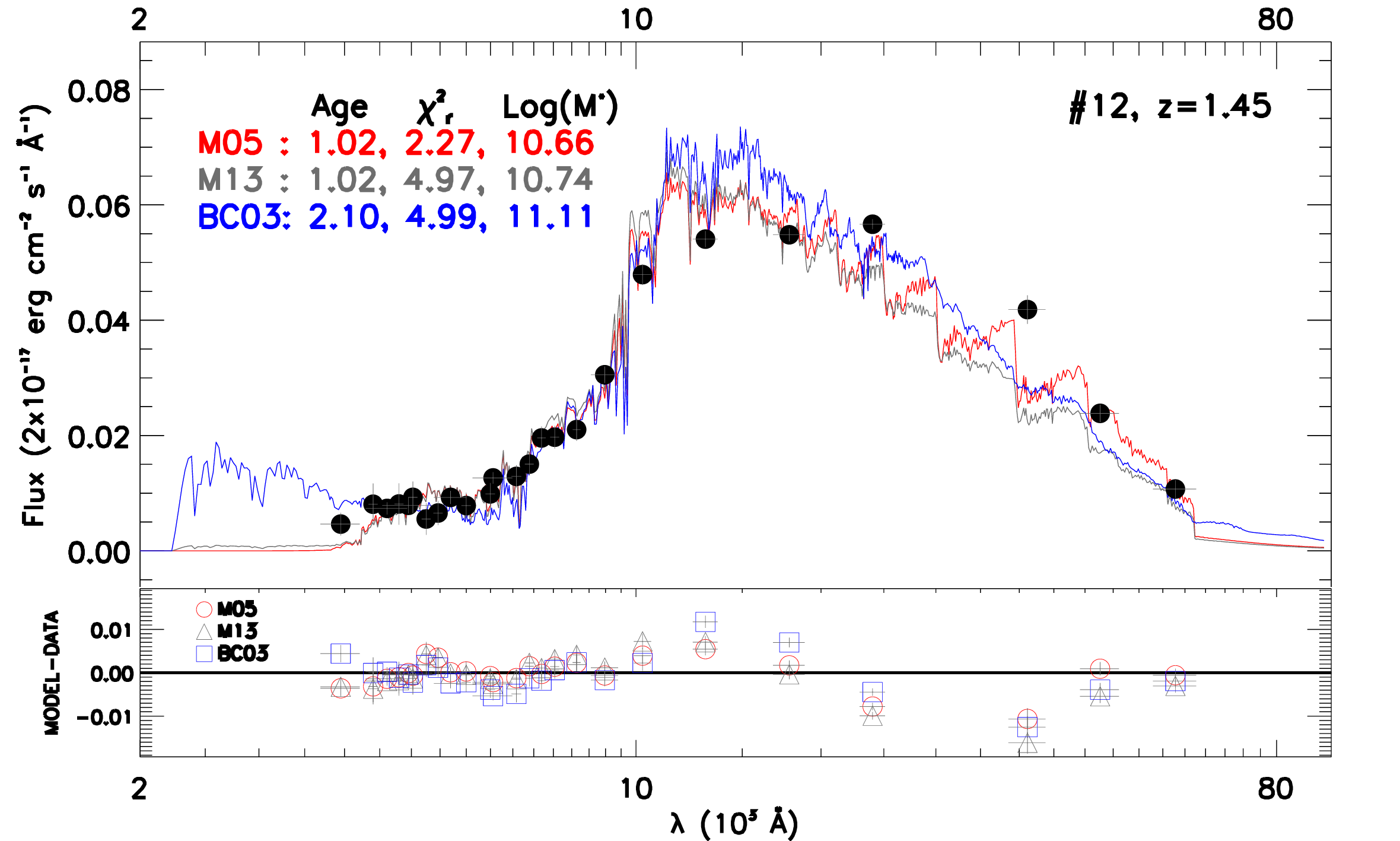}
\includegraphics[width=0.48\textwidth]{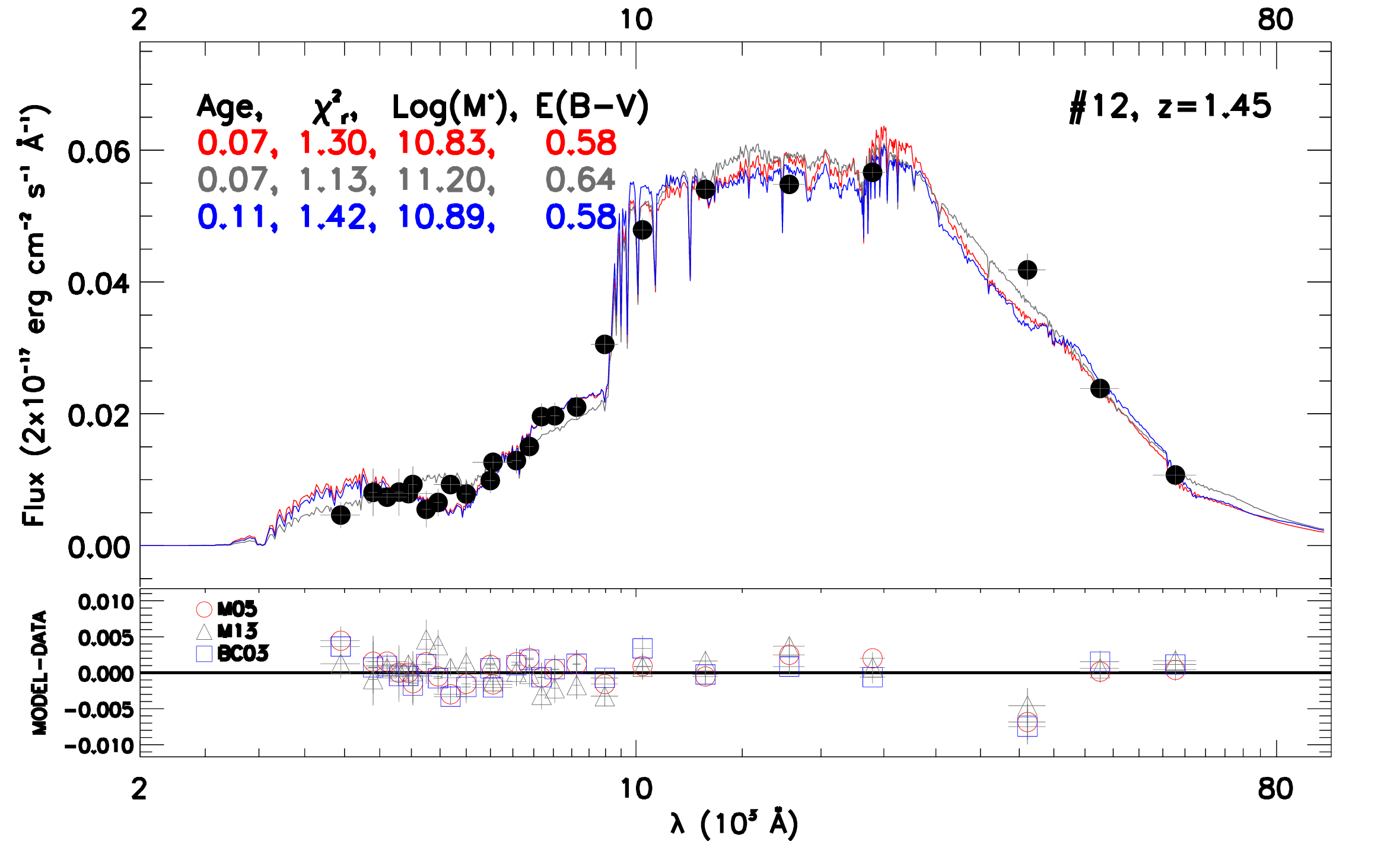}
\caption{Continued.}
\label{fig:Fig1_appB}
\end{figure*}

\addtocounter{figure}{-1}
\begin{figure*}
\centering
\includegraphics[width=0.48\textwidth]{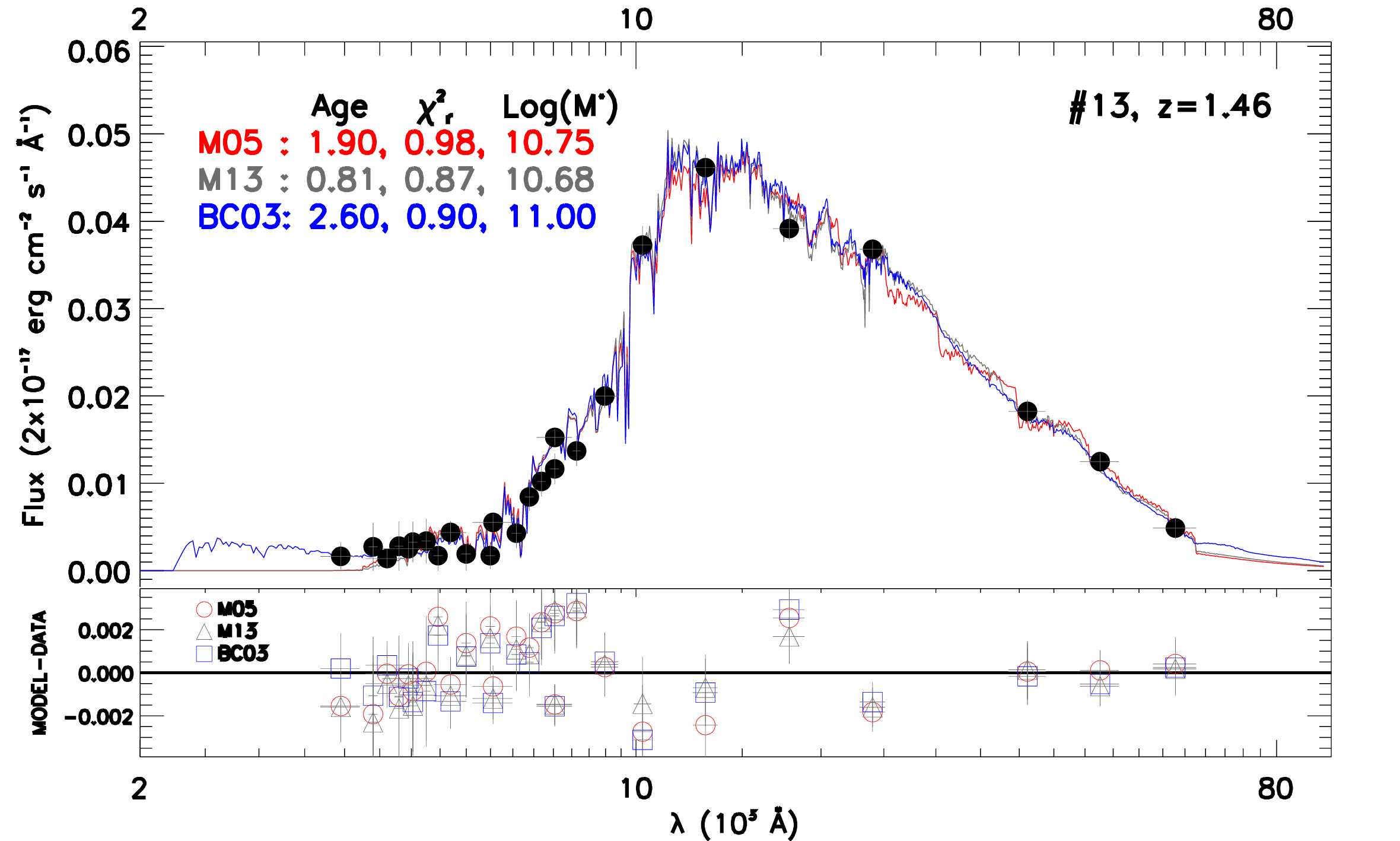}
\includegraphics[width=0.48\textwidth]{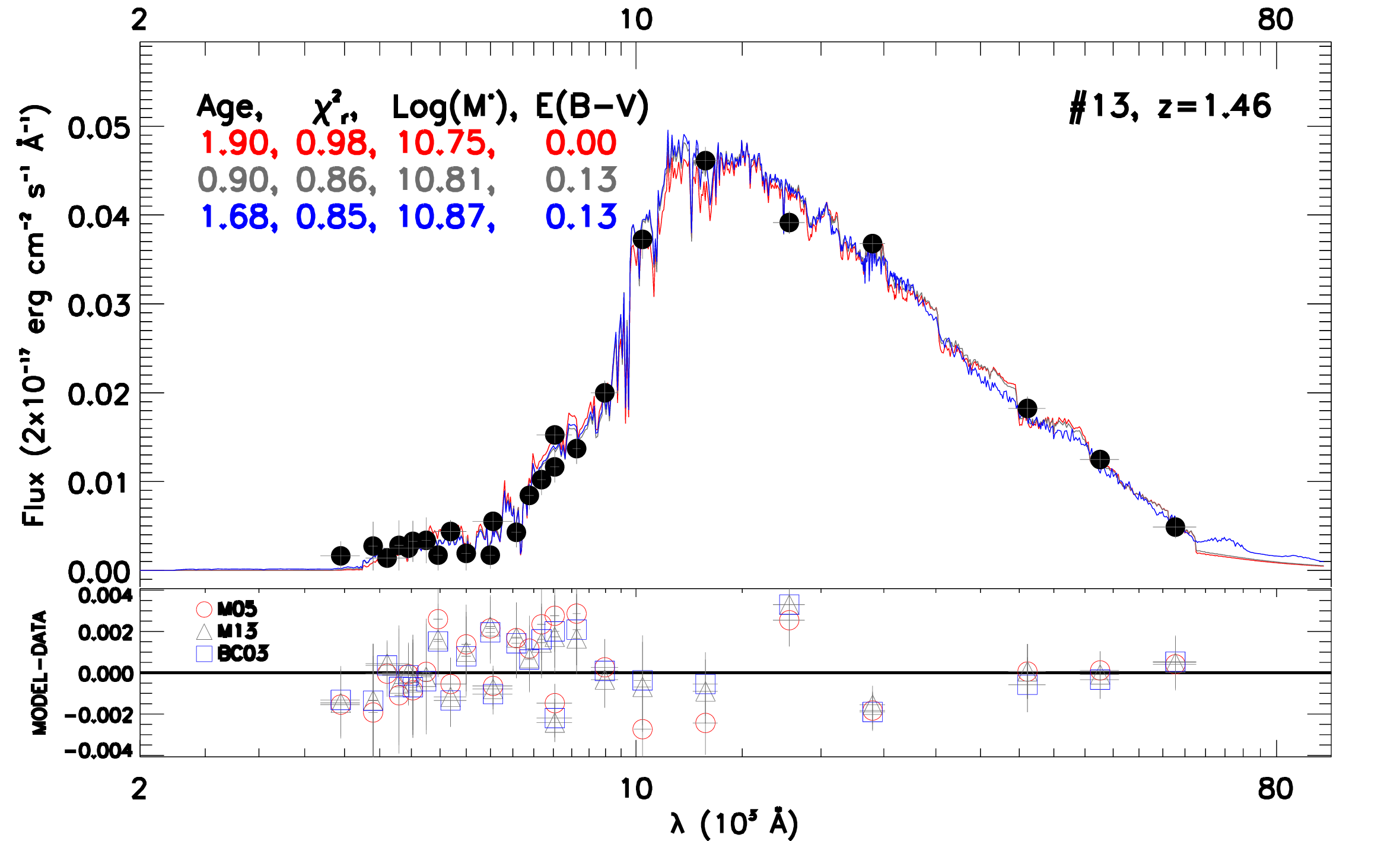}
\includegraphics[width=0.48\textwidth]{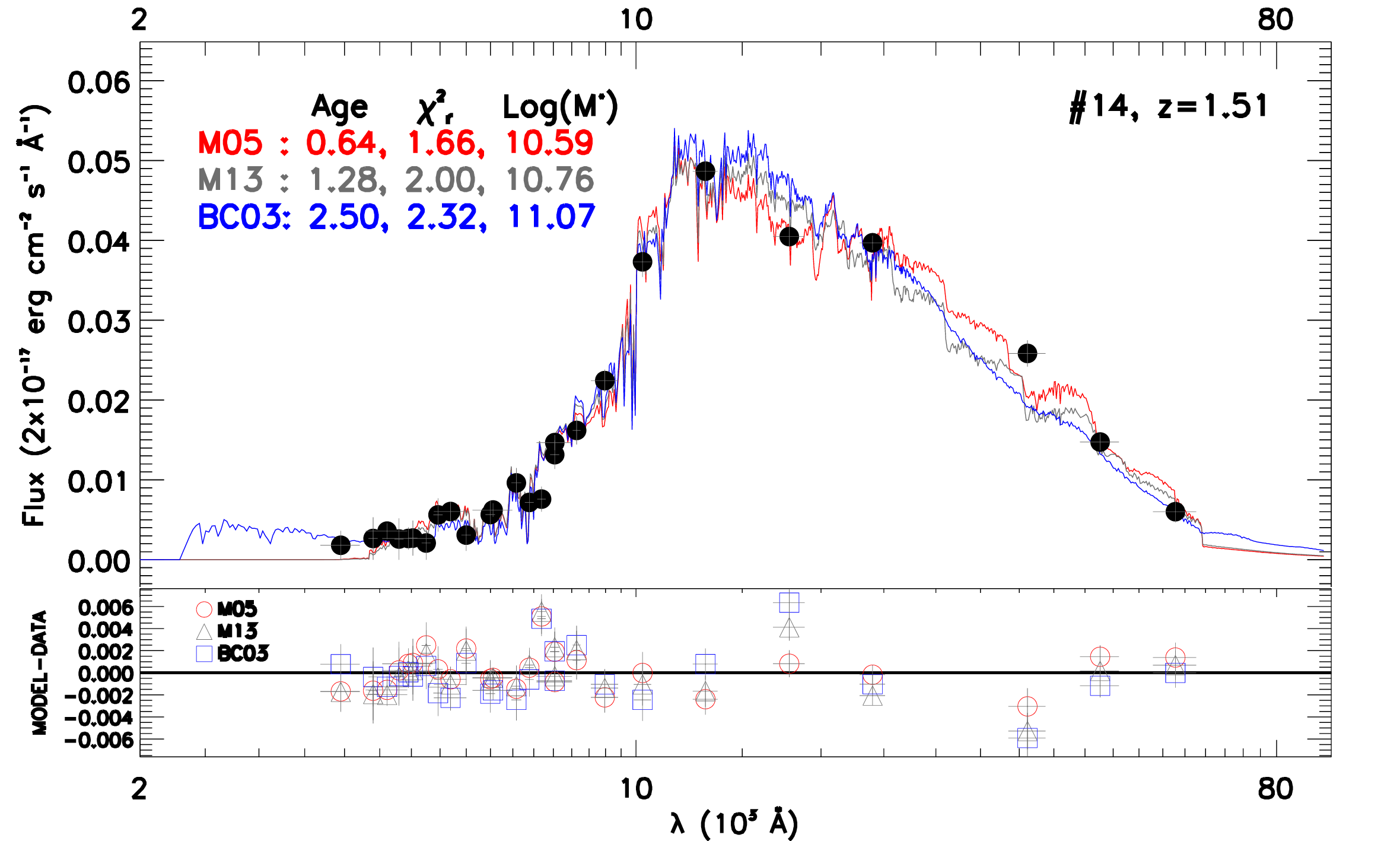}
\includegraphics[width=0.48\textwidth]{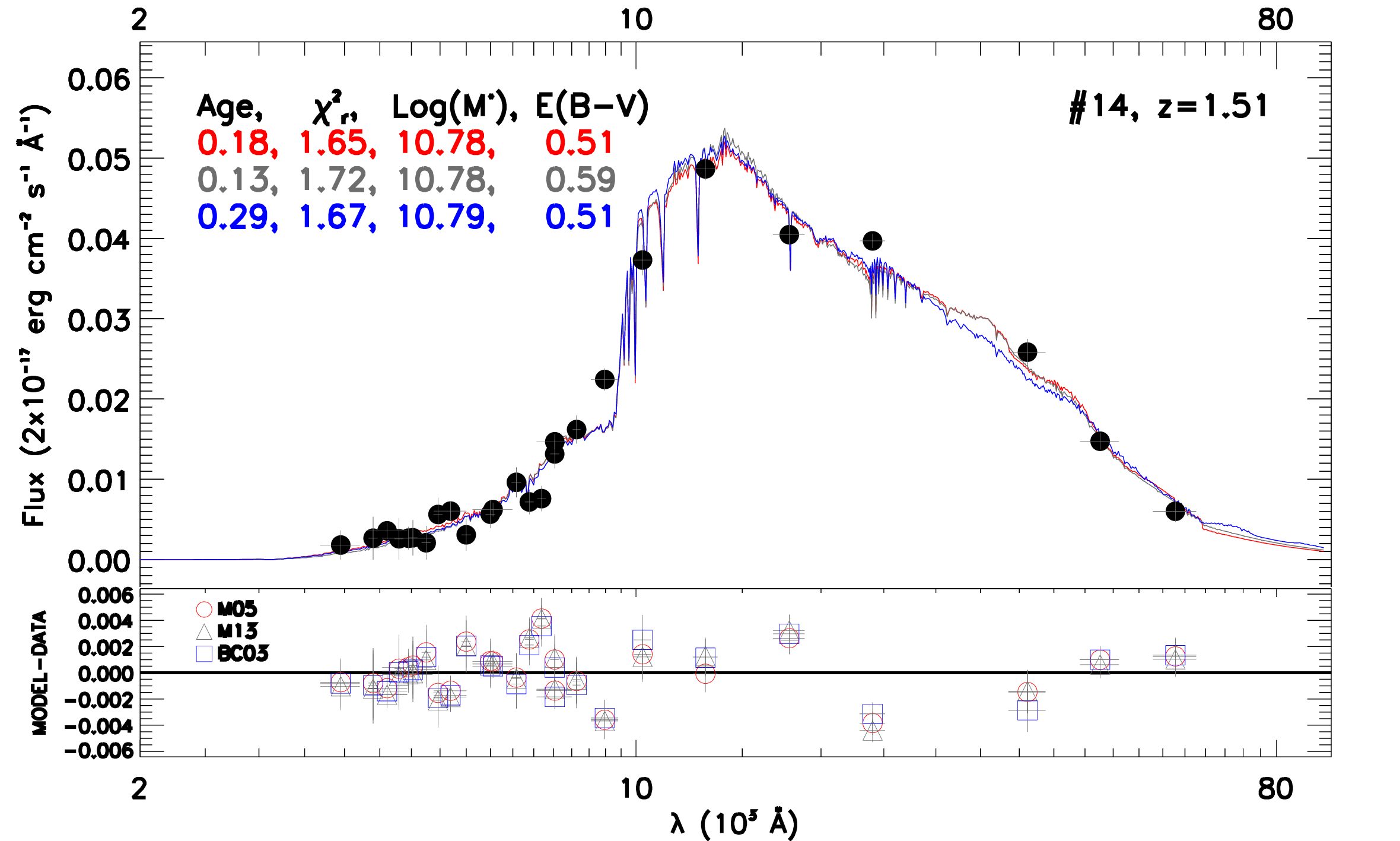}
\includegraphics[width=0.48\textwidth]{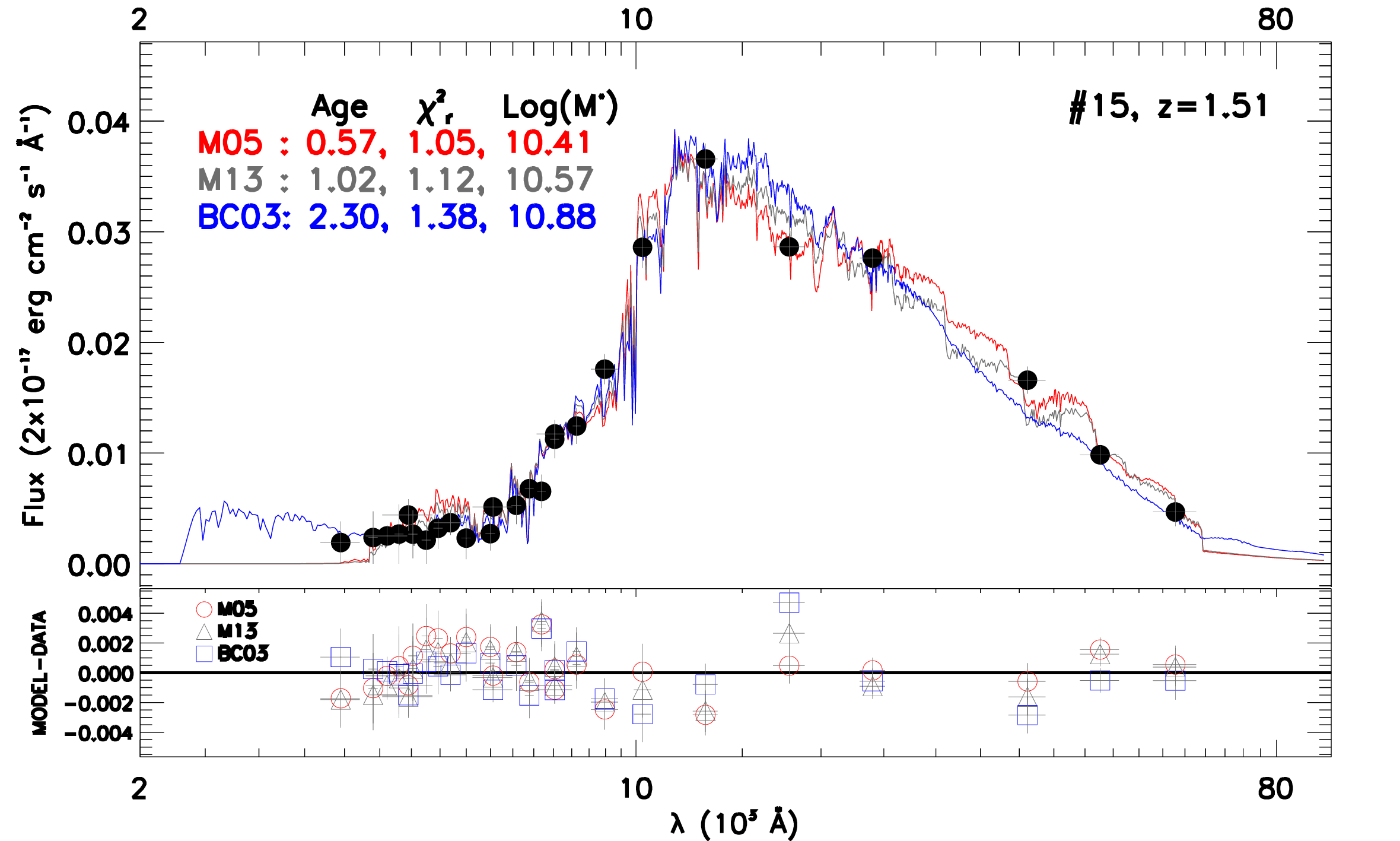}
\includegraphics[width=0.48\textwidth]{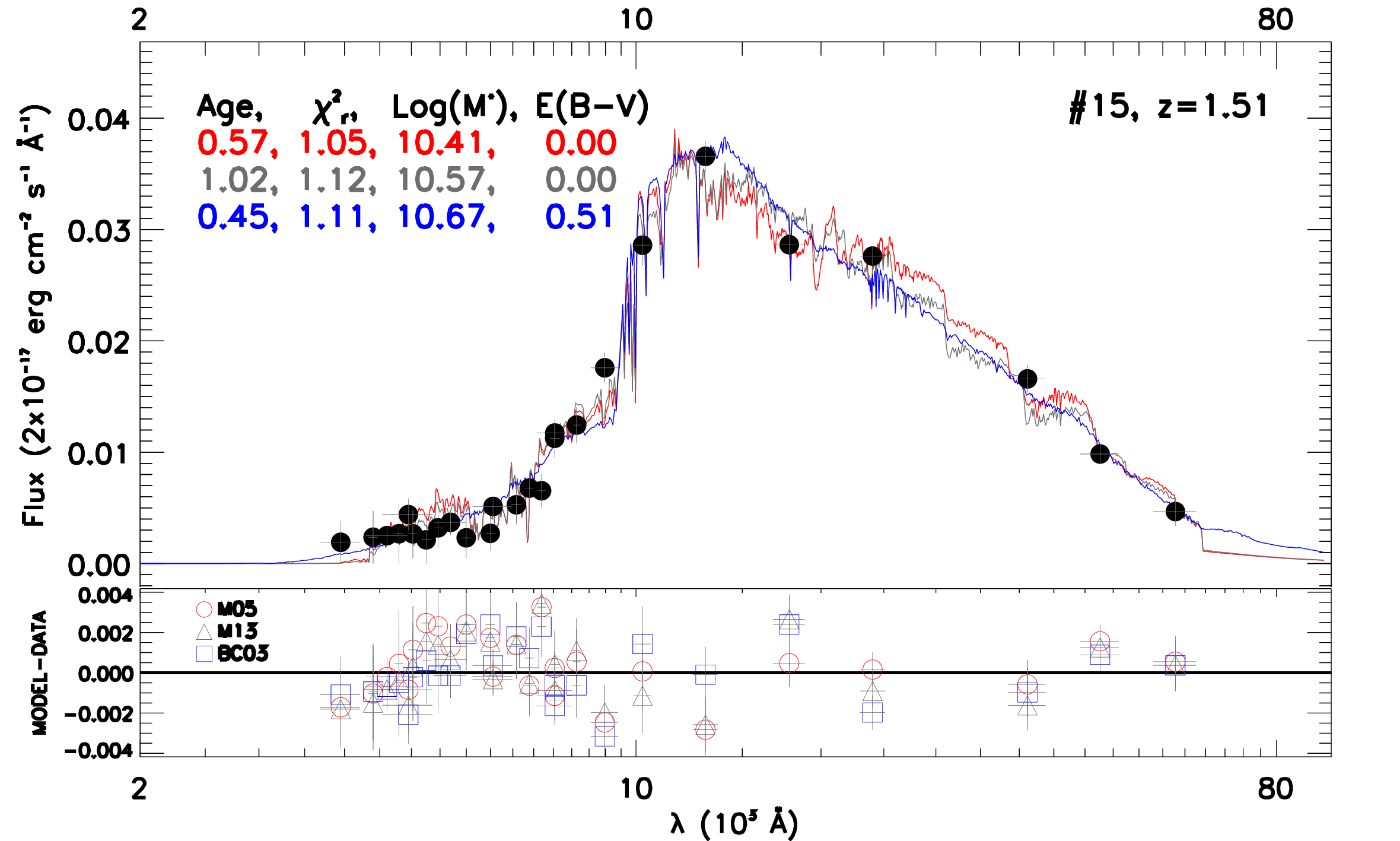}
\includegraphics[width=0.48\textwidth]{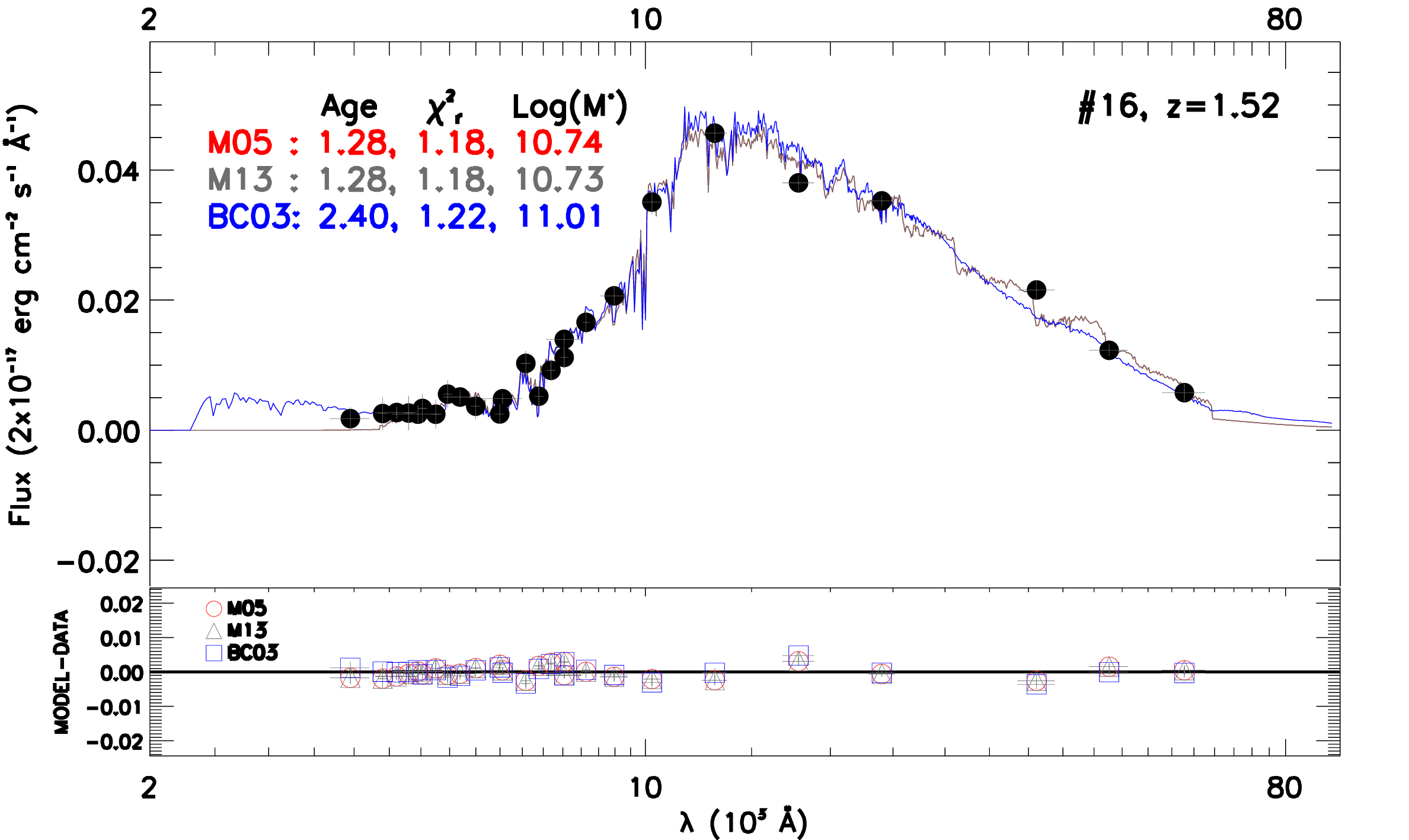}
\includegraphics[width=0.48\textwidth]{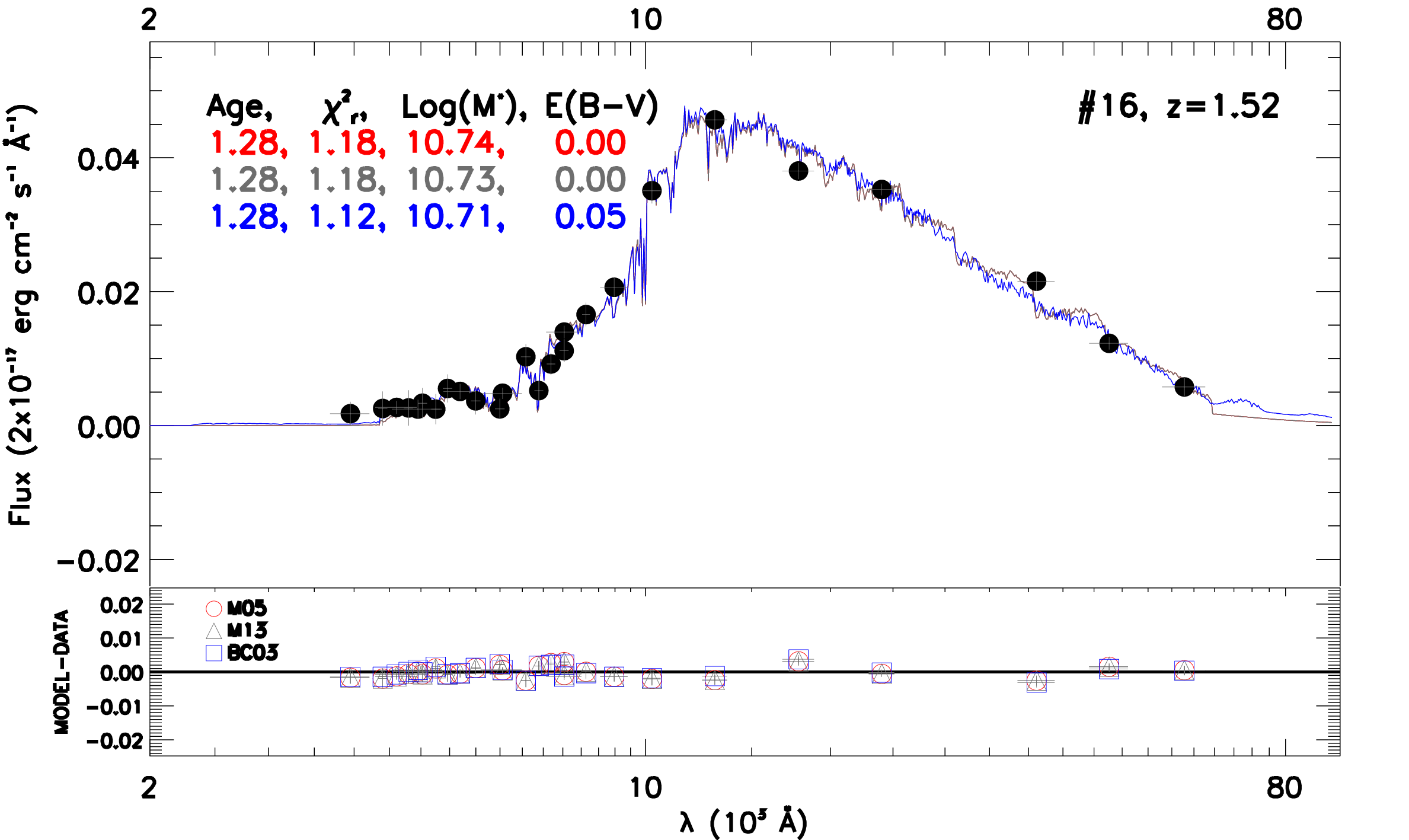}
\caption{Continued.}
\label{fig:Fig1_appB}
\end{figure*}

\addtocounter{figure}{-1}
\begin{figure*}
\centering
\includegraphics[width=0.48\textwidth]{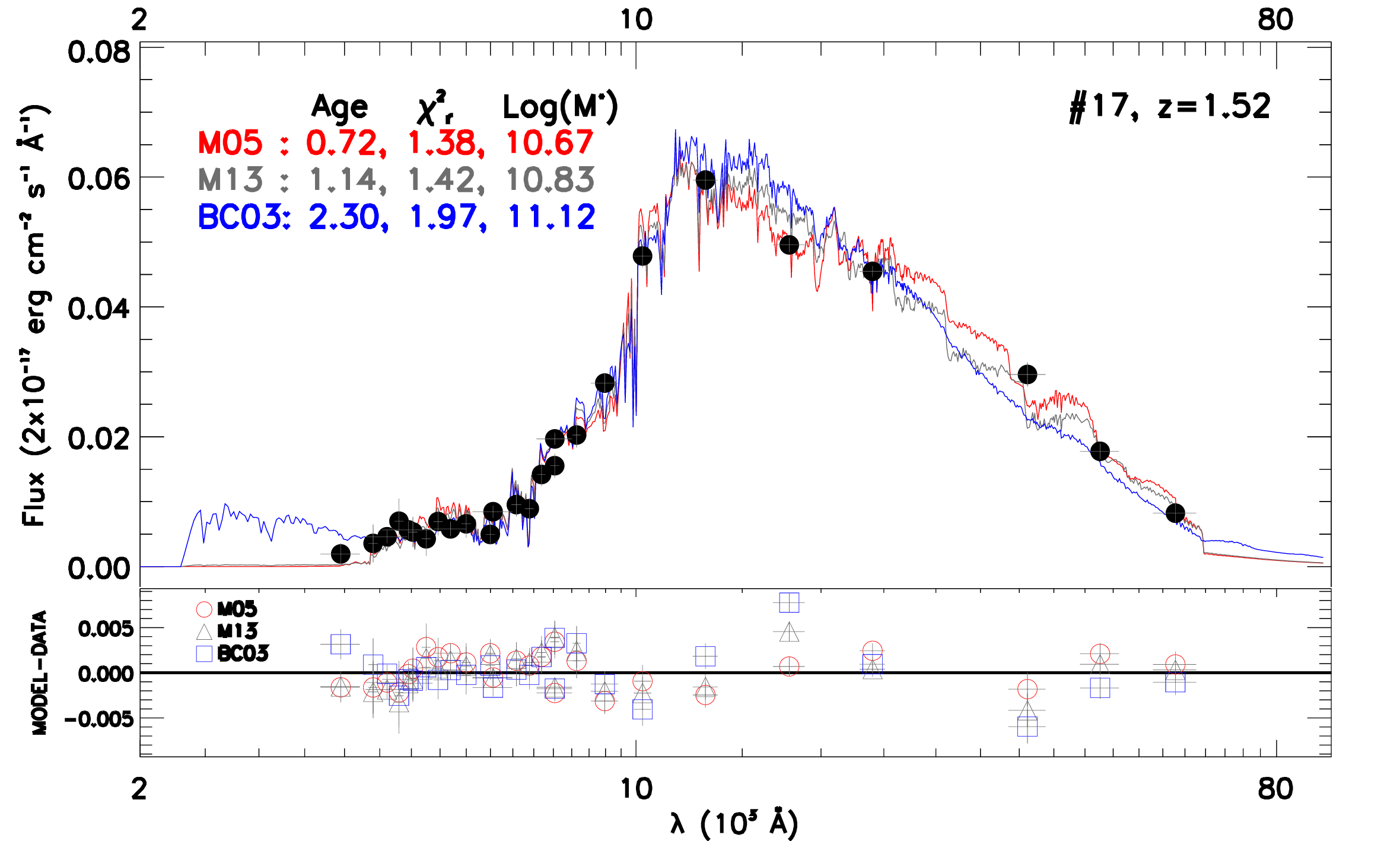}
\includegraphics[width=0.48\textwidth]{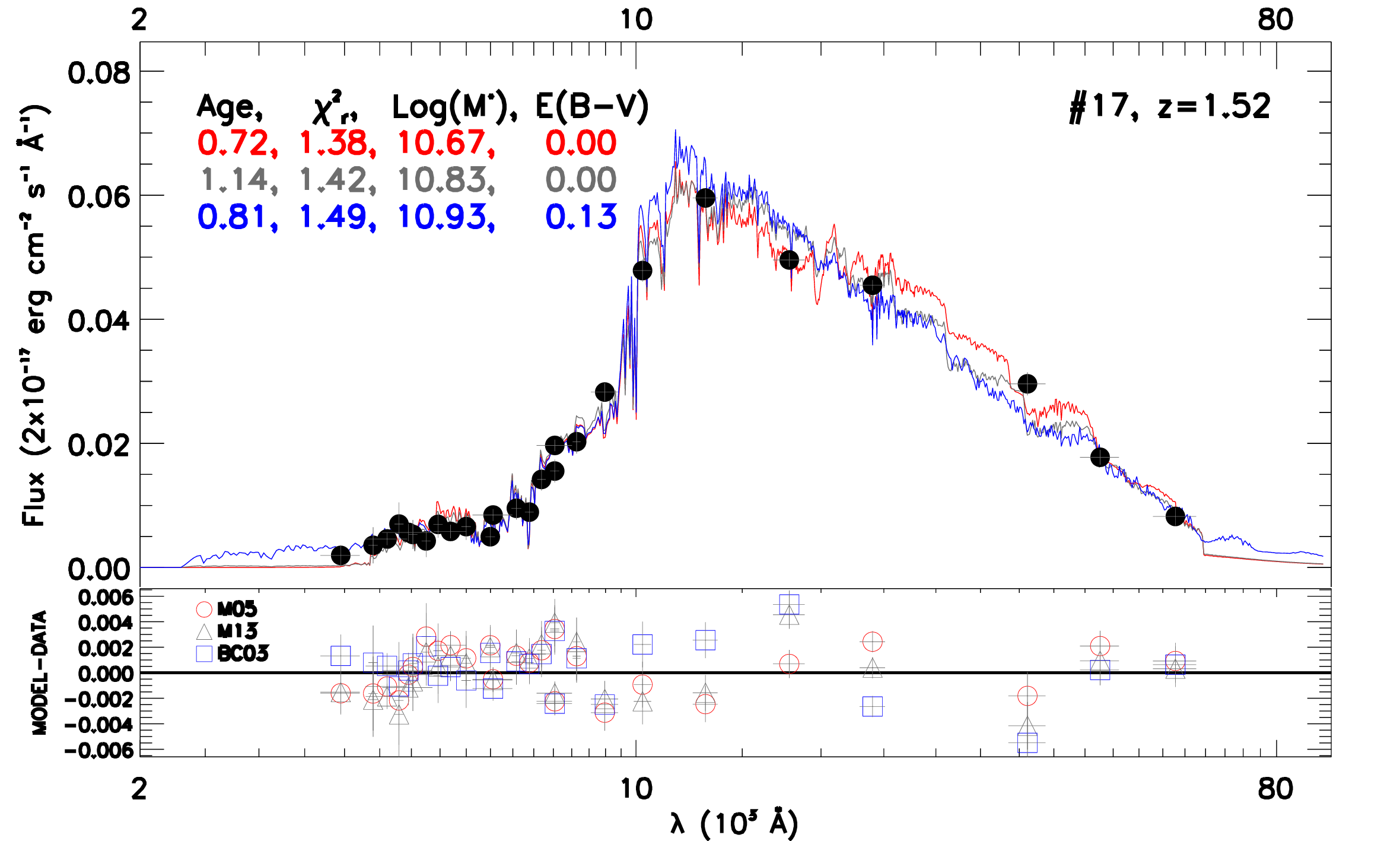}
\includegraphics[width=0.48\textwidth]{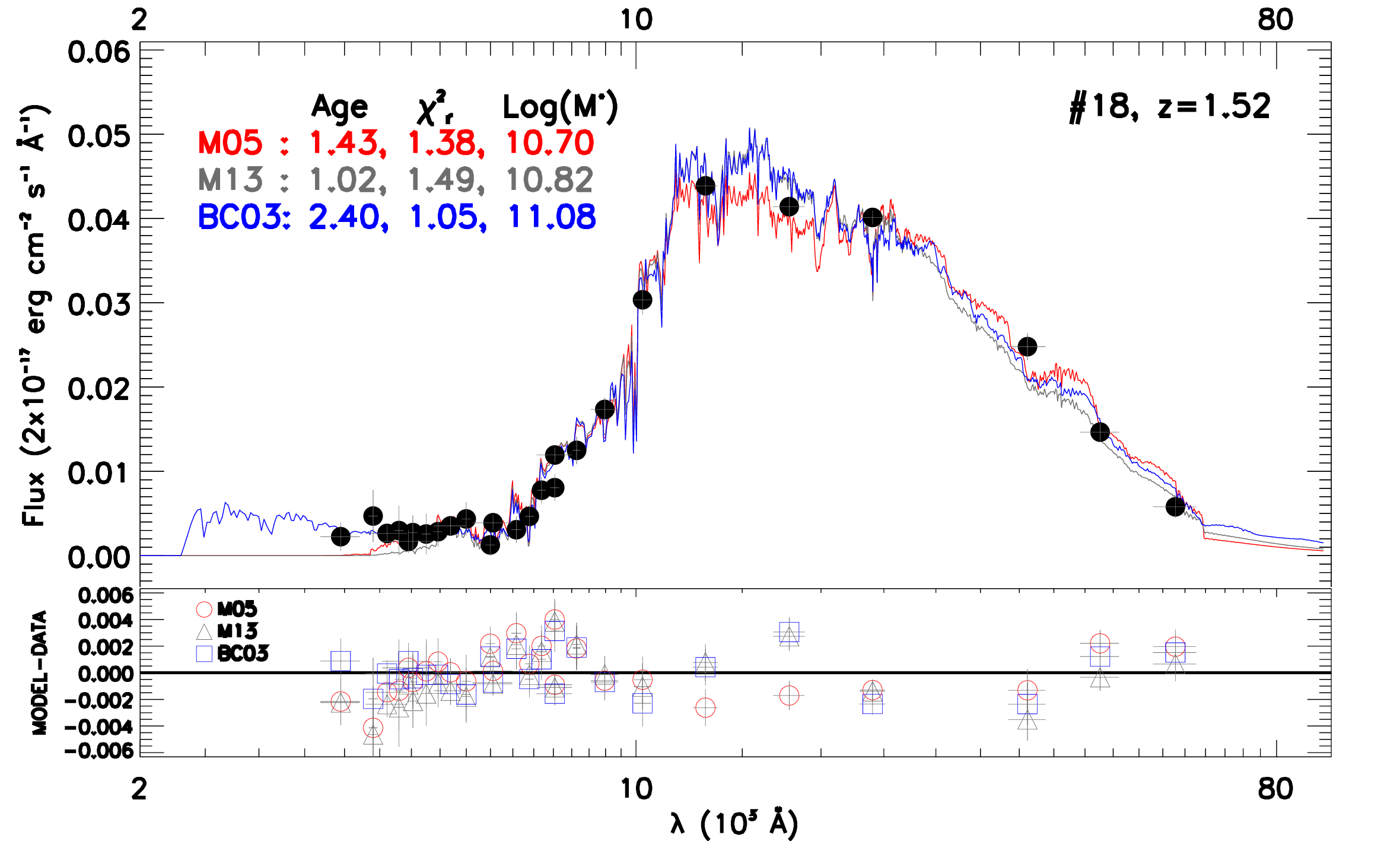}
\includegraphics[width=0.48\textwidth]{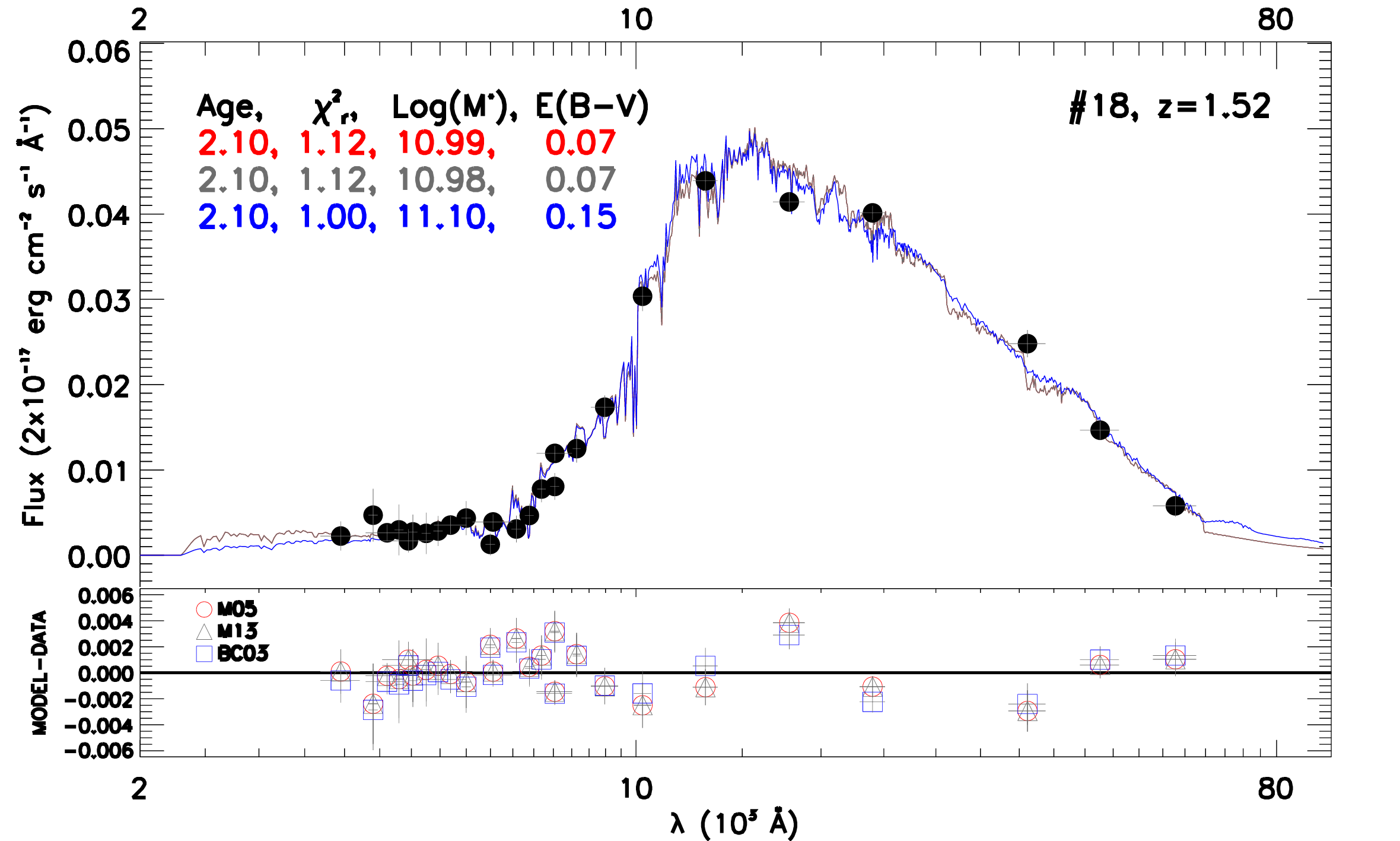}
\includegraphics[width=0.48\textwidth]{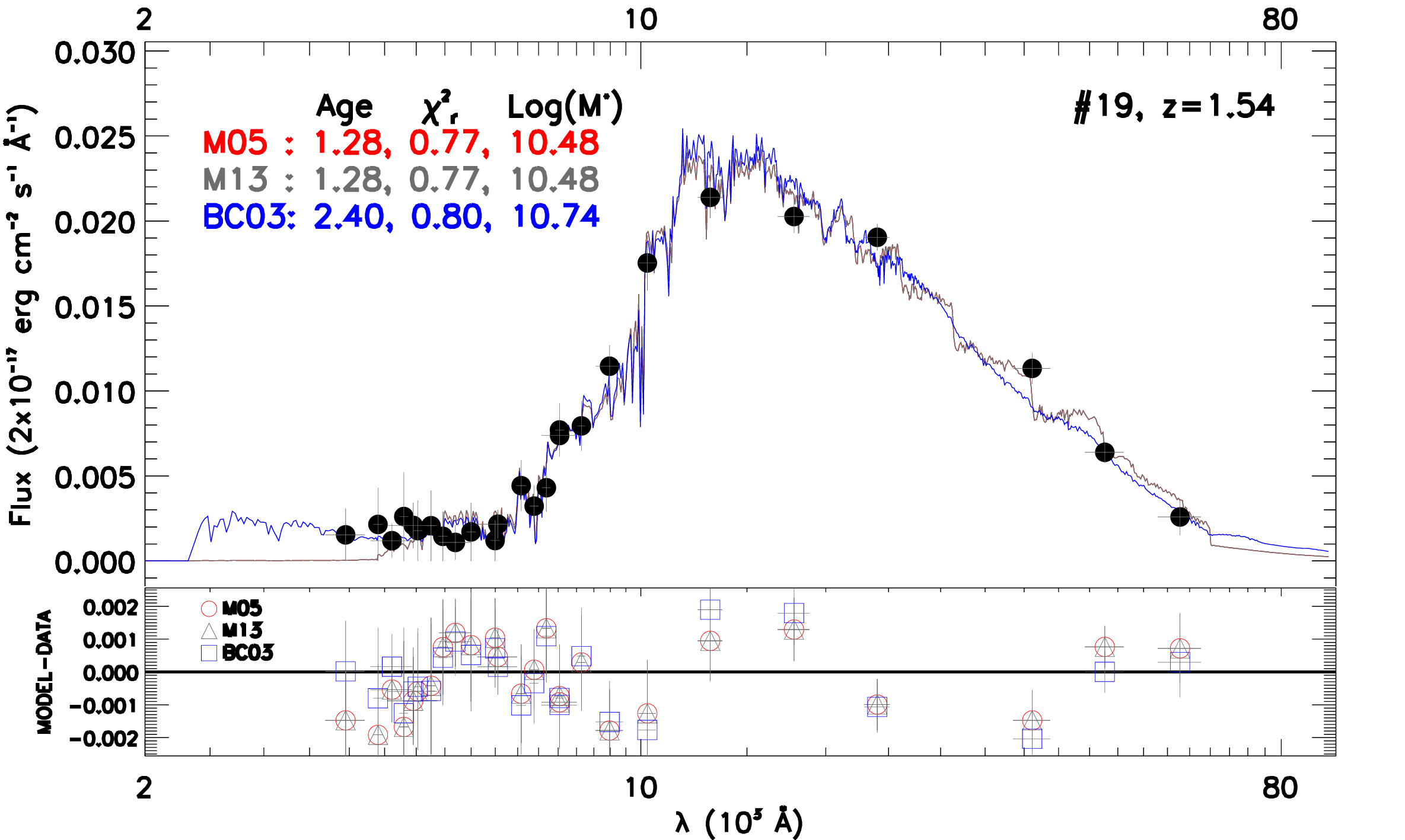}
\includegraphics[width=0.48\textwidth]{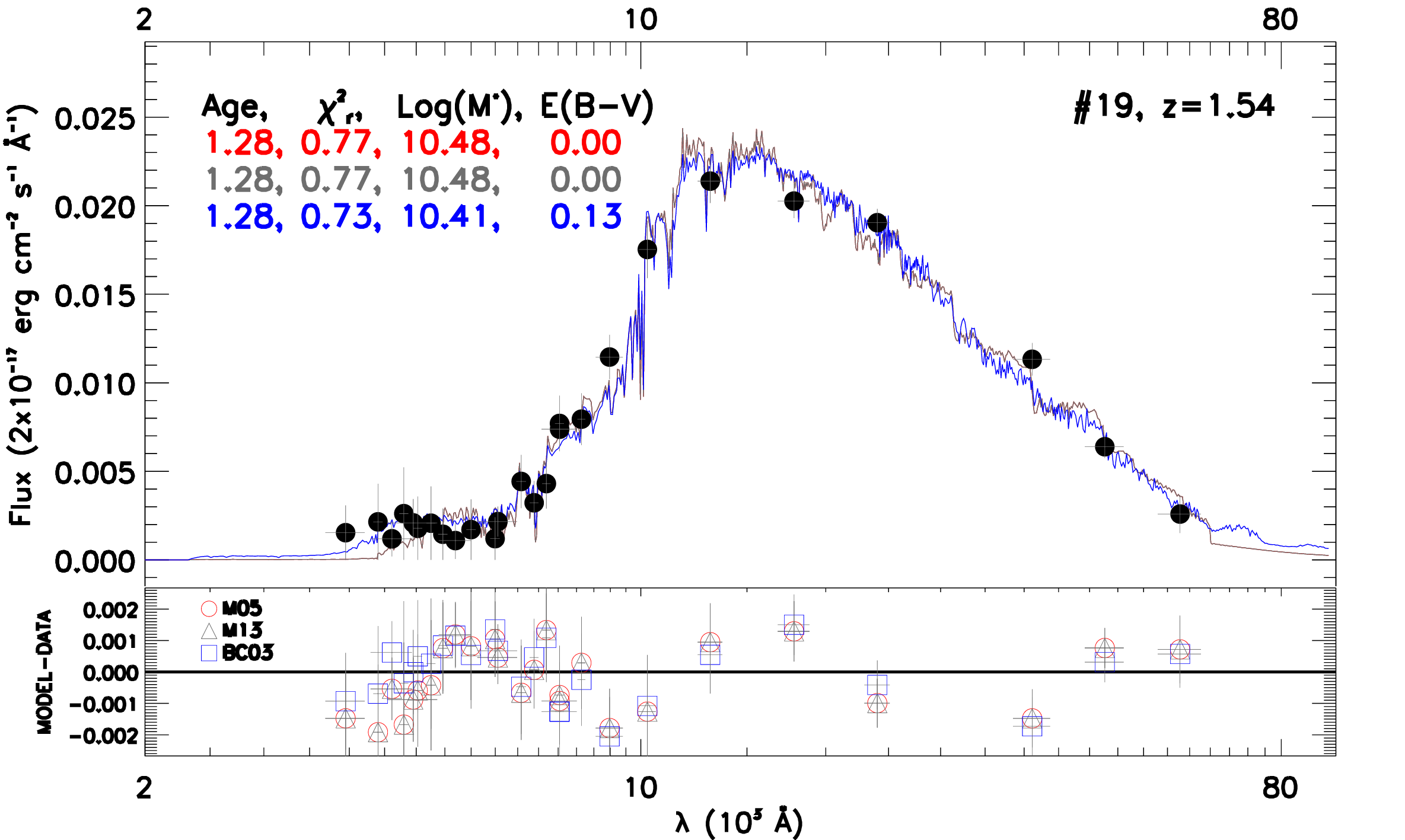}
\includegraphics[width=0.48\textwidth]{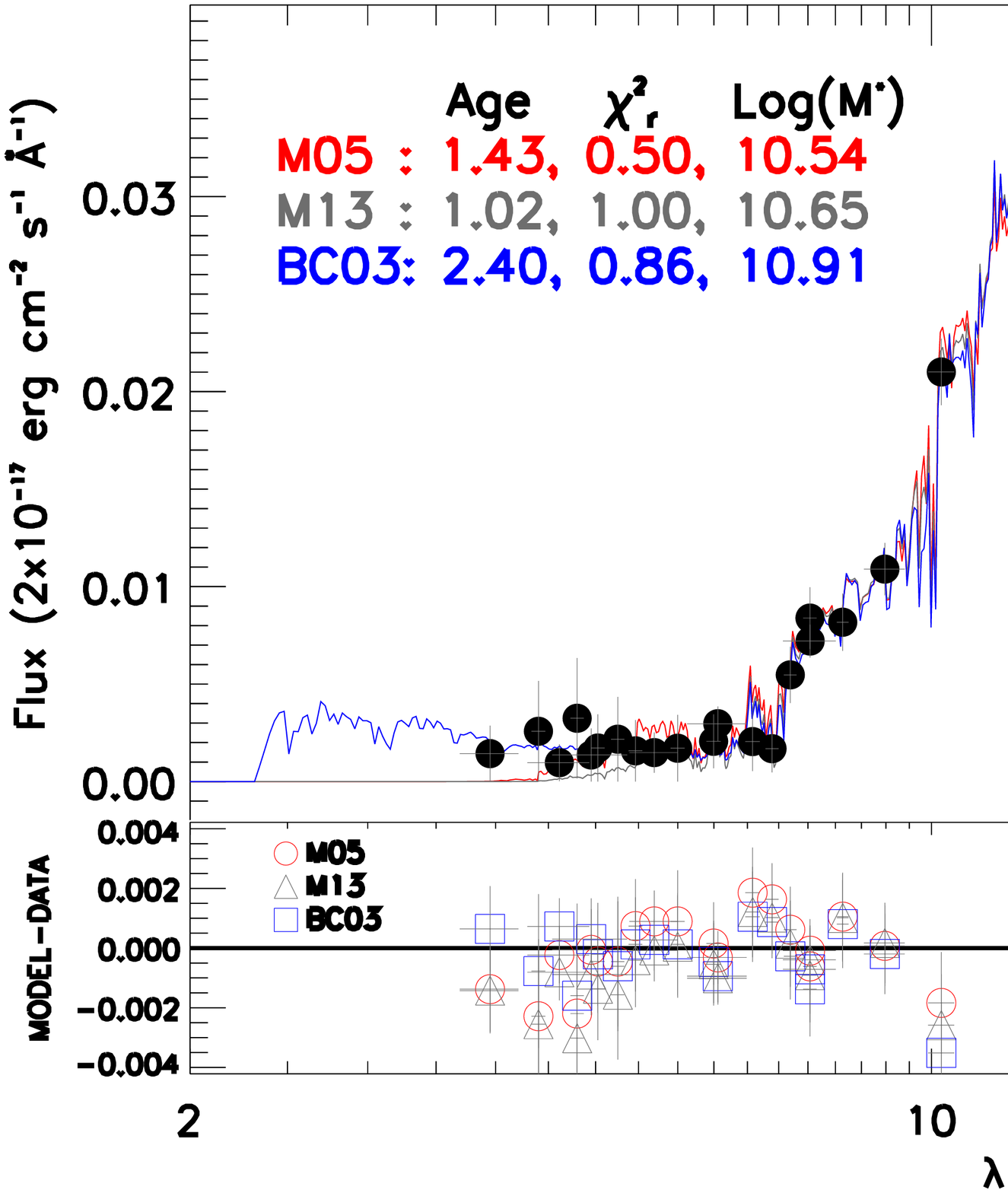}
\includegraphics[width=0.48\textwidth]{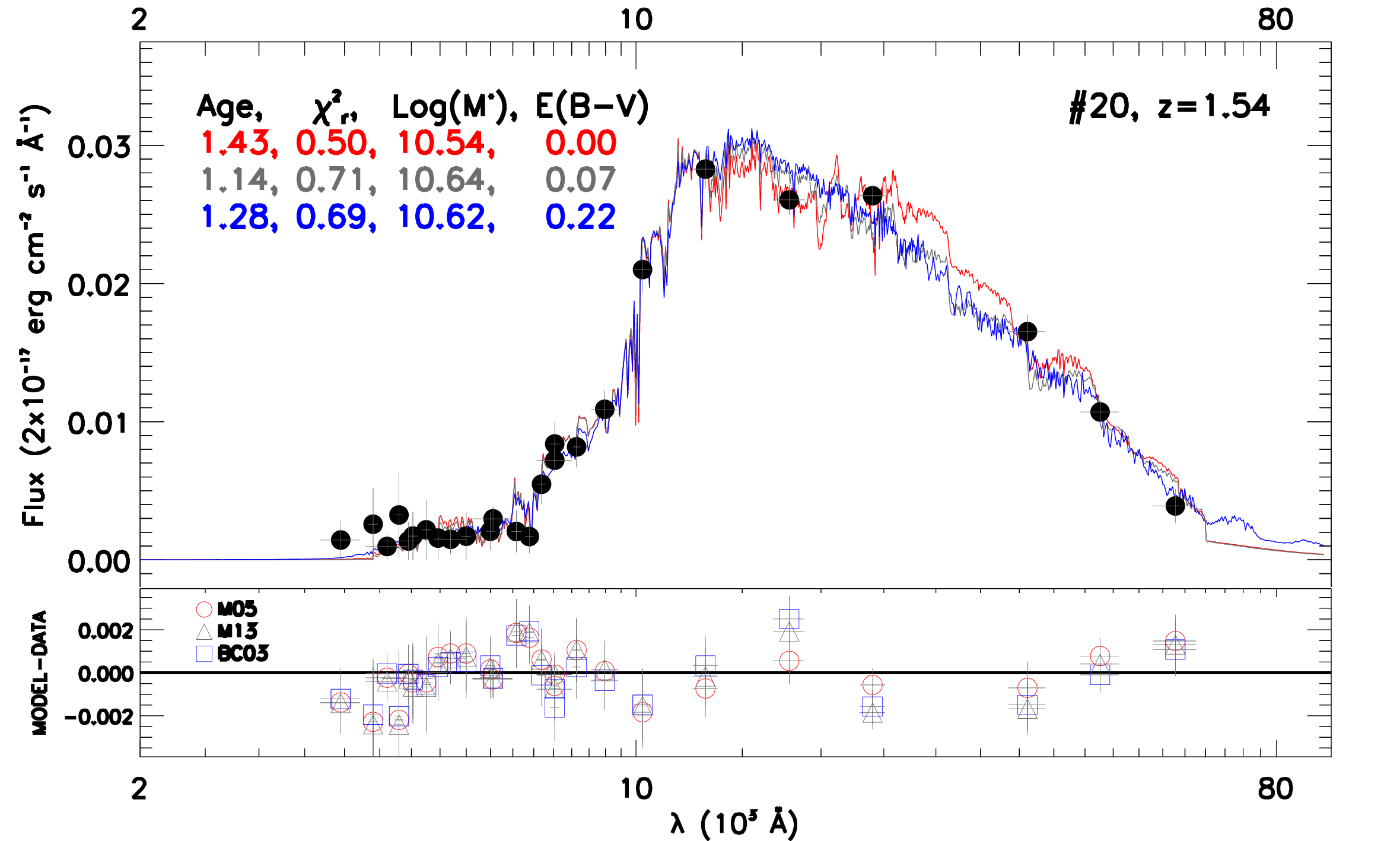}
\caption{Continued.}
\label{fig:Fig1_appB}
\end{figure*}

\addtocounter{figure}{-1}
\begin{figure*}
\centering
\includegraphics[width=0.48\textwidth]{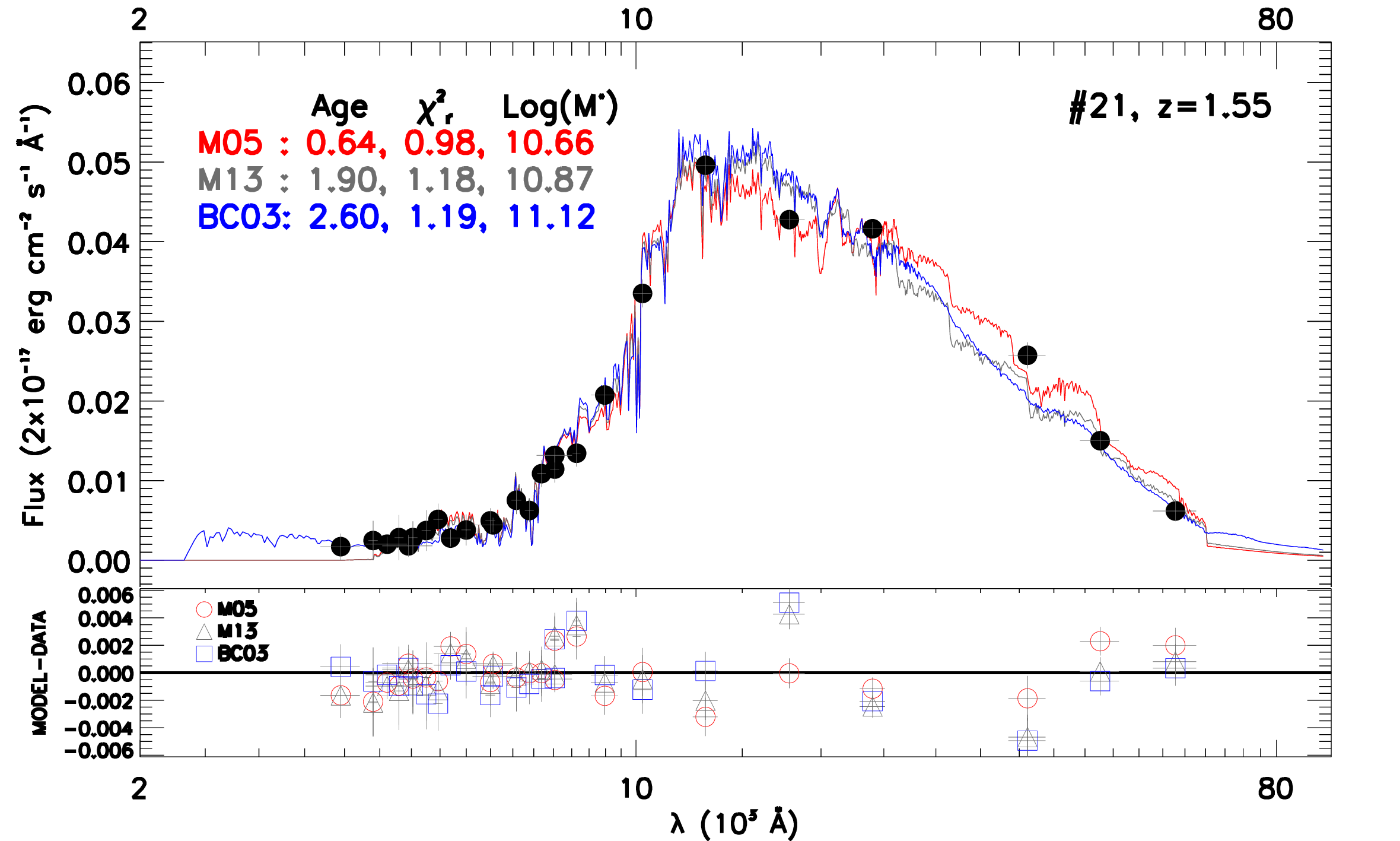}
\includegraphics[width=0.48\textwidth]{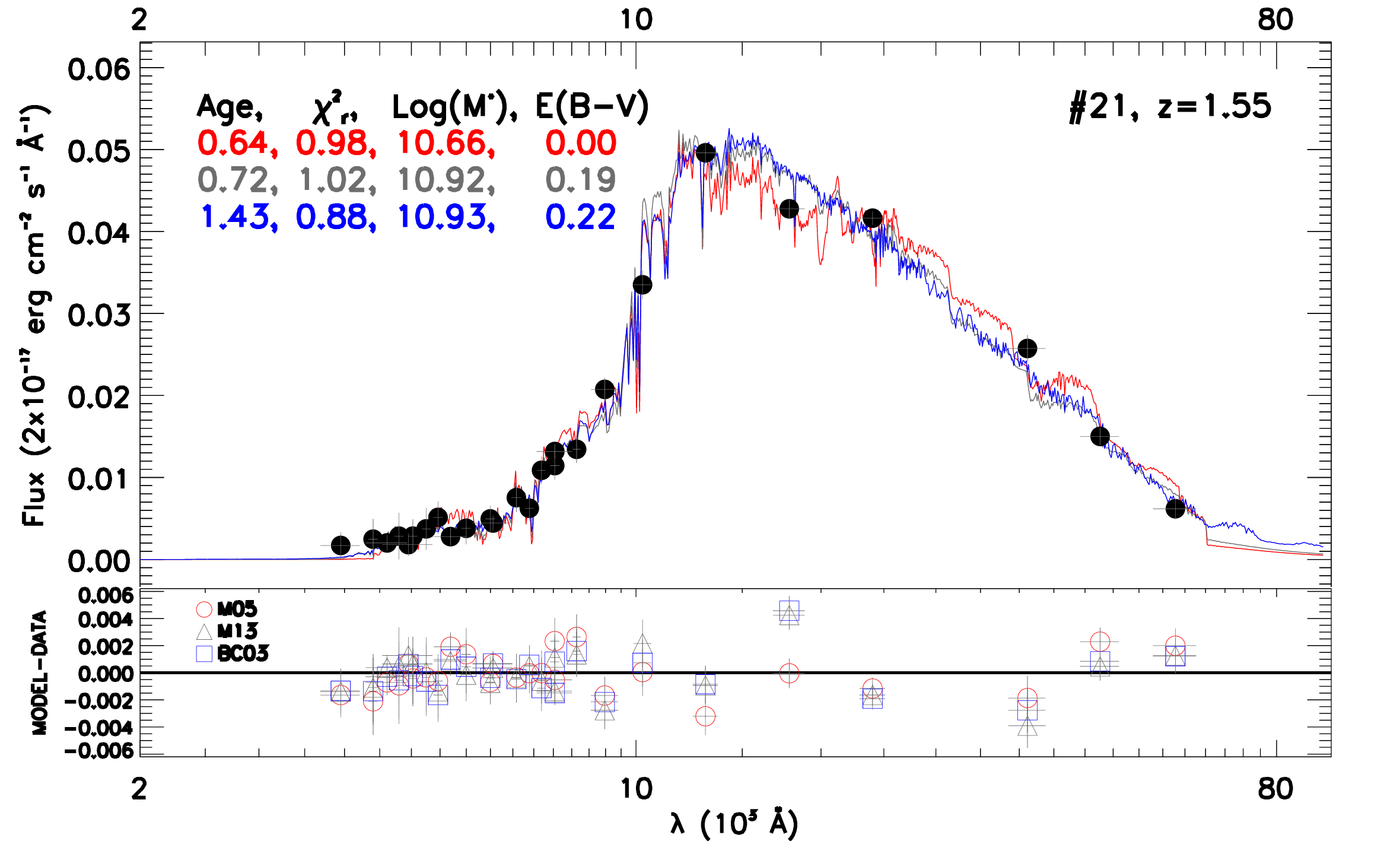}
\includegraphics[width=0.48\textwidth]{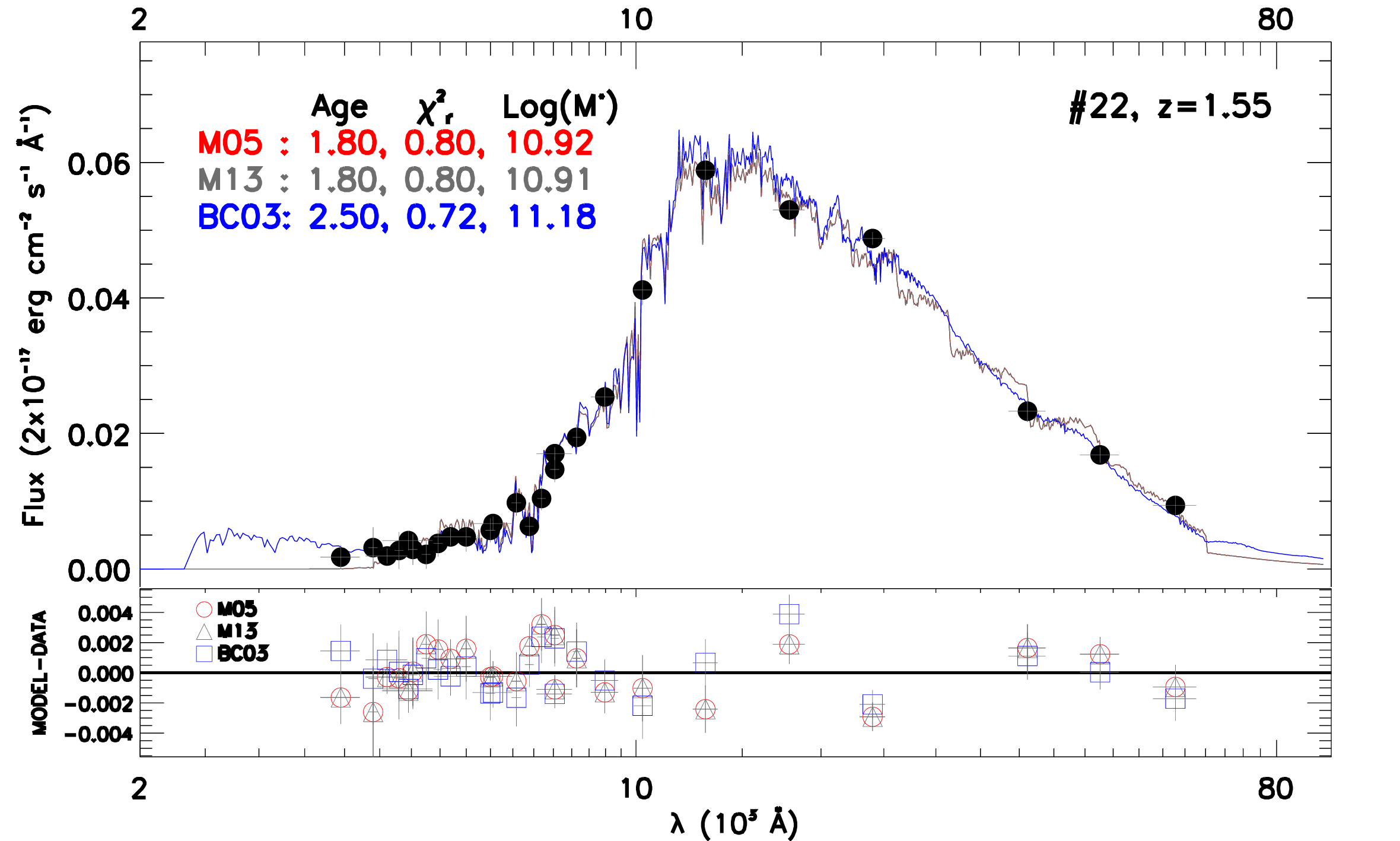}
\includegraphics[width=0.48\textwidth]{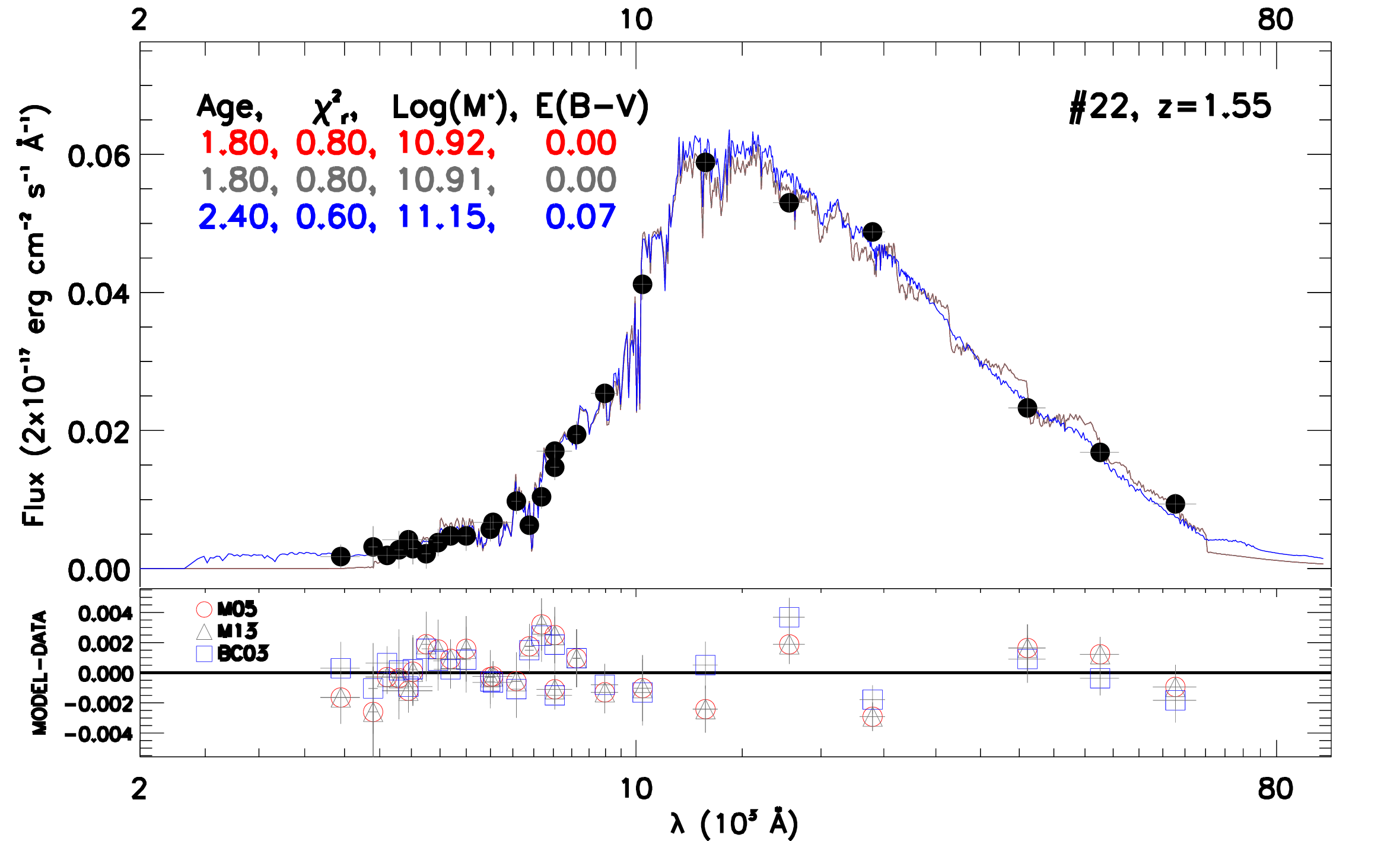}
\includegraphics[width=0.48\textwidth]{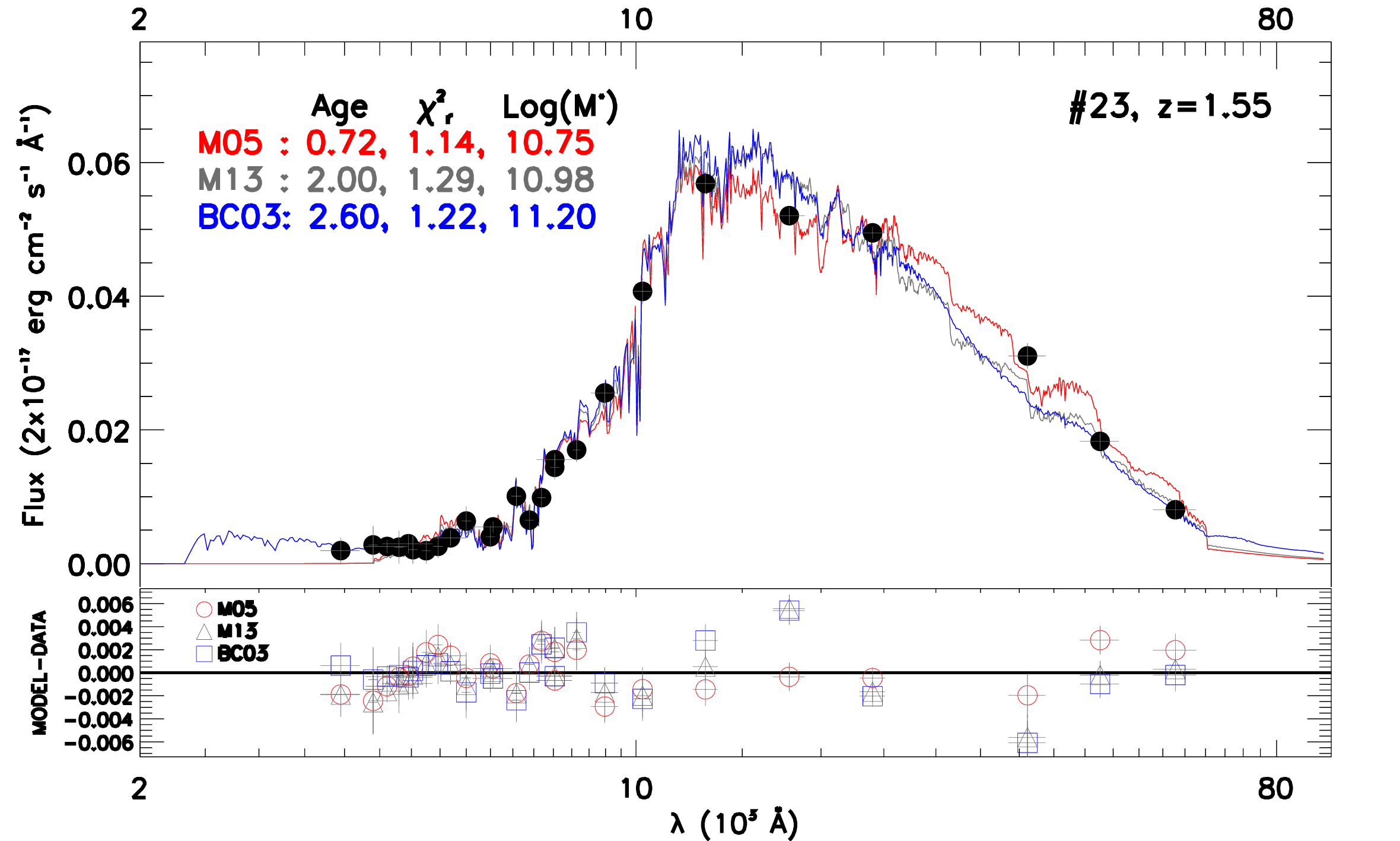}
\includegraphics[width=0.48\textwidth]{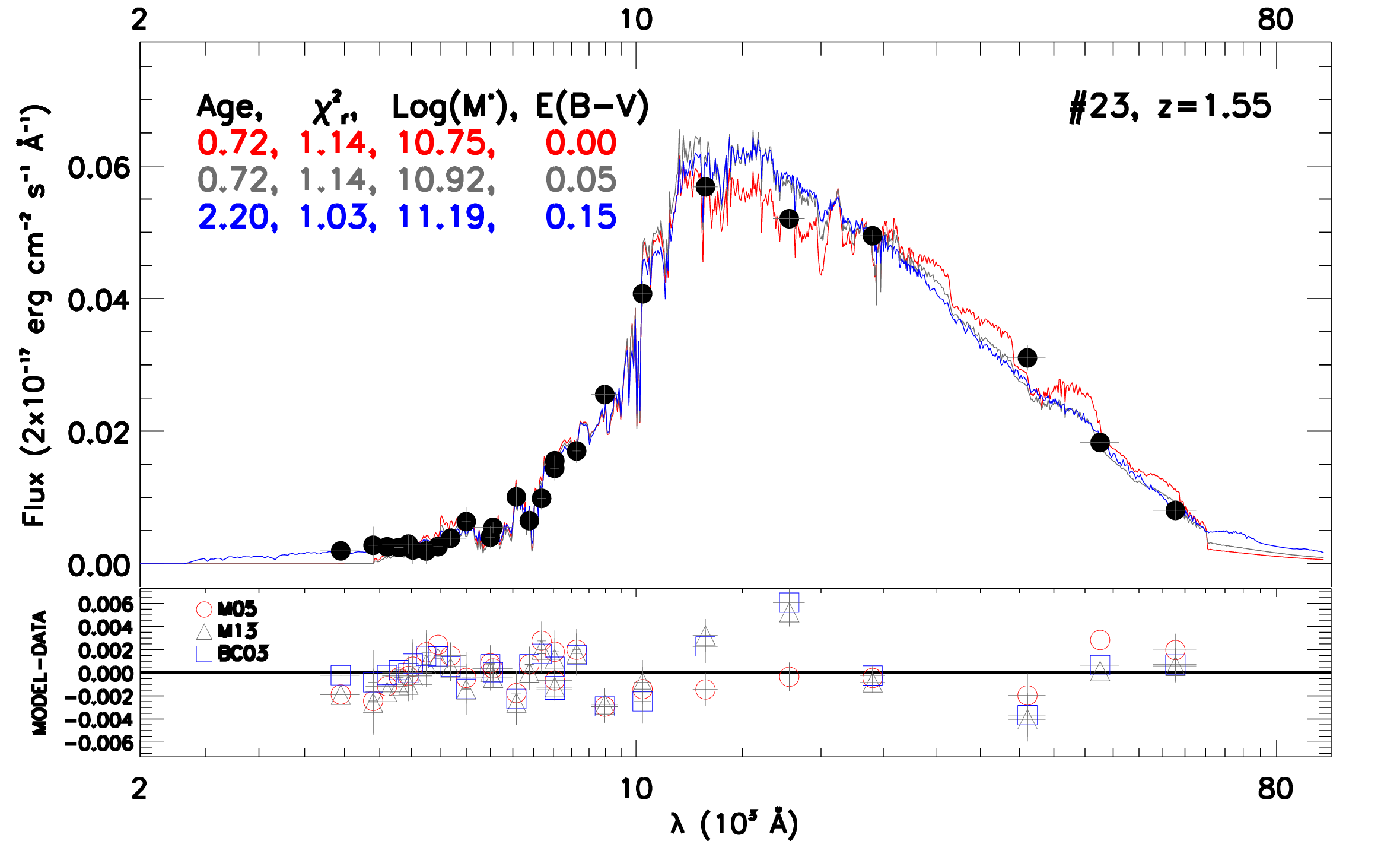}
\includegraphics[width=0.48\textwidth]{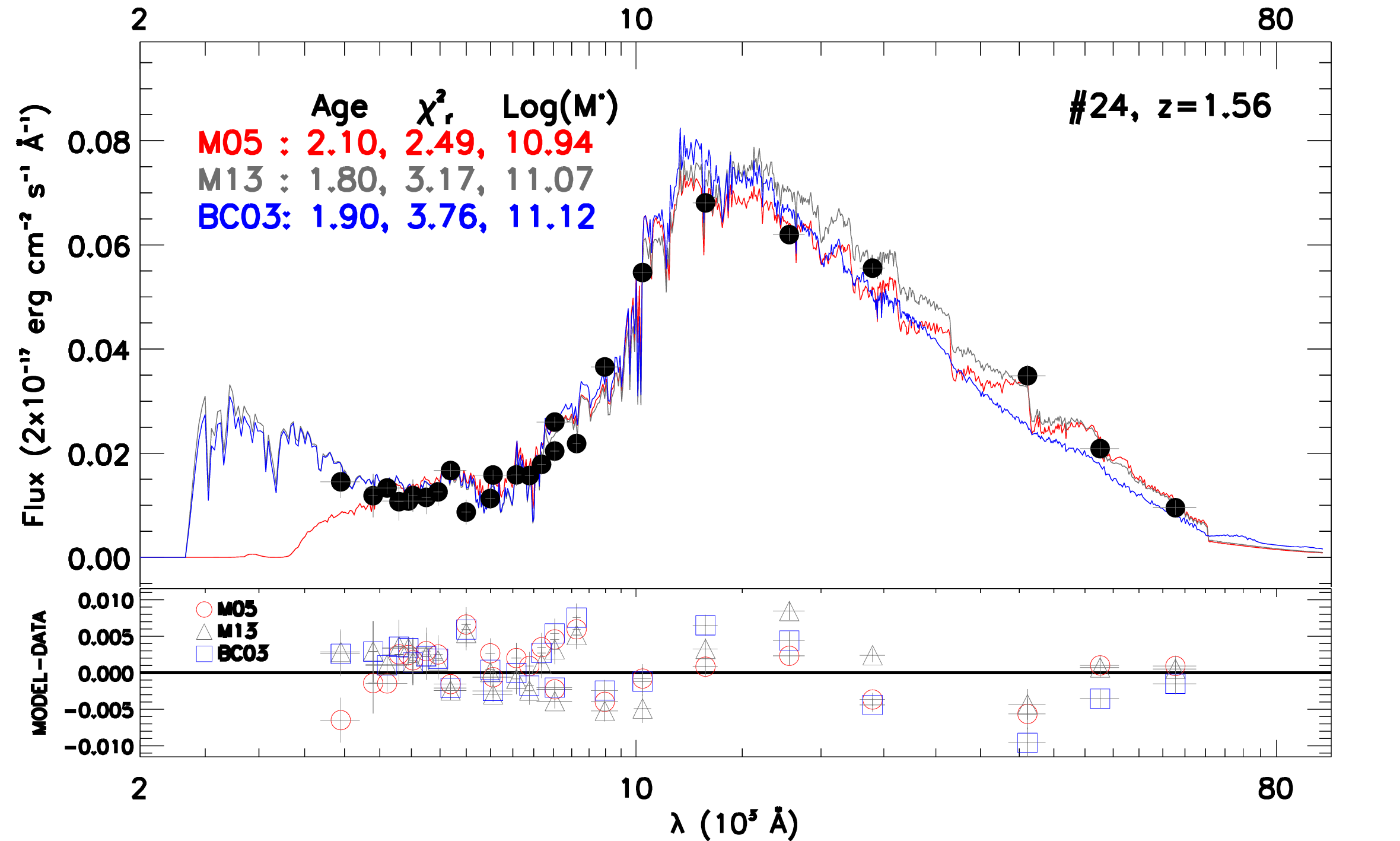}
\includegraphics[width=0.48\textwidth]{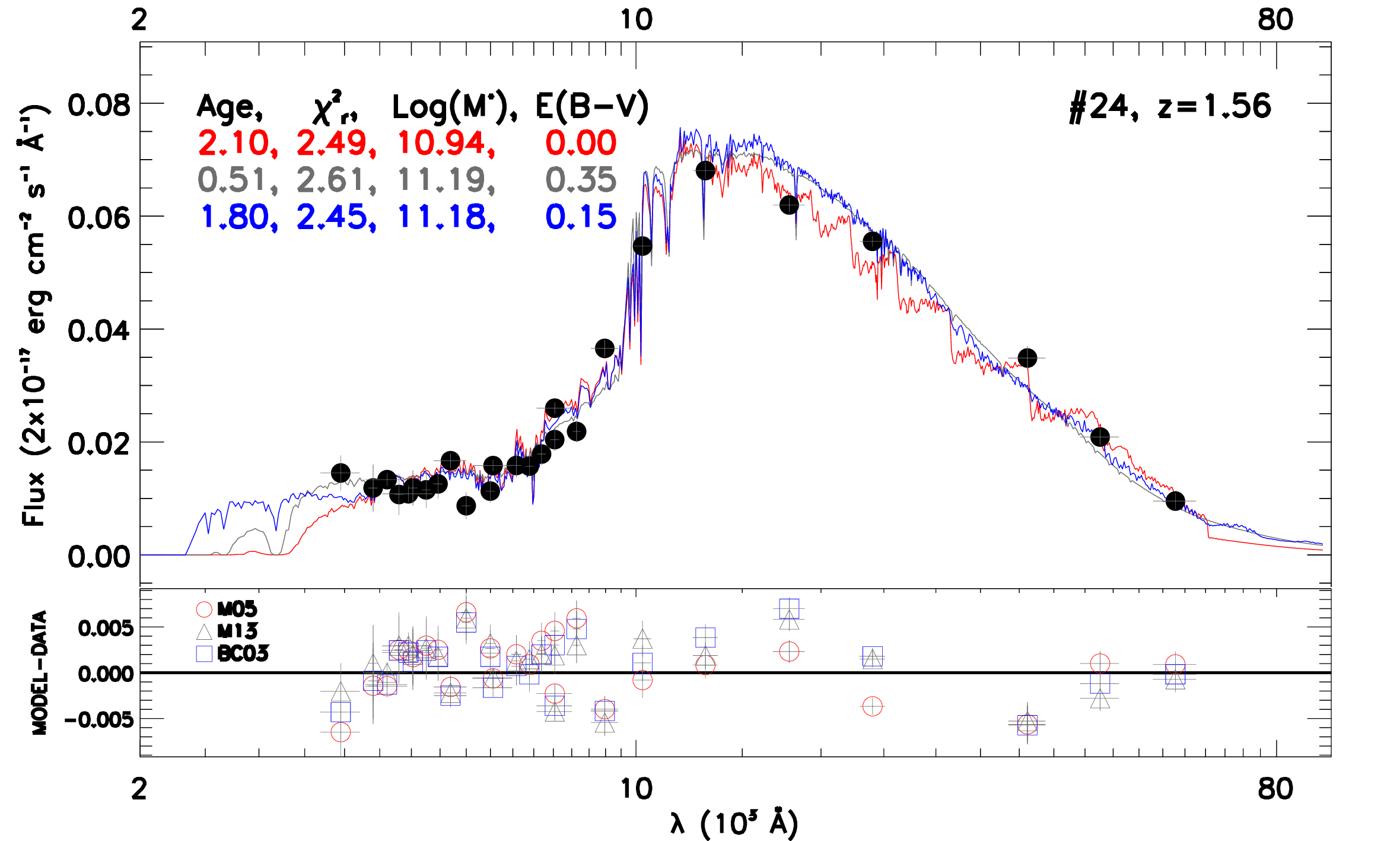}
\caption{Continued.}
\label{fig:Fig1_appB}
\end{figure*}

\addtocounter{figure}{-1}
\begin{figure*}
\centering
\includegraphics[width=0.48\textwidth]{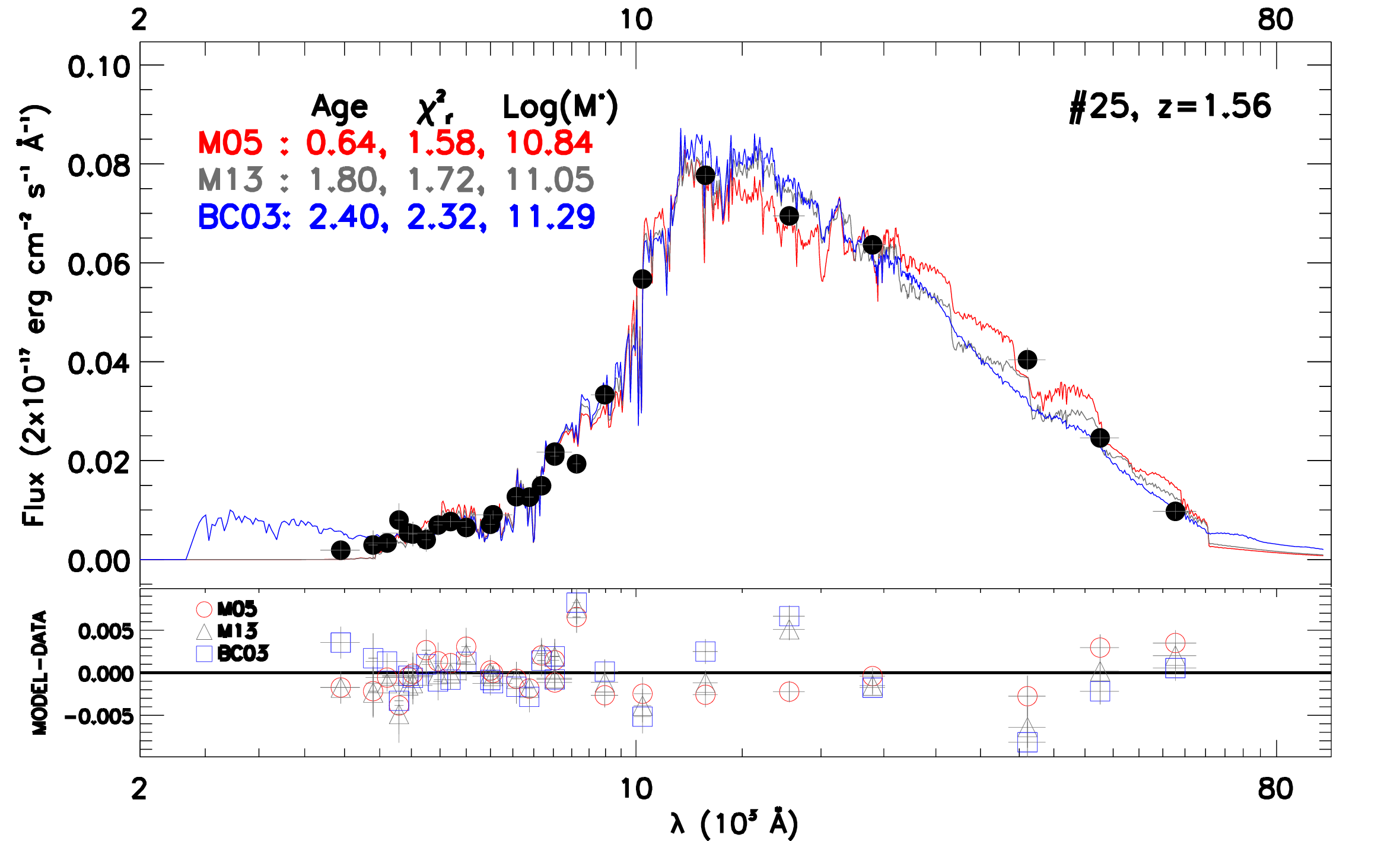}
\includegraphics[width=0.48\textwidth]{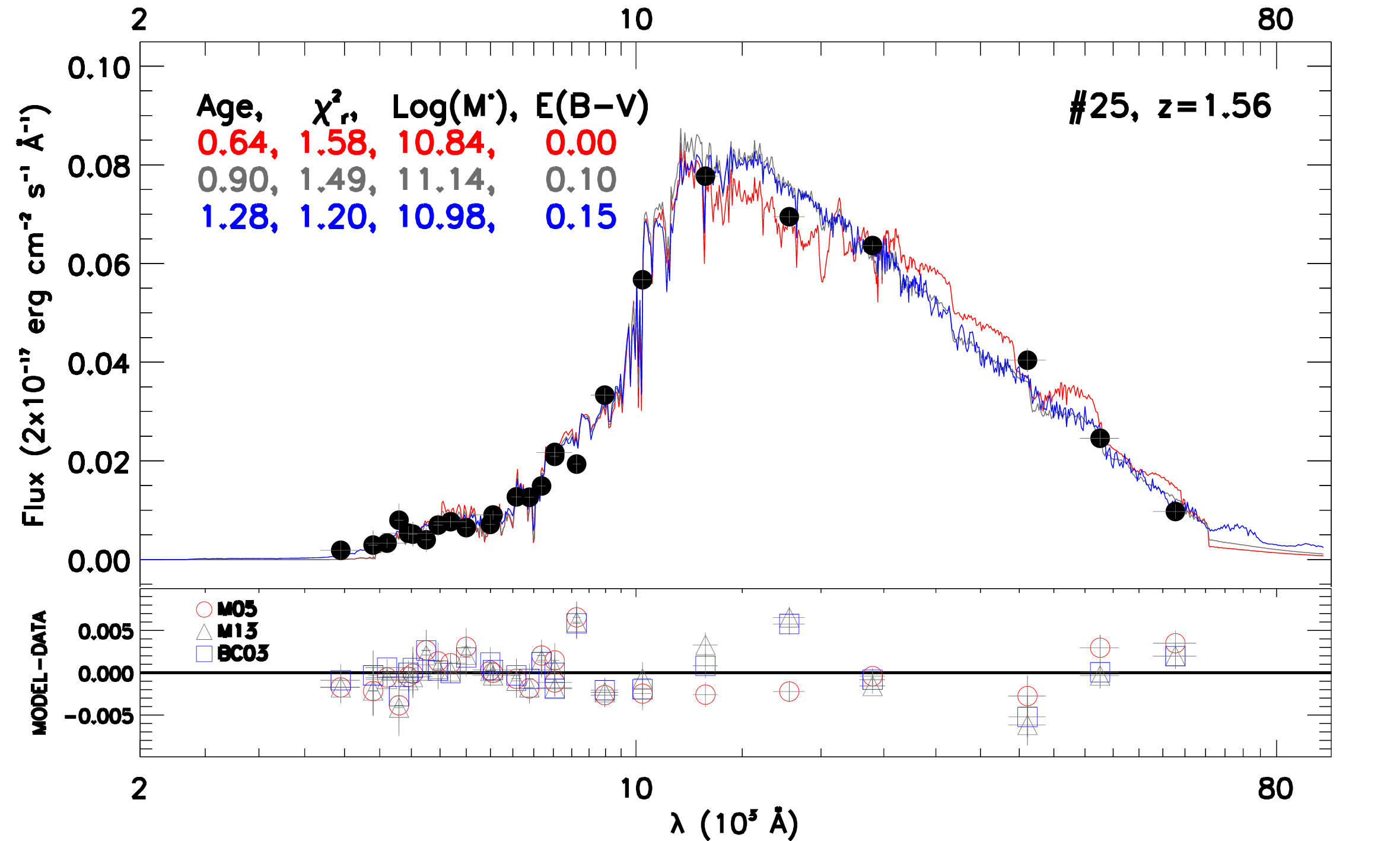}
\includegraphics[width=0.48\textwidth]{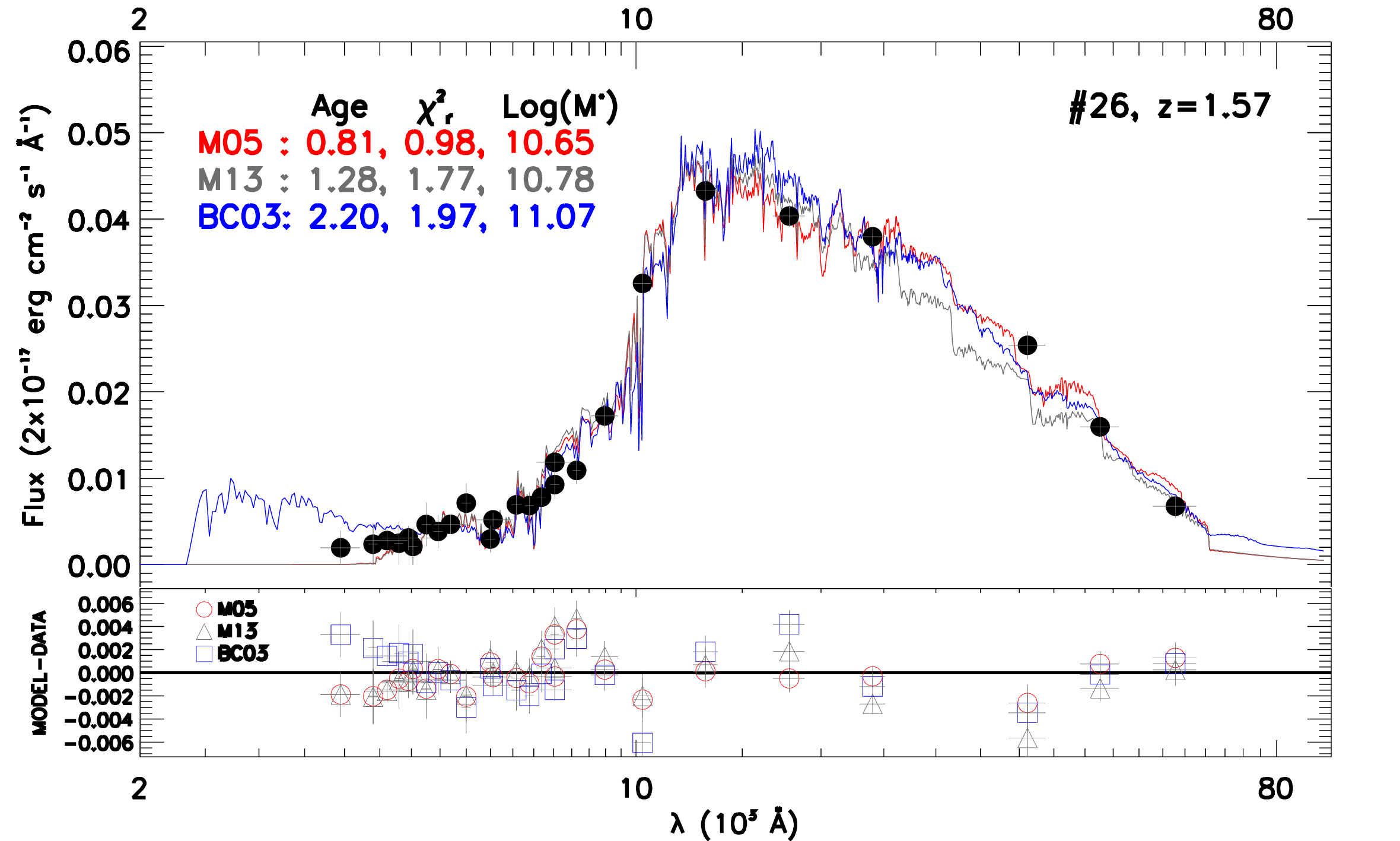}
\includegraphics[width=0.48\textwidth]{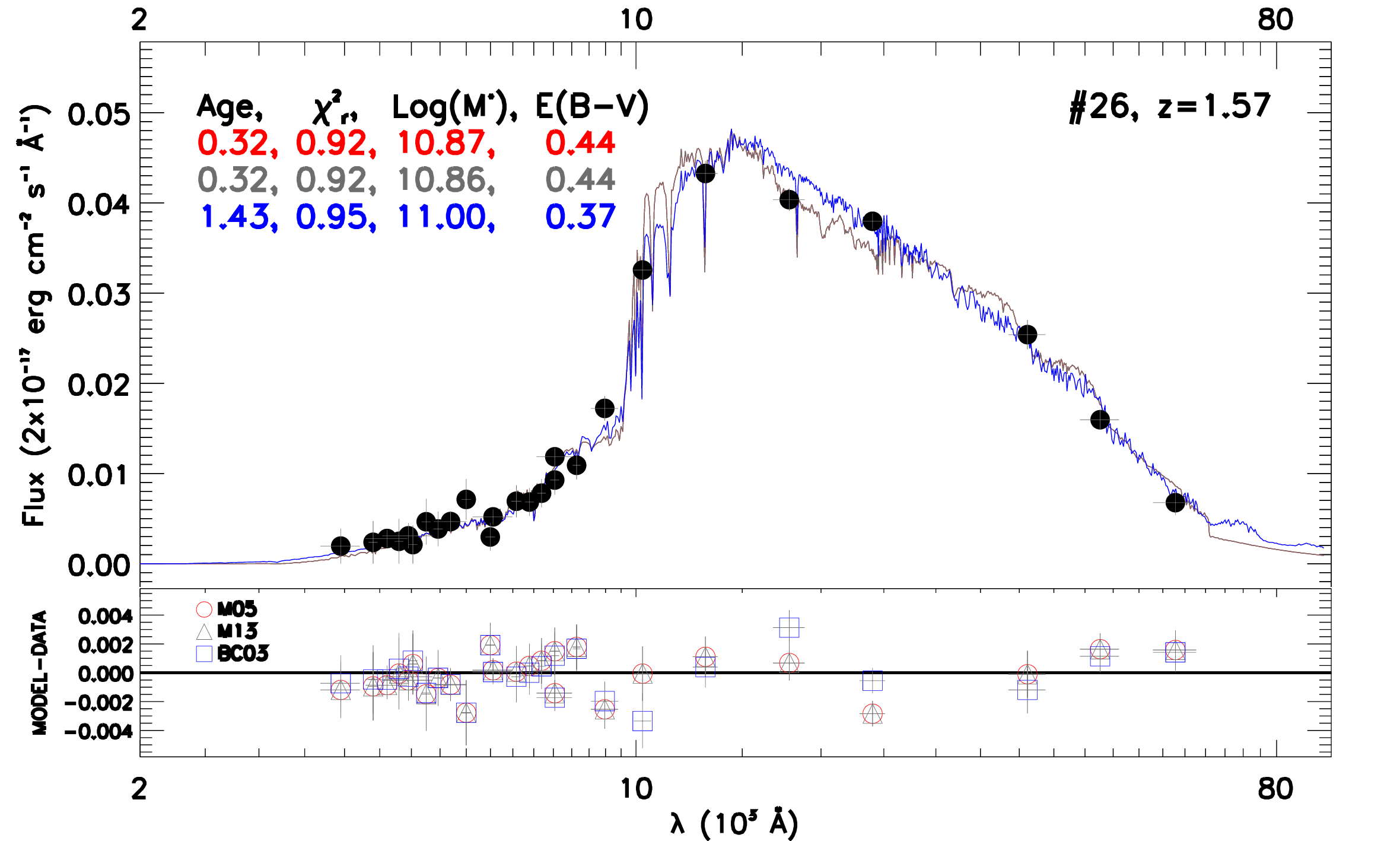}
\includegraphics[width=0.48\textwidth]{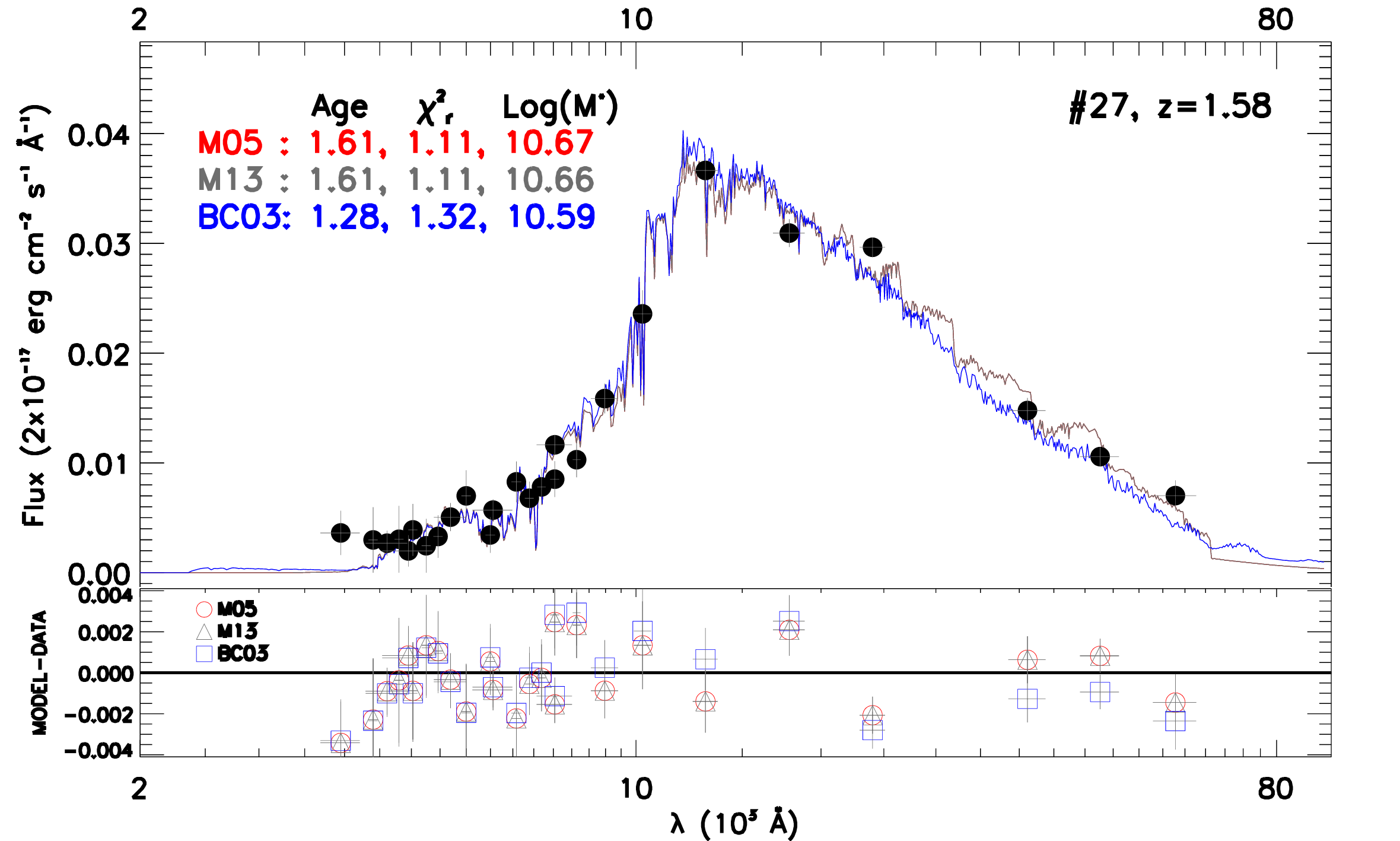}
\includegraphics[width=0.48\textwidth]{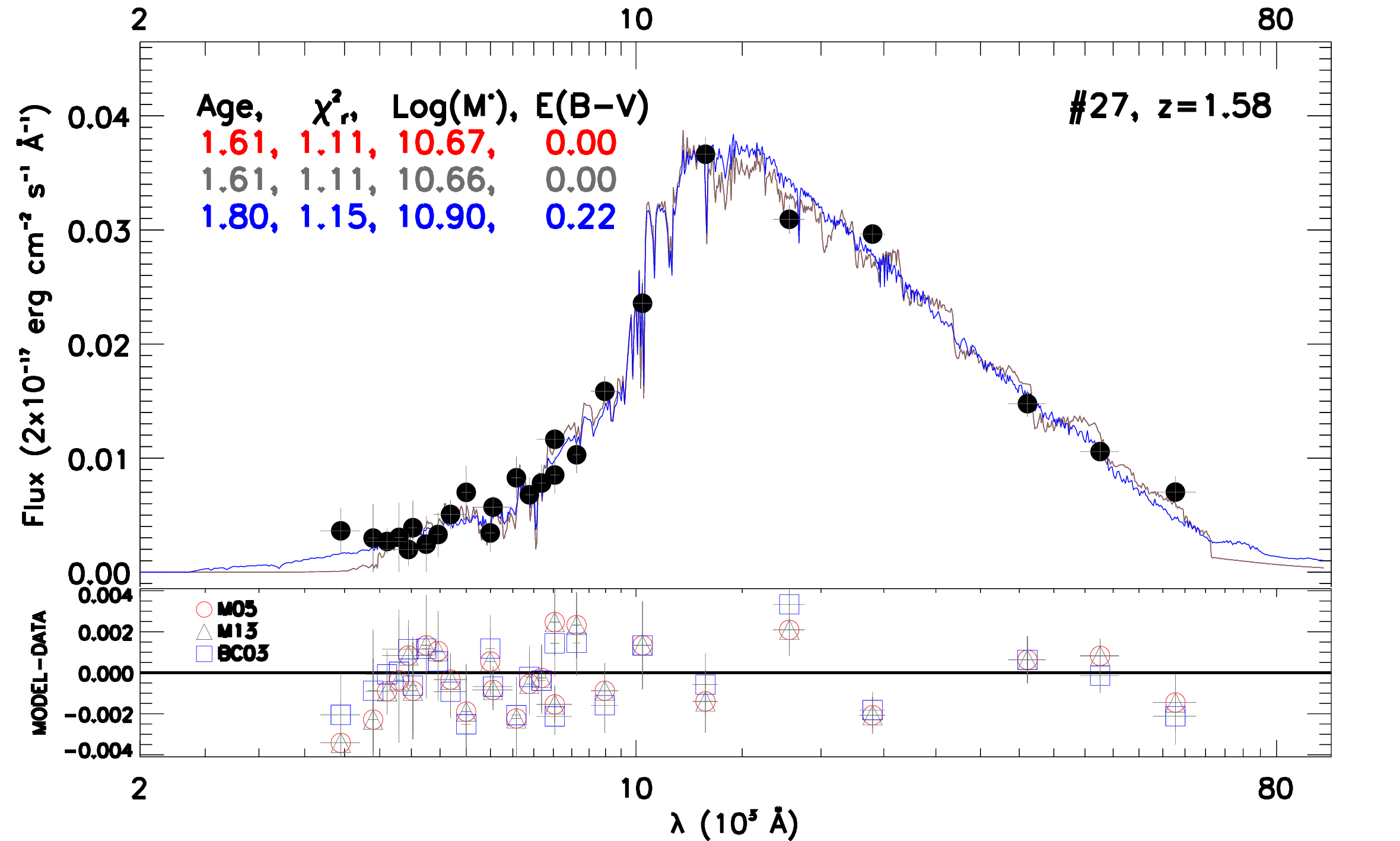}
\includegraphics[width=0.48\textwidth]{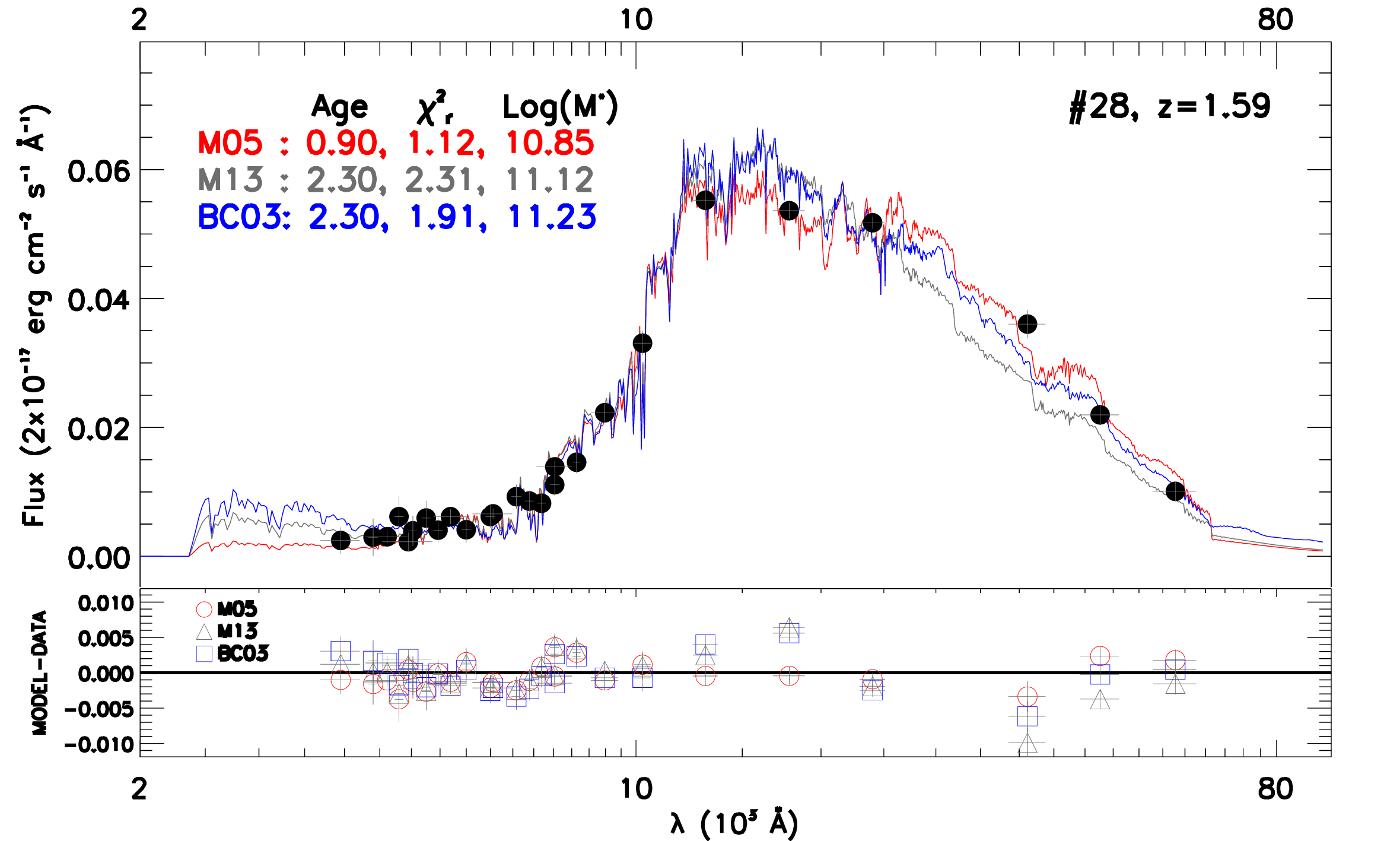}
\includegraphics[width=0.48\textwidth]{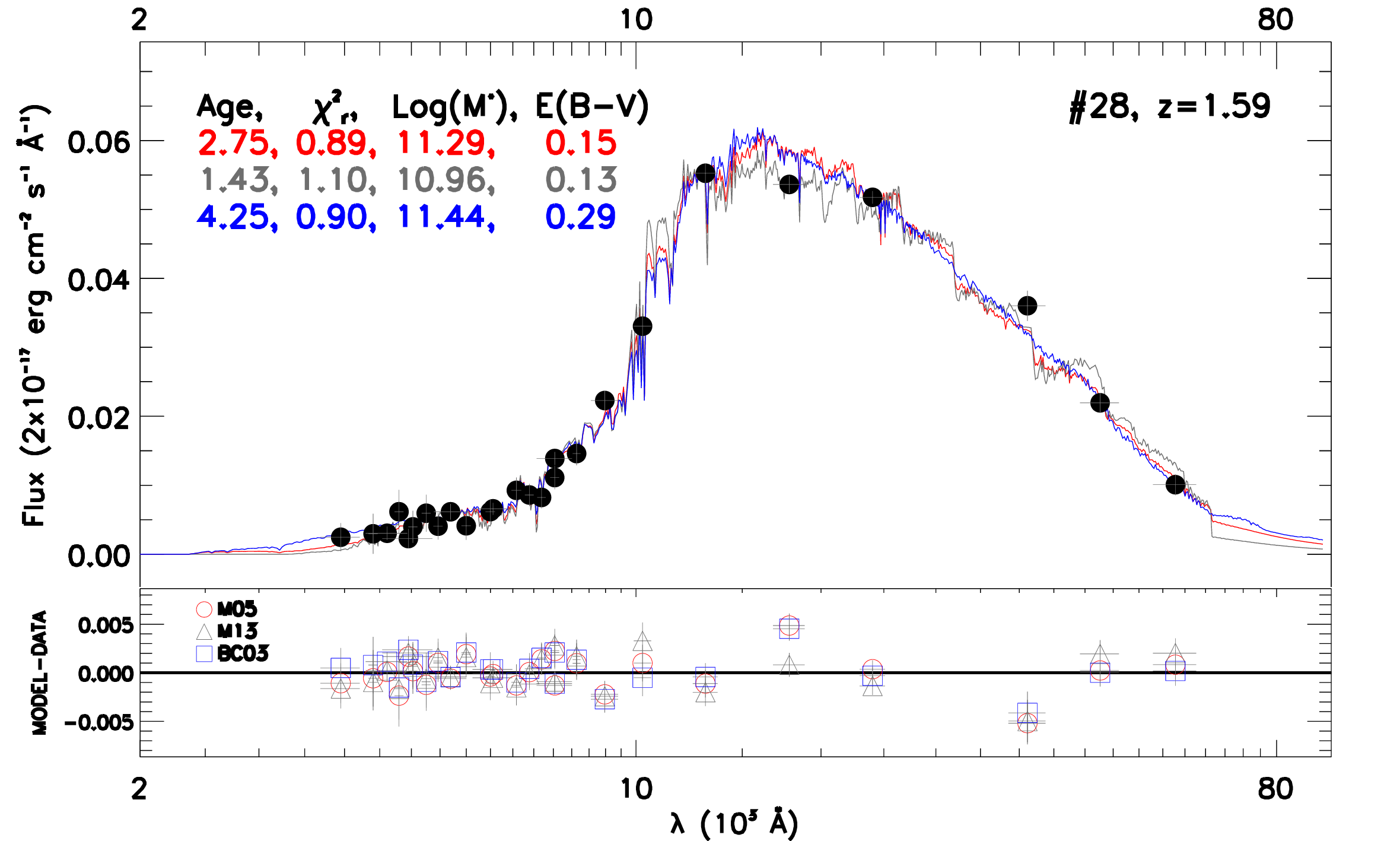}
\caption{Continued.}
\label{fig:Fig1_appB}
\end{figure*}

\addtocounter{figure}{-1}
\begin{figure*}
\centering
\includegraphics[width=0.48\textwidth]{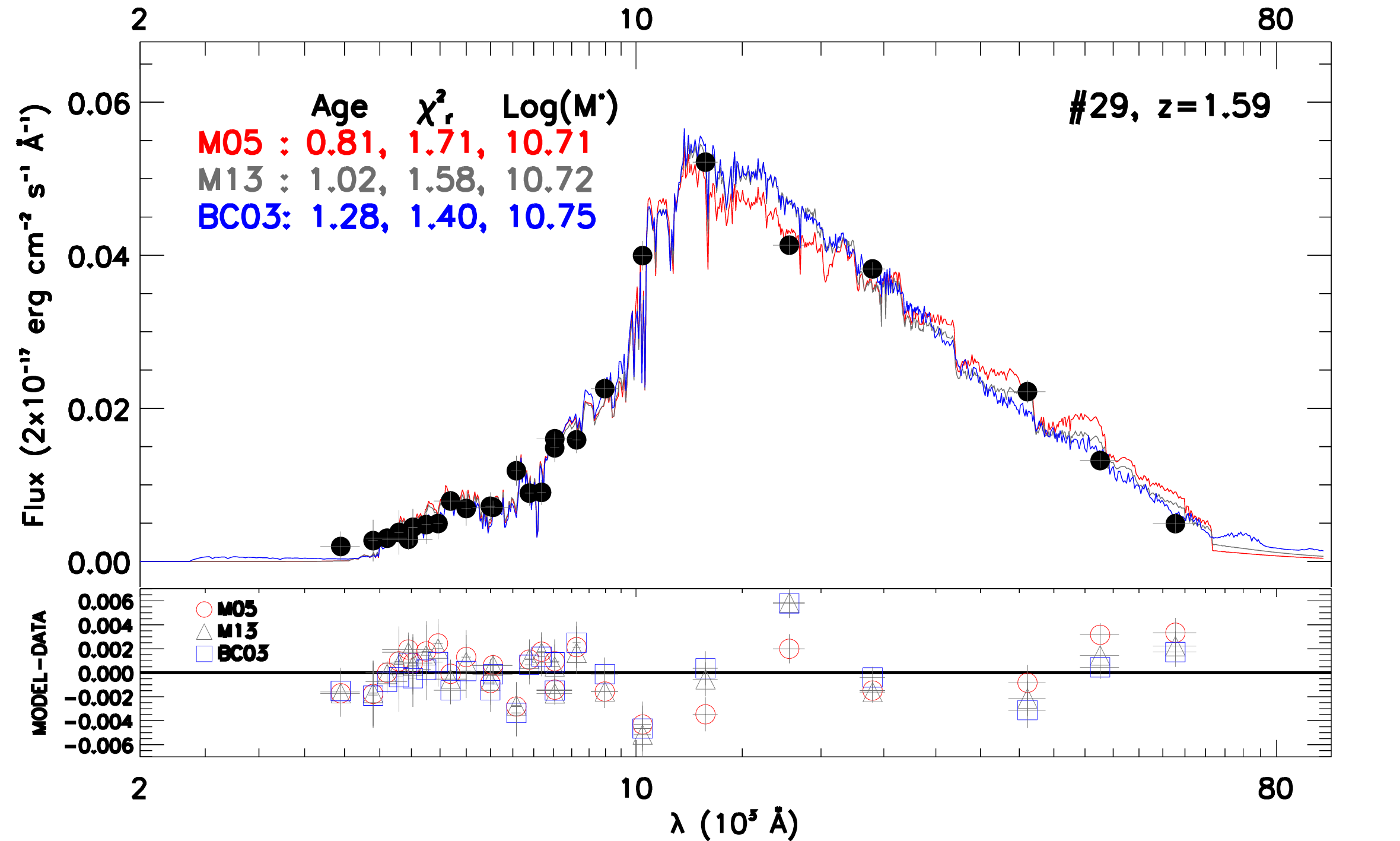}
\includegraphics[width=0.48\textwidth]{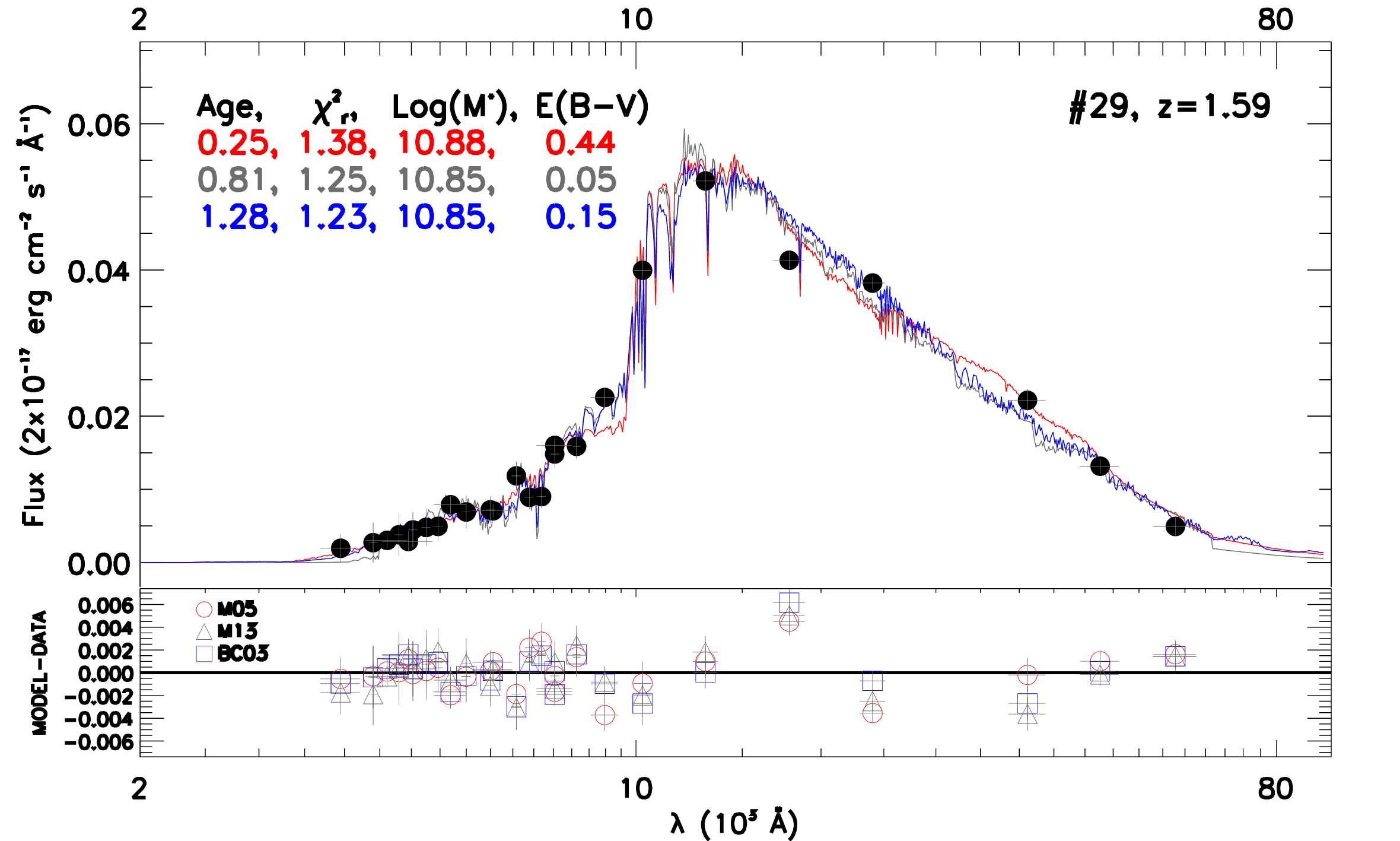}
\includegraphics[width=0.48\textwidth]{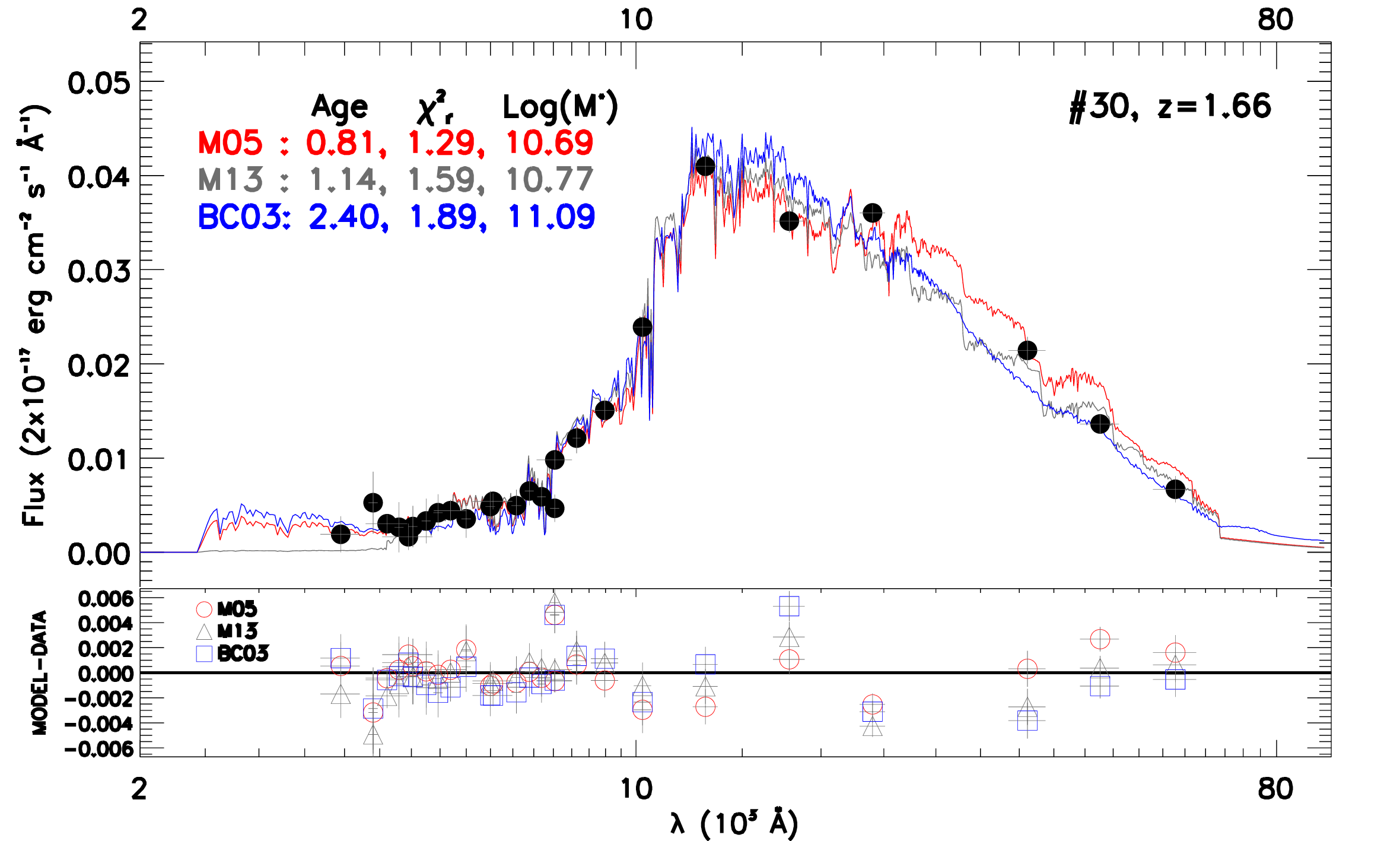}
\includegraphics[width=0.48\textwidth]{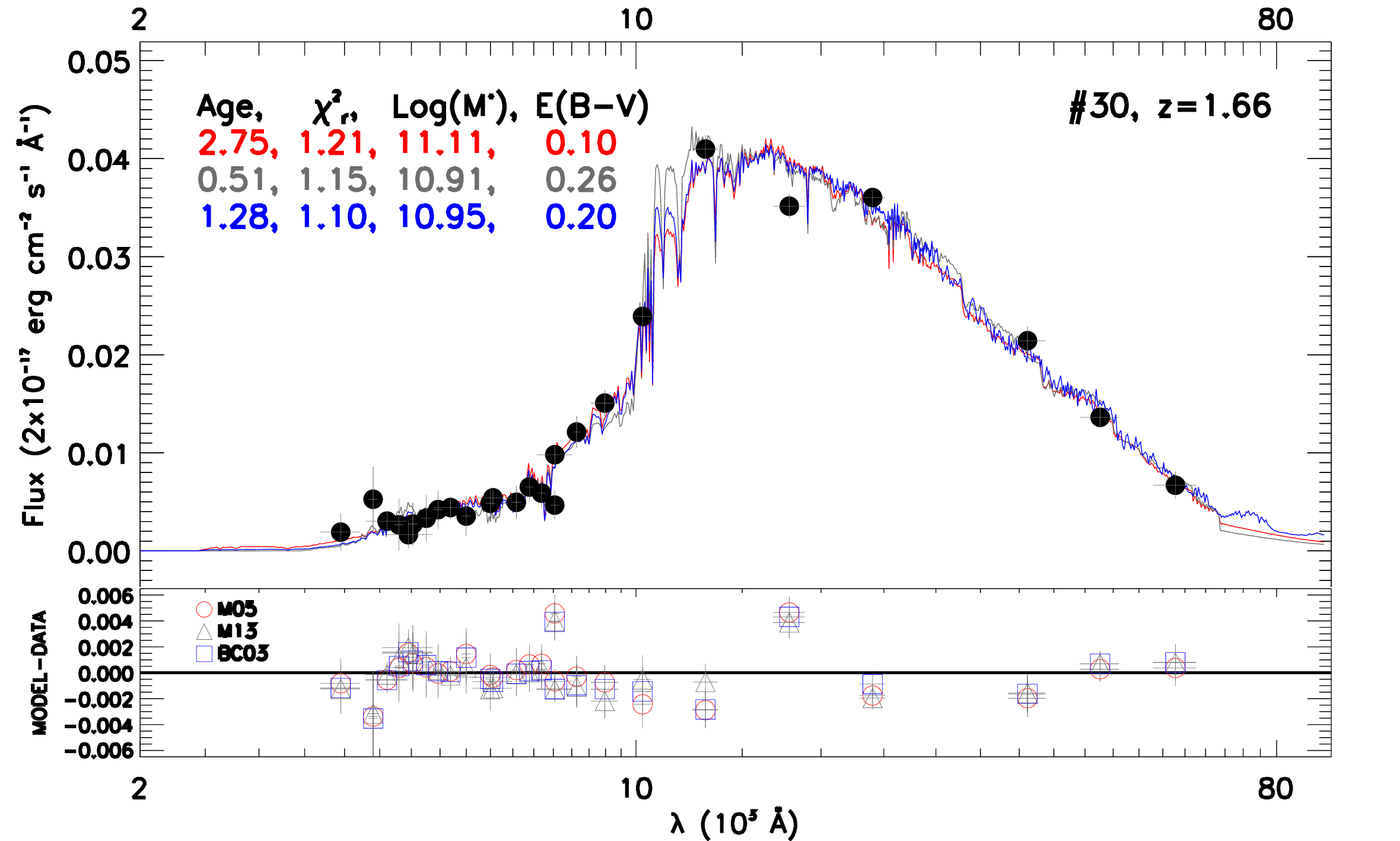}
\includegraphics[width=0.48\textwidth]{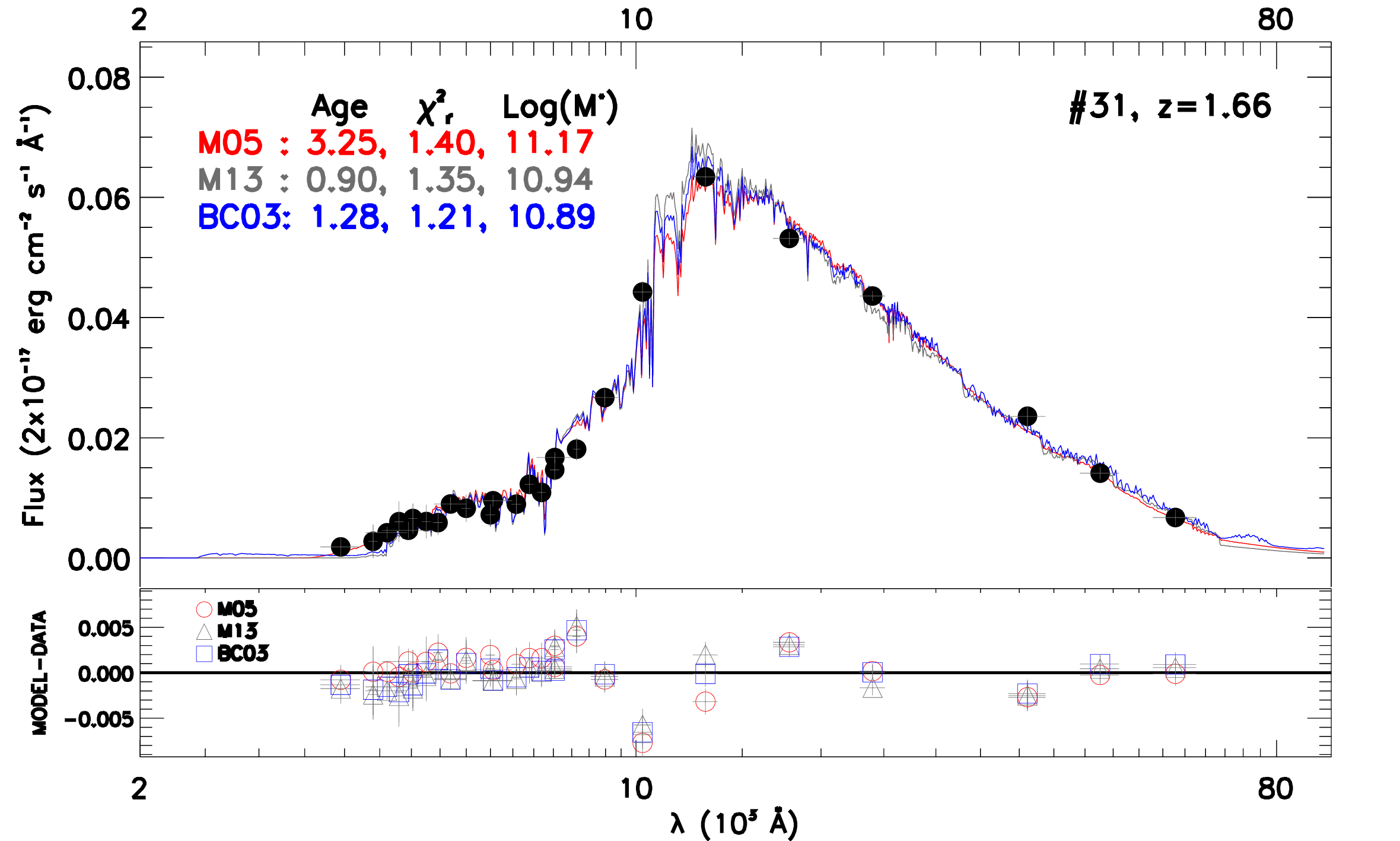}
\includegraphics[width=0.48\textwidth]{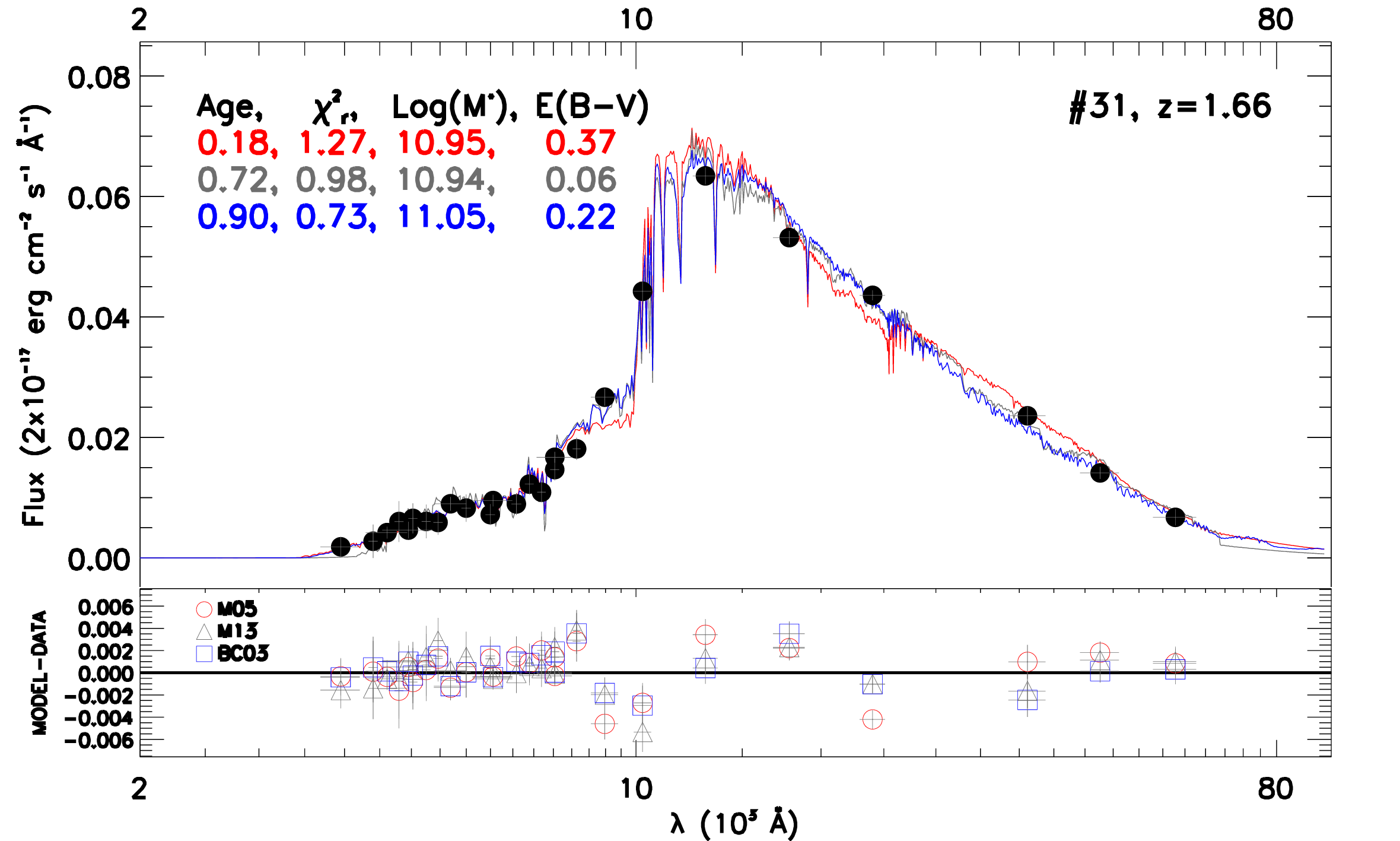}
\includegraphics[width=0.48\textwidth]{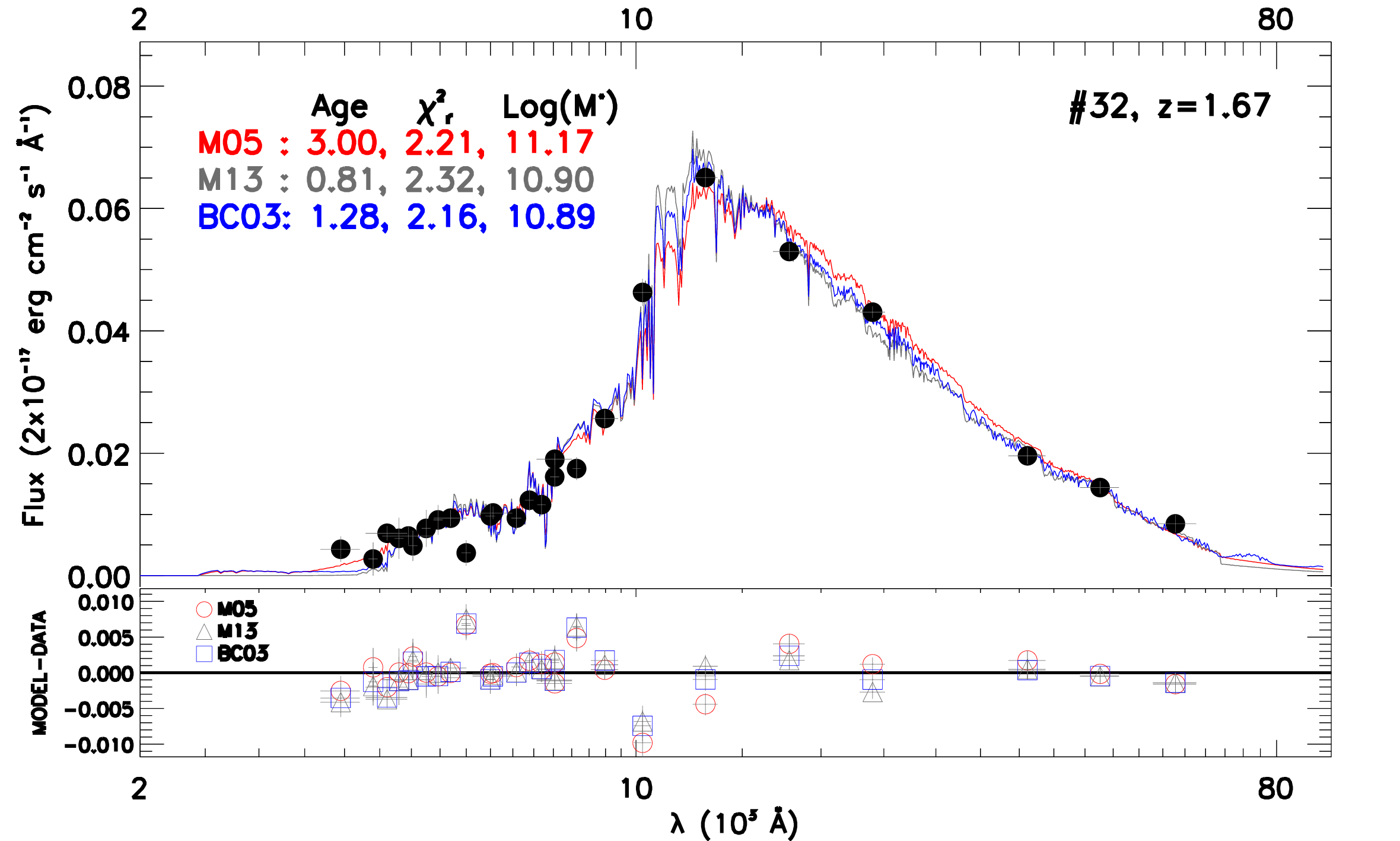}
\includegraphics[width=0.48\textwidth]{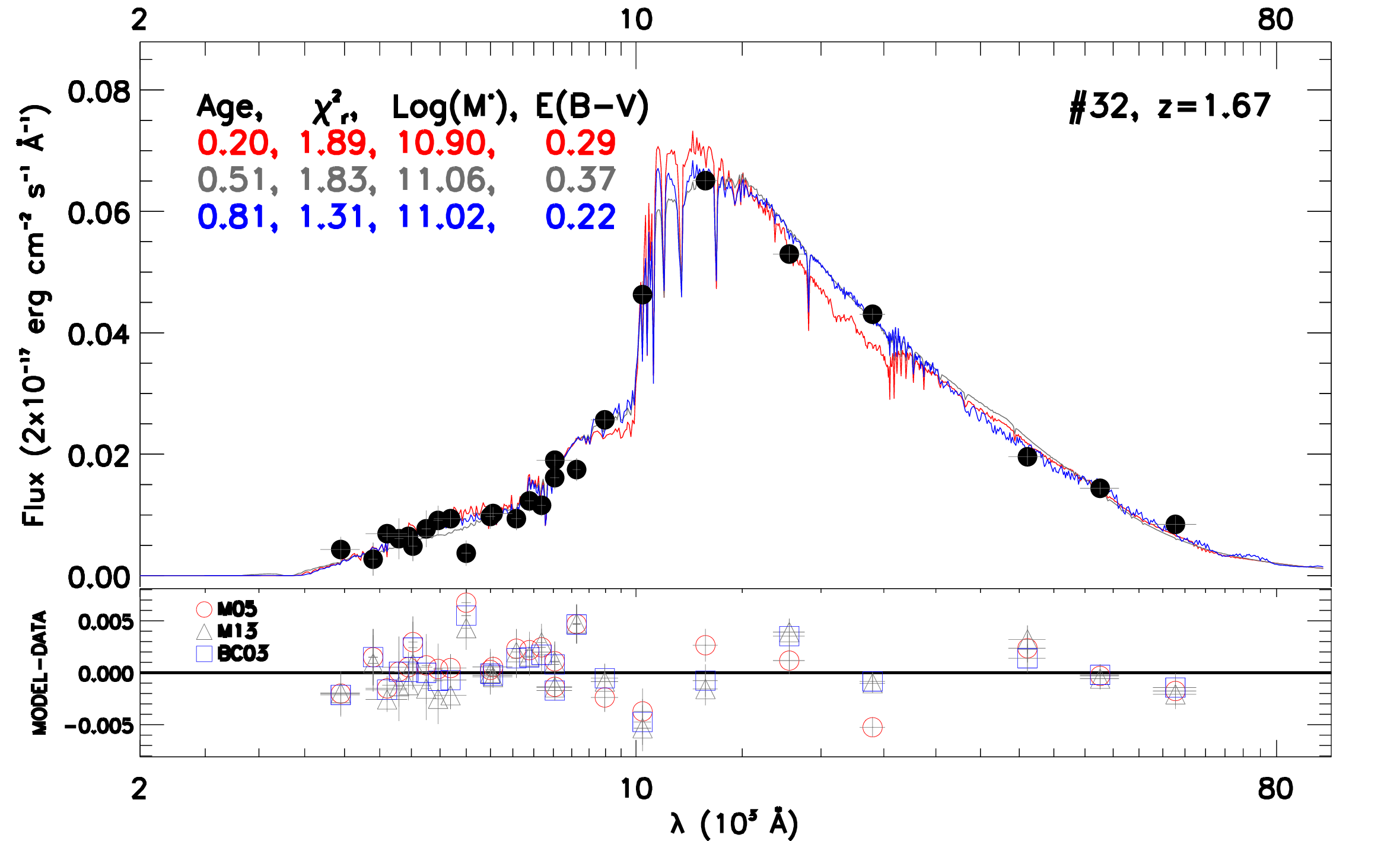}
\caption{Continued.}
\label{fig:Fig1_appB}
\end{figure*}

\addtocounter{figure}{-1}
\begin{figure*}
\centering
\includegraphics[width=0.48\textwidth]{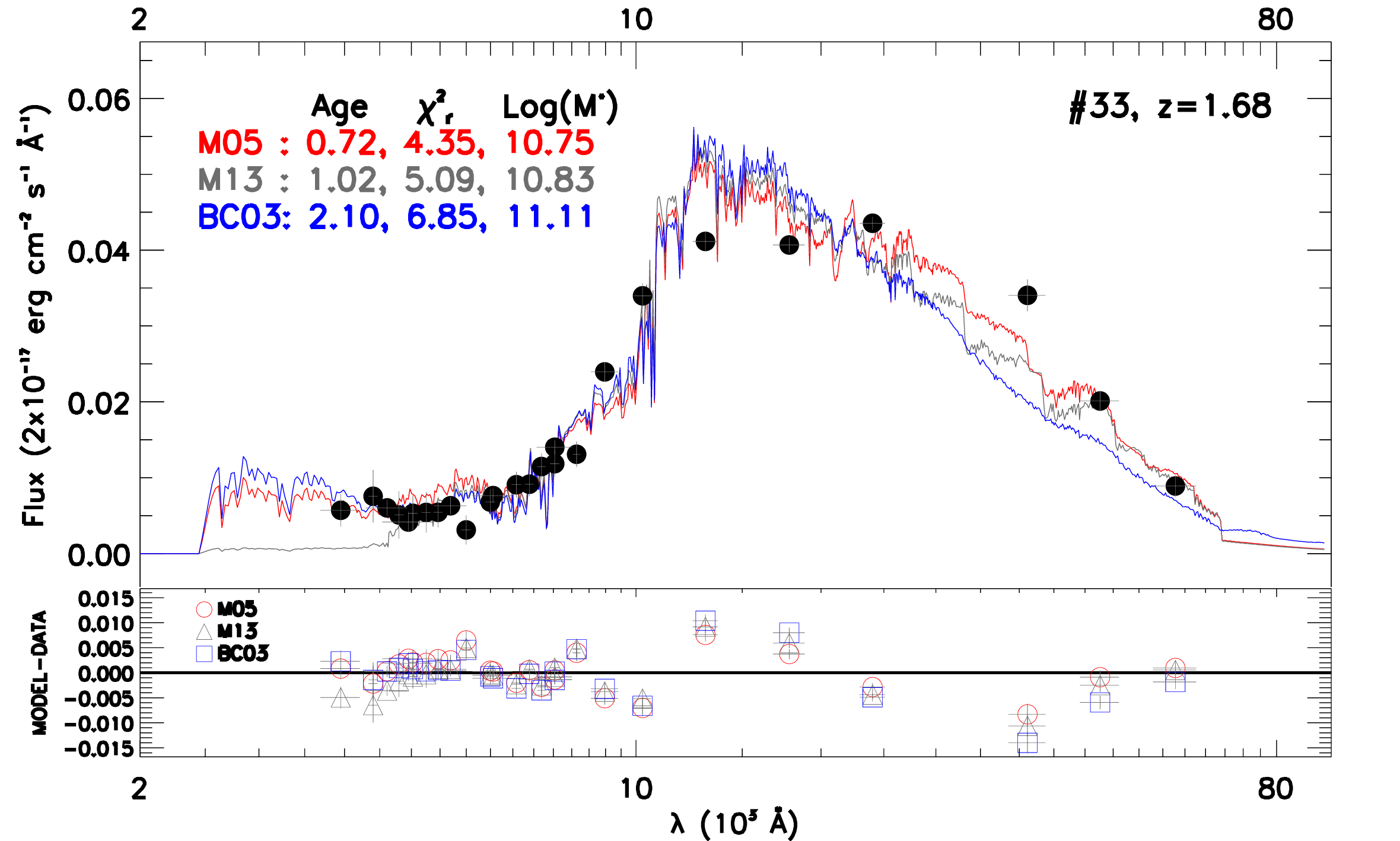}
\includegraphics[width=0.48\textwidth]{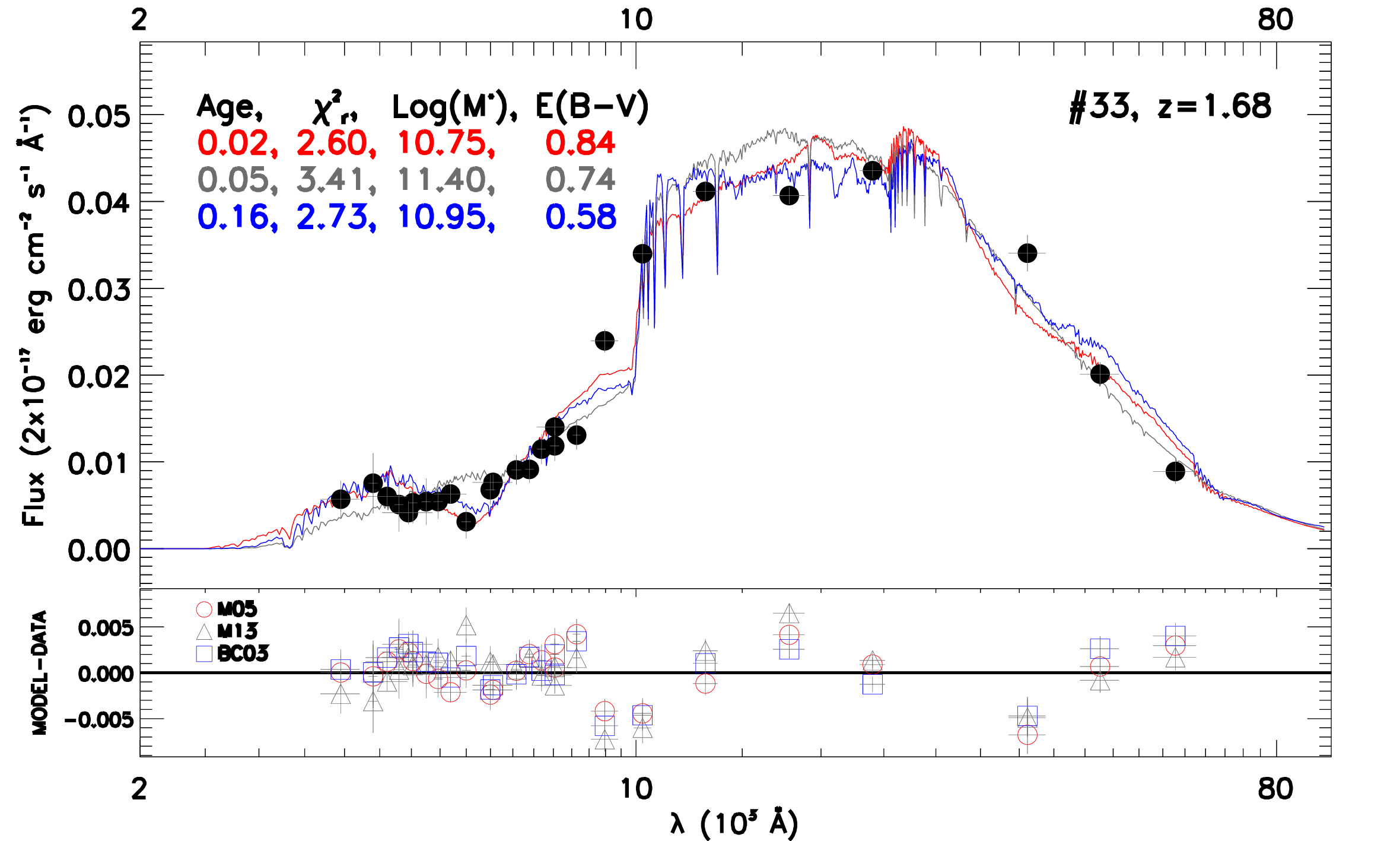}
\includegraphics[width=0.48\textwidth]{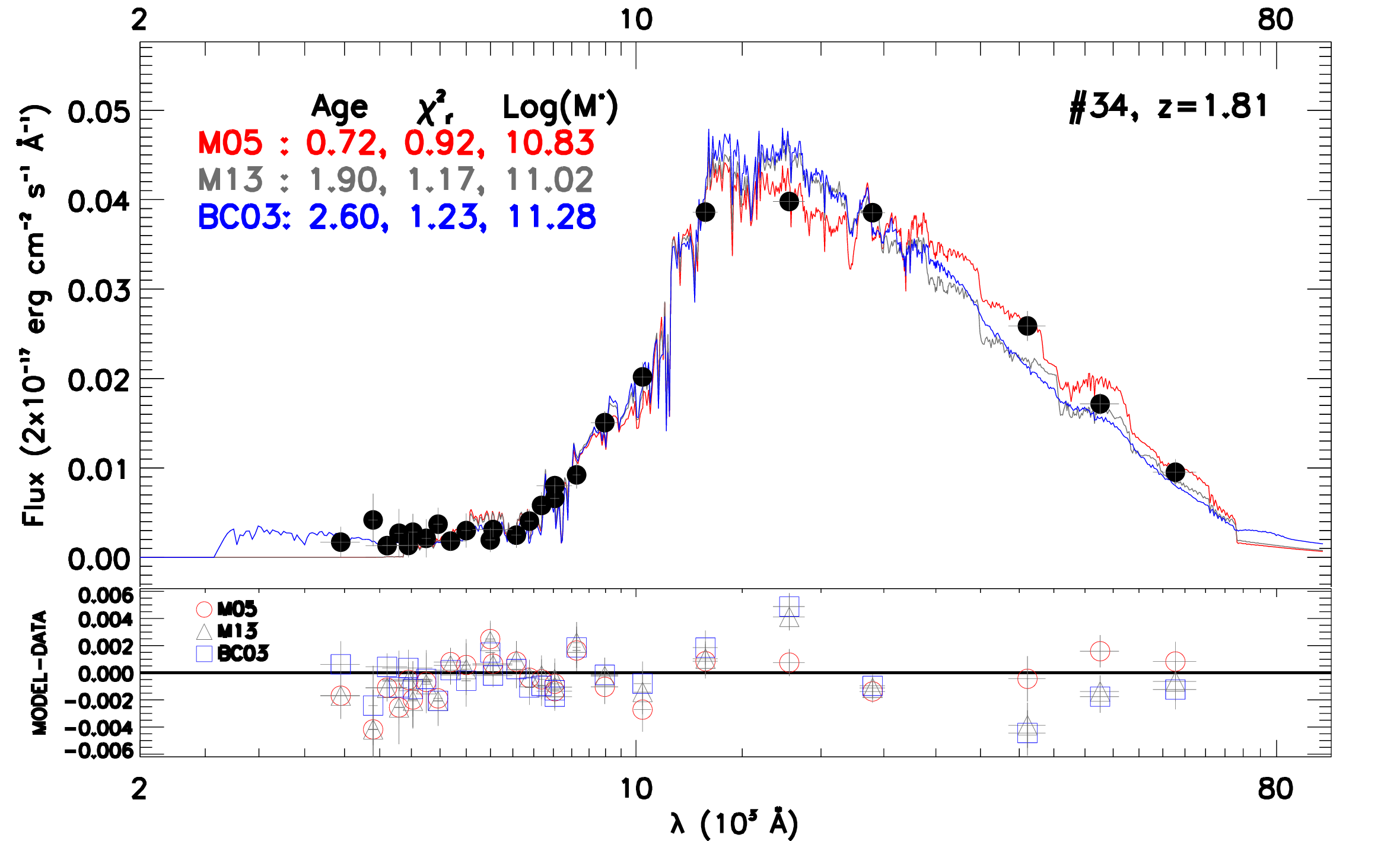}
\includegraphics[width=0.48\textwidth]{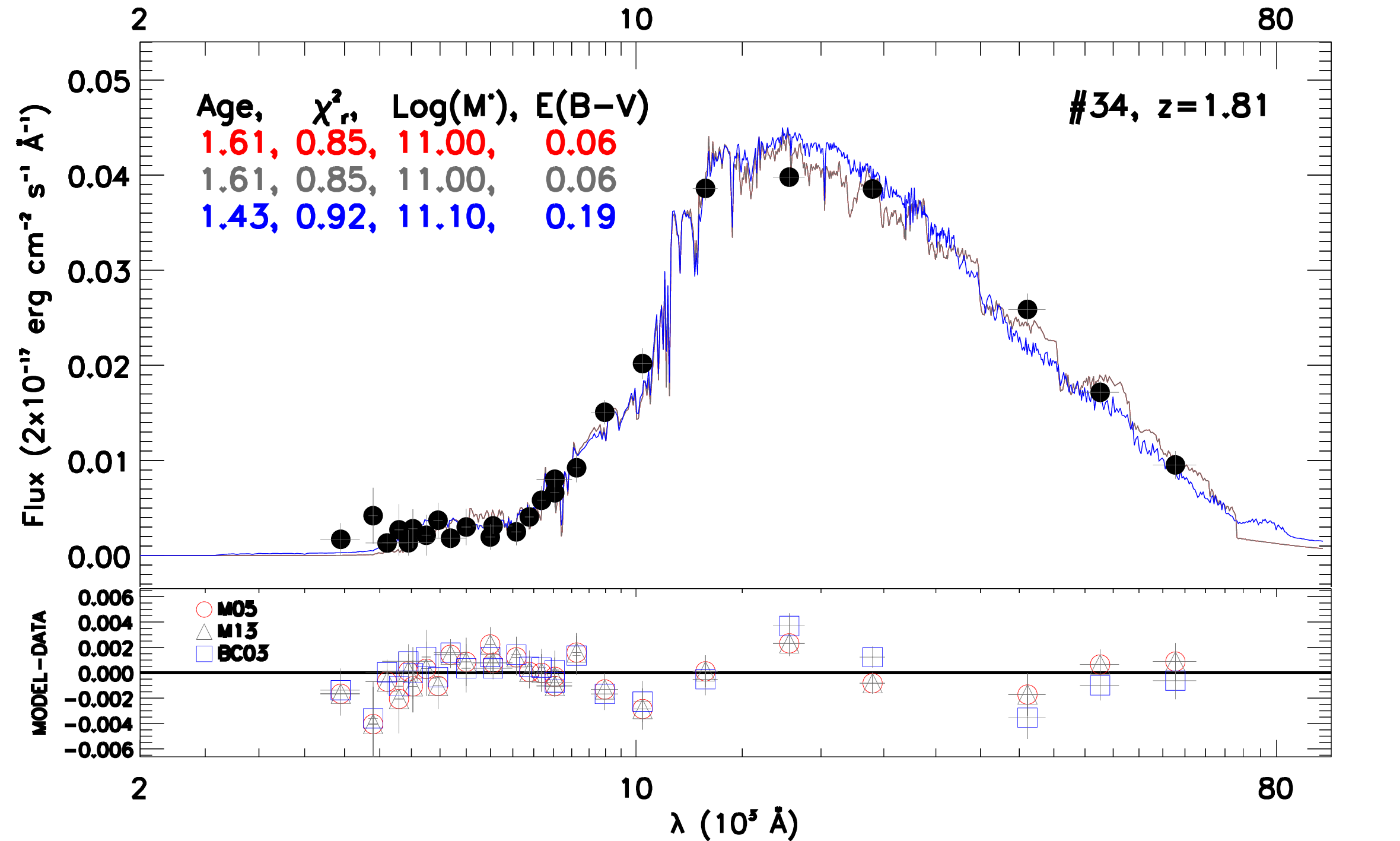}
\includegraphics[width=0.48\textwidth]{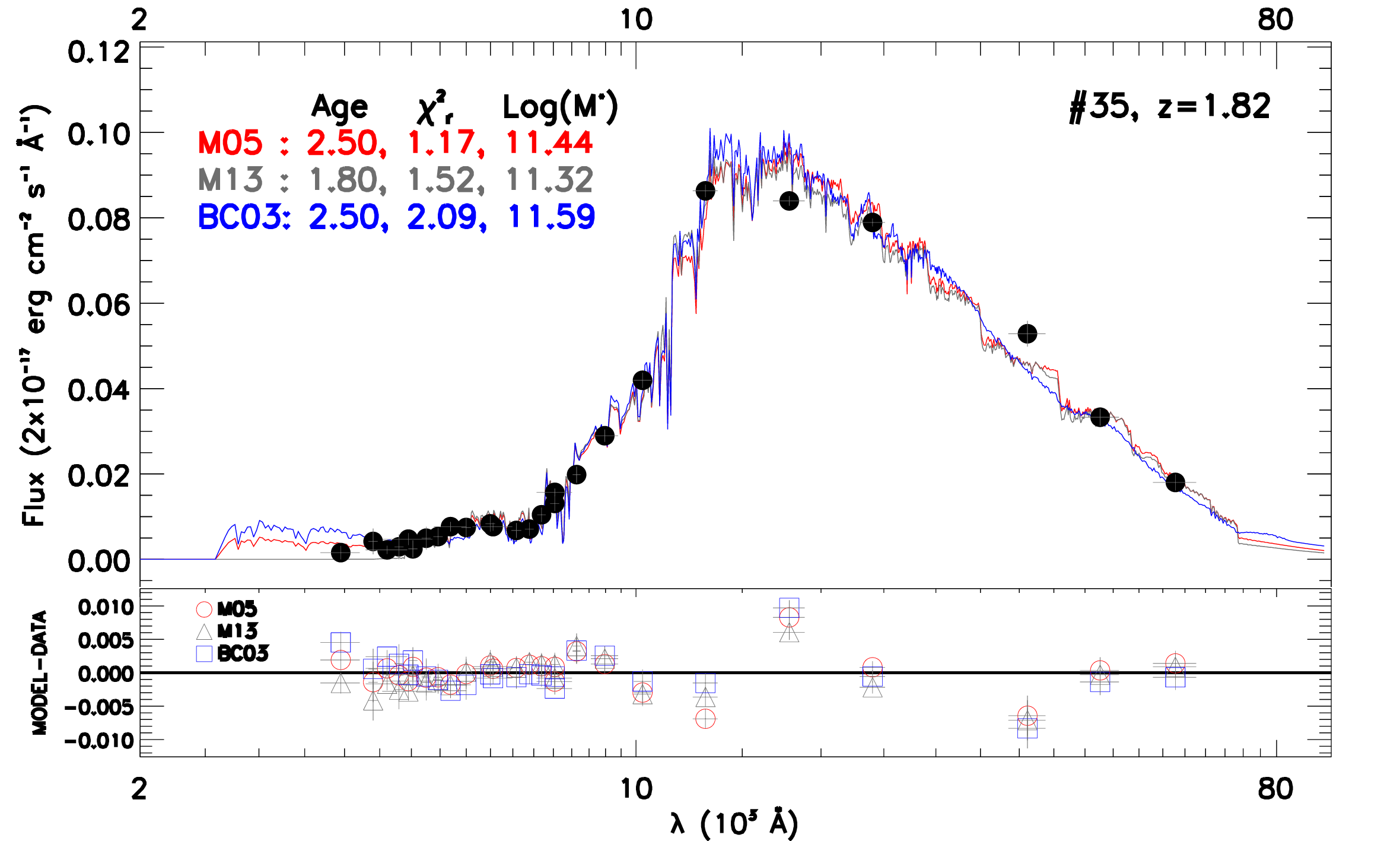}
\includegraphics[width=0.48\textwidth]{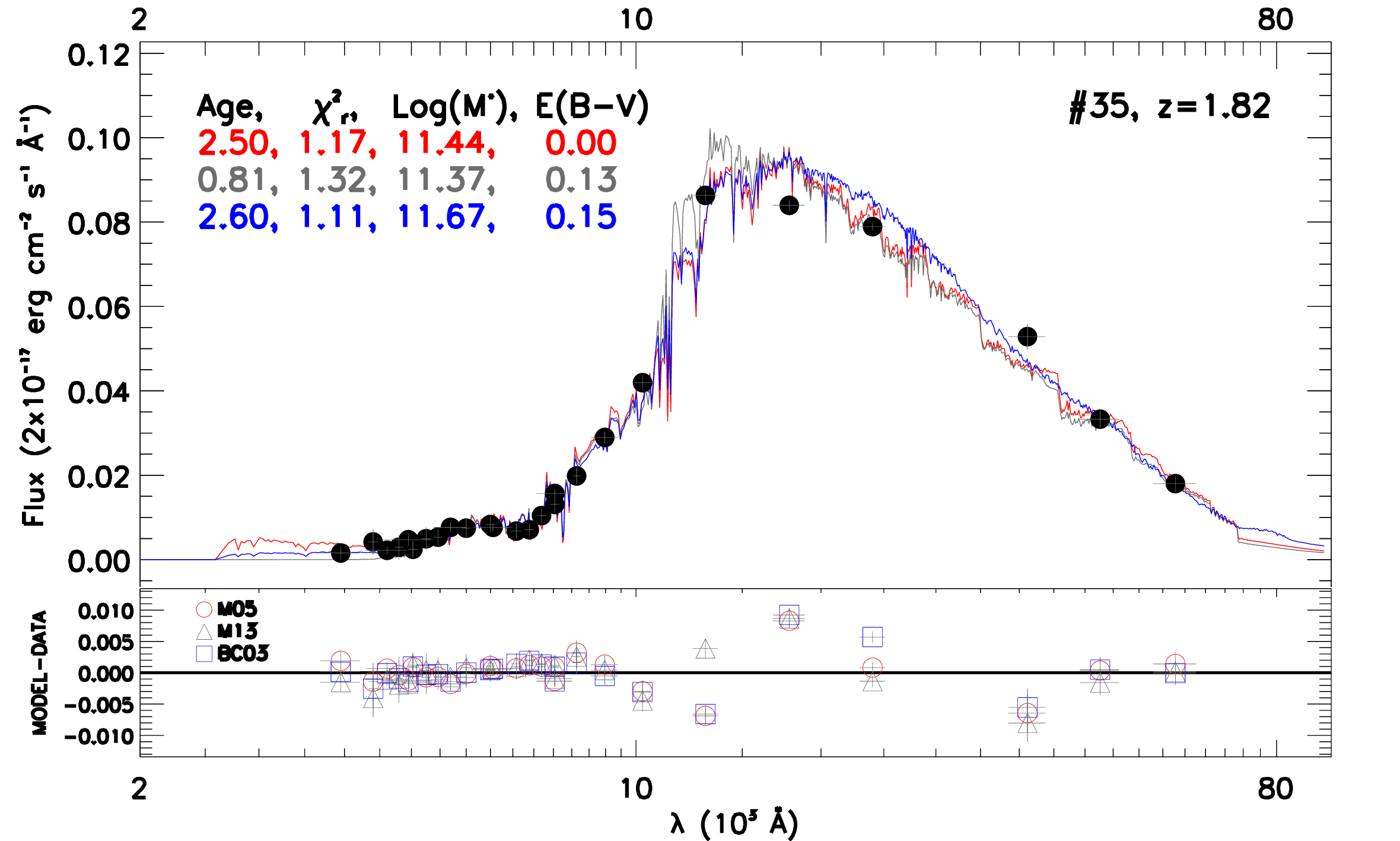}
\includegraphics[width=0.48\textwidth]{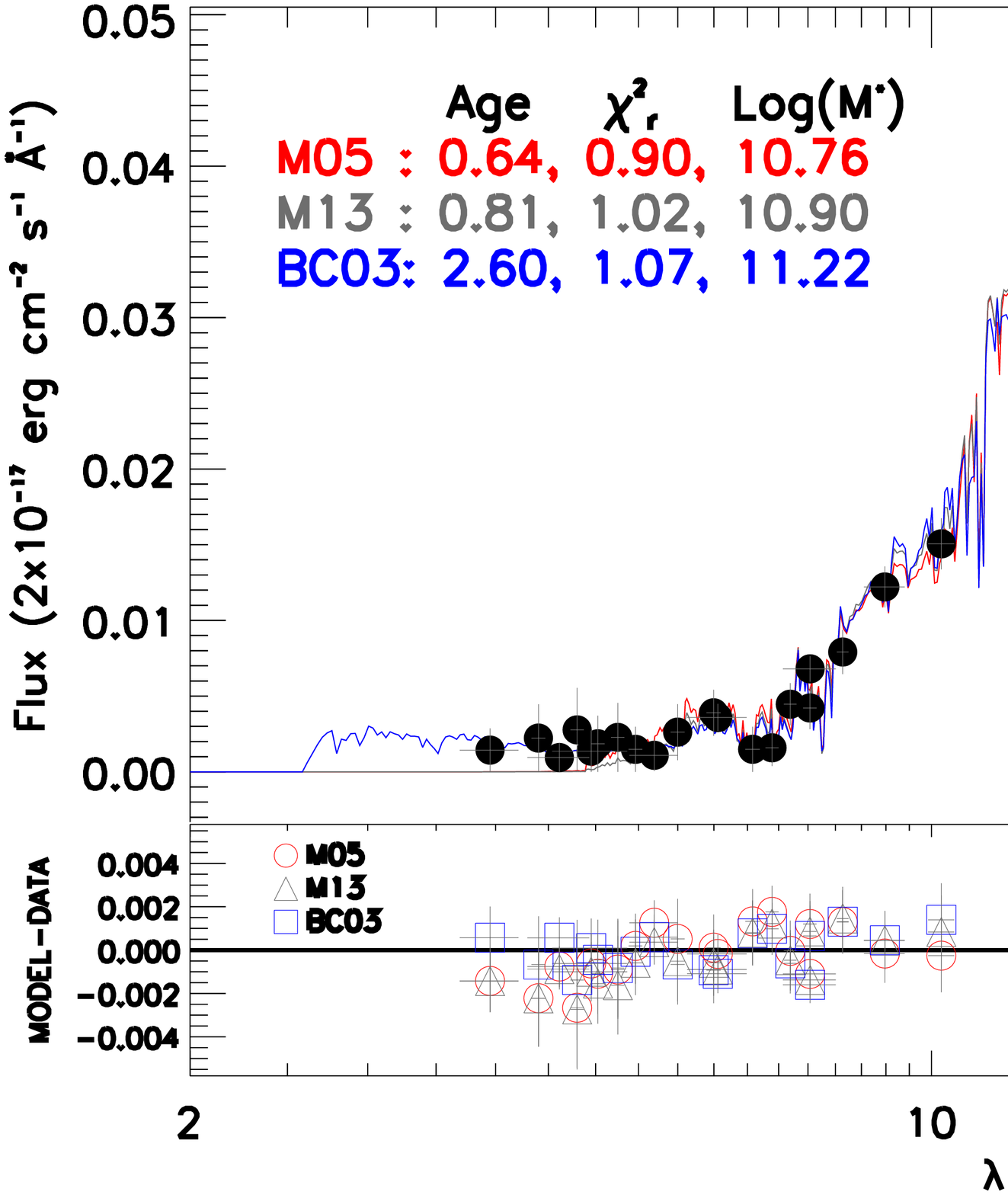}
\includegraphics[width=0.48\textwidth]{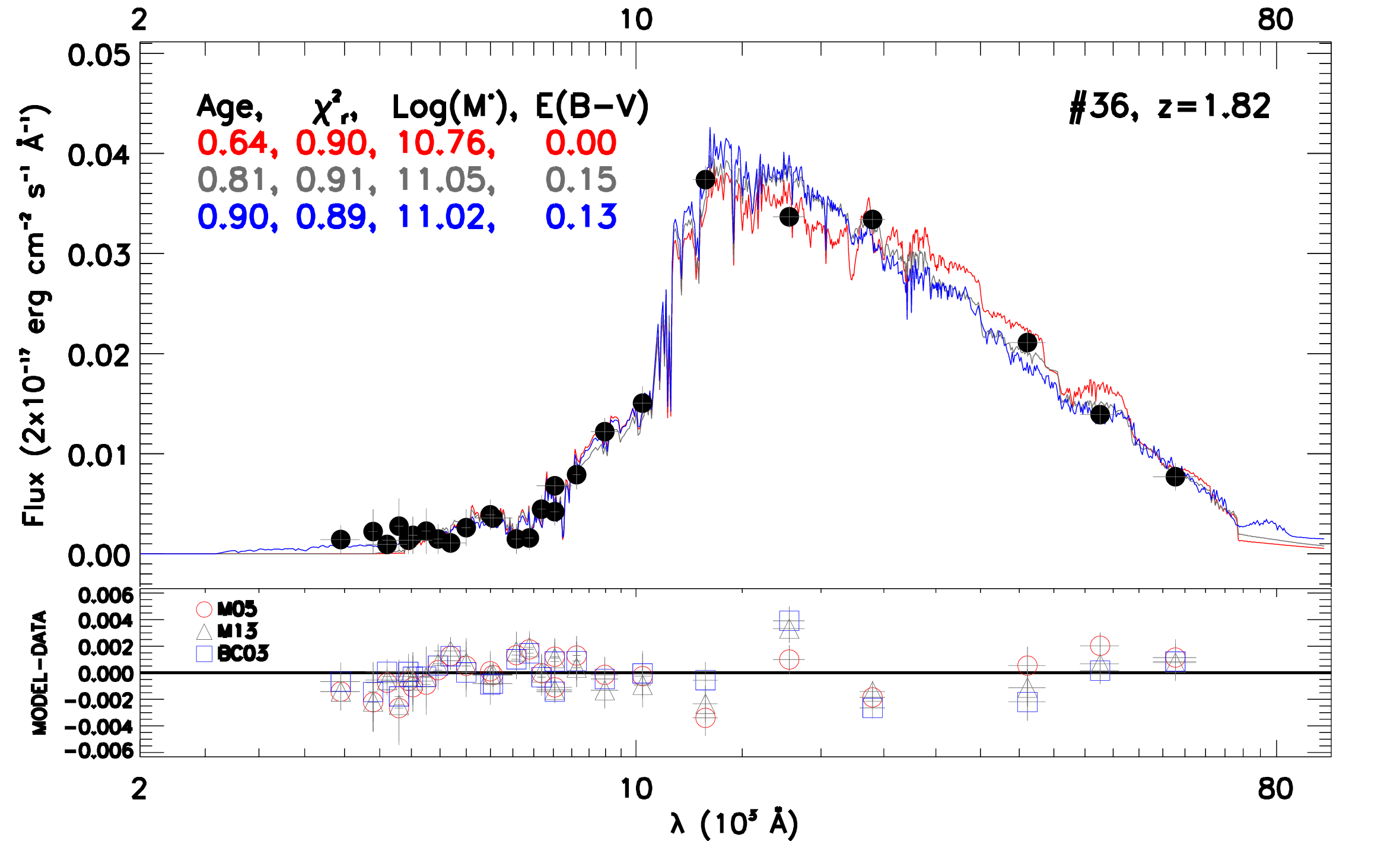}
\caption{Continued.}
\label{fig:Fig1_appB}
\end{figure*}

\addtocounter{figure}{-1}
\begin{figure*}
\centering
\includegraphics[width=0.48\textwidth]{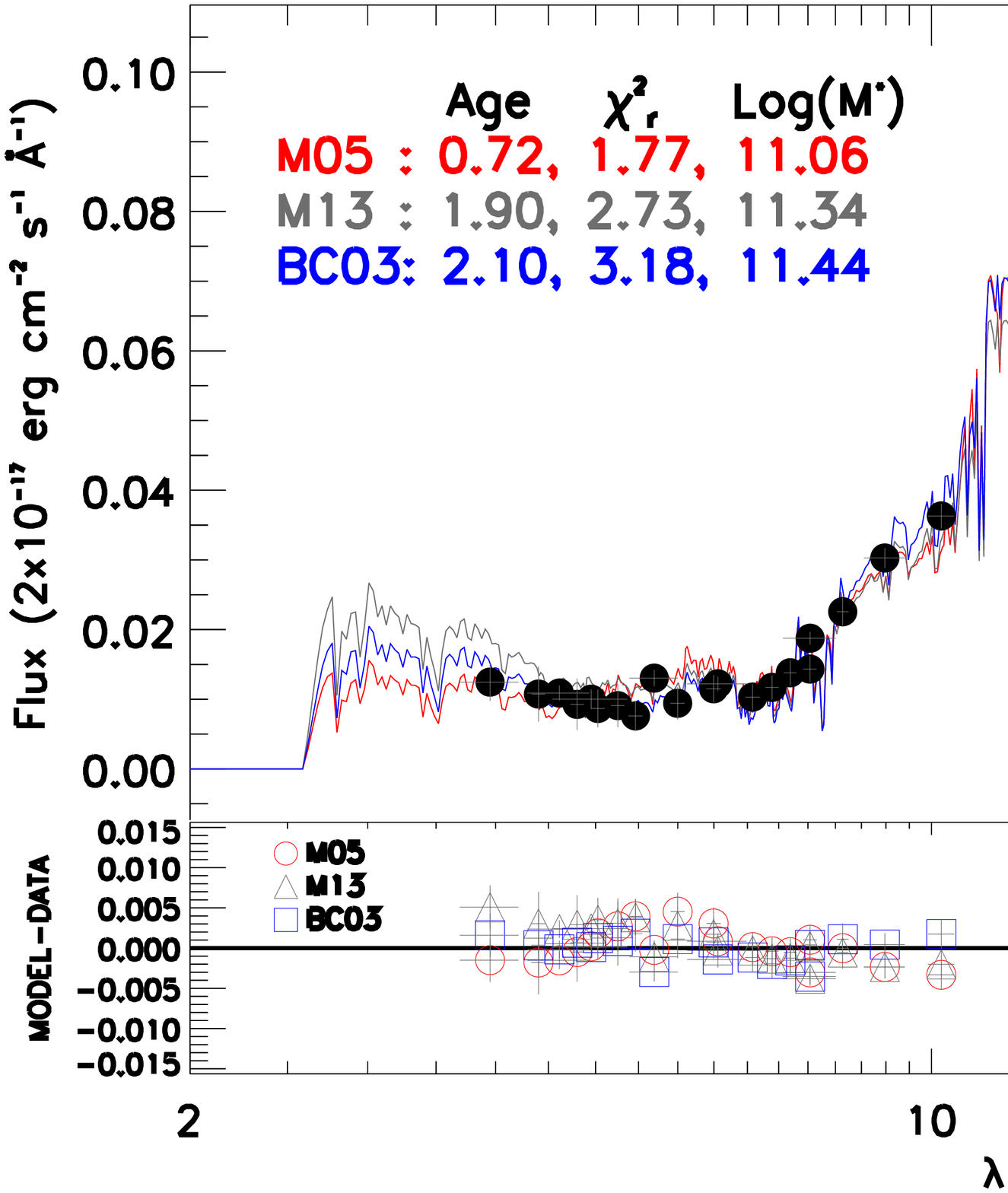}
\includegraphics[width=0.48\textwidth]{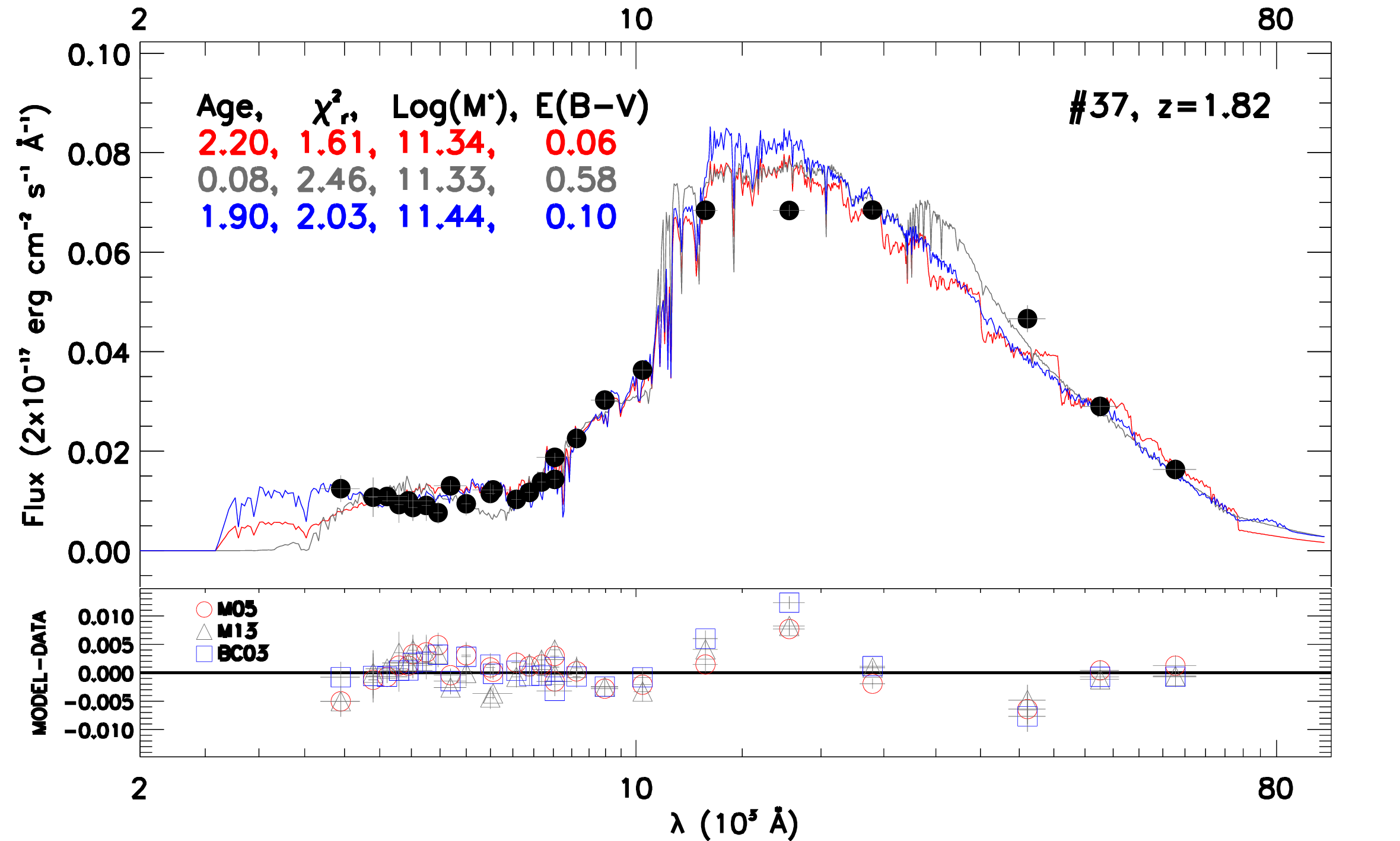}
\includegraphics[width=0.48\textwidth]{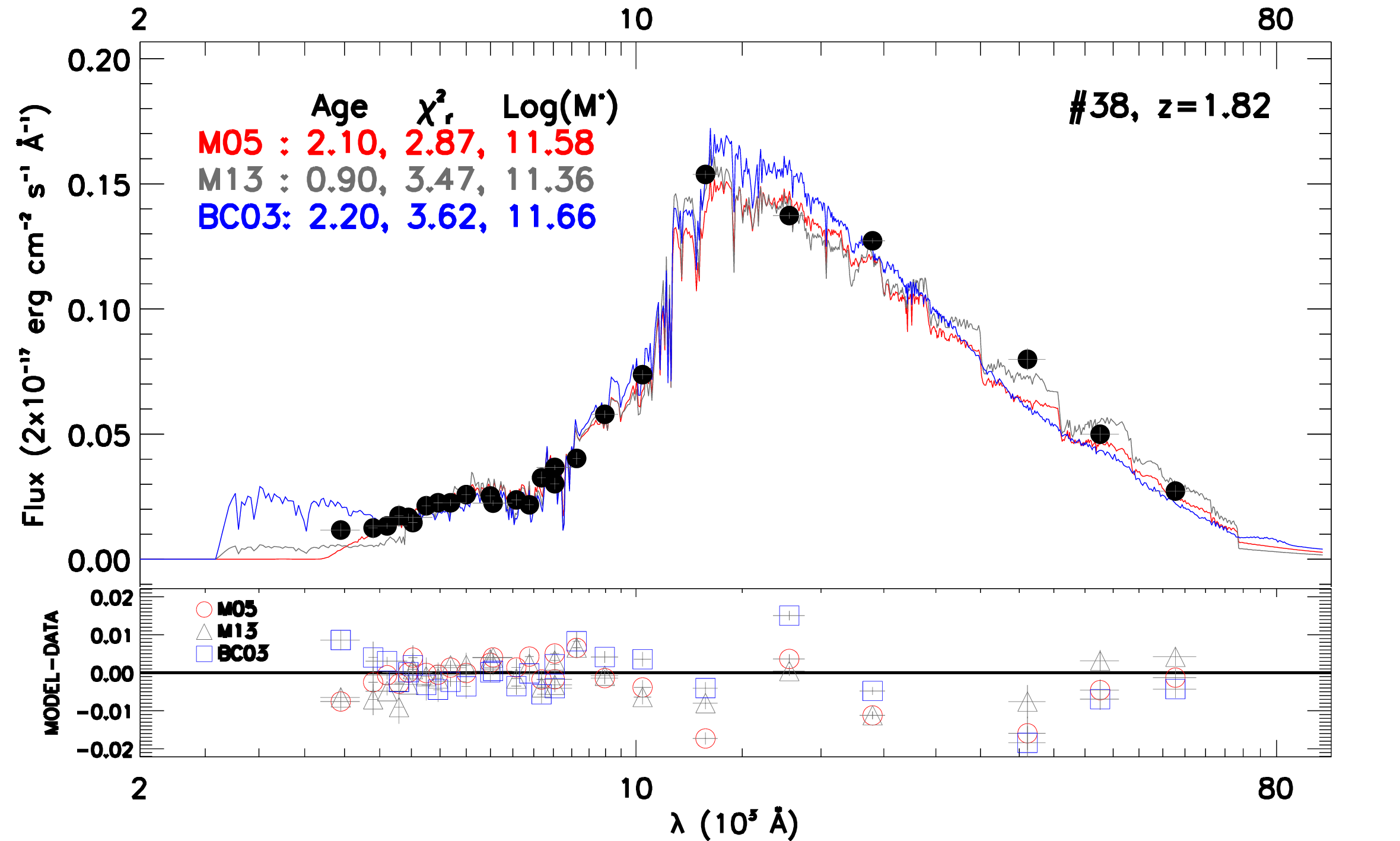}
\includegraphics[width=0.48\textwidth]{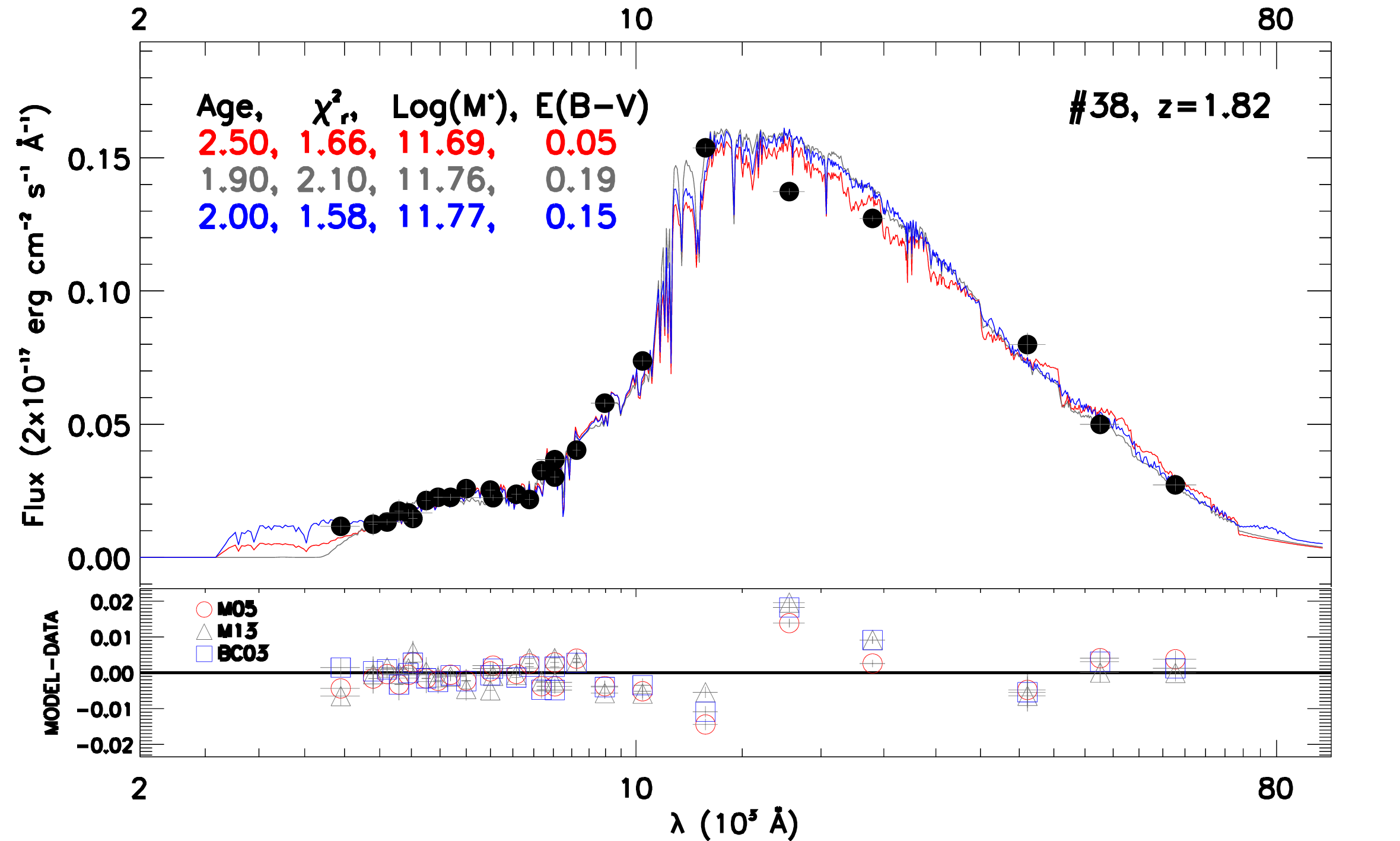}
\includegraphics[width=0.48\textwidth]{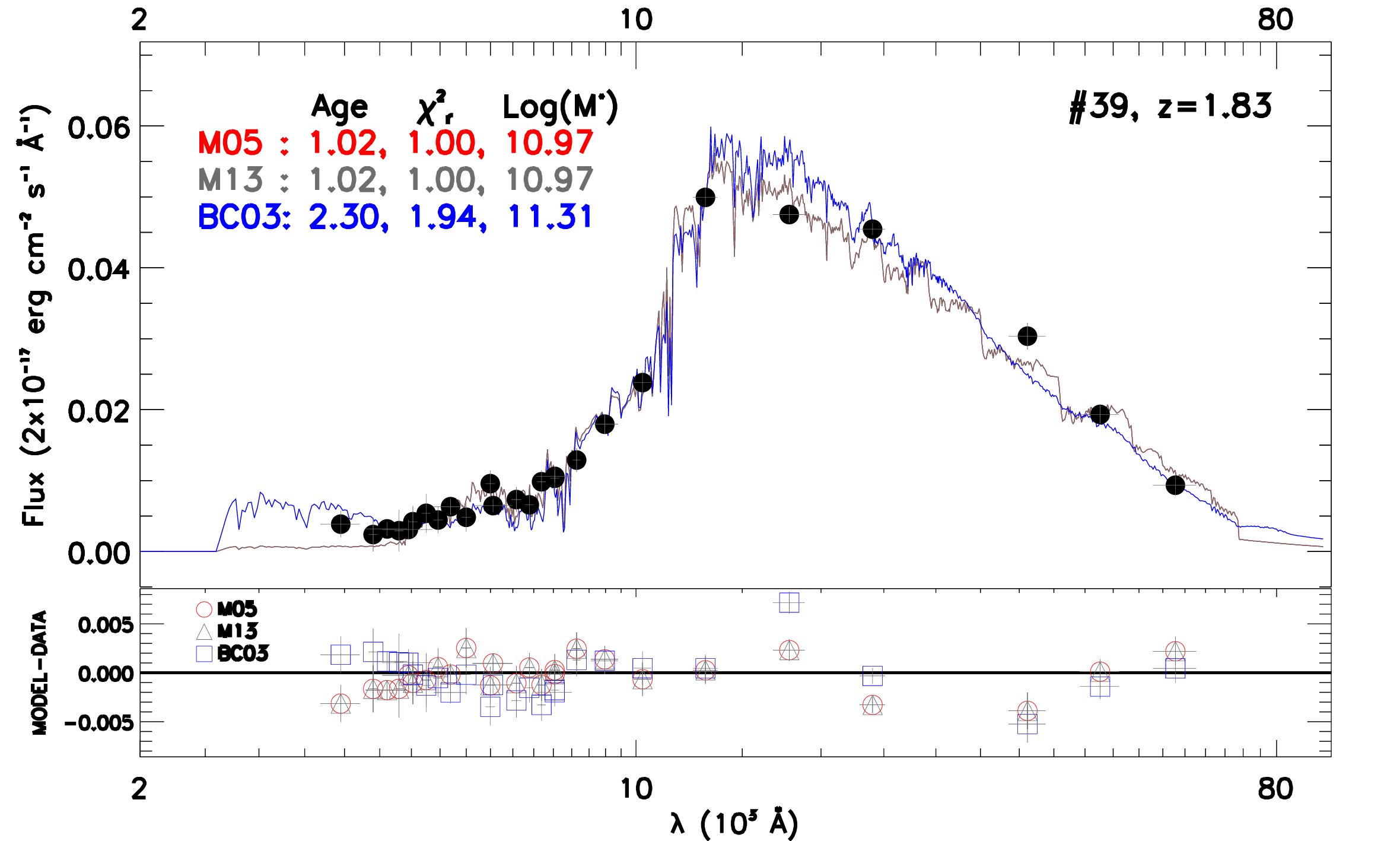}
\includegraphics[width=0.48\textwidth]{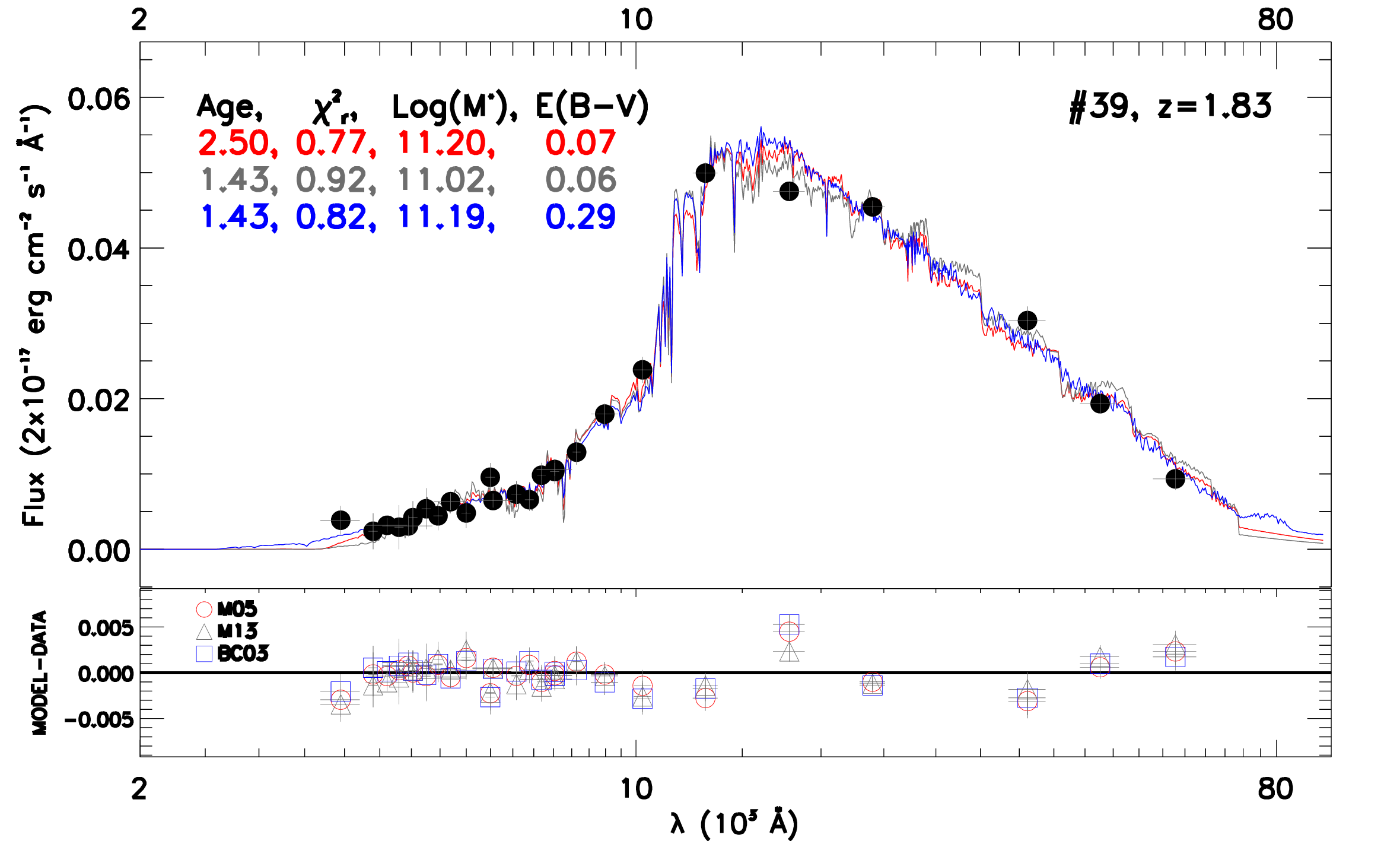}
\includegraphics[width=0.48\textwidth]{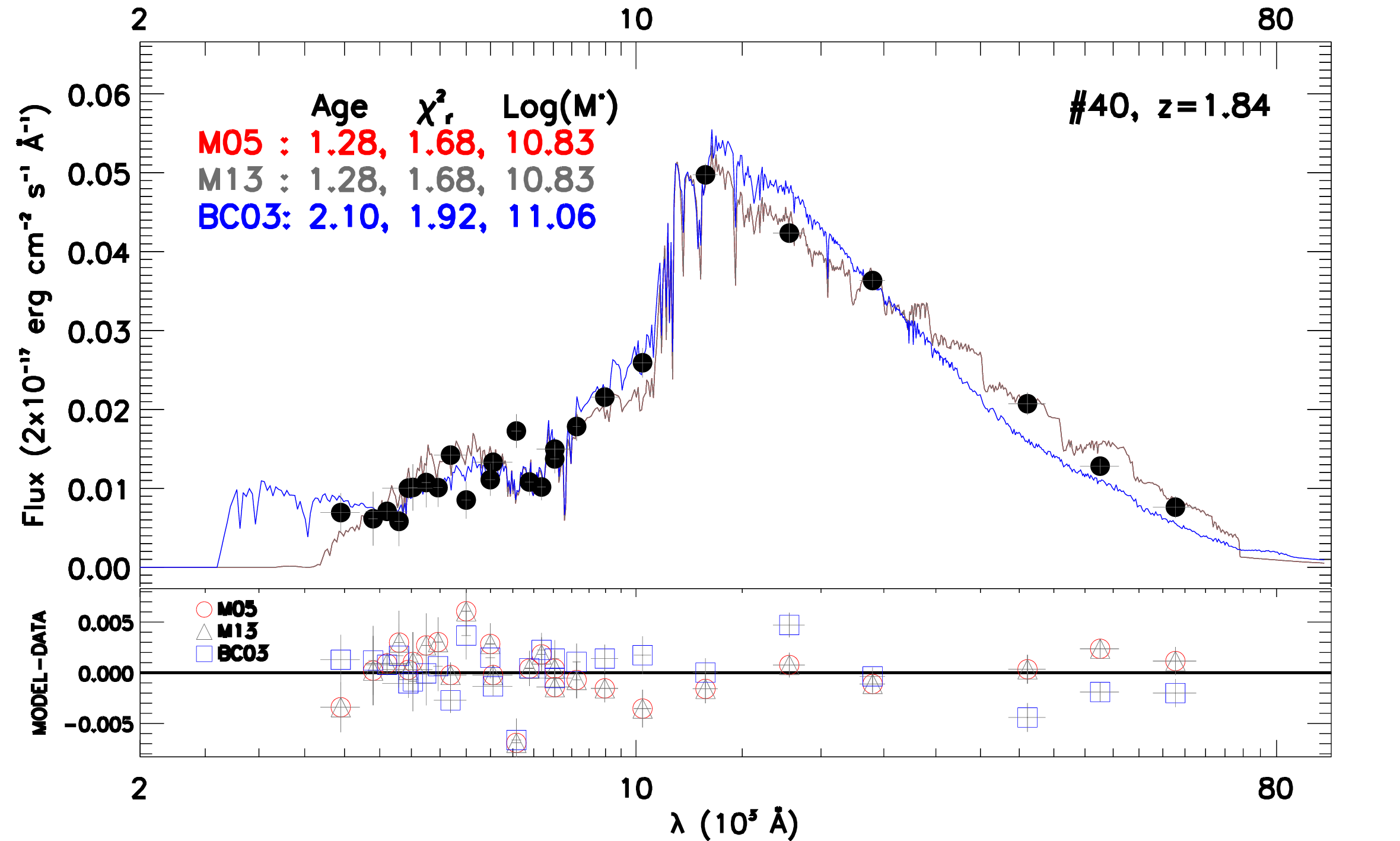}
\includegraphics[width=0.48\textwidth]{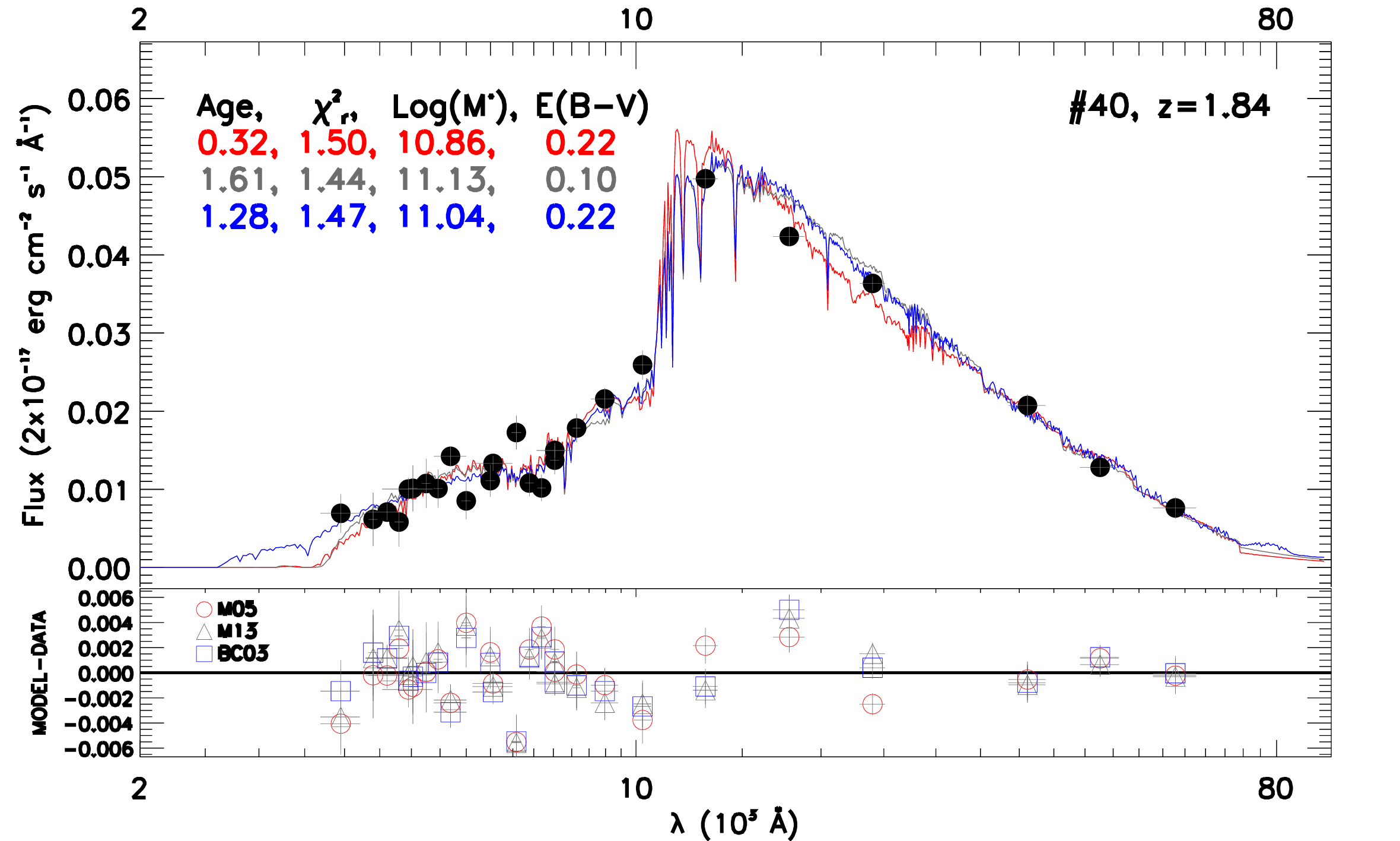}
\caption{Continued.}
\label{fig:Fig1_appB}
\end{figure*}

\addtocounter{figure}{-1}
\begin{figure*}
\centering
\includegraphics[width=0.48\textwidth]{40_SED_nored.ps}
\includegraphics[width=0.48\textwidth]{40_SED_red.ps}
\includegraphics[width=0.48\textwidth]{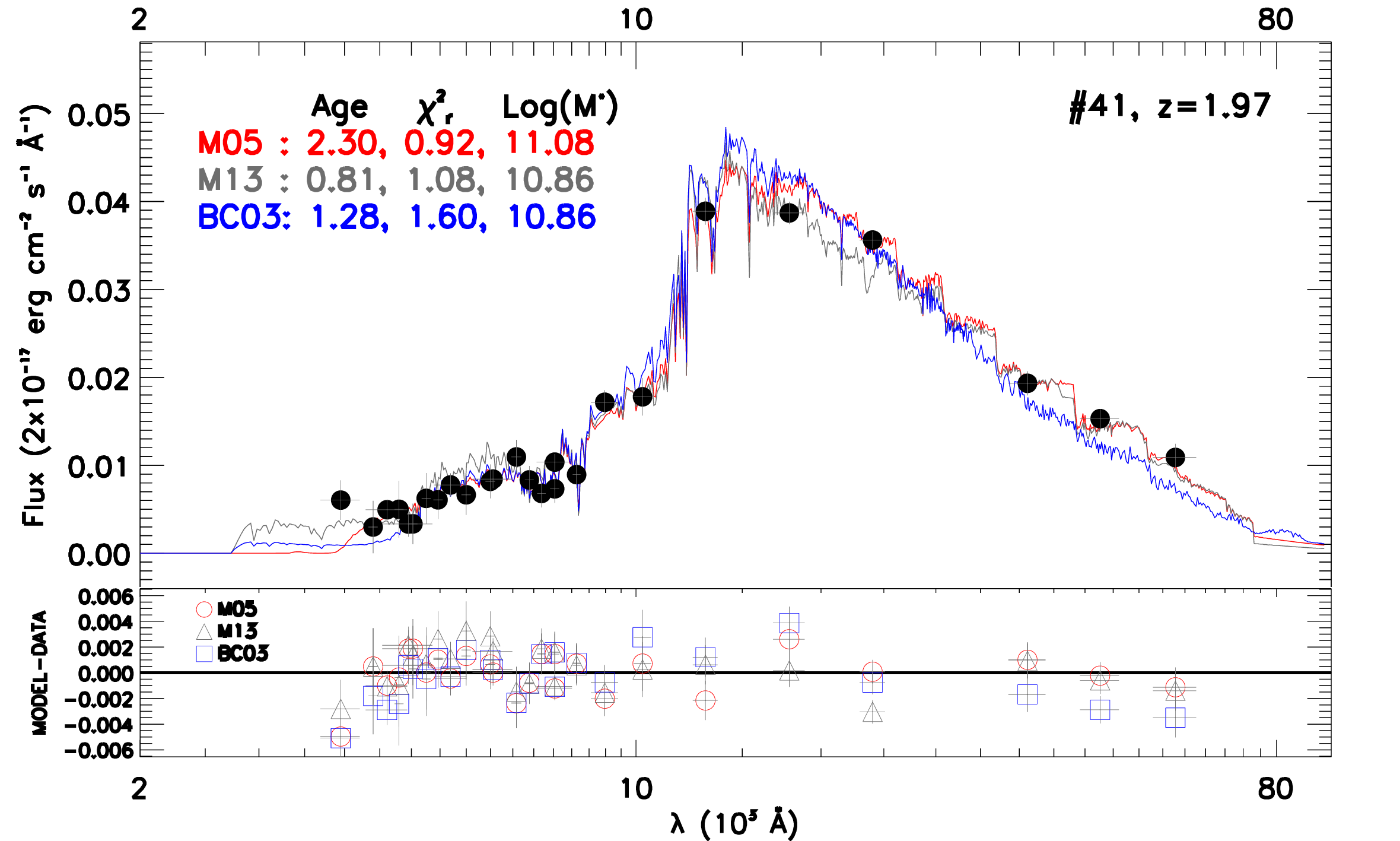}
\includegraphics[width=0.48\textwidth]{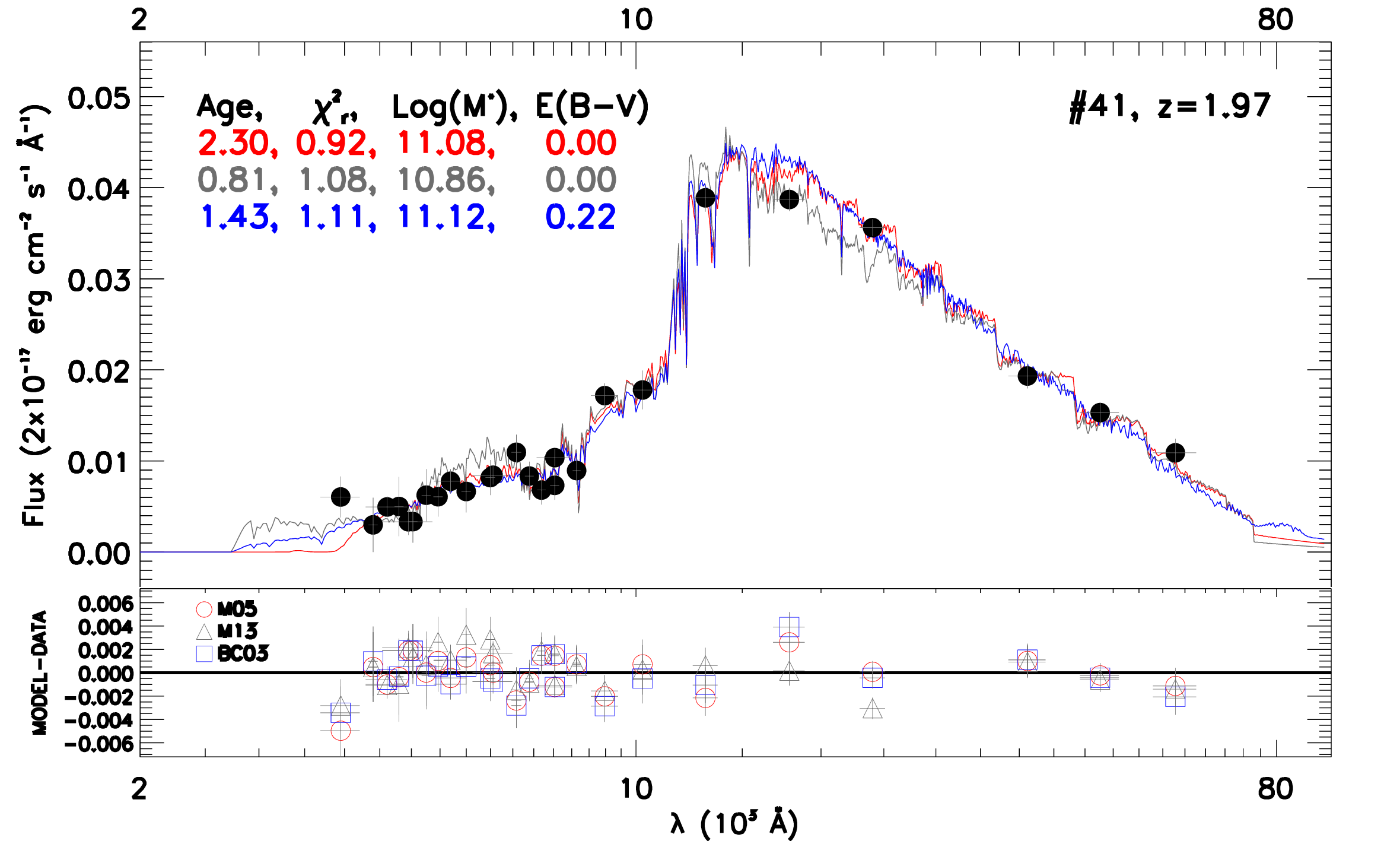}
\includegraphics[width=0.48\textwidth]{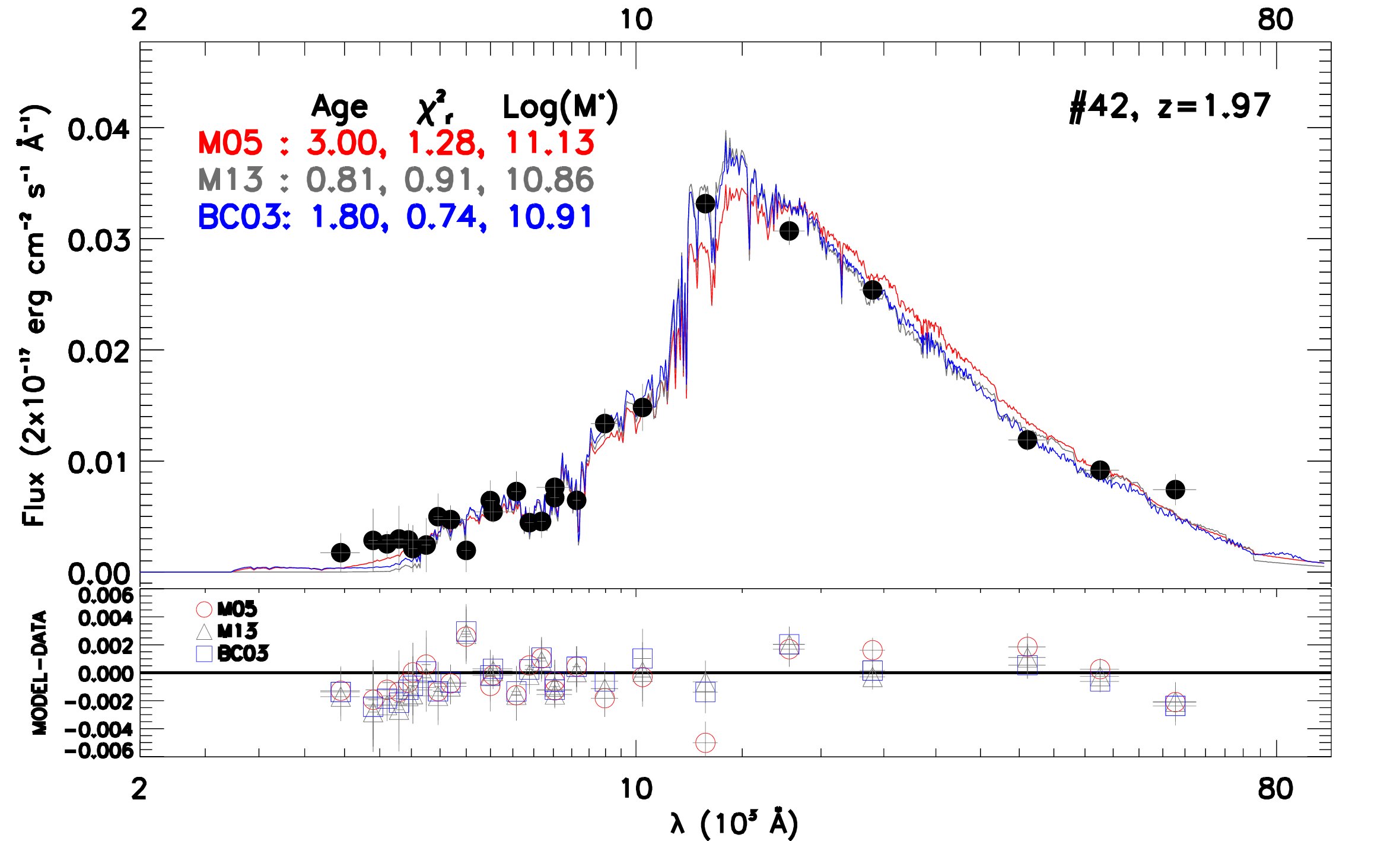}
\includegraphics[width=0.48\textwidth]{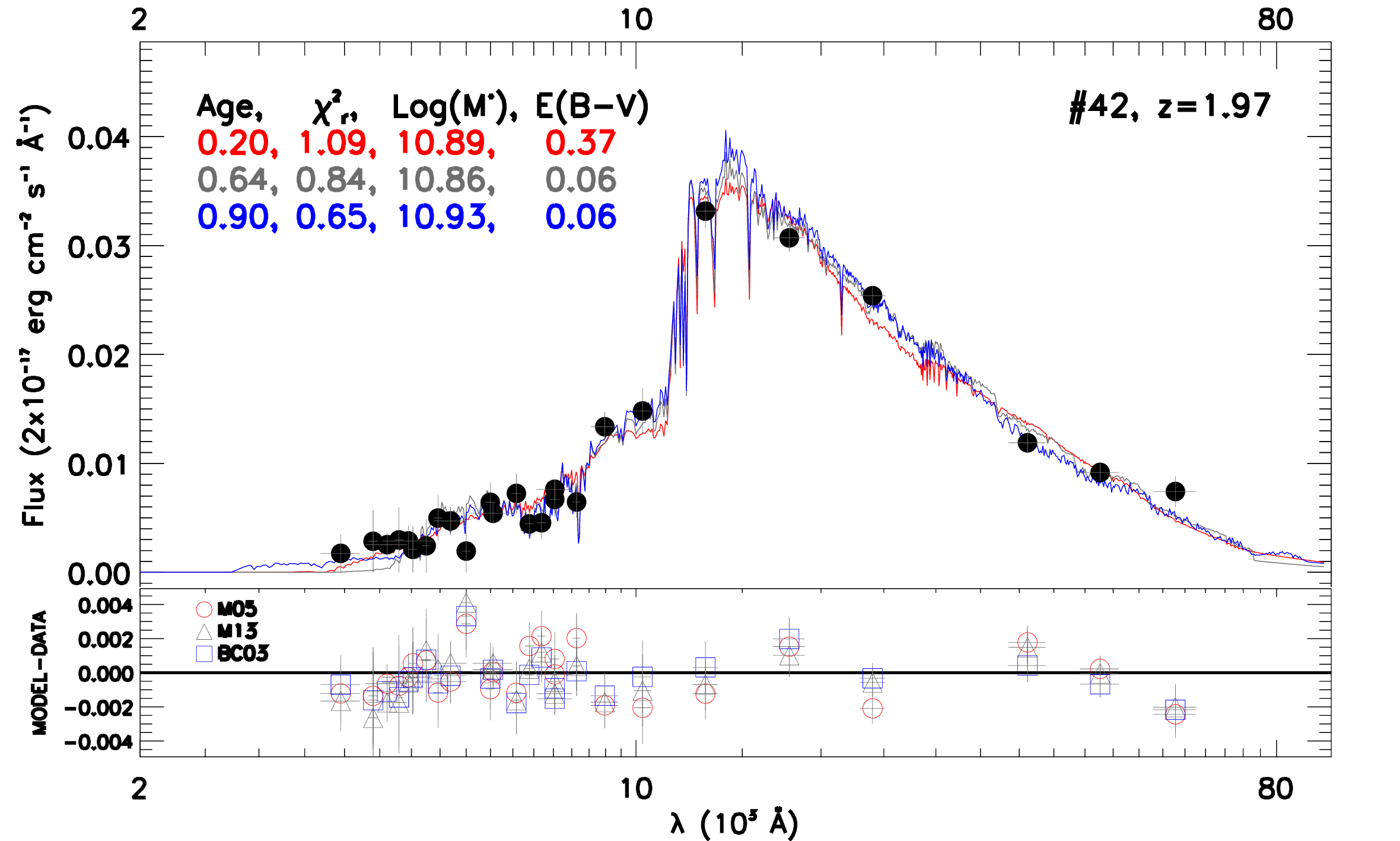}
\includegraphics[width=0.48\textwidth]{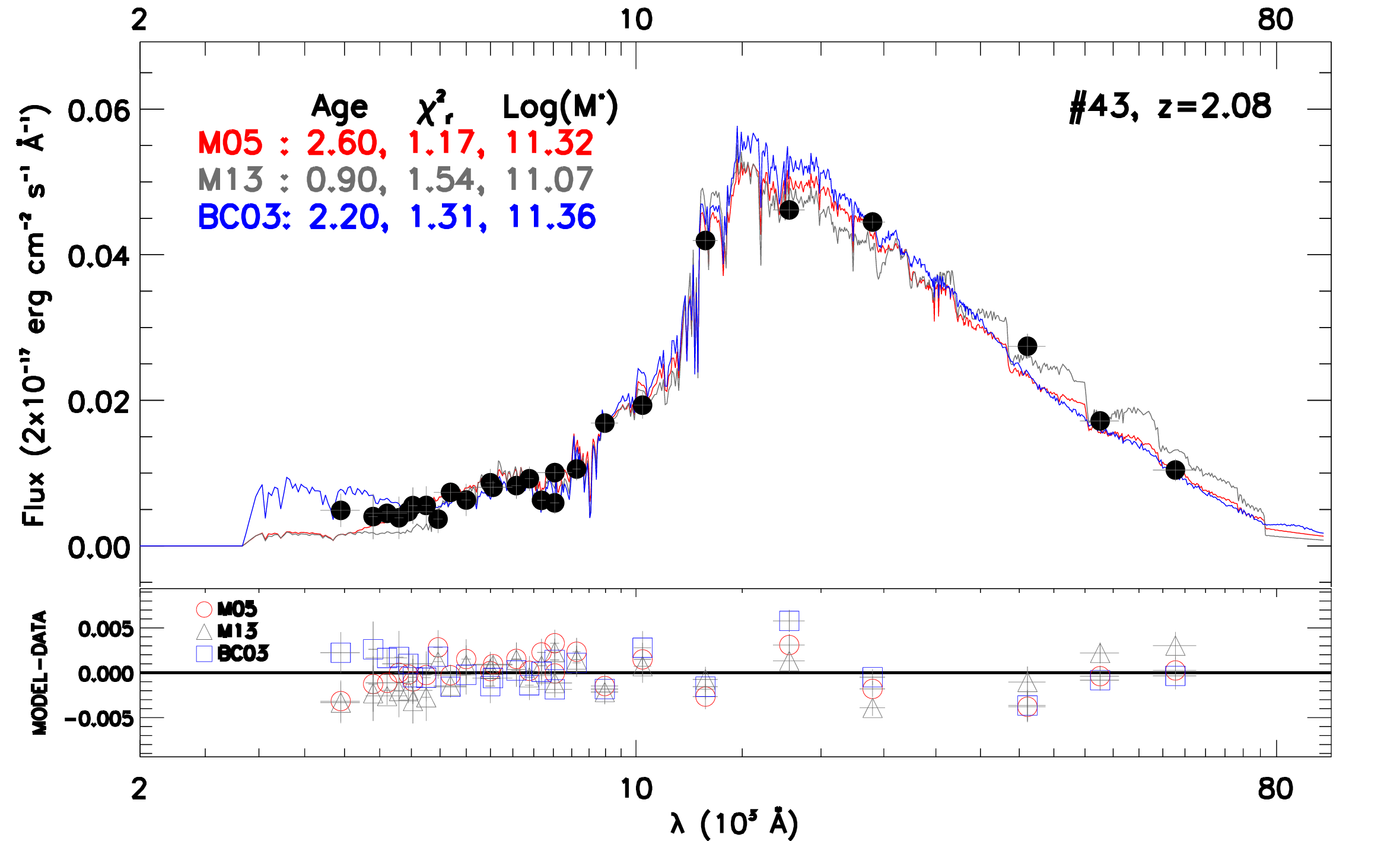}
\includegraphics[width=0.48\textwidth]{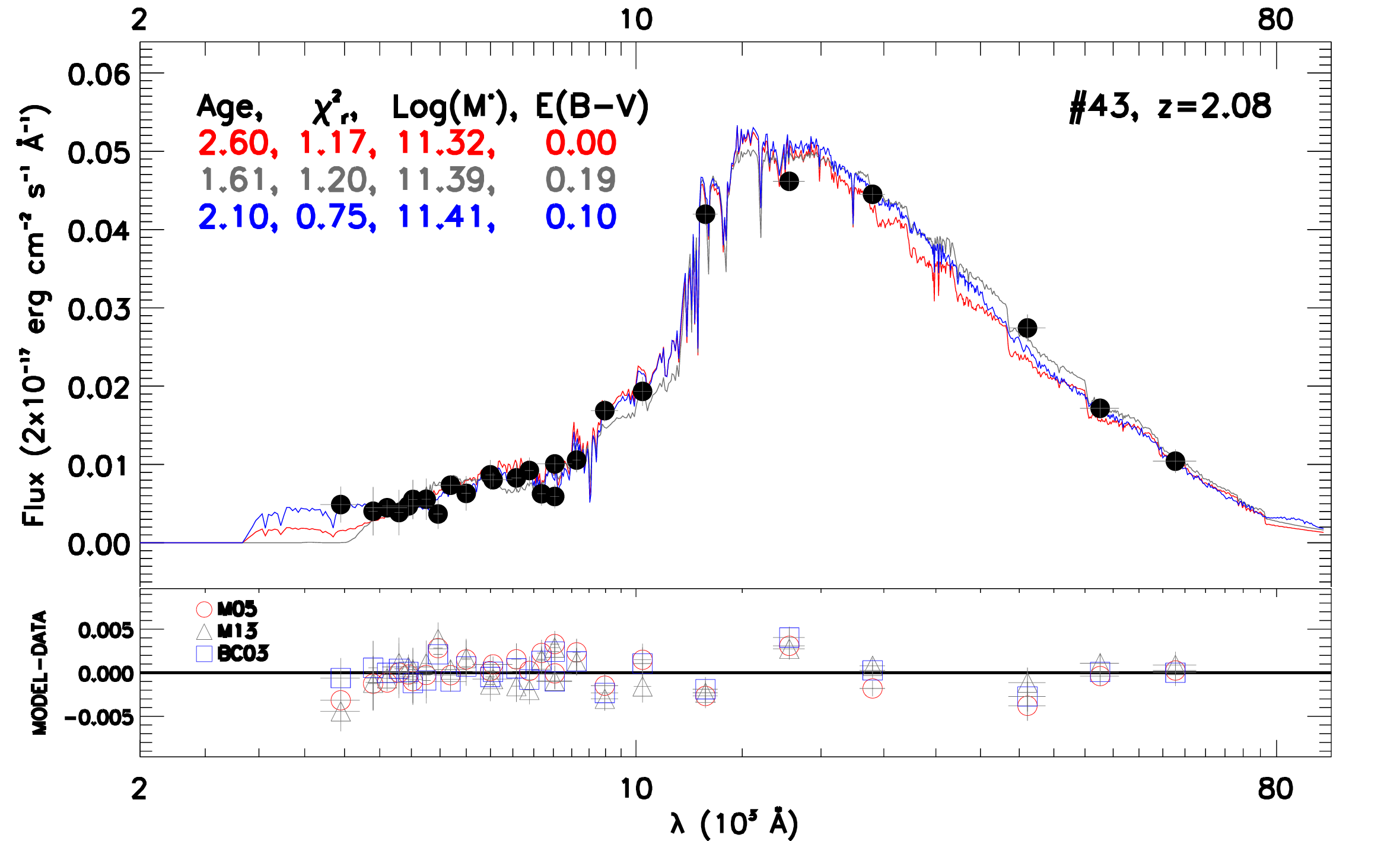}
\caption{Continued.}
\label{fig:Fig1_appB}
\end{figure*}

\addtocounter{figure}{-1}
\begin{figure*}
\centering
\includegraphics[width=0.48\textwidth]{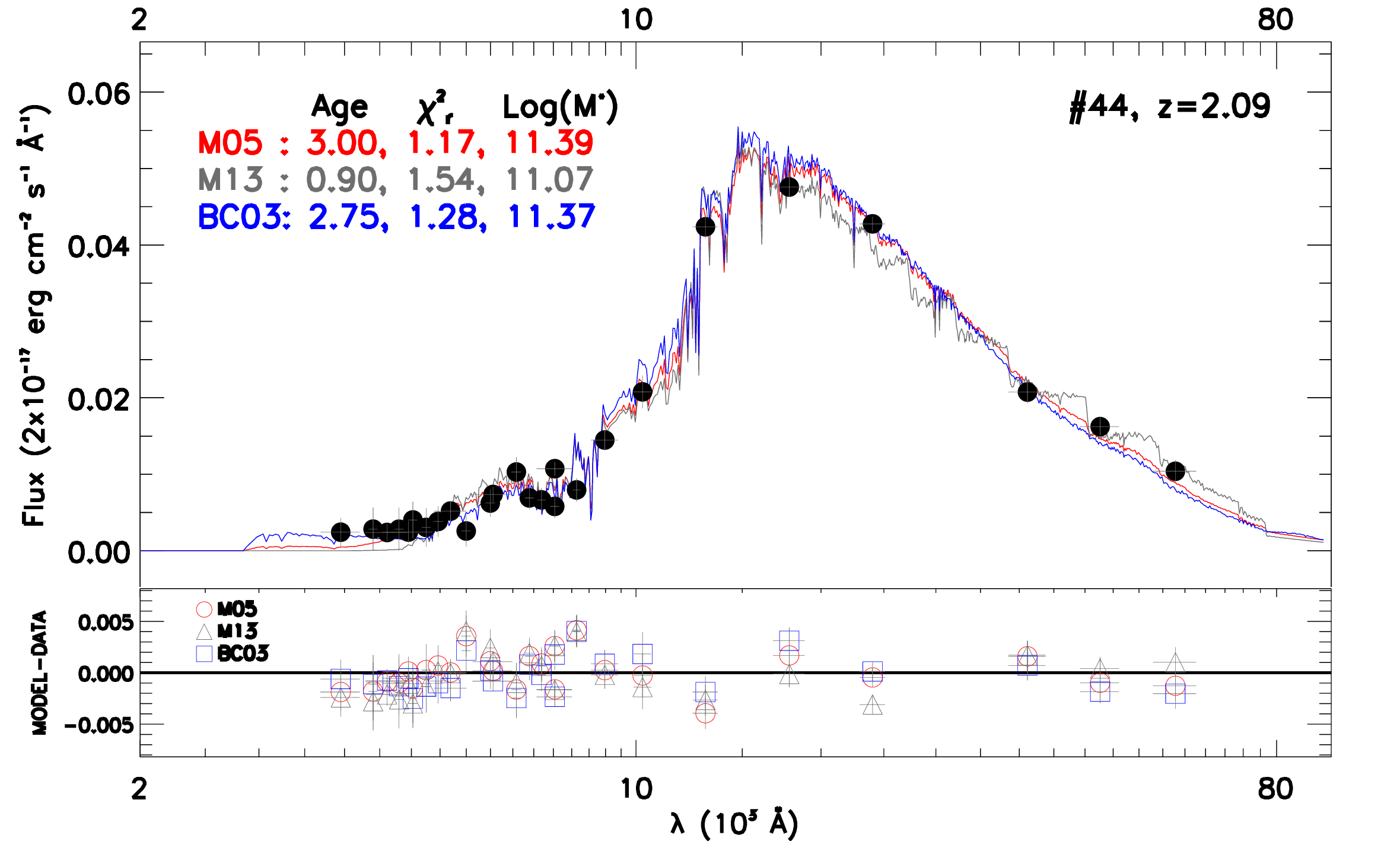}
\includegraphics[width=0.48\textwidth]{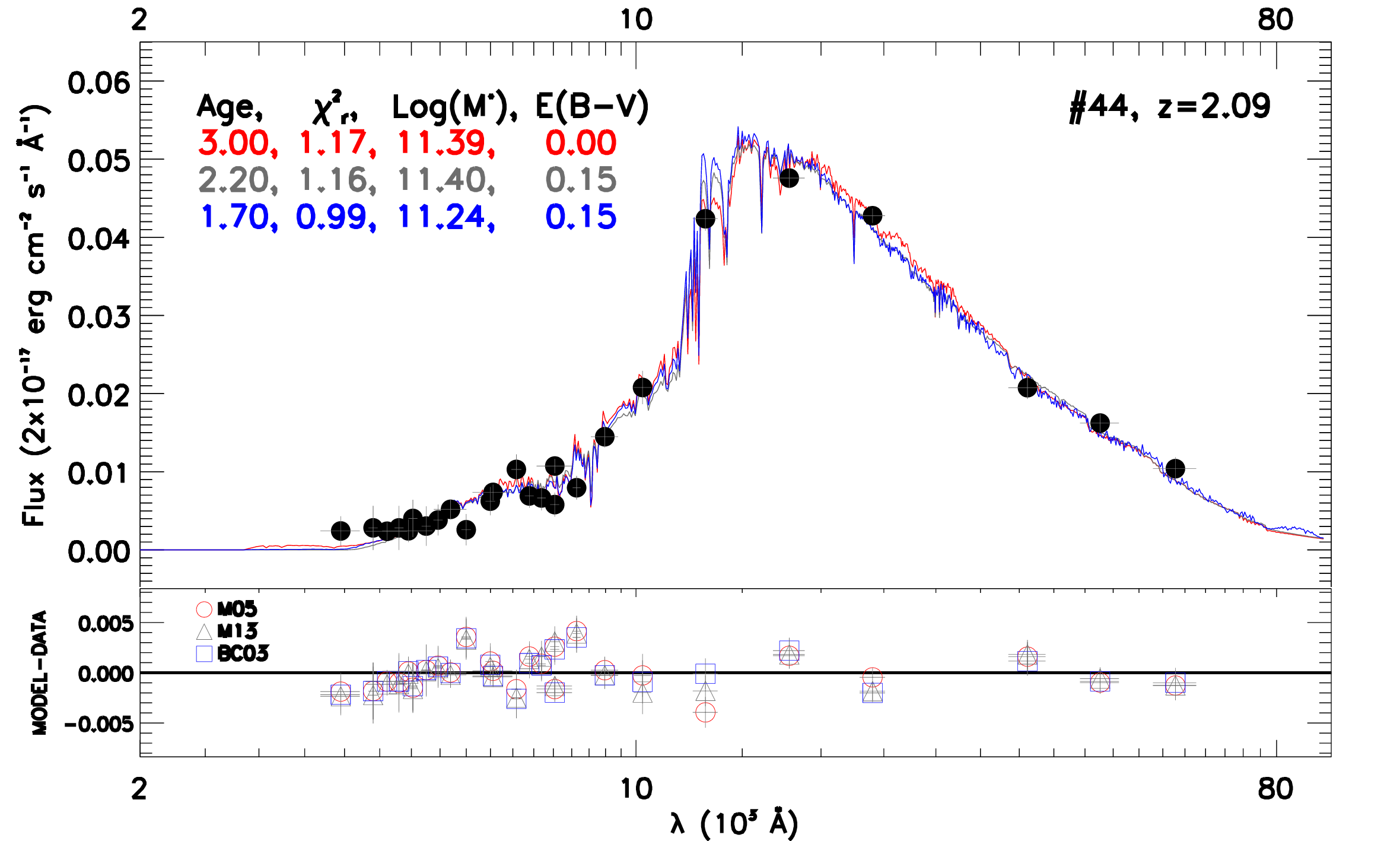}
\caption{Continued.}
\label{fig:Fig1_appB}
\end{figure*}

\begin{figure*}
\centering
\includegraphics[width=0.48\textwidth]{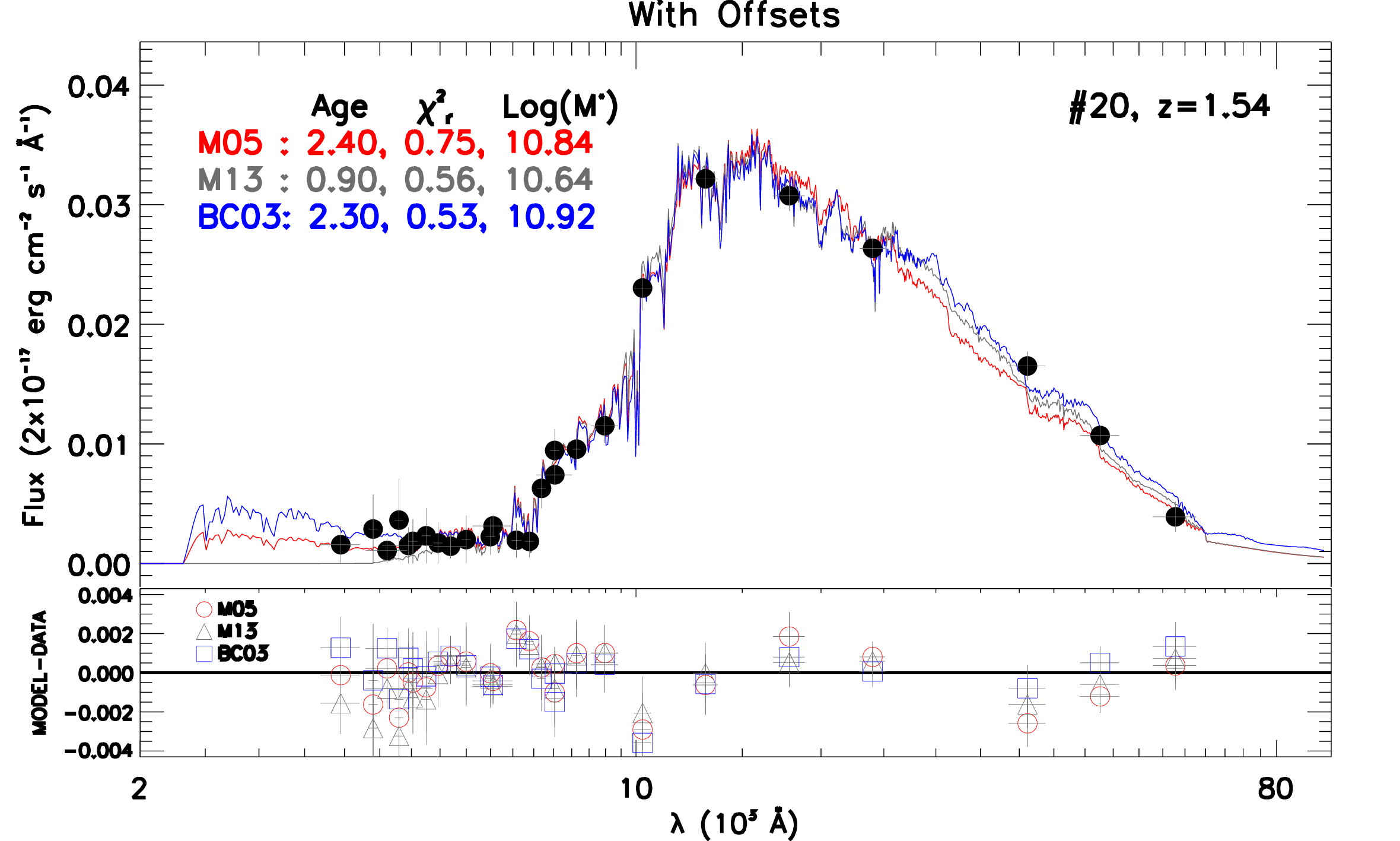}
\includegraphics[width=0.48\textwidth]{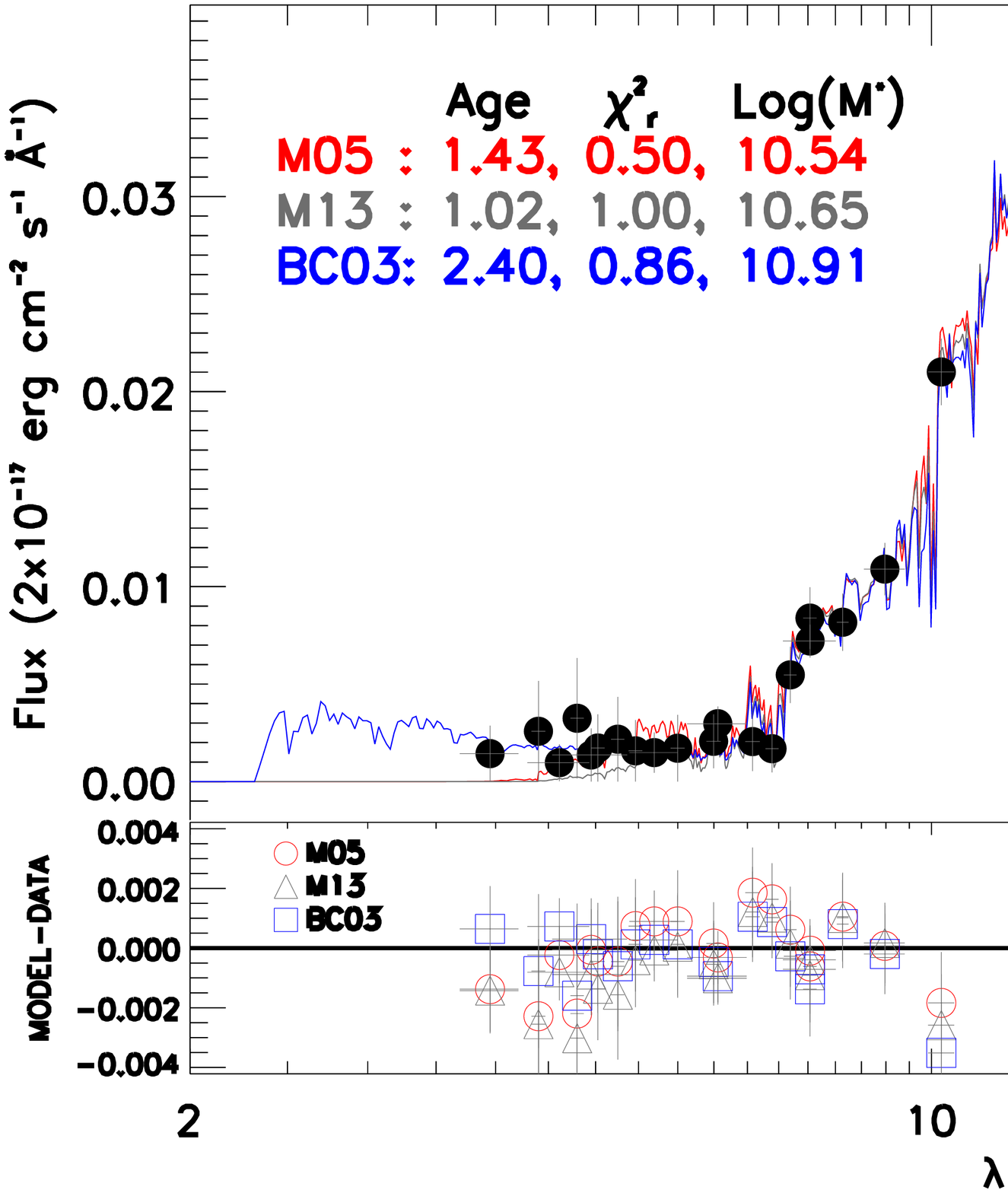}
\includegraphics[width=0.48\textwidth]{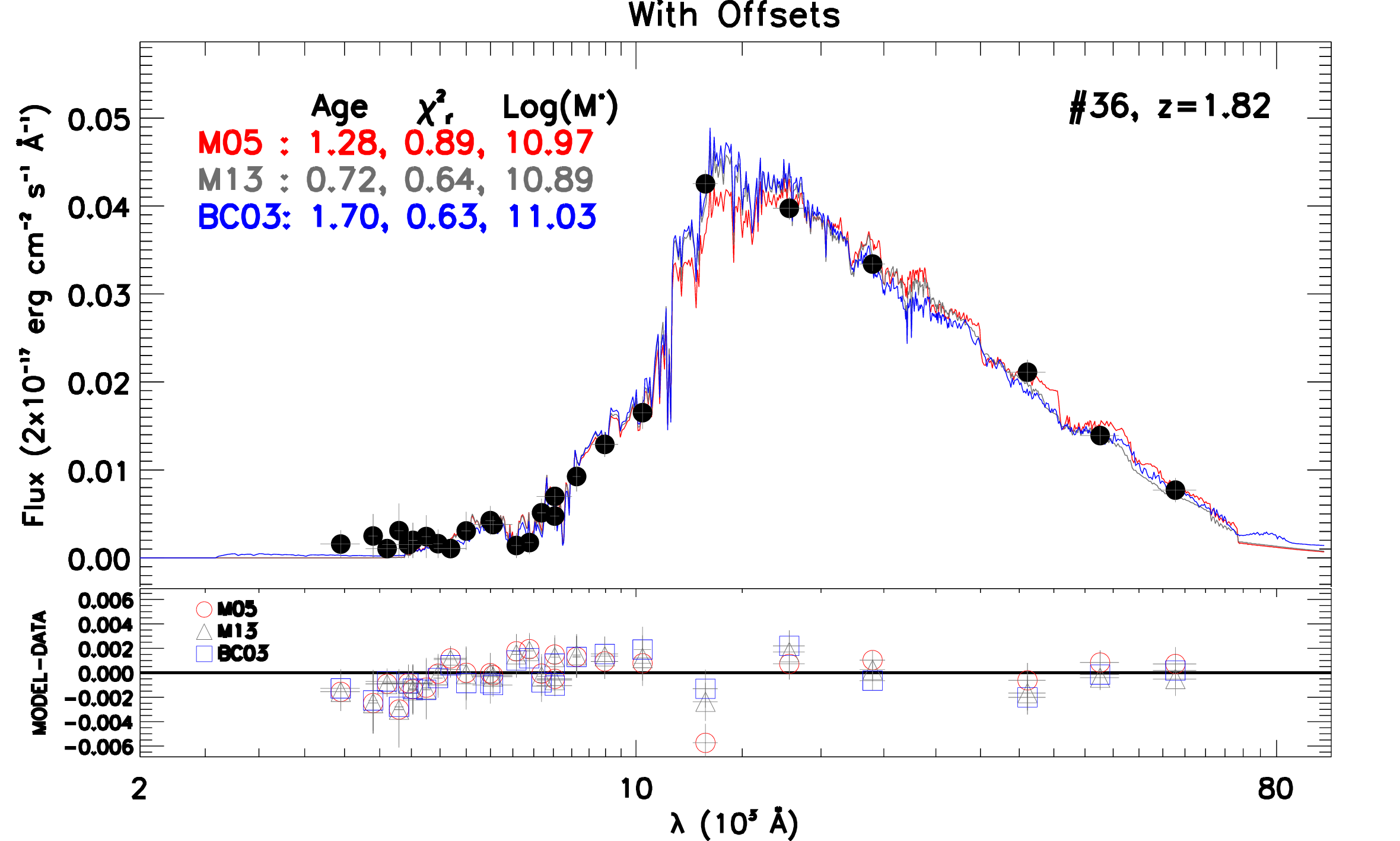}
\includegraphics[width=0.48\textwidth]{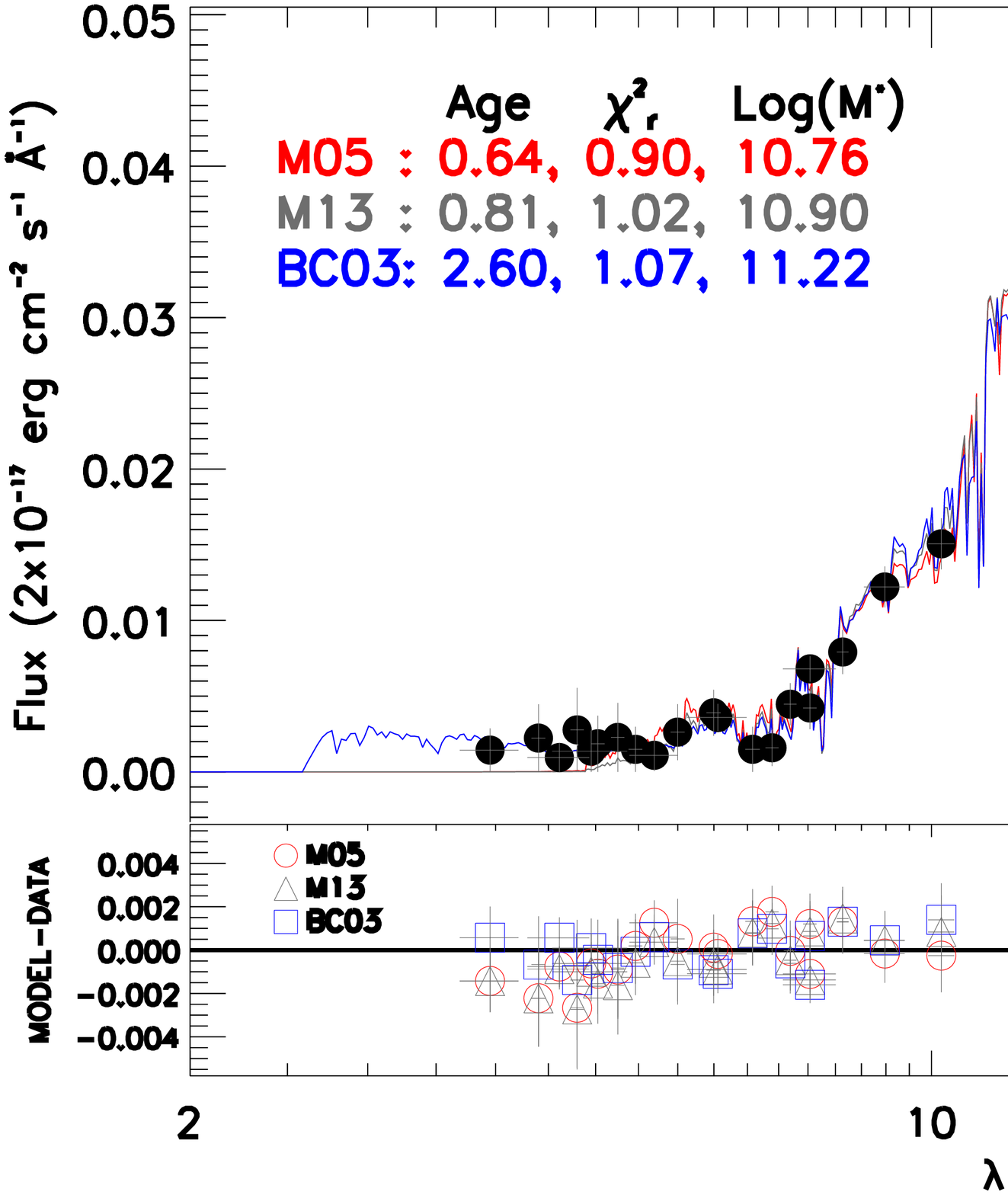}
\includegraphics[width=0.48\textwidth]{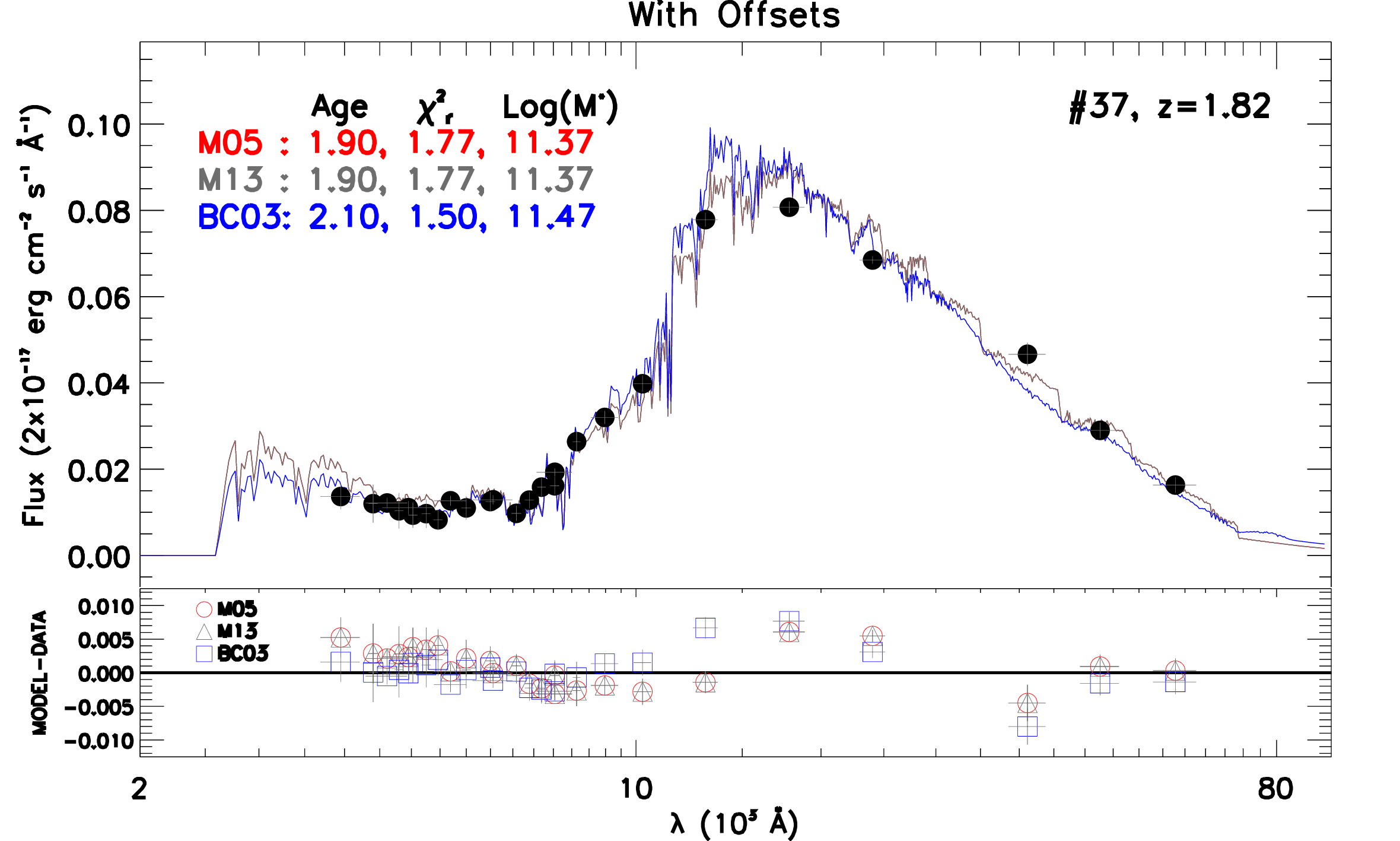}
\includegraphics[width=0.48\textwidth]{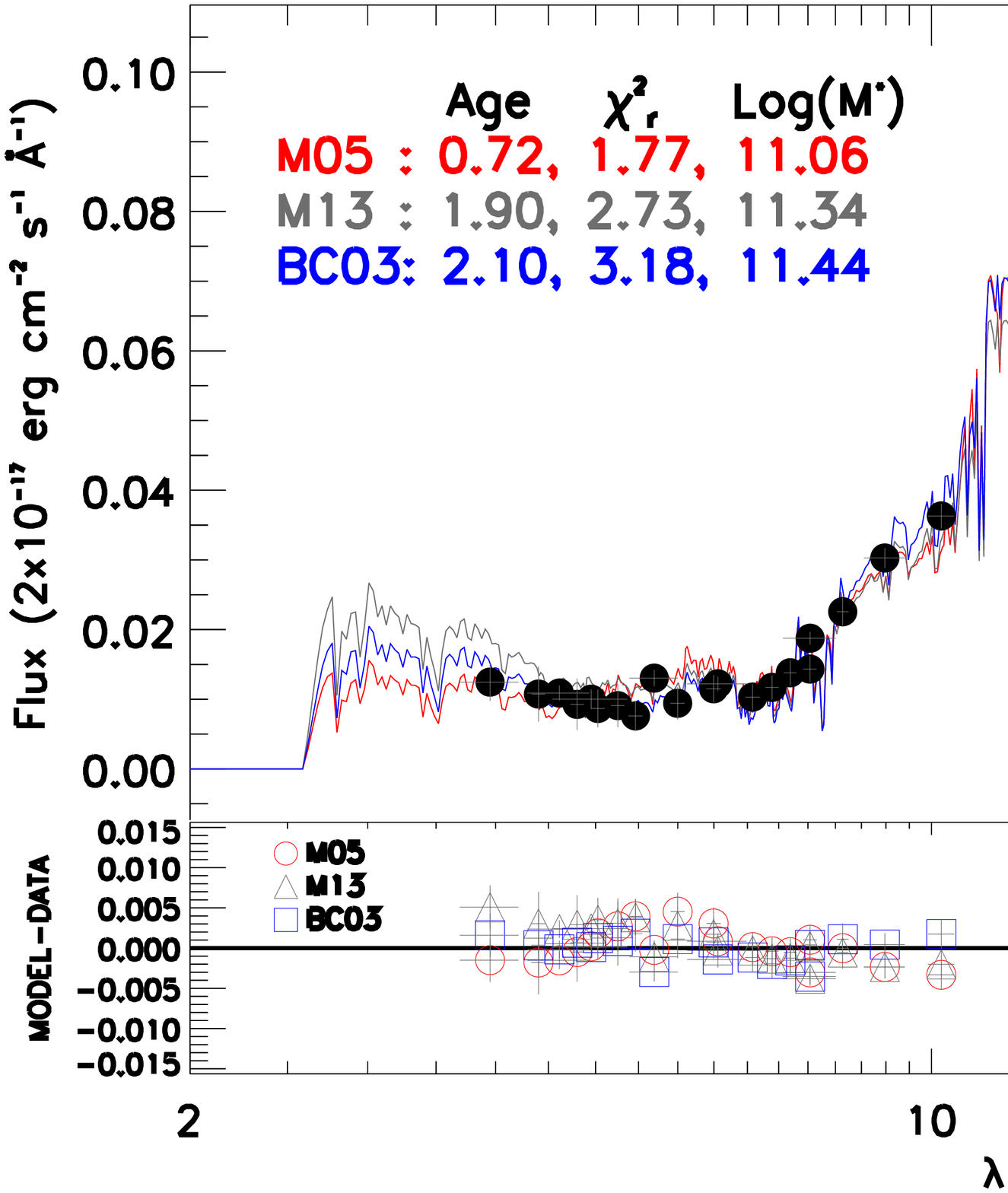}
\caption{SED fits of COSMOS galaxies IDs 20 (two top panels), 36 (two central panels) and 37 (two bottom panels) in absence of reddening for the full-filter-set, showing the sensitivity of the SED fitting to including (left-hand panels) or excluding (right-hand panels) model-dependent corrections for photometric zero-point offsets. Observed fluxes are plotted as full dots over best-fit templates obtained with M05 (red full line), M13 (grey full line) and BC03 (blue full line) models. Flux residuals ($MODEL-DATA$)  are plotted vs. wavelength at the bottom of each panel. }
\label{fig:Fig2_appB}
\end{figure*}

\begin{table*}
\begin{scriptsize}
\begin{center}
\caption{SED fitting solutions obtained after excluding possibly physically unreliable solutions for both HUDF and COSMOS, using the full-filter set and the original photometry in presence of reddening. Columns are as in Tables \ref{tab:Table3} and \ref{tab:Table2_appB}.}
\label{tab:Table3_appB}
\begin{threeparttable} 
\begin{tabular}{cccclcccccl}
\hline
  \multicolumn{1}{c}{\bf ID} &
  \multicolumn{1}{c}{$\mathbf{z_{\rm spec}}$} &
  \multicolumn{1}{c}{\bf Model} &
  \multicolumn{1}{c}{$\mathbf{t}$} &
  \multicolumn{1}{c}{$\mathbf{[Z/H]}$} &
  \multicolumn{1}{c}{\bf SFH} &
  \multicolumn{1}{c}{$\mathbf{\chi^{2}_{\rm r}}$} &
  \multicolumn{1}{c}{$\mathbf{M^{\ast}}$} &
  \multicolumn{1}{c}{$\mathbf{SFR}$} & 
  \multicolumn{1}{c}{$\mathbf{E(B-V)}$} &
  \multicolumn{1}{c}{\bf Reddening Law} \\
  \multicolumn{1}{c}{}&
  \multicolumn{1}{c}{}&
  \multicolumn{1}{c}{}&
  \multicolumn{1}{c}{\rm (Gyr)}&
  \multicolumn{1}{c}{${\rm (Z_{\odot})}$} &
  \multicolumn{1}{c}{}&
  \multicolumn{1}{c}{}&
  \multicolumn{1}{c}{$(10^{11}\ {\rm M_{\odot}})$} &
  \multicolumn{1}{c}{$({\rm M_{\odot}/ yr^{-1}})$} &
  \multicolumn{1}{c}{({\rm mag})}&
  \multicolumn{1}{c}{} \\
  \multicolumn{1}{c}{(1)}&
  \multicolumn{1}{c}{(2)}&
  \multicolumn{1}{c}{(3)}&
  \multicolumn{1}{c}{(4)}&
  \multicolumn{1}{c}{(5)}&
  \multicolumn{1}{c}{(6)}&
  \multicolumn{1}{c}{(7)}&
  \multicolumn{1}{c}{(8)}&
  \multicolumn{1}{c}{(9)}&
  \multicolumn{1}{c}{(10)}&
  \multicolumn{1}{c}{(11)} \\
\hline
\hline
\multirow{3}{*}{1}       &  \multirow{3}{*}{1.3005}	      &  M05   & $1.01$   & $2  $   & $e^{-t/0.1\ {\rm Gyr}}$		& $2.30$	  & $0.60 $ & $<0.1 $	  & $0.00 $	    & NA\tnote{1} \\		    
        	         &				      &  M13   & $2.40$   & $2  $   & $e^{-t/0.3\ {\rm Gyr}}$		& $2.19$	  & $1.30 $ & $0.2  $	  & $0.05 $	    & \citet{Calzetti-2000} \\  	    
        	         &				      &  BC03  & $2.50$   & $2  $   & $e^{-t/0.3\ {\rm Gyr}}$		& $2.02$	  & $1.46 $ & $0.2  $	  & $0.00 $	    & NA\tnote{1} \\  	    
\hline   	      	 					      
\multirow{2}{*}{2}    	 &  \multirow{2}{*}{1.377}	      &  M05   & $1.28$   & $2  $   & $t_{\rm trunc}=1\ {\rm Gyr}$	& $0.74$          & $0.22 $ & $<0.1 $     & $0.00 $	    & NA\tnote{1} \\	
        	      	 &				      &  BC03  & $2.10$   & $1  $   & $e^{-t/0.3\ {\rm Gyr}}$		& $1.07$	  & $0.56 $ & $0.2  $     & $0.05 $	    & \citet{Calzetti-2000} \\
\hline   	      	 	 
\multirow{3}{*}{16273}   &  \multirow{3}{*}{1.39}	      &  M05   & $0.81$   & $1  $   & $t_{\rm trunc}=0.3\ {\rm Gyr}$	& $1.59$	  & $0.21 $ & $<0.1 $	  & $0.10 $	    & \citet{Calzetti-2000} \\  	    
        	      	 &				      &  M13   & $0.81$   & $0.5$   & $t_{\rm trunc}=0.3\ {\rm Gyr}$	& $1.59$	  & $0.21 $ & $<0.1 $	  & $0.10 $	    & \citet{Calzetti-2000} \\    
        	         &				      &  BC03  & $2.00$   & $2  $   & $e^{-t/0.3\ {\rm Gyr}}$		& $7.18$	  & $0.41 $ & $0.2  $	  & $0.00 $	    & NA\tnote{1} \\   
\hline
\multirow{3}{*}{3}    	 &  \multirow{3}{*}{1.3961}	      &  M05   & $1.01$   & $2  $   & $t_{\rm trunc}=0.3\ {\rm Gyr}$	& $1.06$	  & $0.64 $ & $<0.1 $	  & $0.00 $	    & NA\tnote{1} \\    
        	      	 &				      &  M13   & $2.10$   & $2  $   & $t_{\rm trunc}=1\ {\rm Gyr}$	& $0.78$	  & $1.11 $ & $<0.1 $	  & $0.05 $	    & \citet{Calzetti-2000} \\    
        	      	 &				      &  BC03  & $3.50$   & $1  $   & $e^{-t/0.3\ {\rm Gyr}}$		& $0.98$	  & $1.96 $ & $<0.1 $	  & $0.00 $	    & NA\tnote{1} \\    
\hline
\multirow{2}{*}{4}       &  \multirow{2}{*}{1.3965}	      &  M05   & $2.75$   & $1  $   & $e^{-t/0.3\ {\rm Gyr}}$		& $1.58$	  & $1.42 $ & $<0.1 $	  & $0.00 $	    & NA\tnote{1} \\   
        	      	 &				      &  M13   & $1.01$   & $2  $   & SSP				& $1.10$	  & $0.84 $ & $<0.1 $	  & $0.00 $	    & NA\tnote{1} \\   
\hline
\multirow{3}{*}{5}    	 &  \multirow{3}{*}{1.405}	      &  M05   & $0.81$   & $2  $   & $e^{-t/0.1\ {\rm Gyr}}$		& $1.57$	  & $0.87 $ & $0.3  $	  & $0.00 $	    & NA\tnote{1} \\   
        	      	 &				      &  M13   & $2.10$   & $0.2$   & $e^{-t/0.3\ {\rm Gyr}}$		& $2.75$	  & $1.67 $ & $0.7  $	  & $0.00 $	    & NA\tnote{1} \\   
        	      	 &				      &  BC03  & $2.10$   & $2  $   & $e^{-t/0.3\ {\rm Gyr}}$		& $3.55$	  & $2.11 $ & $0.9  $	  & $0.00 $	    & NA\tnote{1} \\   
\hline
\multirow{2}{*}{6}       &  \multirow{2}{*}{1.4072}	      &  M13   & $0.81$   & $2  $   & SSP				& $1.22$	  & $0.5  $ & $<0.1 $	  & $0.00 $	    & NA\tnote{1} \\   
        	      	 &				      &  BC03  & $2.75$   & $1  $   & $e^{-t/0.3\ {\rm Gyr}}$		& $1.16$	  & $1.29 $ & $<0.1 $	  & $0.00 $	    & NA\tnote{1} \\   
\hline
\multirow{3}{*}{7}    	 &  \multirow{3}{*}{1.426}	      &  M05   & $0.90$   & $2  $   & $e^{-t/0.1\ {\rm Gyr}}$		& $1.45$	  & $0.49 $ & $<0.1 $	  & $0.00 $	    & NA\tnote{1} \\   
        	      	 &				      &  M13   & $2.30$   & $0.2$   & $e^{-t/0.3\ {\rm Gyr}}$		& $3.36$	  & $0.92 $ & $0.2  $	  & $0.00 $	    & NA\tnote{1} \\   
        	      	 &				      &  BC03  & $2.30$   & $2  $   & $e^{-t/0.3\ {\rm Gyr}}$		& $2.18$	  & $1.19 $ & $0.2  $	  & $0.00 $	    & NA\tnote{1} \\   
\hline
\multirow{3}{*}{8}    	 &  \multirow{3}{*}{1.428}	      &  M05   & $1.43$   & $2  $   & $t_{\rm trunc}=1\ {\rm Gyr}$	& $2.04$	  & $1.13 $ & $<0.1 $	  & $0.00 $	    & NA\tnote{1} \\   
        	      	 &				      &  M13   & $2.40$   & $0.2$   & $e^{-t/0.3\ {\rm Gyr}}$		& $3.76$	  & $2.10 $ & $0.3  $	  & $0.00 $	    & NA\tnote{1} \\   
        	      	 &				      &  BC03  & $2.40$   & $2  $   & $e^{-t/0.3\ {\rm Gyr}}$		& $3.13$	  & $2.77 $ & $0.4  $	  & $0.00 $	    & NA\tnote{1} \\   
\hline   	      	 	 
\multirow{3}{*}{9}    	 &  \multirow{3}{*}{1.429}	      &  M05   & $0.81$   & $2  $   & SSP				& $2.41$	  & $1.11 $ & $<0.1 $	  & $0.00 $	    & NA\tnote{1} \\   
        	      	 &				      &  M13   & $1.01$   & $2  $   & SSP				& $2.76$	  & $1.50 $ & $<0.1 $	  & $0.00 $	    & NA\tnote{1} \\   
        	      	 &				      &  BC03  & $3.00$   & $1  $   & $e^{-t/0.3\ {\rm Gyr}}$		& $2.99$	  & $3.04 $ & $<0.1 $	  & $0.00 $	    & NA\tnote{1} \\   
\hline
\multirow{2}{*}{10}      &  \multirow{2}{*}{1.4372}	      &  M13   & $1.70$   & $1  $   & $t_{\rm trunc}=1\ {\rm Gyr}$	& $0.52$	  & $0.31 $ & $<0.1 $	  & $0.06 $	    & \citet{Seaton-1979} \\	
        	      	 &				      &  BC03  & $2.30$   & $2  $   & $e^{-t/0.3\ {\rm Gyr}}$		& $0.52$	  & $0.53 $ & $0.1  $	  & $0.00 $	    & NA\tnote{1} \\	
\hline
\multirow{1}{*}{11}      &  \multirow{1}{*}{1.4408}	      &  BC03  & $2.75$   & $1  $   & $e^{-t/0.3\ {\rm Gyr}}$		& $0.71$	  & $0.68 $ & $<0.1 $	  & $0.00 $	    & NA\tnote{1} \\   
\hline
\multirow{3}{*}{12}   	 &  \multirow{3}{*}{1.449}	      &  M05   & $1.01$   & $0.5$   & $t_{\rm trunc}=0.1\ {\rm Gyr}$	& $2.27$	  & $0.46 $ & $<0.1 $	  & $0.00 $	    & NA\tnote{1} \\    
        	      	 &				      &  M13   & $1.01$   & $0.2$   & $e^{-t/0.1\ {\rm Gyr}}$		& $4.97$	  & $0.55 $ & $<0.1 $	  & $0.00 $	    & NA\tnote{1} \\    
        	      	 &				      &  BC03  & $2.10$   & $2  $   & $e^{-t/0.3\ {\rm Gyr}}$		& $4.99$	  & $1.28 $ & $0.5  $	  & $0.00 $	    & NA\tnote{1} \\    
\hline
\multirow{2}{*}{13}      &  \multirow{2}{*}{1.4595}	      &  M13   & $0.81$   & $2  $   & SSP				& $0.87$	  & $0.48 $ & $<0.1 $	  & $0.00 $	    & NA\tnote{1} \\    
        	      	 &				      &  BC03  & $2.60$   & $1  $   & $e^{-t/0.3\ {\rm Gyr}}$		& $0.90$	  & $1.00 $ & $<0.1 $	  & $0.00 $	    & NA\tnote{1} \\    
\hline   	      	 	 
\multirow{3}{*}{14}   	 &  \multirow{3}{*}{1.5126}	      &  M05   & $0.64$   & $2  $   & $t_{\rm trunc}=0.1\ {\rm Gyr}$	& $1.66$	  & $0.39 $ & $<0.1 $	  & $0.00 $	    & NA\tnote{1} \\ 
        	      	 &				      &  M13   & $1.01$   & $0.2$   & $t_{\rm trunc}=0.3\ {\rm Gyr}$	& $1.92$	  & $0.54 $ & $<0.1 $	  & $0.07 $	    & \citet{Prevot-1984,Bouchet-1985} \\ 
        	      	 &				      &  BC03  & $1.28$   & $0.5$   & SSP				& $1.89$	  & $0.56 $ & $<0.1 $	  & $0.07 $	    & \citet{Prevot-1984,Bouchet-1985} \\ 
\hline
\multirow{1}{*}{15}      &  \multirow{1}{*}{1.5145}	      &  BC03  & $1.28$   & $0.2$   & $t_{\rm trunc}=0.1\ {\rm Gyr}$	& $1.14$	  & $0.41 $ & $<0.1 $	  & $0.01 $	    & \citet{Prevot-1984,Bouchet-1985} \\   
\hline
\multirow{1}{*}{17}      &  \multirow{1}{*}{1.521}	      &  BC03  & $1.43$   & $2  $   & $t_{\rm trunc}=1\ {\rm Gyr}$	& $1.73$	  & $0.84 $ & $<0.1 $	  & $0.06 $	    & \citet{Seaton-1979} \\	
\hline
\multirow{1}{*}{18}      &  \multirow{1}{*}{1.523}	      &  BC03  & $2.40$   & $2  $   & $e^{-t/0.3\ {\rm Gyr}}$		& $1.05$	  & $1.20 $ & $0.2  $	  & $0.00 $	    & NA\tnote{1} \\   
\hline
\multirow{1}{*}{20}      &  \multirow{1}{*}{1.542}	      &  BC03  & $1.28$   & $0.5$   & SSP				& $0.76$	  & $0.43 $ & $<0.1 $	  & $0.15 $	    & \citet{Prevot-1984,Bouchet-1985} \\   
\hline
\multirow{2}{*}{21}      &  \multirow{2}{*}{1.549}	      &  M13   & $1.28$   & $0.5$   & SSP				& $1.08$	  & $0.70 $ & $<0.1 $	  & $0.06 $	    & \citet{Seaton-1979} \\	
        	         &				      &  BC03  & $2.20$   & $1  $   & $e^{-t/0.3\ {\rm Gyr}}$		& $1.03$	  & $1.29 $ & $0.4  $	  & $0.07 $	    & \citet{Prevot-1984,Bouchet-1985} \\   
\hline   	      	 	 
\multirow{3}{*}{13586}	 &  \multirow{3}{*}{1.55}	      &  M05   & $1.68$   & $0.5$   & $e^{-t/0.3\ {\rm Gyr}}$		& $1.97$	  & $0.94 $ & $1.5  $	  & $0.00 $	    & NA\tnote{1} \\ 
        	      	 &				      &  M13   & $1.70$   & $0.2$   & $e^{-t/0.3\ {\rm Gyr}}$		& $6.10$	  & $1.24 $ & $1.9  $	  & $0.00 $	    & NA\tnote{1} \\ 
        	      	 &				      &  BC03  & $1.80$   & $2  $   & $e^{-t/0.3\ {\rm Gyr}}$		& $6.42$	  & $1.65 $ & $1.8  $	  & $0.00 $	    & NA\tnote{1} \\ 
\hline
\multirow{1}{*}{23}      &  \multirow{1}{*}{1.5519}	      &  BC03  & $2.30$   & $1  $   & $e^{-t/0.3\ {\rm Gyr}}$		& $1.12$	  & $1.66 $ & $0.3  $	  & $0.07 $	    & \citet{Prevot-1984,Bouchet-1985} \\   
\hline
\multirow{2}{*}{24}      &  \multirow{2}{*}{1.5596}	      &  M13   & $1.80$   & $0.2$   & $e^{-t/0.3\ {\rm Gyr}}$		& $3.17$	  & $1.16 $ & $1.3  $	  & $0.00 $	    & NA\tnote{1} \\    
        	      	 &				      &  BC03  & $1.90$   & $0.5$   & $e^{-t/0.3\ {\rm Gyr}}$		& $2.95$	  & $1.30 $ & $1.0  $	  & $0.05 $	    & \citet{Calzetti-2000} \\    
\hline
\hline
\end{tabular}
\begin{tablenotes}\footnotesize 
\item[1] The SED of this galaxy was best-fitted with $E(B-V)=0$ also when reddening was allowed.
\end{tablenotes}
\end{threeparttable}							    
\end{center}
\end{scriptsize}
\end{table*}

\addtocounter{table}{-1}

\begin{table*}
\begin{scriptsize}
\begin{center}
\caption{Continued.}
\label{tab:Table3_appB}
\begin{threeparttable} 
\begin{tabular}{cccclcccccl}
\hline
  \multicolumn{1}{c}{\bf ID} &
  \multicolumn{1}{c}{$\mathbf{z_{\rm spec}}$} &
  \multicolumn{1}{c}{\bf Model} &
  \multicolumn{1}{c}{$\mathbf{t}$} &
  \multicolumn{1}{c}{$\mathbf{[Z/H]}$} &
  \multicolumn{1}{c}{\bf SFH} &
  \multicolumn{1}{c}{$\mathbf{\chi^{2}_{\rm r}}$} &
  \multicolumn{1}{c}{$\mathbf{M^{\ast}}$} &
  \multicolumn{1}{c}{$\mathbf{SFR}$} & 
  \multicolumn{1}{c}{$\mathbf{E(B-V)}$} &
  \multicolumn{1}{c}{\bf Reddening Law} \\
  \multicolumn{1}{c}{}&
  \multicolumn{1}{c}{}&
  \multicolumn{1}{c}{}&
  \multicolumn{1}{c}{\rm (Gyr)}&
  \multicolumn{1}{c}{${\rm (Z_{\odot})}$} &
  \multicolumn{1}{c}{}&
  \multicolumn{1}{c}{}&
  \multicolumn{1}{c}{$(10^{11}\ {\rm M_{\odot}})$} &
  \multicolumn{1}{c}{$({\rm M_{\odot}/ yr^{-1}})$} &
  \multicolumn{1}{c}{({\rm mag})}&
  \multicolumn{1}{c}{} \\
  \multicolumn{1}{c}{(1)}&
  \multicolumn{1}{c}{(2)}&
  \multicolumn{1}{c}{(3)}&
  \multicolumn{1}{c}{(4)}&
  \multicolumn{1}{c}{(5)}&
  \multicolumn{1}{c}{(6)}&
  \multicolumn{1}{c}{(7)}&
  \multicolumn{1}{c}{(8)}&
  \multicolumn{1}{c}{(9)}&
  \multicolumn{1}{c}{(10)}&
  \multicolumn{1}{c}{(11)} \\
\hline
\hline
\multirow{2}{*}{25}      &  \multirow{2}{*}{1.5642}	      &  M13   & $1.14$   & $0.5$   & SSP				& $1.58$	  & $1.03 $ & $<0.1 $	  & $0.05 $	    & \citet{Calzetti-2000} \\  
        	      	 &				      &  BC03  & $1.28$   & $0.5$   & SSP				& $1.53$	  & $1.00 $ & $<0.1 $	  & $0.07 $	    & \citet{Prevot-1984,Bouchet-1985} \\  
\hline
\multirow{3}{*}{26}   	 &  \multirow{3}{*}{1.5681}	      &  M05   & $0.51$   & $2  $   & SSP				& $0.98$	  & $0.44 $ & $<0.1 $	  & $0.07 $	    & \citet{Prevot-1984,Bouchet-1985} \\  
        	      	 &				      &  M13   & $1.01$   & $0.2$   & $e^{-t/0.1\ {\rm Gyr}}$		& $1.0 $	  & $0.60 $ & $<0.1 $	  & $0.07 $	    & \citet{Prevot-1984,Bouchet-1985} \\  
        	      	 &				      &  BC03  & $2.20$   & $1  $   & $e^{-t/0.3\ {\rm Gyr}}$		& $1.60$	  & $1.24 $ & $0.4  $	  & $0.07 $	    & \citet{Prevot-1984,Bouchet-1985} \\  
\hline
\multirow{1}{*}{27}      &  \multirow{1}{*}{1.5839}	      &  BC03  & $2.10$   & $0.5$   & $e^{-t/0.3\ {\rm Gyr}}$		& $1.18$	  & $0.78 $ & $0.3  $	  & $0.07 $	    & \citet{Prevot-1984,Bouchet-1985} \\  
\hline
\multirow{3}{*}{28}   	 &  \multirow{3}{*}{1.5888}	      &  M05   & $2.20$   & $0.5$   & $e^{-t/0.3\ {\rm Gyr}}$		& $0.98$	  & $1.17 $ & $0.3  $	  & $0.05 $	    & \citet{Calzetti-2000} \\  
        	      	 &				      &  M13   & $1.14$   & $1  $   & $e^{-t/0.1\ {\rm Gyr}}$		& $1.38$	  & $0.91 $ & $<0.1 $	  & $0.06 $	    & \citet{Allen-1976} \\  
        	      	 &				      &  BC03  & $2.00$   & $2  $   & $e^{-t/0.3\ {\rm Gyr}}$		& $1.63$	  & $1.69 $ & $1.0  $	  & $0.07 $	    & \citet{Prevot-1984,Bouchet-1985} \\  
\hline
\multirow{2}{*}{29}      &  \multirow{2}{*}{1.5939}	      &  M05   & $0.81$   & $1  $   & SSP				& $1.71$	  & $0.51 $ & $<0.1 $	  & $0.00 $	    & NA\tnote{1} \\   
        	      	 &				      &  BC03  & $1.28$   & $0.2$   & SSP				& $1.29$	  & $0.53 $ & $<0.1 $	  & $0.07 $	    & \citet{Prevot-1984,Bouchet-1985} \\  
\hline
\multirow{2}{*}{30}      &  \multirow{2}{*}{1.6587}	      &  M13   & $1.43$   & $0.2$   & $t_{\rm trunc}=1\ {\rm Gyr}$	& $1.34$	  & $0.60 $ & $<0.1 $	  & $0.06 $	    & \citet{Allen-1976} \\  
        	      	 &				      &  BC03  & $2.10$   & $0.5$   & $e^{-t/0.3\ {\rm Gyr}}$		& $1.24$	  & $1.20 $ & $0.5  $	  & $0.10 $	    & \citet{Calzetti-2000} \\  
\hline
\multirow{3}{*}{31}      &  \multirow{3}{*}{1.659}	      &  M05   & $3.25$   & $0.2$   & $t_{\rm trunc}=1\ {\rm Gyr}$	& $1.40$	  & $1.47 $ & $<0.1 $	  & $0.00 $	    & NA\tnote{1} \\   
        	      	 &				      &  M13   & $0.90$   & $1  $   & SSP				& $1.34$	  & $0.87 $ & $<0.1 $	  & $0.00 $	    & NA\tnote{1} \\    
        	         &				      &  BC03  & $1.70$   & $0.5$   & $t_{\rm trunc}=1\ {\rm Gyr}$	& $0.92$	  & $1.00 $ & $<0.1 $	  & $0.07 $	    & \citet{Prevot-1984,Bouchet-1985} \\   
\hline
\multirow{3}{*}{32}      &  \multirow{3}{*}{1.6658}	      &  M05   & $3.00$   & $0.2$   & $e^{-t/0.3\ {\rm Gyr}}$		& $2.20$	  & $1.47 $ & $<0.1 $	  & $0.00 $	    & NA\tnote{1} \\   
        	         &				      &  M13   & $0.81$   & $1  $   & SSP				& $2.32$	  & $0.79 $ & $<0.1 $	  & $0.00 $	    & NA\tnote{1} \\   
        	      	 &				      &  BC03  & $2.00$   & $0.2$   & $t_{\rm trunc}=1\ {\rm Gyr}$	& $1.83$	  & $1.03 $ & $<0.1 $	  & $0.07 $	    & \citet{Prevot-1984,Bouchet-1985} \\ 
\hline
\multirow{3}{*}{33}   	 &  \multirow{3}{*}{1.675}	      &  M05   & $1.43$   & $0.2$   & $t_{\rm trunc}=0.1\ {\rm Gyr}$	& $3.91$	  & $0.84 $ & $<0.1 $	  & $0.06 $	    & \citet{Allen-1976} \\ 
        	      	 &				      &  M13   & $0.81$   & $1  $   & $e^{-t/0.1\ {\rm Gyr}}$		& $4.27$	  & $0.78 $ & $0.3  $	  & $0.10 $	    & \citet{Calzetti-2000} \\ 
        	      	 &				      &  BC03  & $1.28$   & $0.2$   & SSP				& $5.90$	  & $0.62 $ & $<0.1 $	  & $0.06 $	    & \citet{Allen-1976} \\ 
\hline
\multirow{1}{*}{10767}   &  \multirow{1}{*}{1.73}	      &  BC03  & $1.28$   & $2  $   & $e^{-t/0.3\ {\rm Gyr}}$		& $3.12$	  & $0.62 $ & $4.0  $	  & $0.05 $	    & \citet{Calzetti-2000} \\ 
\hline
\multirow{3}{*}{12529}	 &  \multirow{3}{*}{1.76}	      &  M05   & $0.40$   & $1  $   & $t_{\rm trunc}=0.1\ {\rm Gyr}$	& $2.72$	  & $0.23 $ & $<0.1 $	  & $0.15 $	    & \citet{Prevot-1984,Bouchet-1985} \\ 
        	      	 &				      &  M13   & $0.51$   & $1  $   & SSP				& $1.80$	  & $0.34 $ & $<0.1 $	  & $0.10 $	    & \citet{Calzetti-2000} \\ 
        	      	 &				      &  BC03  & $1.28$   & $2  $   & $t_{\rm trunc}=1\ {\rm Gyr}$	& $2.10$	  & $0.38 $ & $<0.1 $	  & $0.05 $	    & \citet{Calzetti-2000} \\ 
\hline
\multirow{1}{*}{34}      &  \multirow{1}{*}{1.808}	      &  BC03  & $2.30$   & $1  $   & $e^{-t/0.3\ {\rm Gyr}}$		& $1.07$	  & $1.92 $ & $0.4  $	  & $0.06 $	    & \citet{Seaton-1979} \\ 
\hline
\multirow{2}{*}{35}      &  \multirow{2}{*}{1.82}	      &  M13   & $1.80$   & $0.2$   & $t_{\rm trunc}=1\ {\rm Gyr}$	& $1.52$	  & $2.08 $ & $<0.1 $	  & $0.00 $	    & NA\tnote{1} \\ 
        	      	 &				      &  BC03  & $2.20$   & $0.5$   & $t_{\rm trunc}=1\ {\rm Gyr}$	& $1.89$	  & $2.77 $ & $<0.1 $	  & $0.05 $	    & \citet{Calzetti-2000} \\ 
\hline   	      	 	 
\multirow{2}{*}{36}      &  \multirow{2}{*}{1.82}	      &  M13   & $1.01$   & $1  $   & SSP				& $0.97$	  & $1.15 $ & $<0.1 $	  & $0.10 $	    & \citet{Calzetti-2000} \\ 
        	      	 &				      &  BC03  & $1.28$   & $1  $   & SSP				& $0.99$	  & $1.13 $ & $<0.1 $	  & $0.06 $	    & \citet{Fitzpatrick-1986} \\ 
\hline   	      	 	 
\multirow{3}{*}{37}   	 &  \multirow{3}{*}{1.822}	      &  M05   & $0.72$   & $2  $   & $e^{-t/0.1\ {\rm Gyr}}$		& $1.77$	  & $1.14 $ & $1.1  $	  & $0.00 $	    & NA\tnote{1} \\ 
        	      	 &				      &  M13   & $1.90$   & $0.2$   & $e^{-t/0.3\ {\rm Gyr}}$		& $2.73$	  & $2.16 $ & $1.7  $	  & $0.00 $	    & NA\tnote{1} \\ 
        	      	 &				      &  BC03  & $2.00$   & $0.5$   & $e^{-t/0.3\ {\rm Gyr}}$		& $2.48$	  & $2.44 $ & $1.4  $	  & $0.05 $	    & \citet{Calzetti-2000} \\ 
\hline
\multirow{3}{*}{38}   	 &  \multirow{3}{*}{1.823}	      &  M05   & $2.10$   & $0.2$   & SSP				& $2.87$	  & $3.77 $ & $<0.1 $	  & $0.00 $	    & NA\tnote{1} \\ 
        	      	 &				      &  M13   & $0.91$   & $0.2$   & $e^{-t/0.1\ {\rm Gyr}}$		& $3.47$	  & $2.30 $ & $0.3  $	  & $0.00 $	    & NA\tnote{1} \\ 
        	      	 &				      &  BC03  & $2.20$   & $0.5$   & $e^{-t/0.3\ {\rm Gyr}}$		& $3.6 $	  & $4.52 $ & $1.4  $	  & $0.00 $	    & NA\tnote{1} \\ 
\hline   	      	 	 
\multirow{2}{*}{39}      &  \multirow{2}{*}{1.827}	      &  M05   & $1.01$   & $1  $   & $e^{-t/0.1\ {\rm Gyr}}$		& $1.00$	  & $0.93 $ & $<0.1 $	  & $0.00 $	    & NA\tnote{1} \\ 
        	      	 &				      &  BC03  & $2.30$   & $1  $   & $e^{-t/0.3\ {\rm Gyr}}$		& $1.94$	  & $2.05 $ & $0.4  $	  & $0.00 $	    & NA\tnote{1} \\ 
\hline
\multirow{2}{*}{40}      &  \multirow{2}{*}{1.8368}	      &  M05   & $1.28$   & $1  $   & $t_{\rm trunc}=1\ {\rm Gyr}$	& $1.67$	  & $0.68 $ & $<0.1 $	  & $0.00 $	    & NA\tnote{1} \\ 
        	      	 &				      &  BC03  & $2.10$   & $0.2$   & $e^{-t/0.3\ {\rm Gyr}}$		& $1.92$	  & $1.16 $ & $0.5  $	  & $0.00 $	    & NA\tnote{1} \\ 
\hline
\multirow{2}{*}{12751}   &  \multirow{2}{*}{1.91}	      &  M13   & $0.72$   & $0.5$   & SSP				& $2.68$	  & $0.62 $ & $<0.1 $	  & $0.07 $	    & \citet{Prevot-1984,Bouchet-1985} \\ 
        	      	 &				      &  BC03  & $1.28$   & $2  $   & $t_{\rm trunc}=1\ {\rm Gyr}$	& $1.87$	  & $0.74 $ & $<0.1 $	  & $0.05 $	    & \citet{Calzetti-2000} \\ 
\hline
\multirow{1}{*}{42}      &  \multirow{1}{*}{1.9677}	      &  M05   & $3.00$   & $0.2$   & $e^{-t/0.3\ {\rm Gyr}}$		& $1.28$	  & $1.34 $ & $<0.1 $	  & $0.00 $	    & NA\tnote{1} \\ 
\hline
\multirow{1}{*}{41}      &  \multirow{1}{*}{1.9677}	      &  BC03  & $1.28$   & $0.2$   & SSP				& $1.60$	  & $0.72 $ & $<0.1 $	  & $0.00 $	    & NA\tnote{1} \\ 
\hline
\multirow{1}{*}{12567}   &  \multirow{1}{*}{1.98}	      &  BC03  & $1.28$   & $2  $   & $t_{\rm trunc}=1\ {\rm Gyr}$	& $2.33$	  & $0.36 $ & $<0.1 $	  & $0.06 $	    & \citet{Fitzpatrick-1986} \\ 
\hline
\multirow{1}{*}{43}      &  \multirow{1}{*}{2.0799}	      &  M13   & $0.90$   & $0.2$   & $e^{-t/0.1\ {\rm Gyr}}$		& $1.53$	  & $1.16 $ & $0.2  $	  & $0.00 $	    & NA\tnote{1} \\ 
\hline
\multirow{2}{*}{44}      &  \multirow{2}{*}{2.0892}	      &  M13   & $2.75$   & $0.2$   & SSP				& $1.48$	  & $3.07 $ & $<0.1 $	  & $0.15 $	    & \citet{Prevot-1984,Bouchet-1985} \\ 
        	      	 &				      &  BC03  & $2.30$   & $0.2$   & $e^{-t/0.3\ {\rm Gyr}}$		& $1.03$	  & $2.41 $ & $0.5  $	  & $0.07 $	    & \citet{Prevot-1984,Bouchet-1985} \\ 
\hline
\multirow{3}{*}{11079}	 &  \multirow{3}{*}{2.67}	      &  M05   & $0.72$   & $1  $   & $e^{-t/0.1\ {\rm Gyr}}$		& $3.46$	  & $0.40 $ & $0.4  $	  & $0.00 $	    & NA\tnote{1} \\    
        	      	 &				      &  M13   & $0.64$   & $0.2$   & SSP				& $3.30$	  & $0.71 $ & $<0.1 $	  & $0.20 $	    & \citet{Calzetti-2000} \\
        	      	 &				      &  BC03  & $0.64$   & $1  $   & $e^{-t/0.1\ {\rm Gyr}}$		& $2.32$	  & $0.69 $ & $1.5  $	  & $0.15 $	    & \citet{Calzetti-2000} \\
\hline
\hline
\end{tabular}
\begin{tablenotes}\footnotesize 
\item[1] The SED of this galaxy was best-fitted with $E(B-V)=0$ also when reddening was allowed.
\end{tablenotes}
\end{threeparttable}							    
\end{center}
\end{scriptsize}
\end{table*}

\end{document}